\documentclass[preprint]{aastex63}
\usepackage{hyperref}
\usepackage{color,soul}

\usepackage{graphicx}
\usepackage{subfigure}
\usepackage{tikz} 
\usetikzlibrary{shapes.geometric, arrows}
\usepackage[T1]{fontenc}
\usetikzlibrary{shapes,arrows,calc,fit,backgrounds}
\usetikzlibrary{positioning,decorations.pathreplacing}
\usepackage{tikzpagenodes}\usepackage{mathtools}
\usepackage{natbib}
\usepackage{cleveref}
\usepackage{amsmath, amssymb}
\usepackage{dcolumn}
\usepackage{bm}
\received{~~~~~~}
\revised{~~~~~~}
\accepted{~~~~~~}
\submitjournal{ApJ}

\shorttitle{A Geometric origin for QPOs}
\shortauthors{Rana and Mangalam 2020}

\begin{document}

\title{A Geometric Origin for Quasi-periodic Oscillations in Black Hole X-ray Binaries}

\correspondingauthor{A. Mangalam}
\email{mangalam@iiap.res.in}

\author[0000-0001-6184-5195]{Prerna Rana}
\email{prernarana@iiap.res.in}
\affiliation{Indian Institute of Astrophysics, Sarjapur Road, 2nd Block Koramangala, Bangalore-560034}

\author[0000-0001-9282-0011]{A. Mangalam}
\affiliation{Indian Institute of Astrophysics, Sarjapur Road, 2nd Block Koramangala, Bangalore-560034}
\bibliographystyle{aasjournal}



\begin{abstract}
  We expand the relativistic precession model to include nonequatorial and eccentric trajectories and apply it to quasi-periodic oscillations (QPOs) in black hole X-ray binaries (BHXRBs) and associate their frequencies with the fundamental frequencies of the general case of nonequatorial (with Carter's constant, $Q\neq 0$) and eccentric ($e\neq 0$) particle trajectories, around a Kerr black hole. We study cases with either two or three simultaneous QPOs and extract the parameters \{$e$, $r_p$, $a$, $Q$\}, where $r_p$ is the periastron distance of the orbit, and $a$ is the spin of the black hole. We find that the orbits with $\left[Q=0-4\right]$ should have $e\lesssim 0.5$ and $r_p \sim 2-20$ for the observed range of QPO frequencies, where $a \in [0,1]$, and that the spherical trajectories \{$e=0$, $Q \neq0$\} with $Q=2-4$ should have $r_s \sim 3-20$. We find nonequatorial eccentric solutions for both M82 X-1 and GROJ 1655-40. We see that these trajectories, when taken together, span a torus region and  give rise to a strong QPO signal. For two simultaneous QPO cases, we found equatorial eccentric orbit solutions for XTEJ 1550-564, 4U 1630-47, and GRS 1915+105, and spherical orbit solutions for BHXRBs M82 X-1 and XTEJ 1550-564. We also show that the eccentric orbit solution fits the Psaltis-Belloni-Klis correlation observed in BHXRB GROJ 1655-40. Our analysis of the fluid flow in the relativistic disk edge suggests that instabilities cause QPOs to originate in the torus region. We also present some useful formulae for trajectories and  frequencies of spherical and equatorial eccentric orbits.

\end{abstract}


\section{Introduction} \label{intro}

Black hole X-ray binaries (BHXRBs) are systems with a primary black hole gravitationally bound to a nondegenerate companion star. These systems display transient behavior exhibiting high X-ray luminosities ($L_{\rm X} \sim 10^{38}$ erg s$^{-1}$) during the outburst state, lasting from a few days to many months, followed by a long quiescent state ($L_{\rm X} \sim 10^{30}$ erg s$^{-1}$) \citep{Remillard2006}. The triggering of these X-ray outbursts has been modeled as an instability arising in the accretion disk when the accretion rate is not adequate for the continuous flow of matter to the black hole, and when a critical surface density is reached \citep{Dubus2001}. However, the disk instability model has not been able to explain the outbursts of much shorter or longer time-scales; for example, BHXRB GRS 1915+105 has shown a high X-ray luminosity state for more than 10 yr \citep{FenderBelloni2004AA}. During the outburst phase, the X-ray intensity shows rapid variations with timescales ranging from milliseconds to a few seconds, which are most likely to arise in the proximity to the black hole ($r\sim r_{I}$, where $r_I$ represents the innermost stable spherical orbit (ISSO)). The power density spectrum (PDS) of the X-ray intensity, which is commonly used to probe this fast variability, exhibits distinct features called quasi-periodic oscillations (QPOs) during the outburst period with their peak frequency, $\nu_{0}$, ranging from 0.01 to 450 Hz \citep{Remillard2006,BelloniStella2014}. QPOs can be distinguished from other broad features of the PDS by their high-quality factor $\nu_{0} / {\rm FWHM } \gtrsim 2$. Hence, the study of properties and origin of QPOs in BHXRBs is crucial to understanding the properties of inner accretion flow close to the black hole, where general relativistic effects are ascendant. 

QPOs in BHXRBs are categorized as low-frequency QPOs (LFQPOs) with $\nu_{0} <30$Hz, which are again classified as type A, B, or C based on their various properties, and high-frequency QPOs (HFQPOs) with $\nu_{0} >30$Hz \citep{Motta2016}. These different types of QPOs are also known to show a remarkable association with various spectral states during the outburst phase \citep{FenderBelloni2004,Remillard2006,Fender2012,Motta2016}. The launch of the Rossi X-ray Timing Explorer (RXTE) in 1995 with its high sensitivity significantly increased the detection of BHXRBs, and made it possible to detect HFQPOs in their PDS in the late 1990s \citep{BelloniStella2014}, for example, the detection of 300 Hz and 450 Hz QPOs in GROJ 1655-40 \citep{RemillardMorgan1999, Strohmayer2001a}; QPOs in the range 102-284 Hz, at 188 Hz, 249-276 Hz and near 183 Hz, 283 Hz in XTEJ 1550-564 \citep{Homan2001, Miller2001, Remillard2002}; 67 Hz, 40 Hz, and 170 Hz in GRS 1915+105 \citep{Morgan1997, Strohmayer2001b, Belloni2006}; 250 Hz in XTEJ 1650-500 \citep{Homan2003}; 240 Hz and 160 Hz in H1743-322 \citep{Homan2005, Remillard2006}; and more. Some of these HFQPOs have been detected simultaneously along with their peak frequencies showing nearly 3:2 or 5:3 ratios, indicating a resonance phenomenon \citep{Remillard2006, BelloniStella2014}. There is also an interesting case of BHXRB GROJ 1655-40 which showed three QPOs simultaneously$-$two HFQPOs and one type C LFQPO \citep{Motta2014a}. Understanding of the origin of HFQPOs and their simultaneity has been the prime focus of the observational studies as well as the theoretical models.

The study of general relativistic effects is important for a theoretical understanding of the origin of QPOs and their connection with various spectral states during the X-ray outburst, as these signals appear to emanate very close to the black hole. Several existing models, based on the instabilities in the accretion disk and other geometrical effects, which attempt to explain the origin of LFQPOs and HFQPOs. Most of these models assume that the disk inhomogeneities orbiting in the innermost regions of the accretion disk are the cause of high variability in the X-ray flux, resulting in QPOs in the PDS. A widely accepted model among them is the relativistic precession model (RPM) \citep{Stella1999a, Stella1999b}, which ascribes two simultaneous HFQPOs to the azimuthal, $\nu_{\phi}$, and periastron precession frequencies, $\left( \nu_{\phi}-\nu_{r}\right) $, and a third simultaneous type C LFQPO to the nodal precession frequency, $\left( \nu_{\phi}-\nu_{\theta}\right) $, of a self-emitting blob of matter in the accretion disk. The RPM has been applied to the cases of BHXRBs GROJ 1655-40 \citep{Motta2014a} and XTEJ 1550-564 \citep{Motta2014b} to estimate the spin parameter and mass of the black hole, where they assumed the precession frequencies of nearly circular particle trajectories in the accretion disk around a Kerr black hole. Recently, in contrast with the localized assumption of the RPM, the most frequently detected type C QPOs in BHXRBs have been modeled as the Lense$-$Thirring frequency of a radially extended thick torus precessing as a rigid body \citep{Ingram2009, Ingram2011, Ingram2012}. This model describes the increase in type C QPO frequency with the hard to soft spectral transition during outburst as coincident with the decrease in outer radius of the torus and also shows that the maximum type C QPO frequency should be close to 10$-$30 Hz \citep{Motta2018}. Other models which concentrate on the 3:2 or 5:3 resonance phenomena of simultaneous HFQPOs under the regime of particle approach; for instance, the nonlinear resonance models \citep{ Kato2004b, Kato2008, Torok2005, Torok2011} which explain the phenomenon of simultaneous HFQPOs as an excitation due to the nonlinear resonant coupling between the oscillations within the accretion disk. One such nonlinear resonance phenomenon is the parametric resonance between radial, $\nu_r$, and vertical, $\nu_\theta$,  oscillation frequencies of particles in the accretion disk \citep{Abramowicz2003}. Another explanation of HFQPOs is based on the Keplerian and radial frequencies of the deformation of the clumps of matter that is due to the simulated tidal interactions in the accretion disk \citep{Germana2009}. A recent model involves the study of (magneto)hydrodynamic instabilities, for example, in particular, to understand the 3:2 resonance of HFQPOs using the general relativistic and ray-tracing simulations \citep{Tagger2006,PeggyVarniere2019}.  

The RPM takes into account of the fundamental phenomenon of relativistic precession, which is dominant and inevitable in the strong-field regime around a black hole. Although the emission mechanism for the production of QPOs with strong rms ($\sim$ 20 \%) is hitherto unknown, it explains some important observational relations, for example, the Psaltis$-$Belloni$-$Klis (PBK) \citep{PBK1999}, which is a positive correlation between the HFQPOs and the LFQPOs in different BHXRBs. In a few other BHXRBs, the characteristic frequency of a broad feature (not a QPO) in the PDS during the hard state shows the same correlation with the LFQPOs. This correlation has been explained using the RPM as a variation of the radius of origin around the Kerr black hole, tracing the QPO frequency. 
 
In this paper, we expand the RPM from a restricted study of circular orbits and explore the fundamental frequency range of the nonequatorial eccentric, equatorial eccentric, and spherical particle trajectories around a Kerr black hole and associate them with the properties of QPOs. We call this as the generalized RPM (GRPM). The general trajectory solutions around a Kerr black hole and their corresponding fundamental frequencies have been extensively studied before \citep{Schmidt2002,Fujita2009,RMCQG2019,RMarxiv2019}. The existence of nonequatorial eccentric, equatorial eccentric, and spherical orbits near a rotating black hole is tangible, and hence the relativistic precession of these orbits can also be included in the model for the emission of QPOs. The quadrature form of the general trajectory solution \{$\phi$, $\theta$, $r$, $t$\} around a Kerr black hole \citep{Carter1968} and the corresponding fundamental frequencies \{$\nu_{\phi}$, $\nu_{r}$, $\nu_{\theta}$\} \citep{Schmidt2002} are well known. Later, the complete analytic form for the trajectories and the fundamental frequencies was derived in terms of the Mino time \citep{Mino2003} and the standard elliptic integrals \citep{Fujita2009}. More recently, a more compact, analytic, and numerically faster form was derived, in terms of the standard elliptic integrals, for the particle trajectory solutions and their fundamental frequencies was derived \citep{RMCQG2019,RMarxiv2019}. We use these analytic formulae for the fundamental frequencies via the GRPM for the periastron and nodal precession of nonequatorial eccentric, equatorial eccentric, and spherical trajectories around a Kerr black hole to associate them with the detected QPO frequencies. The RPM was previously predicted for circular \{$e=0$, $Q=0$\} orbits \citep{Stella1999a,Stella1999b}. We now include \{$e\neq0$, $Q\neq0$\} orbits in this paradigm and test the more general model in this paper. Finally, we show that the eccentric trajectory solution also satisfies the PBK correlation for the case of BHXRB GROJ 1655-40.

\begin{figure}[hbt!]
\begin{center}
\vspace{0.5cm}
\scalebox{0.71}{
\tikzstyle{decision} = [diamond, draw,  
    text width=6em, text badly centered, node distance=3cm, inner sep=0pt]
\tikzstyle{block} = [rectangle, draw, 
 minimum width=4.0cm, minimum height=1cm,  text width=14em, text centered, rounded corners]
\tikzstyle{line} = [draw, -latex']
\tikzstyle{process} = [rectangle, draw, minimum width=3cm, minimum height=1cm,   text width=10em, text centered, rounded corners ]
\tikzstyle{cloud} = [draw, ellipse,node distance=3cm,
    minimum height=2em]
\begin{tikzpicture}[node distance = 3cm, auto]
   \small
\node [block, rectangle split,rectangle split parts=2,rectangle split part fill={pink!20,blue!20}] (para1) {{\textbf{Non-equatorial eccentric orbits, $eQ$; \S \ref{eccentricmotivation}, \S \ref{resultseccentric}}} \nodepart{second}$e \neq$0 and $Q \neq$0. };
\node [block, below left of=para1, node distance=6.2cm, rectangle split,rectangle split parts=2,rectangle split part fill={pink!20,blue!20}] (para2) {{\textbf{Spherical orbits, $Q0$; \\ \S \ref{sphericalmotivation}, \S \ref{resultspherical}}}\nodepart{second} $e$=0 and $Q \neq$0.};
\node [block, below of=para1, node distance=8.8cm, rectangle split,rectangle split parts=2,rectangle split part fill={pink!20,blue!20}] (para3) {{\textbf{Circular orbits, $00$}}\nodepart{second} $e$=0 and $Q$=0,\\
 Previously applied to QPOs \citep{Motta2014a,Motta2014b}.};
\node [block, below right of=para1, node distance=6.2cm, rectangle split,rectangle split parts=2,rectangle split part fill={pink!20,blue!20}] (para4) {{\textbf{Equatorial eccentric orbits, $e0$; \S \ref{eccentricmotivation}, \S \ref{resultseccentric}} }\nodepart{second} $e\neq$0 and $Q$=0.};

\draw[->,thick] (para1)--(para2);
\draw[->,thick] (para1)--(para4);
\draw[->,thick] (para2)--(para3);
\draw[->,thick] (para4)--(para3);
 \end{tikzpicture}
}
\end{center}
\caption{Flowchart of various Kerr orbits (with the nomenclature used here of nonequatorial eccentric ($eQ$), spherical ($Q0$), eccentric equatorial ($e0$), and circular ($00$) orbits) studied to explore QPO frequencies using the GRPM in various sections of this paper, where the most specialized case of circular orbits was previously studied \citep{Motta2014a,Motta2014b}. Clearly, the GRPM is valid strictly only when $e \neq 0$.}
\label{orbitflowchart}
\end{figure}
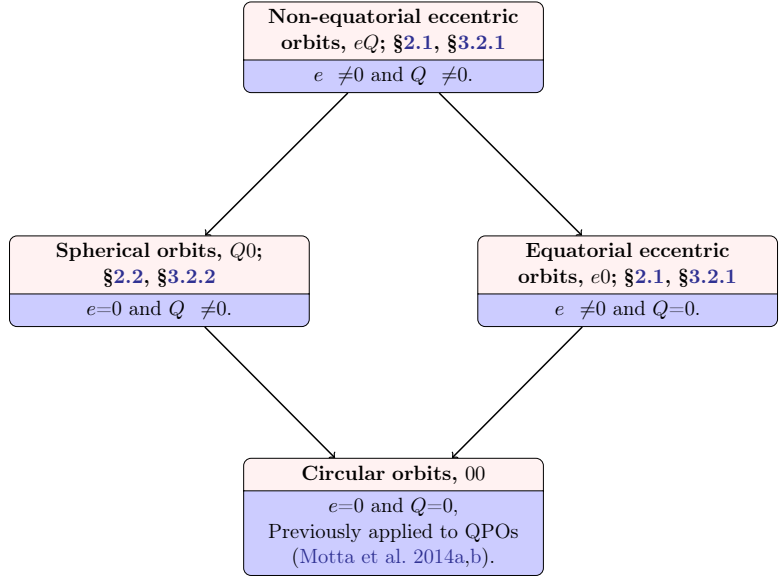

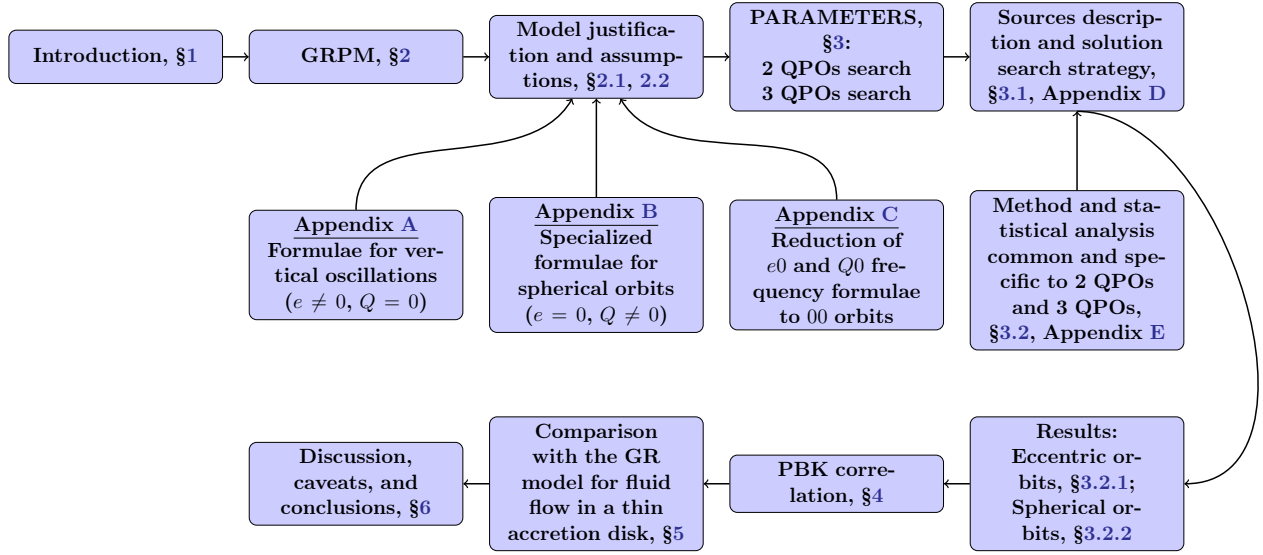
\begin{figure}
\begin{center}
\scalebox{0.71}{
\tikzstyle{decision} = [diamond, draw,  
    text width=6em, text badly centered, node distance=3cm, inner sep=0pt]
\tikzstyle{block} = [rectangle, draw, 
 minimum width=4.0cm, minimum height=1cm,  text width=9.0em, text centered, rounded corners]
\tikzstyle{line} = [draw, -latex']
\tikzstyle{process} = [rectangle, draw, minimum width=3cm, minimum height=1cm,   text width=11em, text centered, rounded corners ]
\tikzstyle{cloud} = [draw, ellipse,node distance=3cm,
    minimum height=2em]
\begin{tikzpicture}[node distance = 3cm, auto]
   \small
\node [block, rectangle, fill=blue!20] (para1) {{\textbf{Introduction, \S \ref{intro}}}};
\node [block, right of=para1, node distance=4.5cm, rectangle, fill=blue!20] (para2) {{\textbf{GRPM, \S \ref{motivation}}}};
\node [block, right of=para2, node distance=4.5cm, rectangle, fill=blue!20] (para3) {{\textbf{Model justification and assumptions, \S \ref{eccentricmotivation}, \ref{sphericalmotivation}}}};
\node [block, below of=para2, node distance=3.9cm, rectangle, fill=blue!20] (para32) {{\textbf{\underline{Appendix \ref{nuthetaderivation}} Formulae for vertical oscillations ($e\neq0$, $Q=0$)}}};
\node [block, right of=para3, node distance=4.5cm, rectangle, fill=blue!20] (para4) {{\textbf{PARAMETERS, \S \ref{parameterestimation}: \\ 2 QPOs search \\ 
3 QPOs search }}};
\node [block, below of=para3, node distance=3.9cm, rectangle, fill=blue!20] (para33) {{\textbf{\underline{Appendix \ref{sphericalorbitsderivations}} Specialized formulae for spherical orbits ($e=0$, $Q\neq0$)}}};
\node [block, right of=para4, node distance=4.5cm, rectangle, fill=blue!20] (para5) {{\textbf{ Sources description and solution search strategy, \S \ref{samplesel}, Appendix \ref{sourcehistory}}}};
\node [block, below of=para4, node distance=3.9cm, rectangle, fill=blue!20] (para34) {{\textbf{\underline{Appendix \ref{reducecircular}} \\ Reduction of $e0$ and $Q0$ frequency formulae to $00$ orbits}}};
\node [block, below of=para5, node distance=4.0cm, rectangle, fill=blue!20] (para6) {{\textbf{Method and statistical analysis common and specific to 2 QPOs and 3 QPOs, \S \ref{method}, Appendix \ref{methodsection} }}};
\node [block, below of=para6, node distance=4.0cm, rectangle, fill=blue!20] (para7) {{\textbf{Results: \\ Eccentric orbits, \S \ref{resultseccentric}; \\
Spherical orbits, \S \ref{resultspherical}}}};
\node [block, left of=para7, node distance=4.5cm, rectangle, fill=blue!20] (para8) {{\textbf{PBK correlation, \S \ref{PBK}}}};
\node [block, left of=para8, node distance=4.5cm, rectangle, fill=blue!20] (para9) {{\textbf{Comparison with the GR model for fluid flow in a thin accretion disk, \S \ref{gasflowmodel}}}};
\node [block, left of=para9, node distance=4.5cm, rectangle, fill=blue!20] (para11) {{\textbf{Discussion, caveats, and conclusions, \S \ref{discconcl} }}};

\draw[->,thick] (para1)--(para2);
\draw[->,thick] (para2)--(para3);
\draw[->,thick] (para3)--(para4);
\draw[->,thick] (para4)--(para5);
\draw[->,thick] (para6)--(para5); 
\draw[->,thick] (para5.south) to [out=0,in=360] (para7.east);
\draw[->,thick] (para32) to [out=90,in=240] (para3);
\draw[->,thick] (para33) to [out=90,in=270] (para3);
\draw[->,thick] (para34) to [out=90,in=300] (para3);
\draw[->,thick] (para7)--(para8); 
\draw[->,thick] (para8)--(para9); 
\draw[->,thick] (para9)--(para11);
 \end{tikzpicture}
}
\caption{Concept flowchart of the paper.}
\end{center}
\label{conceptflowchart}
\end{figure}
The paper is structured as follows. We first motivate the association of fundamental frequencies of the general eccentric and spherical trajectories with the QPOs in BHXRBs assuming the GRPM in \S \ref{eccentricmotivation} and \S \ref{sphericalmotivation}; see Figure \ref{orbitflowchart} for the terminology used for $eQ$ (general case), $Q0$ (spherical), $e0$ (eccentric equatorial), and $00$ (circular orbits). We then take up the cases of BHXRBs M82 X-1, GROJ 1655-40, XTEJ 1550-564, 4U 1630-47, and GRS 1915+105, where HFQPOs have been discovered before. We discuss their observation history in Appendix \ref{sourcehistory}, and we discuss observations of each BHXRB that we use for our analysis in \S \ref{samplesel}. Using the observed QPO frequencies in these BHXRBs, we calculate the corresponding orbital parameters. The method for the parameter estimation and its corresponding errors are discussed in \S \ref{method} and in Appendix \ref{methodsection}. We discuss the results for general eccentric trajectories in \S \ref{resultseccentric}, and those corresponding to the spherical orbit in \S \ref{resultspherical}. We also show in \S \ref{PBK} that the PBK correlation is well explained by the eccentric trajectory solutions found in the case of BHXRB GROJ 1655-40. In \S \ref{gasflowmodel}, we compare our model with another model for the fluid flow in the general-relativistic thin accretion disk. We finally discuss and conclude our results in \S \ref{discconcl}. A glossary of symbols used in this article is given in Table \ref{gloss}, and a concept flowchart of the paper is given in Figure \ref{conceptflowchart}.

\begin{deluxetable}{c l c l}[h!]
\tablecaption{Glossary of Symbols Used.\label{gloss}}
\tabletypesize{\scriptsize}
\tablehead{
& & & \\
\colhead{{\bf Common physical parameters }}  & & &}
\startdata
$c$& Speed of light  & $G$ & Gravitational constant \\
$M_{\bullet}$& Mass of the black hole  &  $a$ & Spin of the black hole scaled by $\left(G M^2_{\bullet}/c\right)$  \\
$\mathcal{M}$ & $M_{\bullet}/M_{\odot}$ & & \\
\hline
{\bf Orbital parameters} & &&\\
\hline
  $E$ & Energy per unit rest mass of the  & $L_z$ & z component of Angular momentum  \\
  & particle, scaled by $m c^2$ & & per unit rest mass of the particle, \\
  & & &  scaled by $\left(G M_{\bullet}/c\right)$ \\
  $L$ & Angular momentum per unit rest &$Q$ & Carter's constant scaled by $\left(G M^2_{\bullet}/c\right)^2$ \\
   & mass of the particle, scaled by $\left(G M_{\bullet}/c\right)$ & & \\
$r_a$ & Apastron distance of the orbit & $r_p$ & Periastron distance of the orbit \\
&  scaled by $\left(G M_{\bullet}/c^2\right)$ & &  scaled by $\left(G M_{\bullet}/c^2\right)$ \\
 $e$ & Eccentricity parameter & $\mu$ & Inverse latus-rectum parameter \\
  $r_s$ & Radius of spherical orbit scaled & $r_I$ & ISSO radius scaled by $\left(G M_{\bullet}/c^2\right)$ \\
&   by $\left(G M_{\bullet}/c^2\right)$ & &  \\
\hline
{\bf Fundamental frequencies} & &&\\
\hline
$\nu_{\phi}$ & Azimuthal frequency & $\nu_{\rm np}$ & Nodal precession frequency, $\left( \nu_{\phi}-\nu_{\theta}\right)$  \\
$\nu_{\theta}$ & Vertical oscillation frequency & $\nu_{\rm pp}$ & Periastron precession frequency, $\left( \nu_{\phi}-\nu_{r}\right)$ \\
$\nu_{r}$ & Radial frequency &$\nu_{0}$ & Centroid frequency of the QPO\\
$\bar{\nu}$ & Frequency scaled by the factor $\left( c^3 / G M_{\bullet}\right)$  &  &  \\
\hline
{\bf Probability analysis for  estimating }& {\bf parameter errors} & &\\
\hline
$P$ & Probability density (space) & $\mathcal{P}$ & Normalized probability density (space) \\
$\mathcal{N}$ & Normalization factor & $\mathcal{J}_l$ & Jacobian of transformation from frequency \\
& & &  to parameter space \\
 \enddata
\end{deluxetable}
\section{Generalized Relativistic Precession Model (GRPM)}
\label{motivation}
\begin{figure}
\hspace{1.0cm}
\mbox{ \subfigure[]{
\includegraphics[scale=0.25]{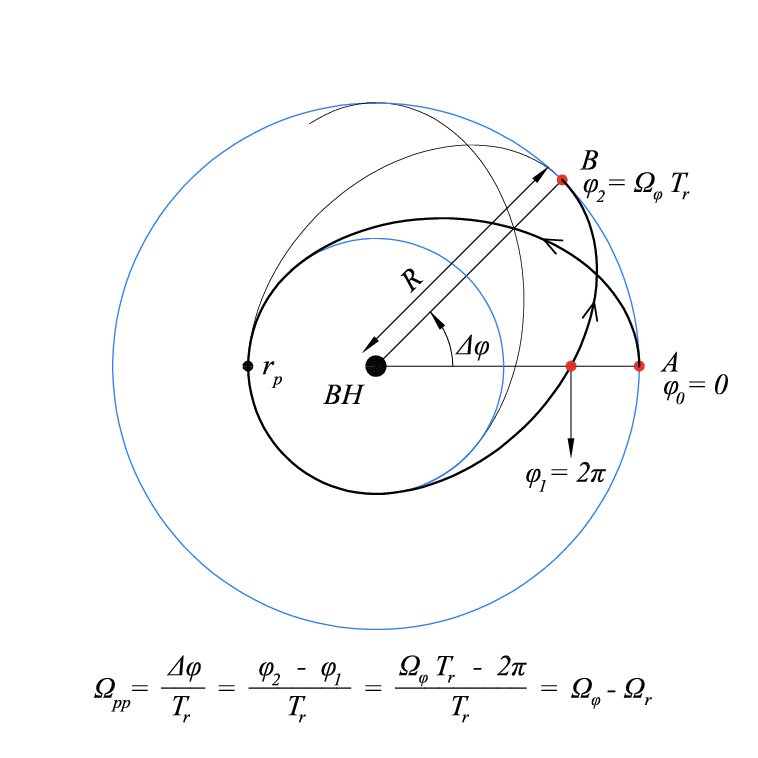}}
\hspace{0.9cm}
\subfigure[]{
\includegraphics[scale=0.255]{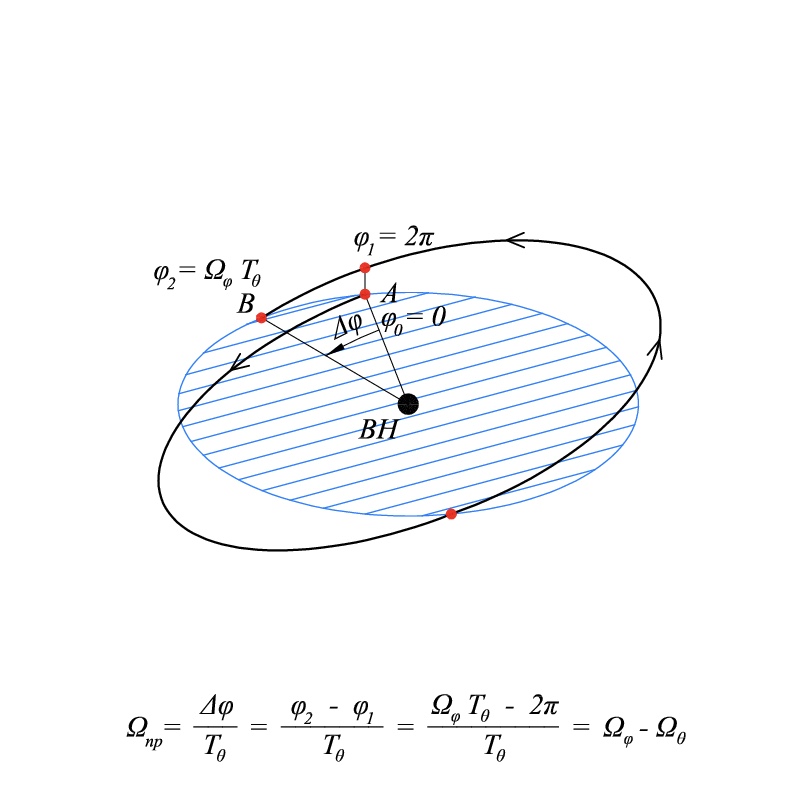}}}
 \caption{\label{precessionplots}Generalized relativistic precession phenomenon for $Q\neq0$, near a black hole (BH) at the center, rotating anticlockwise, where $\Omega_{\rm pp}$ represents the periastron precession and $\Omega_{\rm np}$ represents the nodal precession frequency. The initial point of the trajectory is indicated by point A, from where the particle follows an eccentric trajectory before completing one (a) radial or (b) vertical oscillation to reach point B. The particle sweeps an extra $\Delta \phi$ azimuthal angle during one (a) radial or (b) vertical oscillation because the azimuthal motion is faster than the radial or vertical motion causing the periastron or nodal precession. }
 \end{figure}

The relativistic precession is a phenomenon that is due to strong gravity near a rotating black hole, and its consequence for QPOs originating very close to the black hole is studied. We motivate the association of QPOs in BHXRBs with the fundamental frequencies of general nonequatorial bound particle trajectories around a Kerr black hole through the GRPM. Figure \ref{precessionplots} shows the periastron and nodal precession of an eccentric particle trajectory near the equatorial plane of a rotating black hole. 

\begin{figure}[h!]
\mbox{ \subfigure[]{
\includegraphics[scale=0.32]{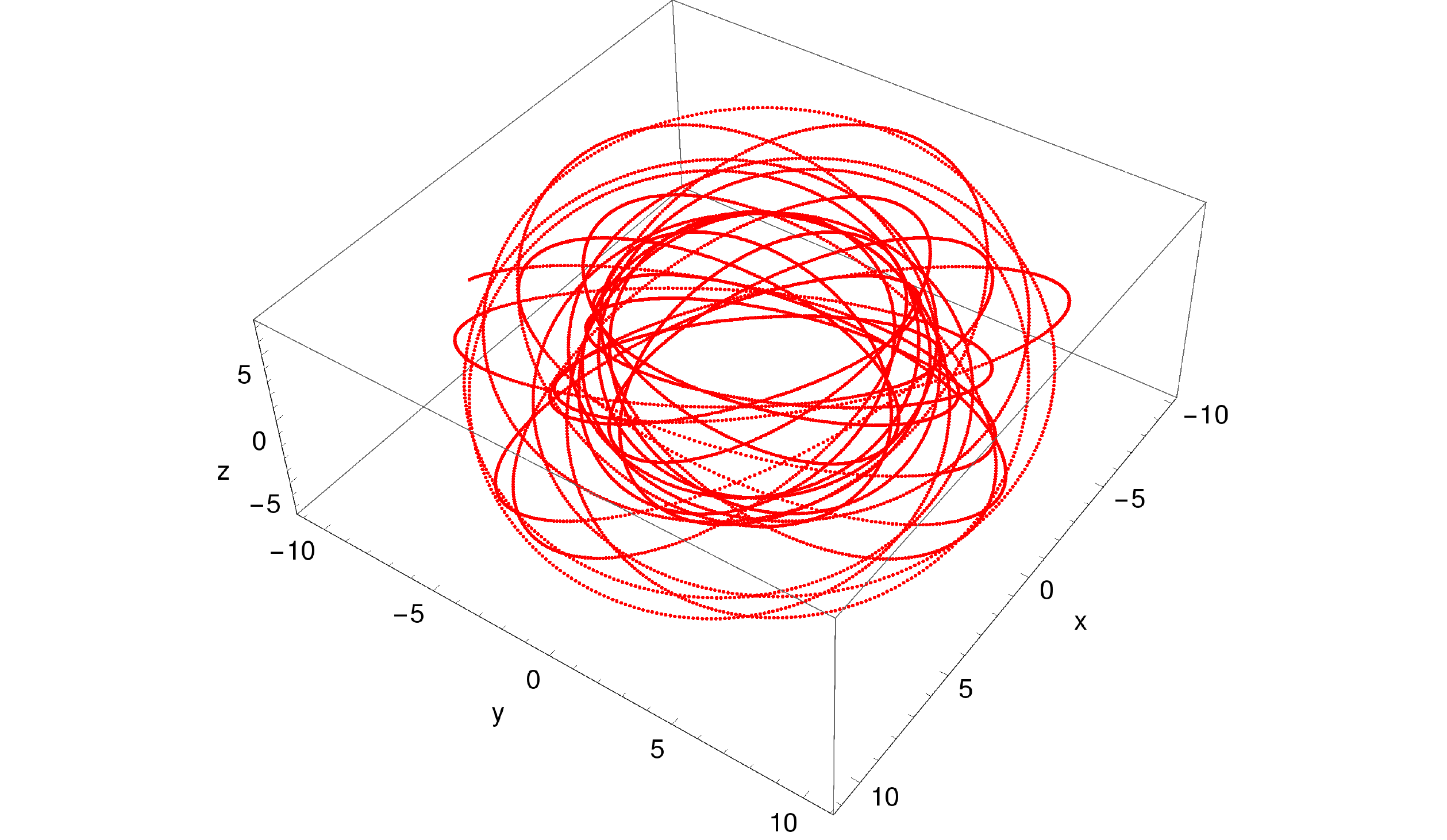}}
\hspace{-0.5cm}
\subfigure[]{
\includegraphics[scale=0.36]{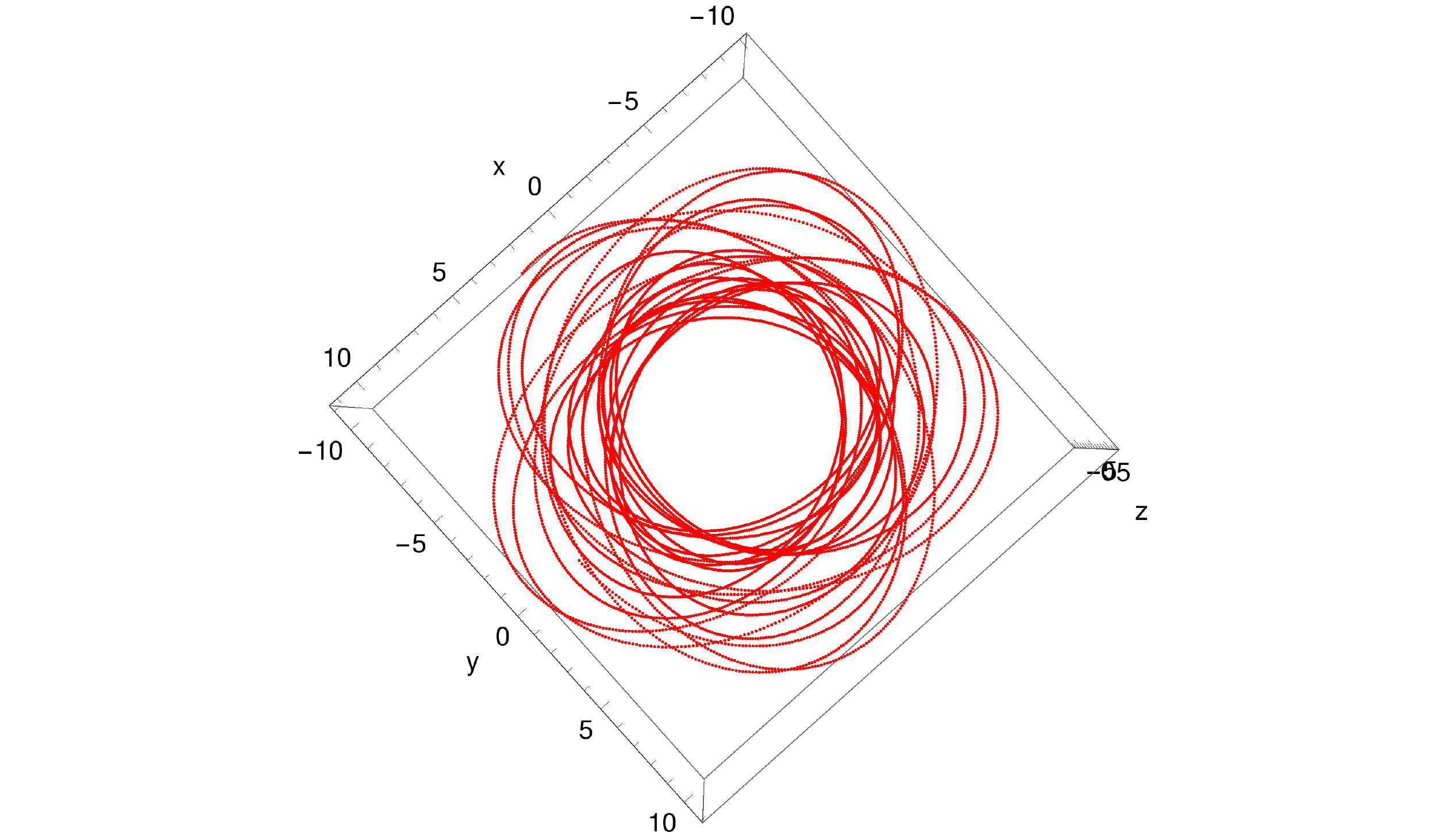}}}
\caption{\label{trajectory}Example of $eQ$ trajectory with parameters $\{ e=0.3, \ r_p=5.917, \ a=0.5, \ Q=5 \}$ around a Kerr black hole at the origin, with its spin pointing in the positive $z$-direction: (a) shows the side view of the orbit representing the nodal precession phenomenon of the orbital plane about the spin axis of the black hole; (b) top view of the orbit showing the periastron precession phenomenon.}
\end{figure}

 We suggest that the instabilities in the inner region close to the rotating black hole might provide a radiating plasma cloud (it could be a blob or a torus with the collection of such trajectories degenerate in the parameter space) with enough energy and angular momentum to attain an eccentric ($e\neq0$) trajectory, or a nonequatorial trajectory ($Q \neq 0$, Carter's constant, \cite{Carter1968}), or both simultaneously ($e\neq0$, $Q\neq0$). The Carter's constant can be roughly interpreted as representative of the residual of the angular momentum in the $x-y$ plane, $Q \propto L^2 -L^2_{z}$, so we have $Q=0$ for the equatorial orbits where $L=L_z$. We first try to find the suitable range for the parameters, $\{e, r_p , a, Q \}$, of these orbits that produce the fundamental frequencies to compare with the observed range of QPO frequencies in BHXRBs, where $r_p$ represents the periastron point of the orbit and $a$ represents the spin of the black hole. We divide our study of the trajectories into three categories (see Figure \ref{orbitflowchart}), where a particle follows one of these:
 \begin{enumerate}
\item A nonequatorial eccentric trajectory ($e\neq0$, $Q\neq0$) called $eQ$.
\item An equatorial eccentric trajectory ($e\neq0$, $Q=0$) called $e0$.
\item A nonequatorial and noneccentric, also called a spherical trajectory ($e=0$, $Q\neq0$), called $Q0$.
 \end{enumerate} 
  We are using dimensionless parameters ($G=c=M_{\bullet}=1$) as the convention in this article for simplicity, so that $r_p \rightarrow r_p/ \left(G M_{\bullet}/c^2\right),\ r_a \rightarrow r_a/ \left(G M_{\bullet}/c^2\right) ,\ a \rightarrow J/ \left(G M^2_{\bullet}/c\right)$, and $Q \rightarrow Q/ \left(G M^2_{\bullet}/c\right)^2$, where $J$ is the angular momentum and $M_{\bullet}$ is the mass of the black hole, and $r_a$ is the apastron point of the bound orbit, while $e=\left(r_a - r_p\right) /\left(r_a + r_p\right)$, the eccentricity parameter, is dimensionless by definition (see Table \ref{gloss}). We also define another mass parameter $\mathcal{M}=M_{\bullet}/M_{\odot}$ scaled by solar mass for convenience. The most general nonequatorial trajectory ($eQ$) around a Kerr black hole comprises of periastron precession in the orbital plane, superimposed on the precession of the orbital plane about the spin axis of the rotating black hole. Figure \ref{trajectory} shows one such trajectory around a Kerr black hole centered at the origin. 
 
\begin{figure}[htb!]
\mbox{ 
\hspace{1.2cm}
\subfigure[]{
\includegraphics[scale=0.37]{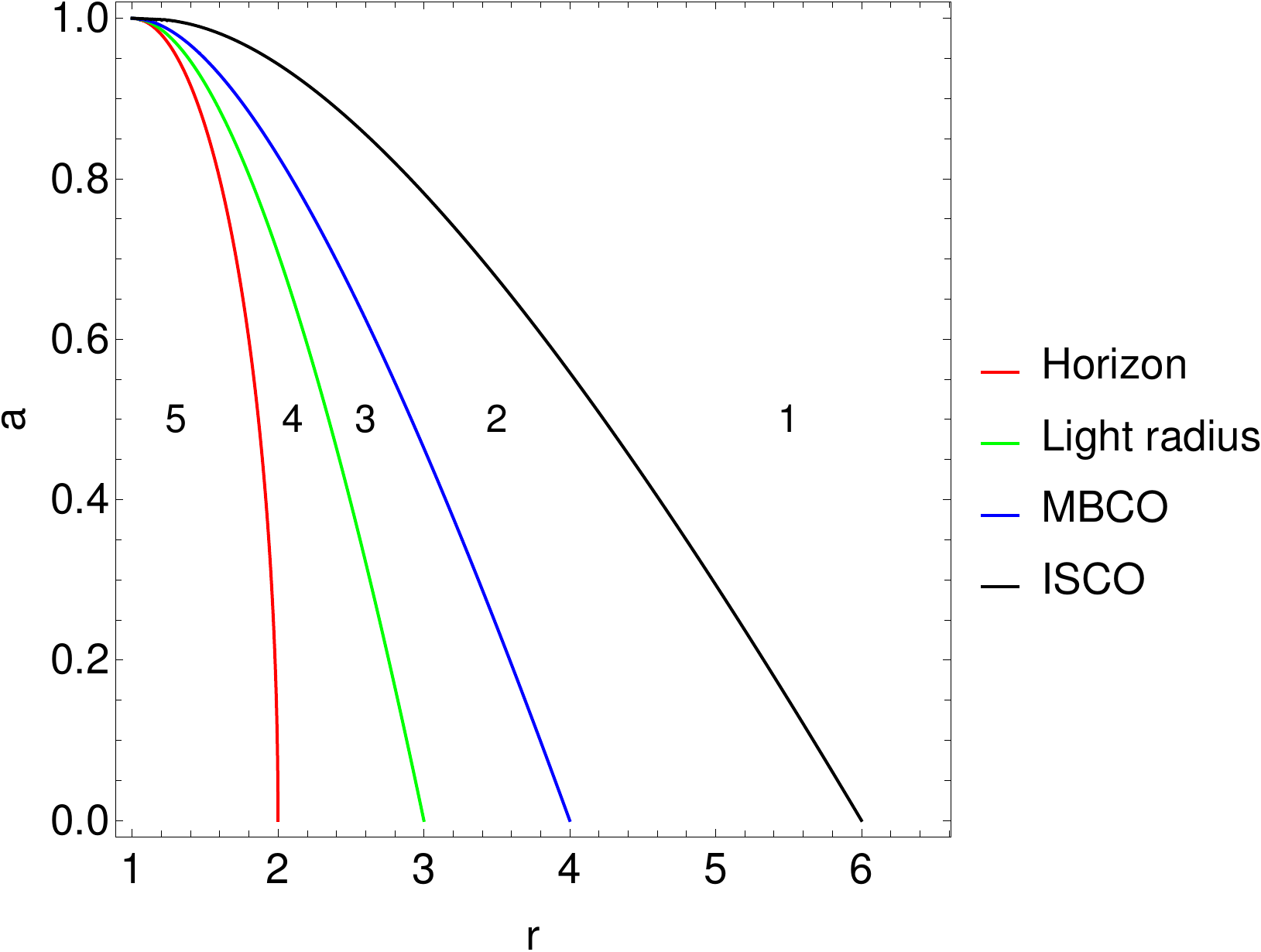}\label{radiiQ0}}
\hspace{1.9cm}
\subfigure[]{
\includegraphics[scale=0.37]{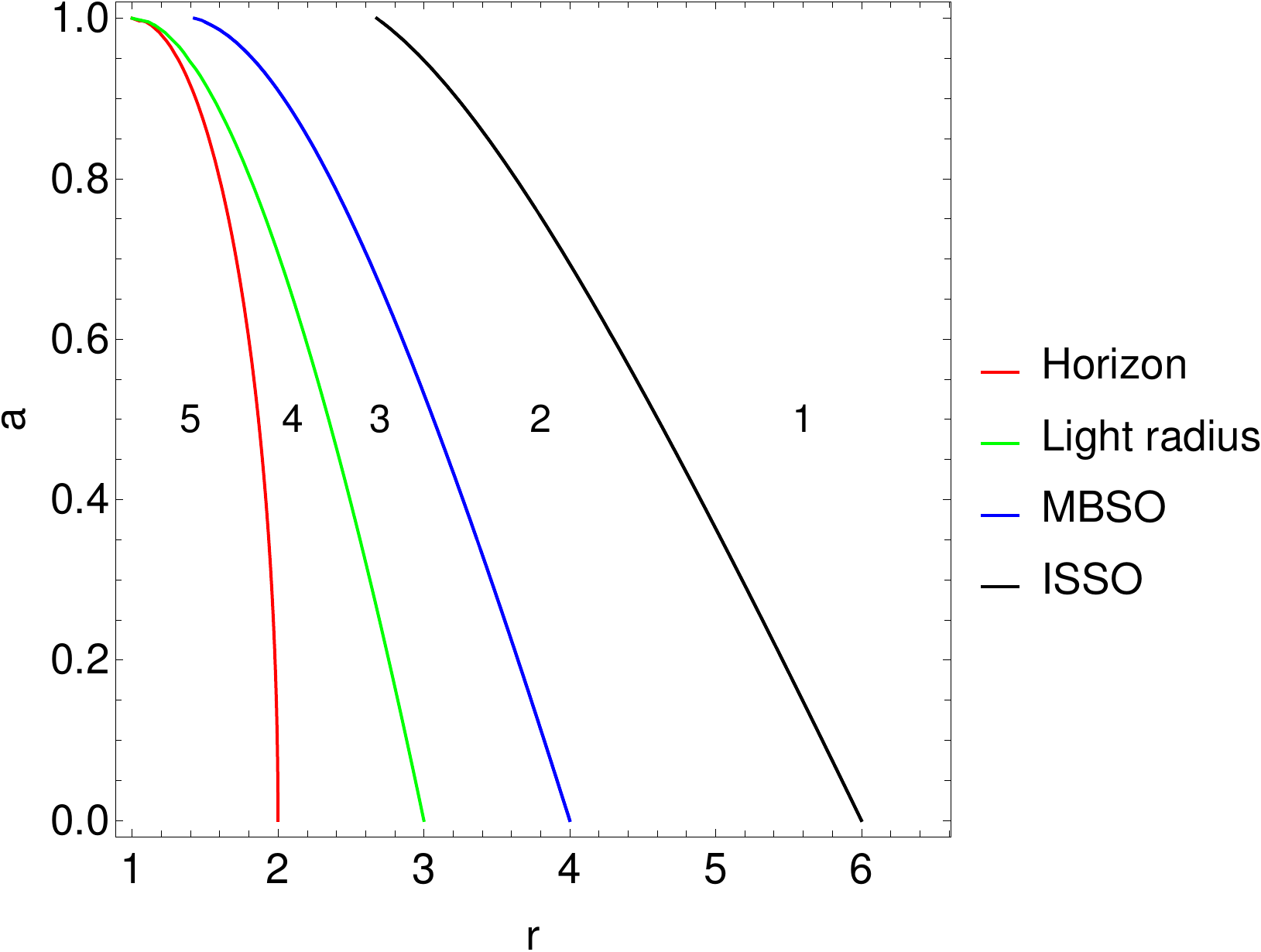}\label{radiiQ4}}}
 \caption{\label{radii}Important radii: the ISCO (ISSO), MBCO (MBSO), light radius, and the horizon. These radii separate various kinds of orbits outside a Kerr black hole in the ($r$, $a$) plane, indicated by different regions that are depicted by numbers, for (a) the equatorial orbits with $Q=0$, and (b) nonequatorial orbits with $Q=4$. }
 \end{figure}

 There are a variety of bound Kerr orbits, for example, nonequatorial eccentric, separatrix, zoom-whirl, and spherical orbits, that have been systematically studied before [e.g. \cite{RMCQG2019,RMarxiv2019} and references within]. Hence, here we first discuss the distribution of these orbits in the parameter space and then isolate the most plausible type of orbits, which should give us the observed range of QPO frequencies assuming the GRPM. A complete description of various types of trajectories is given in Table \ref{trajectorytab}, where MBSO(MBCO) is the marginally bound spherical (circular) orbit, and ISCO is the innermost stable circular orbit. These bound orbits are distributed in particular regions in the parameter space and into different parameter ranges for different types of orbits. In Figure \ref{radii}, we show how this distribution belongs in different regions in the ($r$, $a$) plane, where $r=R/R_g$ represents distance from the black hole, and $R_g=\left(G M_{\bullet}/c^2\right)$. These regions are separated by important radii, which are shown as various curves for the equatorial ($Q=0$) and nonequatorial ($Q=4$) trajectories in Figure \ref{radii}, where we see that the (un)stable bound orbits are found in regions 1, 2, and 3. Region 4 is beyond the light radius, which extends down to the horizon radius [$r_{+}=\left(1+ \sqrt{1-a^2} \right)$], where bound particle orbits are not present, which means any particle in this region would plunge into the black hole, and region 5 is inside the horizon surface. Hence, we restrict our exploration search of suitable parameters for required QPO frequencies to the regions 1 and 2, where stable circular (spherical), equatorial (nonequatorial) eccentric, zoom-whirl, and separatrix orbits are found.
 \begin{deluxetable}{l l c}[h!]
\tablecaption{Various Types of Trajectories around a Kerr Black Hole with Their Description and the Region in the ($r$, $a$) Plane Where They Are Found, as Shown in Figure \ref{radii}.\label{trajectorytab}}
\tabletypesize{\footnotesize}
\tablehead{
\colhead{{\bf Type of Orbit or Radius}}  & \colhead{{\bf Description }} & \colhead{{\bf Region or Curve\tablenotemark{a} }} }
\startdata
 Eccentric (1), $eQ$ or $e0$ & $\bullet$ Stable eccentric bound orbits. & 1 and 2 \\
Separatrix (1), (2), $eQ$ or $e0$ & $\bullet$ They are the intermediate case between bound and  & 2  \\
&  plunge orbits, while their periastron points correspond & \\
&  to an unstable spherical (or circular) orbit, where a & \\ 
& particle reaches asymptotically. & \\
& $\bullet$ The eccentricity of a separatrix orbit increases as its & \\
&  periastron moves closer to the black hole for a given $a$. & \\
& $\bullet$ The $r_p$ of a separatrix orbit with a given eccentricity & \\
 & defines the innermost radial limit for the eccentric bound & \\
 & orbits having the same eccentricity. & \\
Zoom-whirl (1), (3), $eQ$ or $e0$ & $\bullet$ Represent an extreme form of the periastron & 1 and 2  \\
& precession in the strong-field regime. & \\
& $\bullet$ A particle spends enough time near the periastron & \\
& to make finite spherical (or circular) revolutions before  & \\
& zooming out to the apastron point. &  \\
& $\bullet$ Found near and outside the separatrices. & \\
Stable spherical (circular) (1), $Q0$ ($00$) & $\bullet$ Have a constant radius with the precession of & 1  \\
&  orbital plane partially spanning the surface of a sphere & \\
&  around the black hole. & \\
& $\bullet$ Found outside ISSO (ISCO). & \\
Unstable spherical (circular) (1), $Q0$ ($00$)& $\bullet$ Have a constant radius like stable spherical &  2 and 3 \\
&  (circular) orbits. & \\
& $\bullet$ Found outside MBSO (MBCO). & \\
ISSO (ISCO) (1), $Q0$ ($00$)& $\bullet$ Innermost stable spherical (circular) orbit. & Black curve  \\
& $\bullet$ Defined by Equation (22) of \cite{RMarxiv2019}. & \\
MBSO (MBCO) (1), $Q0$ ($00$)& $\bullet$ Marginally bound spherical (circular) orbit.  & Blue curve  \\
& $\bullet$ Defined by Equation (23) of \cite{RMarxiv2019}. & \\
Light radius (1), $Q0$ or $00$ & $\bullet$ Only a photon orbit can exist at this radius. &  Green curve \\
& $\bullet$ Defined by Equation (24) of \cite{RMarxiv2019}. & \\
& $\bullet$ Innermost boundary for the unstable spherical & \\
&  (circular) particle orbits. & \\ 
 \enddata
\tablenotetext{a}{The regions for $e0$ and $00$ orbits are shown in Figure \ref{radiiQ0}, whereas $eQ$ or $Q0$ orbits are shown in Figure \ref{radiiQ4}.}
\tablerefs{(1) \cite{RMCQG2019,RMarxiv2019}; (2) \cite{Levin2009}, \cite{Perez-Giz2009}; (3)\cite{Glampedakis2002}.}
\end{deluxetable}
 
\begin{figure}[h!]
 \begin{center}
 \mbox{ 
\subfigure[]{
 \includegraphics[scale=0.37]{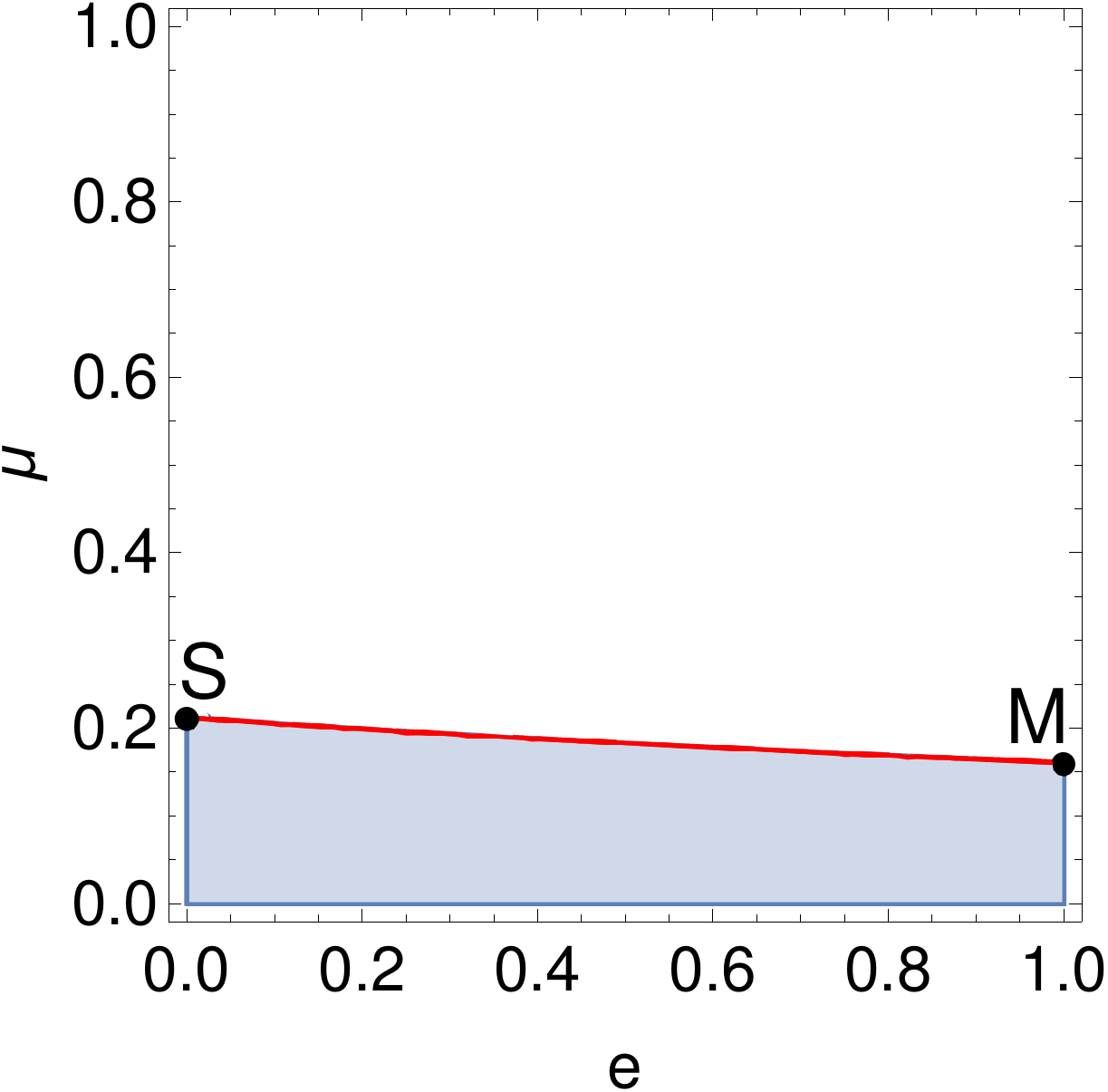}}
 \subfigure[]{
 \includegraphics[scale=0.32]{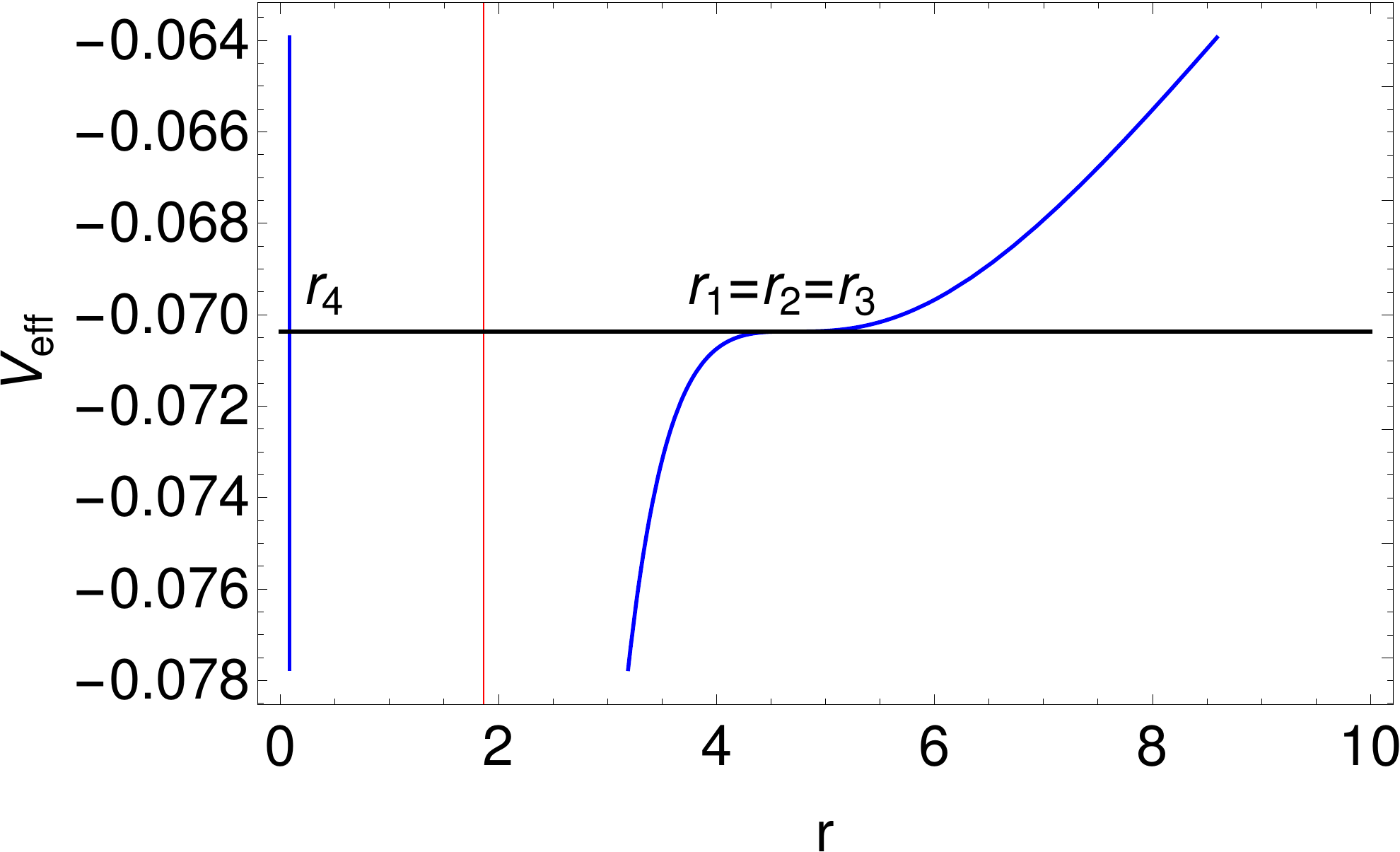}}
 \subfigure[]{
 \includegraphics[scale=0.32]{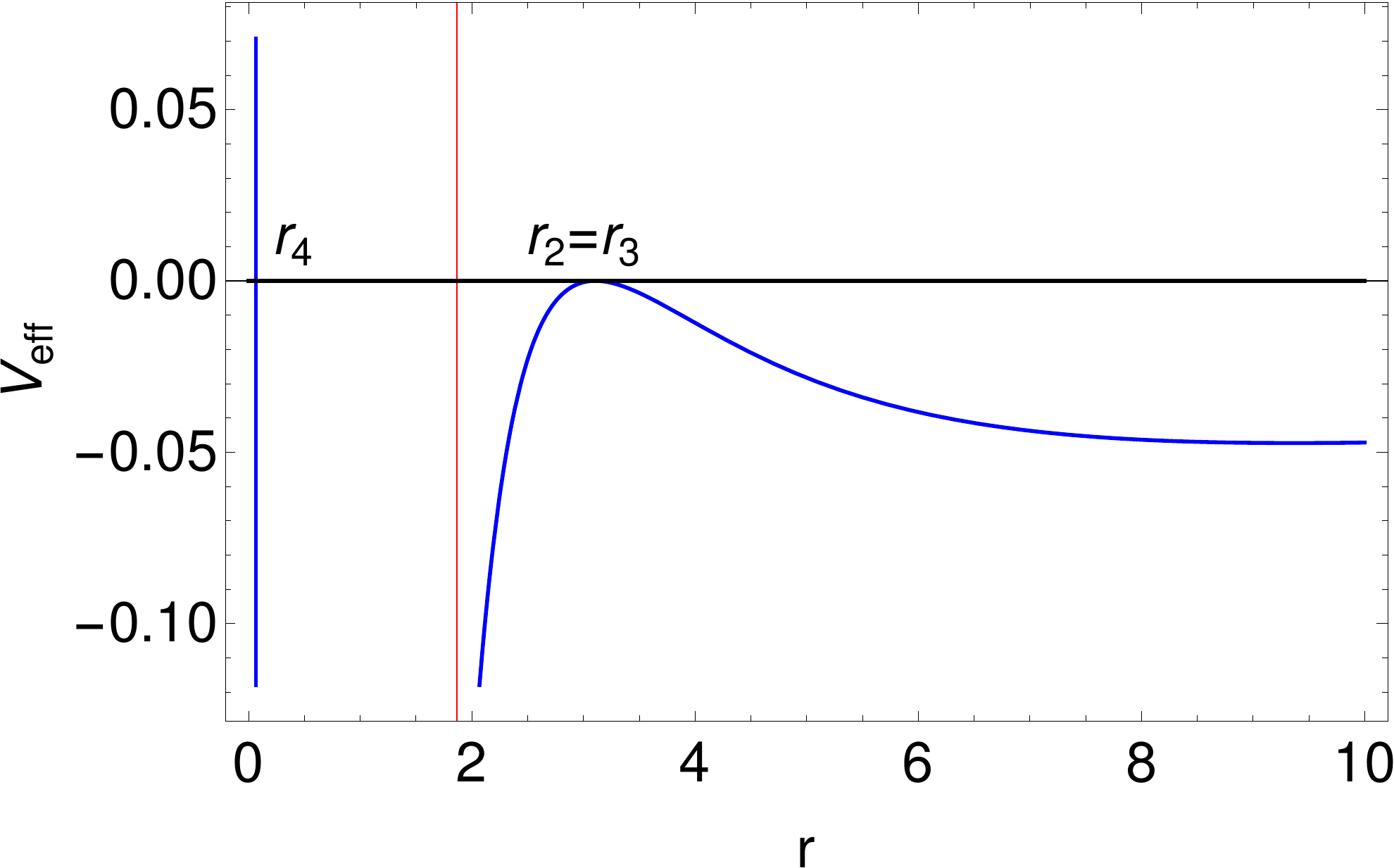}}}
 \caption{\label{bndregn}(a) The shaded region represents all possible bound orbits in the ($e$, $\mu$) plane for \{$a=0.5$, $Q=5$\}, where $S$ depicts the ISSO and $M$ depicts  the MBSO radius, and the red curve represents separatrix orbits (see \cite{RMCQG2019}, Figure (2a)); the corresponding effective potential diagrams are shown as a function of $r$ for (b) ISSO and (c) MBSO, where the horizontal black curve represents $\left(E^2-1\right) /2$ and the vertical red curve represents the horizon radius, and \{$r_1$, $r_2$, $r_3$, $r_4$\} are four roots of the effective potential, which are also the turning points of a trajectory, and where $r_1=\infty$ for MBSO. }
 \end{center}
 \end{figure}

\begin{figure}
\mbox{
 \subfigure[]{
\includegraphics[scale=0.4]{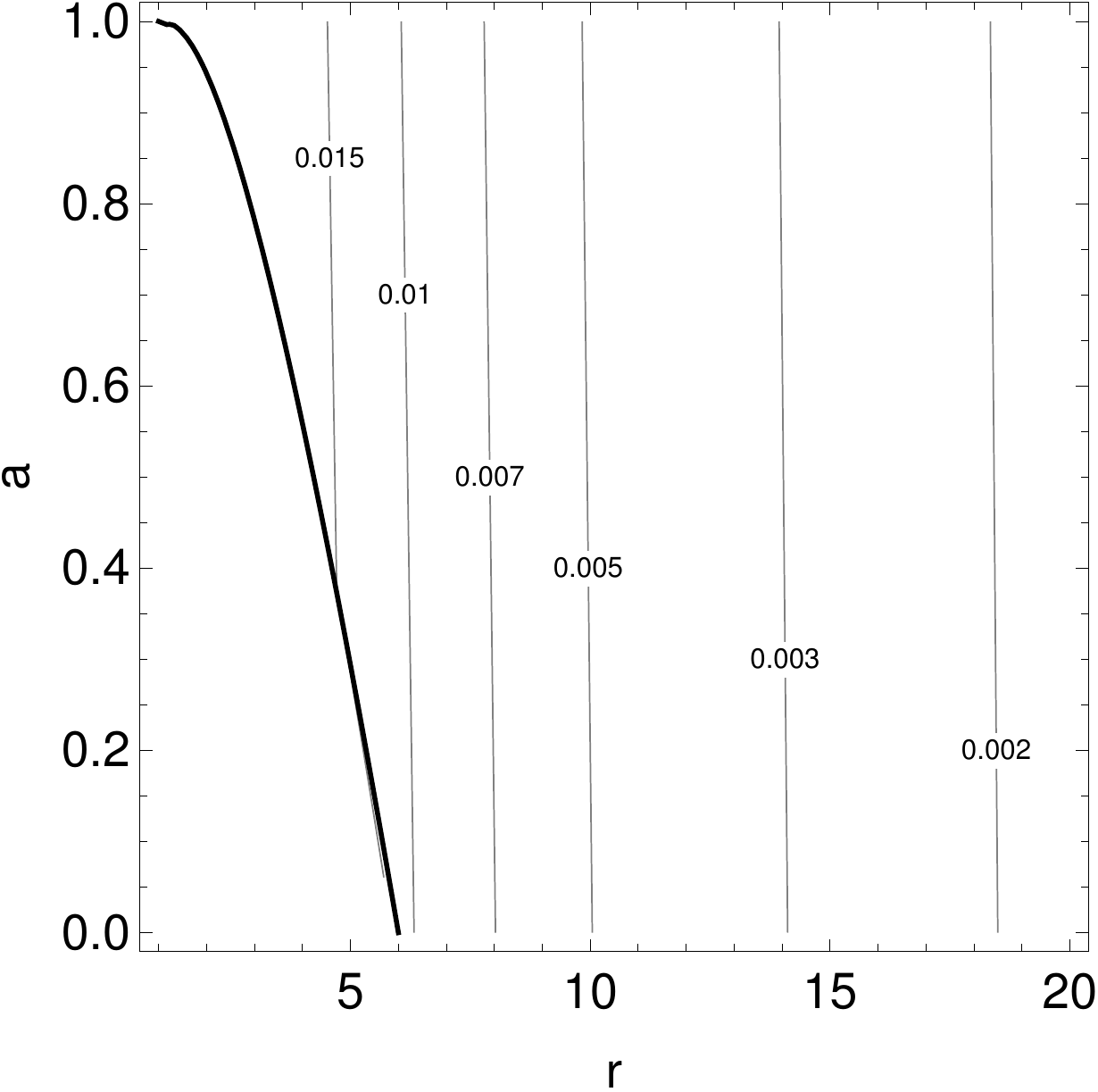}}
\hspace{0.65cm}
\subfigure[]{
\includegraphics[scale=0.4]{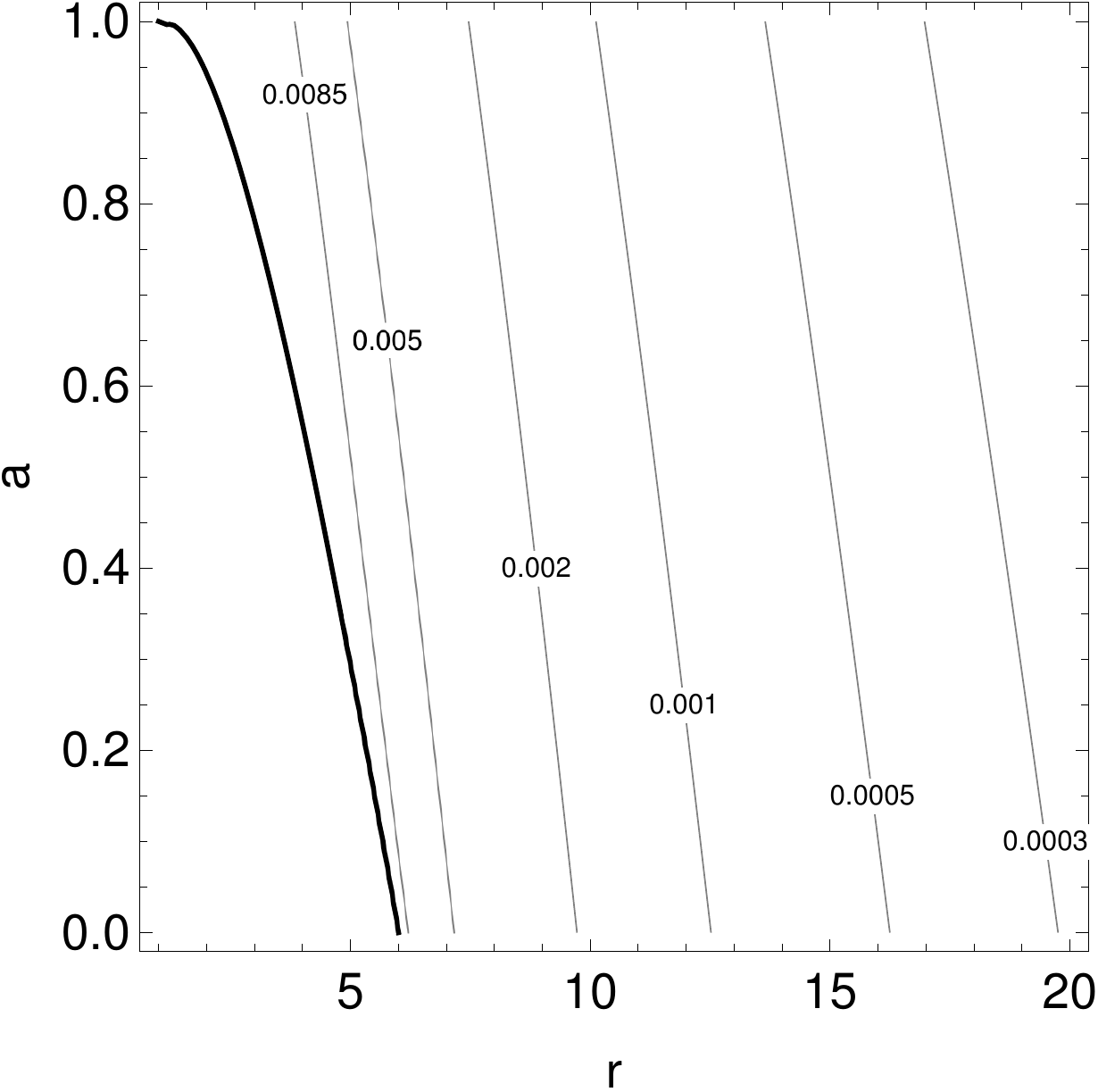}}
\hspace{0.65cm}
 \subfigure[]{
\includegraphics[scale=0.4]{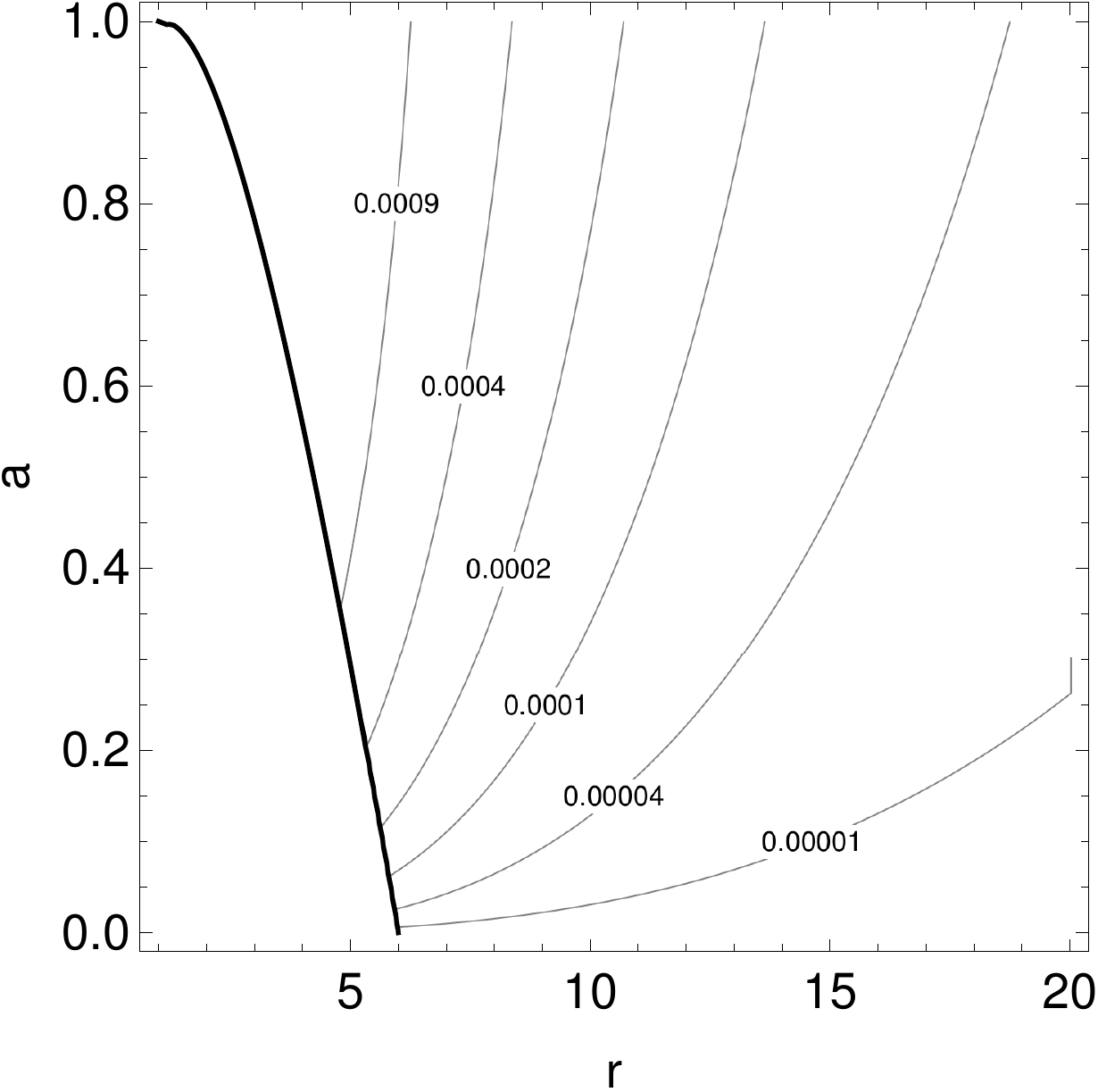}}}
\caption{\label{nucirccontrs}Dimensionless frequency contours are shown for circular and equatorial trajectories ($00$), using Equations (\ref{nuphicirc}$-$\ref{nuthetacirc}), in the ($r$, $a$) plane outside the ISCO radius, which is indicated by a thick black contour as also depicted in Figure \ref{radii}, for HFQPOs (a) $\bar{\nu}_{\phi}$, (b) $\bar{\nu}_{\rm pp}$, and type C LFQPO (c) $\bar{\nu}_{\rm np}$ assuming the RPM.}
\end{figure}

 These bound orbits can also be shown as a region in the ($e$, $\mu$) space, which is defined as
  \begin{equation}
  e=\dfrac{r_a-r_p}{r_a+r_p}, \ \ \ \ \ \ \ \ \ \  \mu=\dfrac{r_a+r_p}{2 r_a r_p},
\end{equation} 
 where $r_a$ is the apastron point of the orbit. This bound orbit region is shown as a shaded region in Figure \ref{bndregn}. The condition for these bound orbits is given by \citep{RMCQG2019,RMarxiv2019}
\begin{equation}
\left[ \mu^3 a^2 Q \left(1+e \right)^2 +\mu^2 \left( \mu a^2 Q -x^2 -Q\right) \left(3-e \right) \left(1+e \right) +1 \right]\geq 0, \label{boundcondition}
\end{equation} 
where $\mu$ can also be written as $\mu=1/\left[r_p \left( 1+ e\right) \right]$, where the equality sign corresponds to the separatrix trajectories. This bound orbit region shown in Figure \ref{bndregn} only includes regions 1 and 2 of the ($r_p$, $a$) plane shown in Figure \ref{radii}. 
 The RPM has been applied to two cases of BHXRBs, assuming the precession of nearly circular orbits (negligible eccentricity \footnote{as there is no periastron precession for $e=0$.}) in the equatorial plane of a Kerr black hole \citep{Motta2014a, Motta2014b}. In general, the observed range of HFQPOs in BHXRBs is 40-500 Hz, whereas that of type C LFQPOs is 10 mHz to 30 Hz \citep{Remillard2006,BelloniStella2014}. The formulae for fundamental particle frequencies of nearly circular and equatorial orbits are given by \cite{Bardeen1972} and \cite{Wilkins1972}; see Appendix \ref{reducecircular} for the derivation of these formulae from the general frequency formulae of $e0$ (Equation \eqref{eqfreq}) and $Q0$ (Equation \eqref{sphfreq}) orbits:
 \begin{subequations}
\begin{eqnarray}
\nu_{\phi}\left( r, a\right)=&& \frac{c^3}{2 \pi G M_{\bullet}}\frac{1}{\left( r^{3/2} + a\right)}, \ \  \bar{\nu}_{\phi}\left( r, a\right)=\frac{\nu_{\phi}}{\left( c^3 /G M_{\bullet} \right)}=\frac{1}{2\pi \left( r^{3/2} + a\right)}, \label{nuphicirc}  \\
\nu_{r}\left( r, a\right)=&& \nu_{\phi}\left( 1- \dfrac{6}{r} - \dfrac{3 a^2}{r^2} + \dfrac{8a}{r^{3/2}}\right)^{1/2}, \ \ \bar{\nu}_{r}\left( r, a\right)=\frac{\nu_{r}}{\left( c^3 /G M_{\bullet} \right)}, \label{nurcirc} \\
 \nu_{\theta}\left( r, a\right)=&& \nu_{\phi} \left( 1+ \dfrac{3 a^2}{r^2} - \dfrac{4a}{r^{3/2}}\right)^{1/2}, \ \  \bar{\nu}_{\theta}\left( r, a\right)=\frac{\nu_{\theta}}{\left( c^3 /G M_{\bullet} \right)}, \label{nuthetacirc}
\end{eqnarray}
\label{circfreq}
 \end{subequations}
 where $\{ \bar{\nu}_{\phi} , \ \bar{\nu}_{r} , \ \bar{\nu}_{\theta} \}$ are the dimensionless frequencies, where we use the convention $a>0$ for the prograde and $a<0$ for the retrograde orbits in this article. Using these formulae and assuming the RPM, it was retrodicted for BHXRB GROJ 1655-40 and XTEJ 1550-564 that these signals originated very close to and outside the ISCO radius, at nearly $r=5.677\pm 0.035$ for GROJ 1655-40 and $r=5.47\pm0.12$ for XTE J1550-564 \citep{Motta2014a, Motta2014b}. We show that the expected QPO frequency range associated with the $00$ orbits in the RPM $ \{ \nu_{\phi}$, $\nu_{\rm pp}\equiv(\nu_{\phi}-\nu_{r})$, $\nu_{\rm np}\equiv(\nu_{\phi}-\nu_{\theta}) \}$ is valid for a wide range of $r$, where $\nu_{\phi}$, $\nu_{\rm pp}$, and $\nu_{\rm np}$ correspond to the HFQPO-1, HFQPO-2, and type C LFQPO, respectively \footnote{where ${\rm pp}$ and ${\rm np}$ represent the periastron and nodal precession frequencies, respectively.}. To illustrate this, we present a mass-independent model of these frequencies. In Table \ref{QPOfreqtable}, we have shown the observed range of the HFQPO and LFQPO frequencies in BHXRBs along with a typical range in dimensionless values \{$\bar{\nu}_{\phi}$, $\bar{\nu}_{\rm pp}$, $\bar{\nu}_{\rm np}$\}, obtained by scaling the observed frequencies of HFQPOs in BHXRBs using the corresponding known value of the black hole mass $\mathcal{M}\sim 5-10$ (given in Table \ref{sourcelist}). For a BHXRB, the typical frequency range of the type C QPOs is 10 mHz to 30 Hz, and we have scaled this frequency range with $\mathcal{M}=10$ (a typical mass value for BHXRB) to obtain the dimensionless frequency range. This provides an expected range of the geometrical orbital parameters independent of the black hole mass that implies largely a range of $r_p$. Figure \ref{nucirccontrs} shows the contours of $\bar{\nu}_{\phi}$, $\bar{\nu}_{\rm pp}$, and $\bar{\nu}_{\rm np}$ for the $00$ orbits, using Equations (\ref{nuphicirc}$-$\ref{nuthetacirc}), in the ($r$, $a$) plane outside the ISCO radius (region 1 of Figure \ref{radiiQ0}). We find the following: 
 \begin{enumerate}
 \item The expected range of simultaneous QPO frequencies corresponds to a wide range of $r\sim 5-15 $ for the $00$ orbits, which is typically the inner region of the accretion disk. 
 \item The simultaneous QPOs, if associated with the $00$ orbits, should originate very near to the ISCO radius.
 \item We expect much higher QPO frequency values $\{ \bar{\nu}_{\phi} \gtrsim 0.015 , \ \bar{\nu}_{\rm pp} \gtrsim 0.009, \ \bar{\nu}_{\rm np} \gtrsim 0.001 \}$ for the $00$ orbits near the ISCO radius for $a\gtrsim 0.5$, as seen in Figure \ref{nucirccontrs}, which are outside the observed QPO frequency range.
\end{enumerate} 
 \begin{deluxetable}{c c c c}
\tablecaption{Summary of the Observed QPO Frequency Range in BHXRBs and Their Corresponding Dimensionless Values Derived from Data Given in Table \ref{sourcelist}.
\label{QPOfreqtable}}
\tablewidth{400pt}
\tabletypesize{\footnotesize}
\tablehead{
\colhead{$\ \ $Type of QPO  $\ \ \ $} & \colhead{$\ \ \ $ QPOs in the$\ \ \ $} & 
\colhead{$\ \ \ $ Observed QPO Frequency  $\ \ $} & \colhead{Dimensionless Frequency Range} \\ 
\colhead{} & \colhead{RPM and GRPM} & \colhead{Range in Hz} & \colhead{$ \ \ \ \bar{\nu}= \nu \cdot 10^{-3} / \left( c^3 / G M_{\bullet} \right) \ \ $} 
} 
\startdata
HFQPO-1 & $\nu_{\phi}$	&	$100-500$	& $2-15$	\\
HFQPO-2 & $\nu_{\rm pp}$ & $40-300$ &  $1-9$\\
 Type C LFQPO & $\nu_{\rm np}$ & $10^{-2}-30$ &  $10^{-5}-1$\\
 \enddata
\end{deluxetable}

Now, with this, we can explore the frequency range of the nonequatorial eccentric, equatorial eccentric, and spherical orbits using a similar approach assuming the GRPM in the regions 1 and 2 of Figure \ref{radii} (shaded region of Figure \ref{bndregn}).

\subsection{Nonequatorial and Equatorial Eccentric Orbits: $eQ$ and $e0$ }
\label{eccentricmotivation}
We first discuss the useful formulae of the fundamental frequencies for the nonequatorial and equatorial eccentric particle trajectories derived in \cite{RMCQG2019,RMarxiv2019}. Later, we use these formulae to explore the required frequency range  for QPOs in BHXRBs, based on the GRPM, and determine the corresponding parameter range \{$e$, $r_p$, $a$, $Q$\} associated with these trajectories. 

As shown in Figure \ref{trajectory}, the orbital plane of a nonequatorial eccentric trajectory oscillates with respect to the spin axis of the black hole, along with the phenomenon of periastron precession taking place in the orbital plane. A complete analytic trajectory solution and the fundamental frequencies for such trajectories around a Kerr black hole were derived in terms of \{$e$, $\mu$, $a$, $Q$\} parameters \citep{RMCQG2019,RMarxiv2019}, where $\mu$ is the inverse latus rectum of the orbit, and it can be written in terms of \{$e, r_p$\} as $\mu=1/\left[ r_p \left(1+e \right)\right]$. The expressions of dimensionless fundamental frequencies for these trajectories are given by \citep{RMCQG2019,RMarxiv2019}
\begin{subequations}
 \begin{eqnarray}
\bar{\nu}_{\phi}\left(   e, r_p , a, Q \right)&&=\frac{ \left[ - I_1\left( e, r_p, a, Q \right) - 2 L_z I_8\left(  e, r_p, a, Q \right)\right] F\left( \frac{\pi}{2},\frac{z_{-}^{2}}{z_{+}^{2}}\right)  + 2 L_z I_8\left(  e, r_p, a, Q \right) \Pi\left( z_{-}^2, \frac{\pi}{2},\frac{z_{-}^{2}}{z_{+}^{2}}\right) }{2 \pi \left\lbrace \left[I_2\left(  e, r_p, a, Q \right) + 2 a^2 z_{+}^2 E I_8\left(e, r_p, a, Q \right) \right] F\left( \frac{\pi}{2},\frac{z_{-}^{2}}{z_{+}^{2}}\right) -  2a^2 z_{+}^2 E I_8\left(  e, r_p, a, Q \right) K\left( \frac{\pi}{2},\frac{z_{-}^{2}}{z_{+}^{2}}\right)  \right\rbrace }, \nonumber \\
 \label{nuphi} 
 \end{eqnarray}
\begin{eqnarray}
\bar{\nu}_r \left(   e, r_p, a, Q \right)&&= \frac{F\left( \frac{\pi}{2},\frac{z_{-}^{2}}{z_{+}^{2}}\right) }{\left\lbrace \left[ I_2 \left( e, r_p, a, Q \right)  + 2 a^2 z_{+}^2 E I_8 \left( e, r_p, a, Q \right)  \right] F\left( \frac{\pi}{2},\frac{z_{-}^{2}}{z_{+}^{2}}\right)- 2 a^2 z_{+}^2 E I_8 \left( e, r_p, a, Q \right) K\left( \frac{\pi}{2},\frac{z_{-}^{2}}{z_{+}^{2}}\right) \right\rbrace  }, \nonumber \\
\label{nur}  \\
 \bar{\nu}_{\theta} \left(  e, r_p, a, Q \right)&&=  \frac{a \sqrt{1- E^2} z_{+} I_8 \left(  e, r_p, a, Q \right)}{2\left\lbrace \left[ I_2\left(  e, r_p, a, Q \right) + 2 a^2 z_{+}^2 E I_8\left( e, r_p, a, Q \right)  \right] F\left( \frac{\pi}{2},\frac{z_{-}^{2}}{z_{+}^{2}}\right)- 2 a^2 z_{+}^2 E I_8 \left(  e, r_p, a, Q \right) K\left( \frac{\pi}{2},\frac{z_{-}^{2}}{z_{+}^{2}}\right)  \right\rbrace }, \nonumber \\ 
 \label{nutheta}
 \end{eqnarray}
 \label{genfreq}
\end{subequations}
where $L_z$ is the $z$-component of a particle's angular momentum and $E$ is its energy per unit rest mass, which can be explicitly expressed as the functions of \{$e$, $\mu$, $a$, $Q$\} parameters [see Equations (5a$-$5e) in \cite{RMCQG2019}]. Here, $I_1\left( e, r_p, a, Q \right)$, $I_2\left( e, r_p, a, Q \right)$, and $I_8\left( e, r_p, a, Q \right)$ are the radial integrals of motion given in their simplest analytic forms, along with the constants involved, by Equations (6a$-$6h), (7a$-$7l), (8a$-$8c), and (9d) in \cite{RMCQG2019}; $F\left( \varphi , p^2 \right) $, $K\left( \varphi , p^2 \right) $, and $\Pi\left(q^2 , \varphi , p^2 \right) $ used in Equations (\ref{nuphi}$-$\ref{nutheta}) are the standard elliptic integrals \citep{Grad}.

Next, in the case of equatorial eccentric orbits ($e0$), the expressions for the azimuthal and radial fundamental frequencies can be further reduced to a form simpler than Equations (\ref{nuphi}$-$\ref{nutheta}), which are given by \citep{RMCQG2019,RMarxiv2019}
\begin{subequations}
\begin{equation}
\bar{\nu}_{\phi}\left( e, r_p, a\right) =\frac{  a_1  \Pi  \left( -p_{2}^2, \frac{\pi}{2}, m^2\right) + b_1 \Pi\left( -p_{3}^2, \frac{\pi}{2}, m^2\right) }{ 2 \pi \left\lbrace \splitfrac{  \frac{a_2}{2\left(1+p_{1}^2 \right)} \left[ - F\left( \frac{\pi}{2}, m^2\right)  +  \frac{p_{1}^2K\left( \frac{\pi}{2}, m^2\right) }{ \left( m^2 + p_{1}^2\right) } \right] + \Pi\left( -p_{1}^2 , \frac{\pi}{2} , m^2\right)  \left[ a_2  \frac{\left[ p_{1}^4 +2p_{1}^2 \left( 1+m^2\right) + 3m^2\right] }{2\left( 1+p_{1}^2\right) \left( m^2 +p_{1}^2\right)} +b_2\right]}{  + c_2 \Pi\left( -p_{2}^2 , \frac{\pi}{2} , m^2\right) + d_2 \Pi\left( -p_{3}^2 , \frac{\pi}{2} , m^2\right)  } \right\rbrace},   \label{eqnuphi}
\end{equation}
\begin{equation}
\bar{\nu}_{r}\left( e, r_p, a\right)=\frac{1}{2 \left\lbrace \splitfrac{  \frac{a_2}{2\left(1+p_{1}^2 \right)} \left[ - F\left( \frac{\pi}{2}, m^2\right)  +  \frac{p_{1}^2K\left( \frac{\pi}{2}, m^2\right) }{ \left( m^2 + p_{1}^2\right) } \right] + \Pi\left( -p_{1}^2 , \frac{\pi}{2} , m^2\right)  \left[ a_2  \frac{\left[ p_{1}^4 +2p_{1}^2 \left( 1+m^2\right) + 3m^2\right] }{2\left( 1+p_{1}^2\right) \left( m^2 +p_{1}^2\right)} +b_2\right]}{  + c_2 \Pi\left( -p_{2}^2 , \frac{\pi}{2} , m^2\right) + d_2 \Pi\left( -p_{3}^2 , \frac{\pi}{2} , m^2\right)  } \right\rbrace }, \label{eqnur}
\end{equation}
\begin{equation}
\bar{\nu}_{\theta}\left( e, r_p, a\right)=\frac{2 \bar{\nu}_{r}\left( e, \mu, a\right) \mu^{1/2} \sqrt{\left( x^2+a^2+ 2a E x \right) } \cdot F\left( \dfrac{\pi}{2}, m^2\right)}{\pi \left[ 1- \mu^2 x^2 \left( 3- e^2 -2e\right) \right]^{1/2} }, \label{eqnutheta}
\end{equation}
\label{eqfreq}
\end{subequations}
where $x=\left( L_z - a E\right)$, and \{$p_{1}^2$, $p_{2}^2$, $p_{3}^2$\} are given by Equation (7k) of \cite{RMCQG2019}, while $m^2$ is given by Equation (13c) of \cite{RMCQG2019}, for the $e0$ orbits. See Appendix \ref{nuthetaderivation} for the derivation of Equation \eqref{eqnutheta}, which is a novel reduced form for $\bar{\nu}_{\theta}$.
\begin{figure}
\mbox{
 \subfigure[]{
 \hspace{-0.8cm}
\includegraphics[scale=0.4]{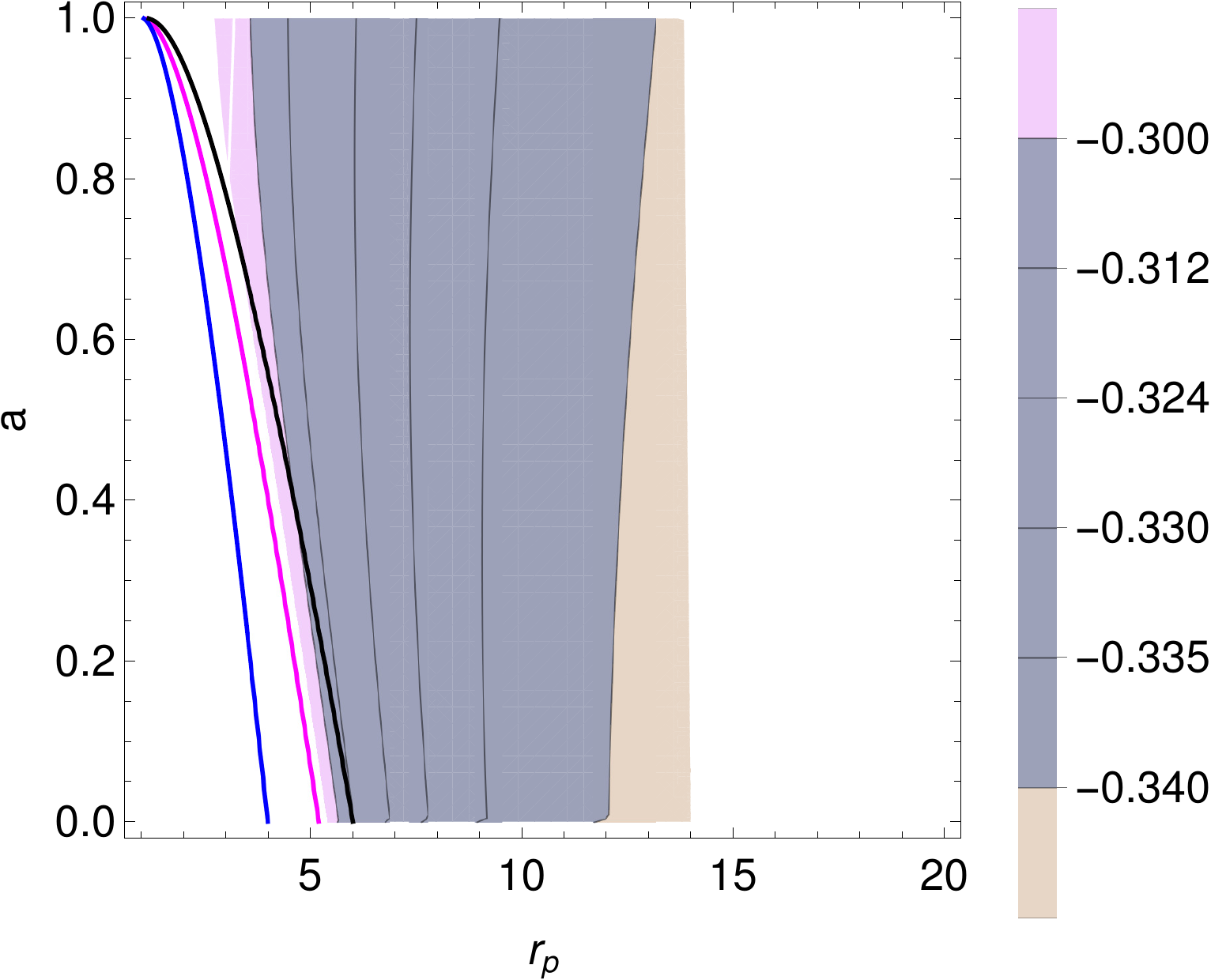}\label{nuphie25Q0}}
\hspace{0.1cm}
 \subfigure[]{
\includegraphics[scale=0.4]{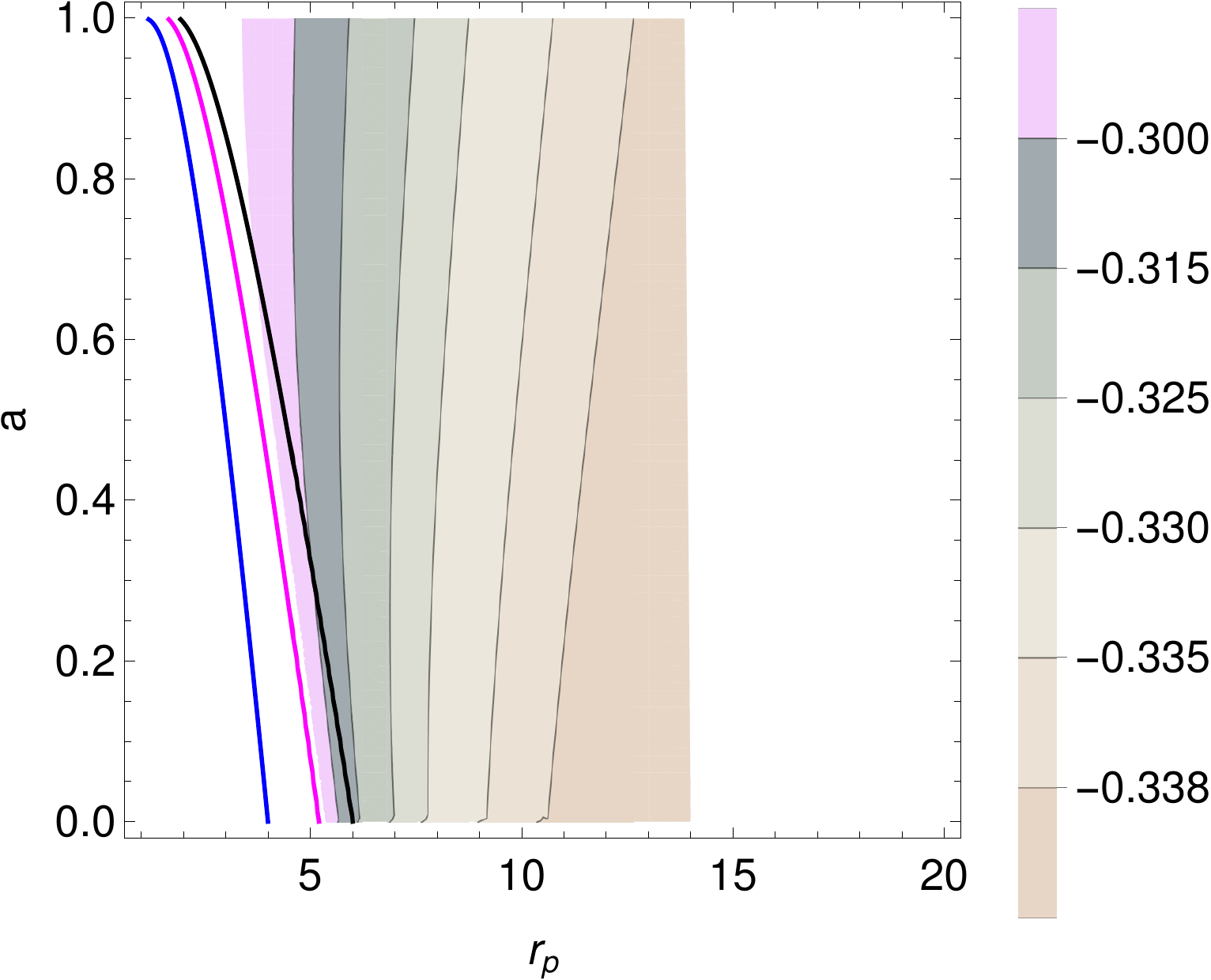}\label{nuphie25Q2}}
\hspace{0.1cm}
 \subfigure[]{
\includegraphics[scale=0.4]{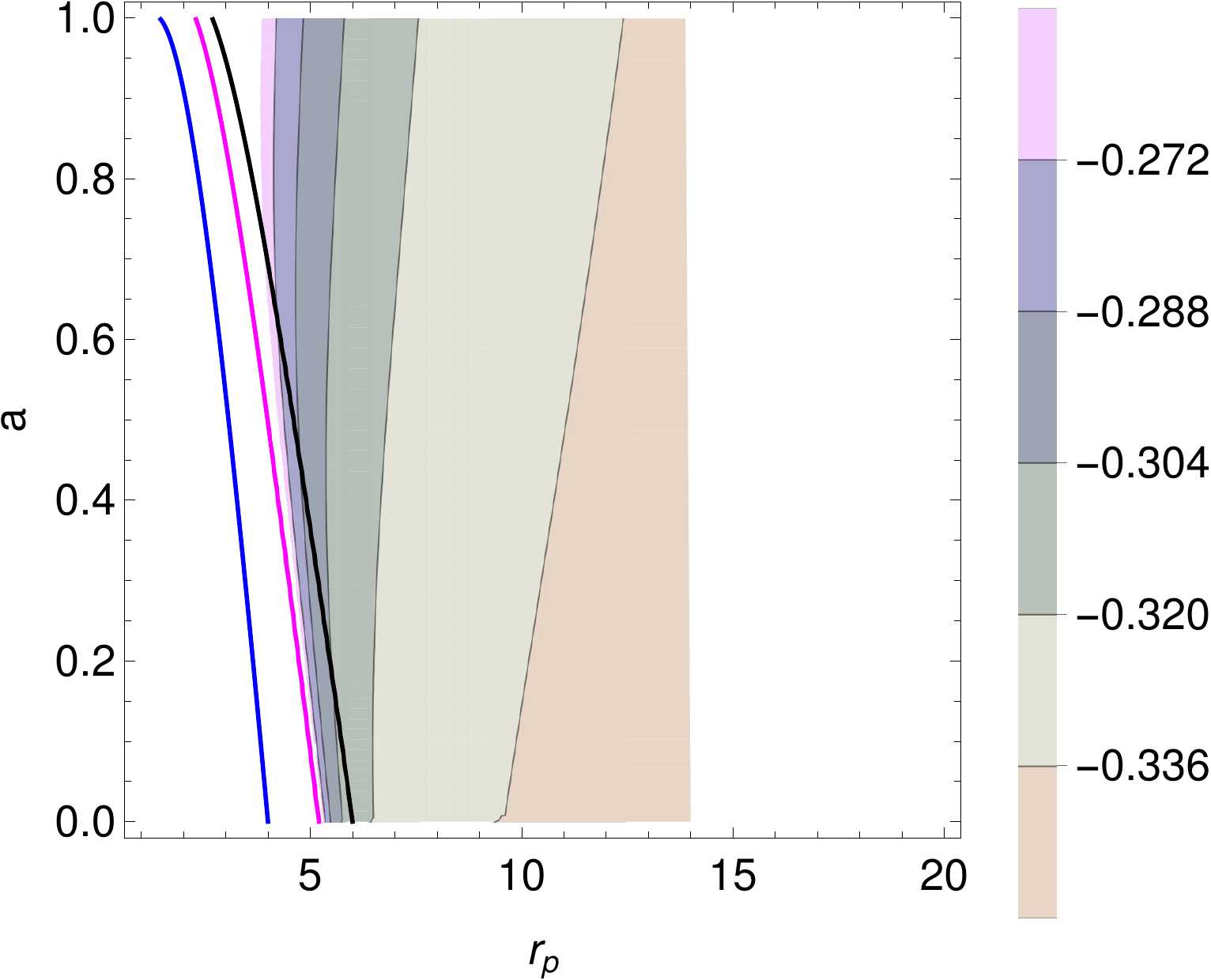}\label{nuphie25Q4}}}
\mbox{
 \hspace{-0.8cm}
 \subfigure[]{
\includegraphics[scale=0.4]{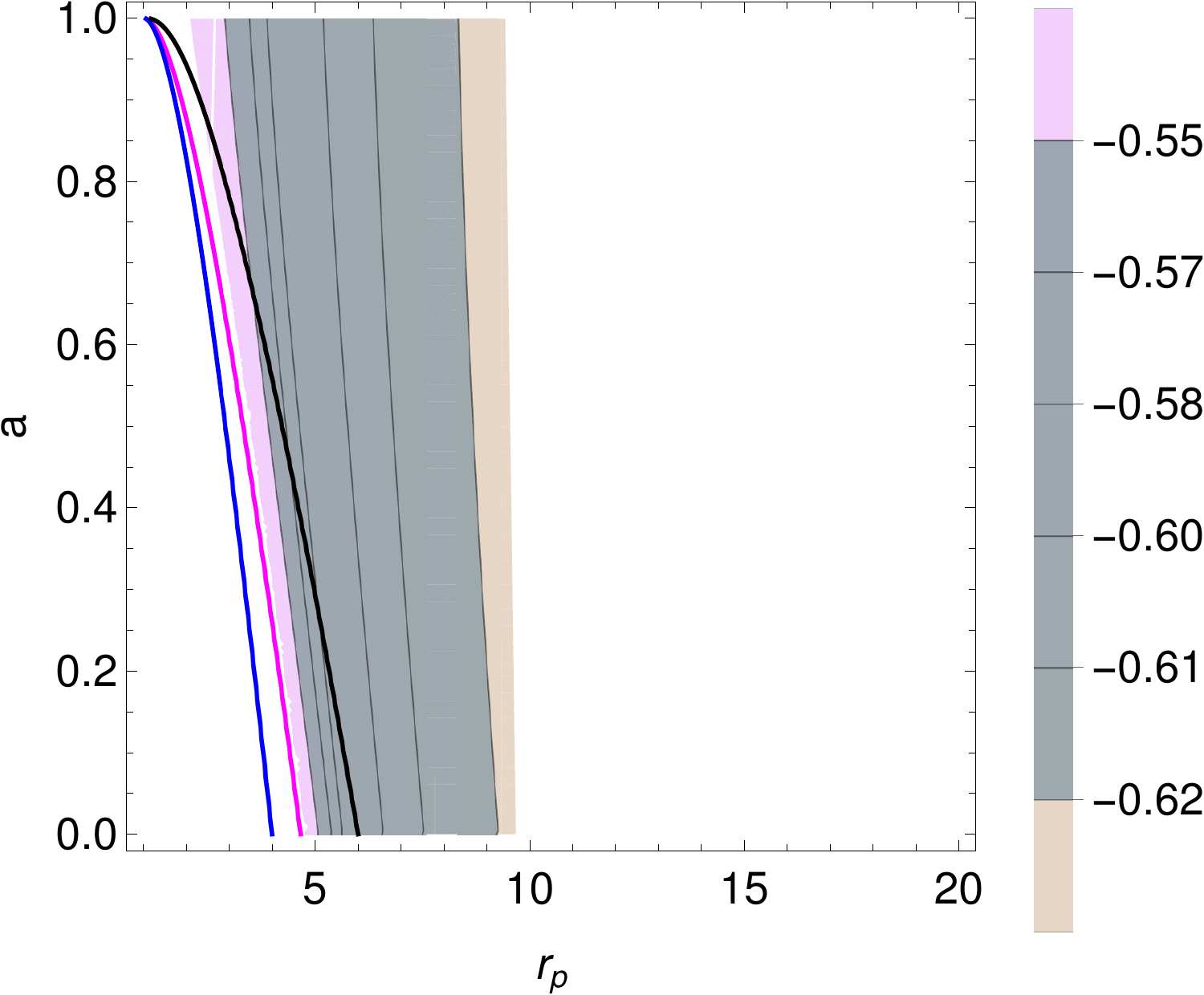}\label{nuphie5Q0}}
\hspace{0.1cm}
 \subfigure[]{
\includegraphics[scale=0.4]{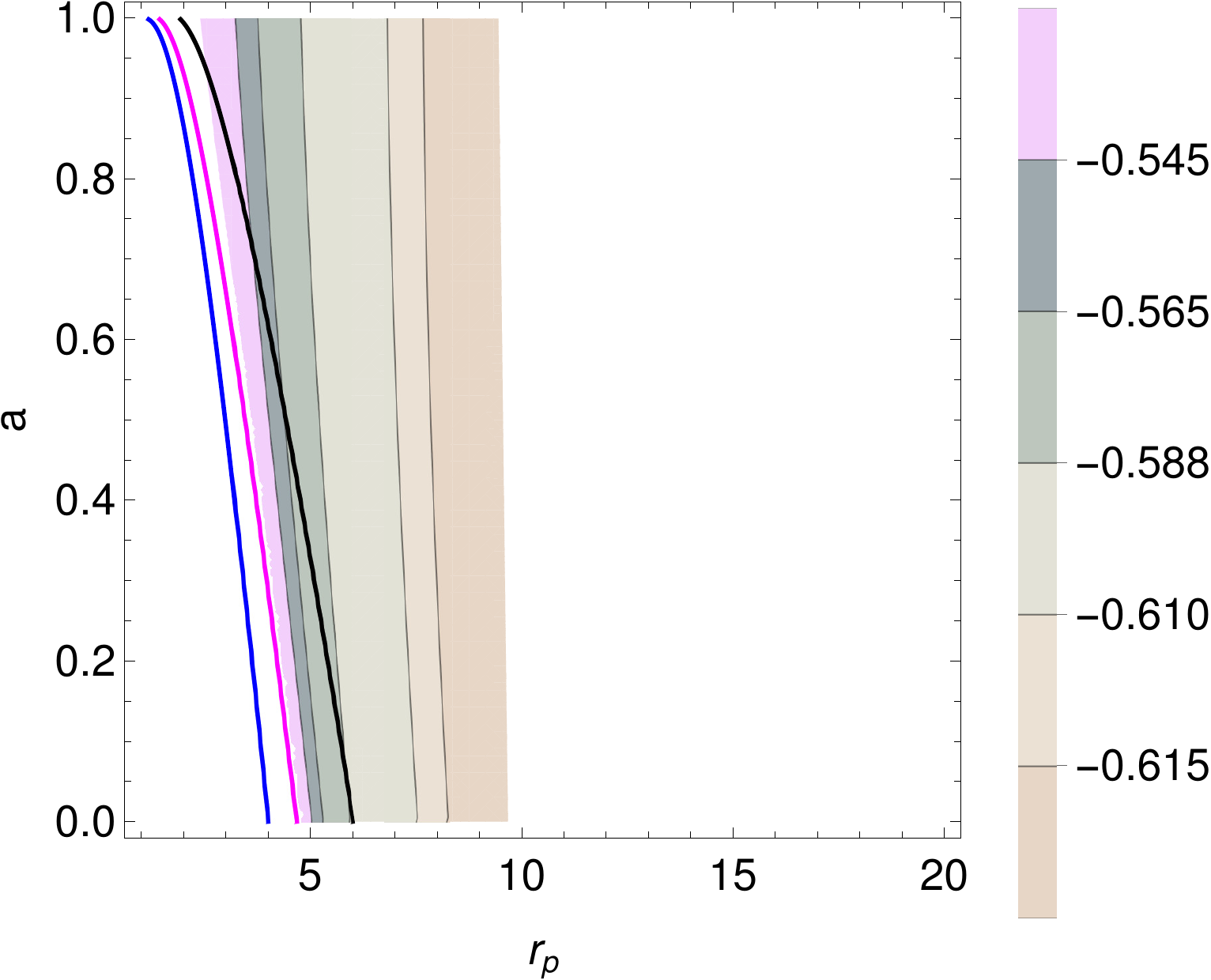}\label{nuphie5Q2}}
\subfigure[]{
\hspace{0.1cm}
\includegraphics[scale=0.4]{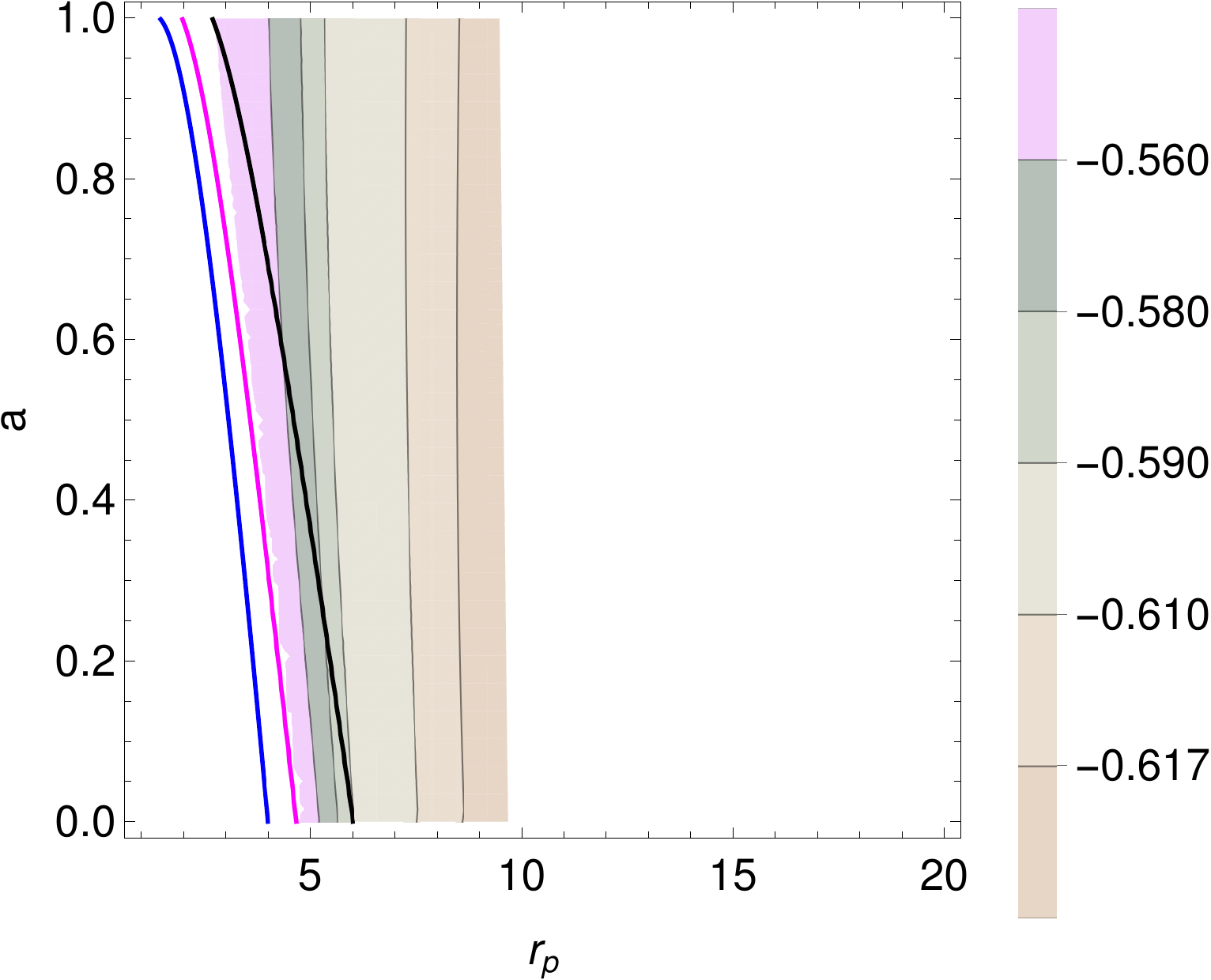}\label{nuphie5Q4}}}
\caption{\label{nuphieccentriccontrs} The contours of $\delta_{\phi}\left(e, r_p, a, Q \right)$ are shown in the ($r_p$, $a$) plane for eccentric orbits around a Kerr black hole, where the parameter combinations are (a) \{$e=0.25, Q=0$\}, (b) \{$e=0.25, Q=2$\}, (c) \{$e=0.25, Q=4$\}, (d) \{$e=0.5, Q=0$\}, (e) \{$e=0.5, Q=2$\}, and (f) \{$e=0.5, Q=4$\}.}
\end{figure}

\begin{figure}
\mbox{
 \subfigure[]{
  \hspace{-0.8cm}
\includegraphics[scale=0.4]{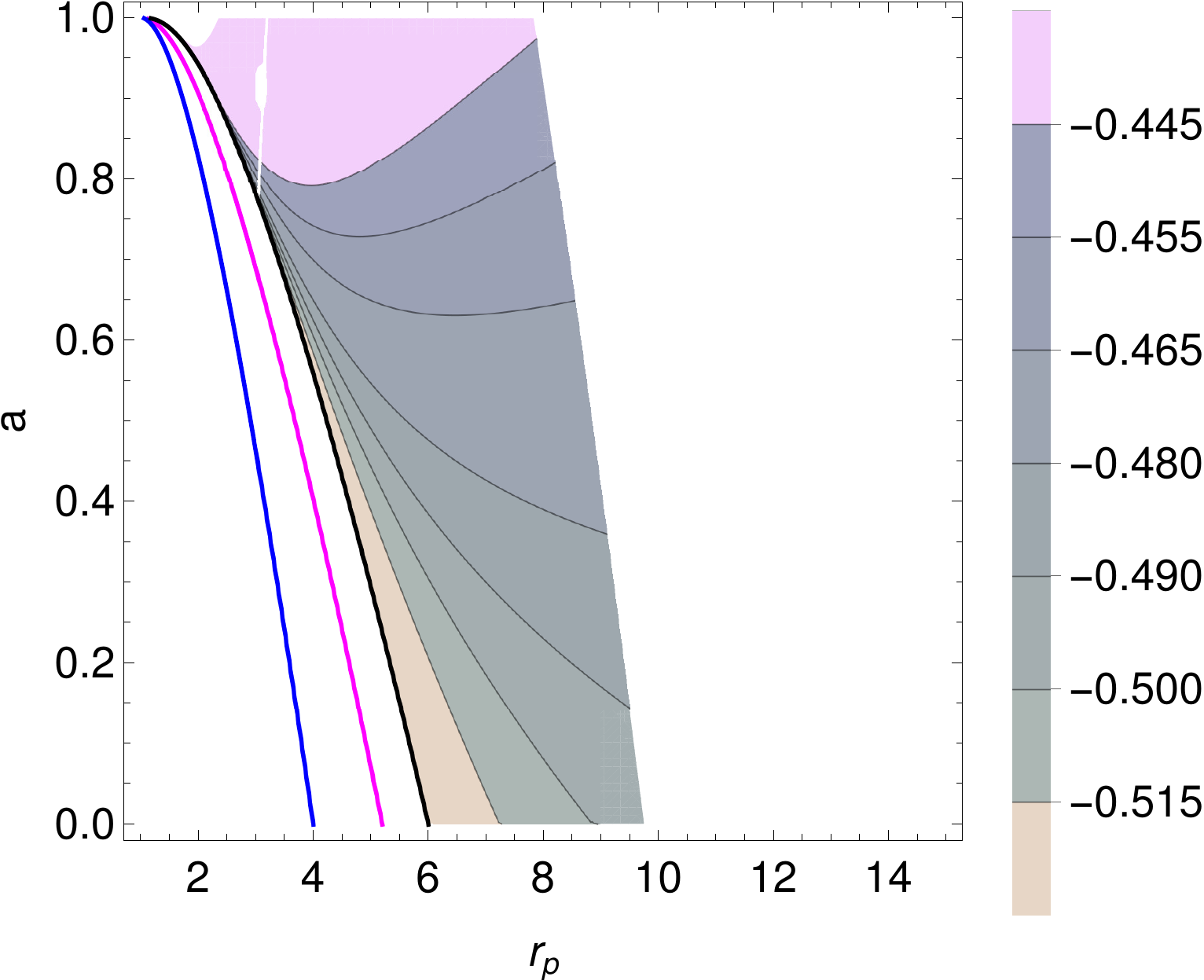}\label{nupere25Q0}}
\hspace{0.1cm}
 \subfigure[]{
\includegraphics[scale=0.4]{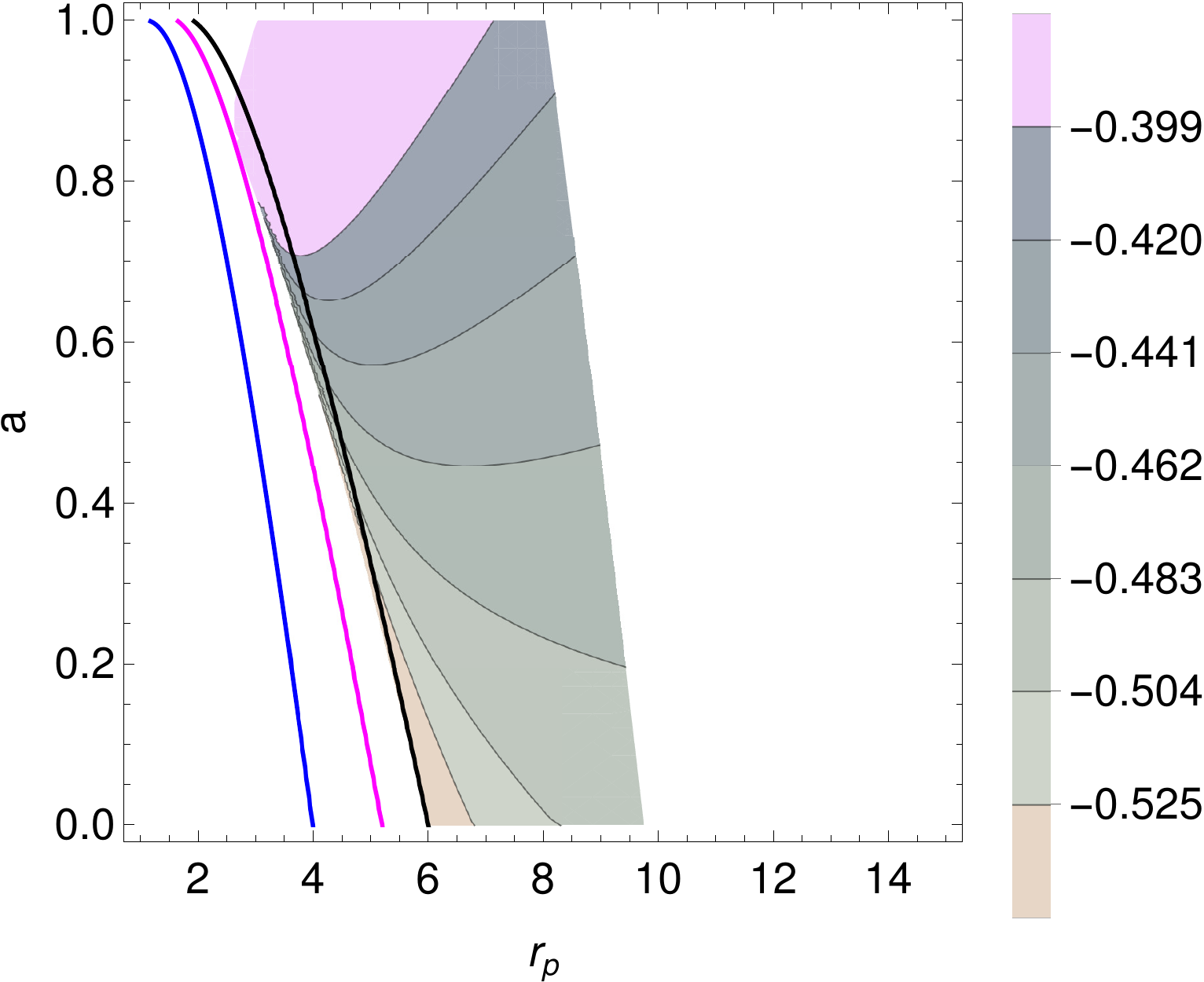}\label{nupere25Q2}}
\hspace{0.1cm}
\subfigure[]{
\includegraphics[scale=0.4]{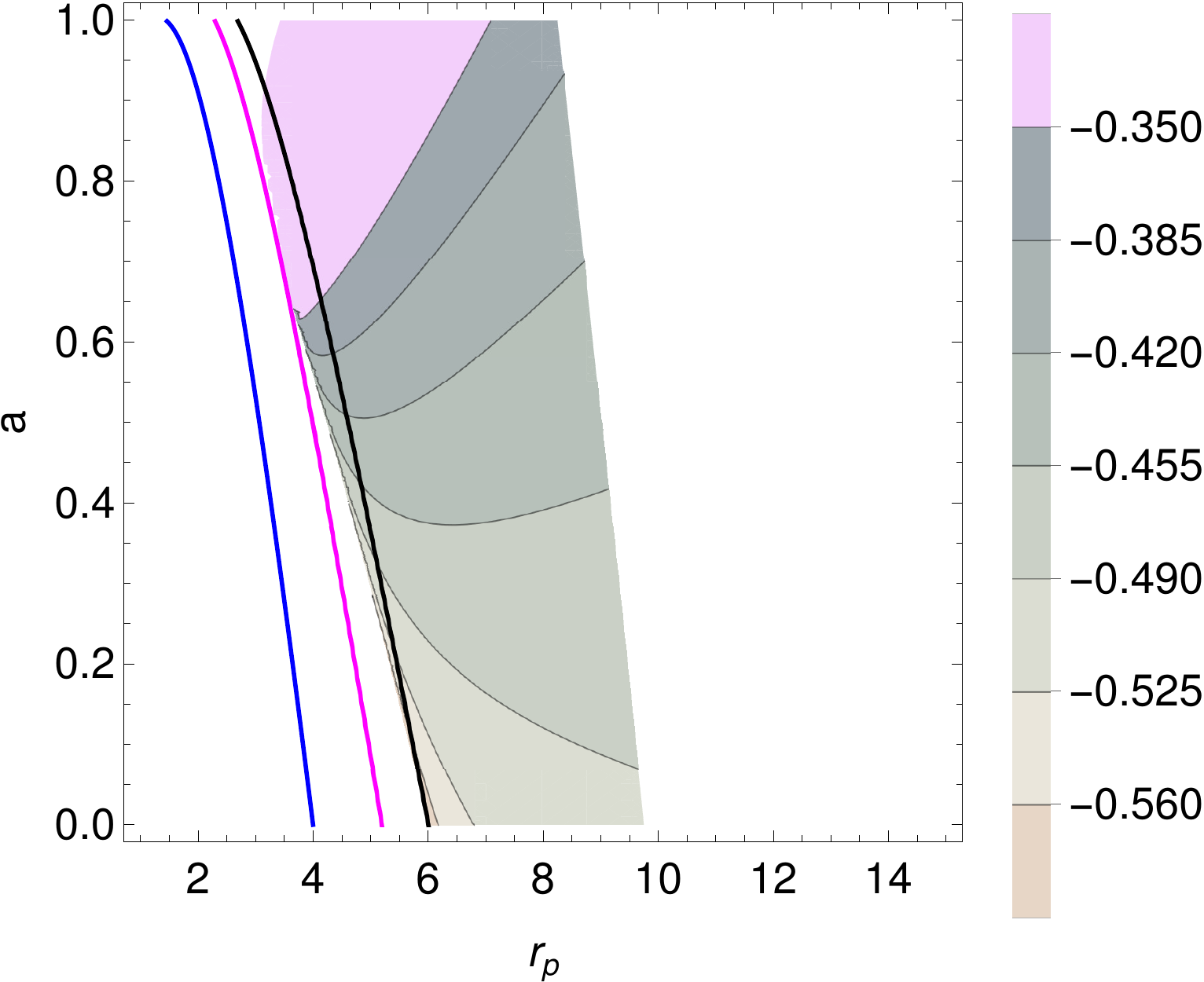}\label{nupere25Q4}}}
\mbox{
 \subfigure[]{
  \hspace{-0.8cm}
\includegraphics[scale=0.4]{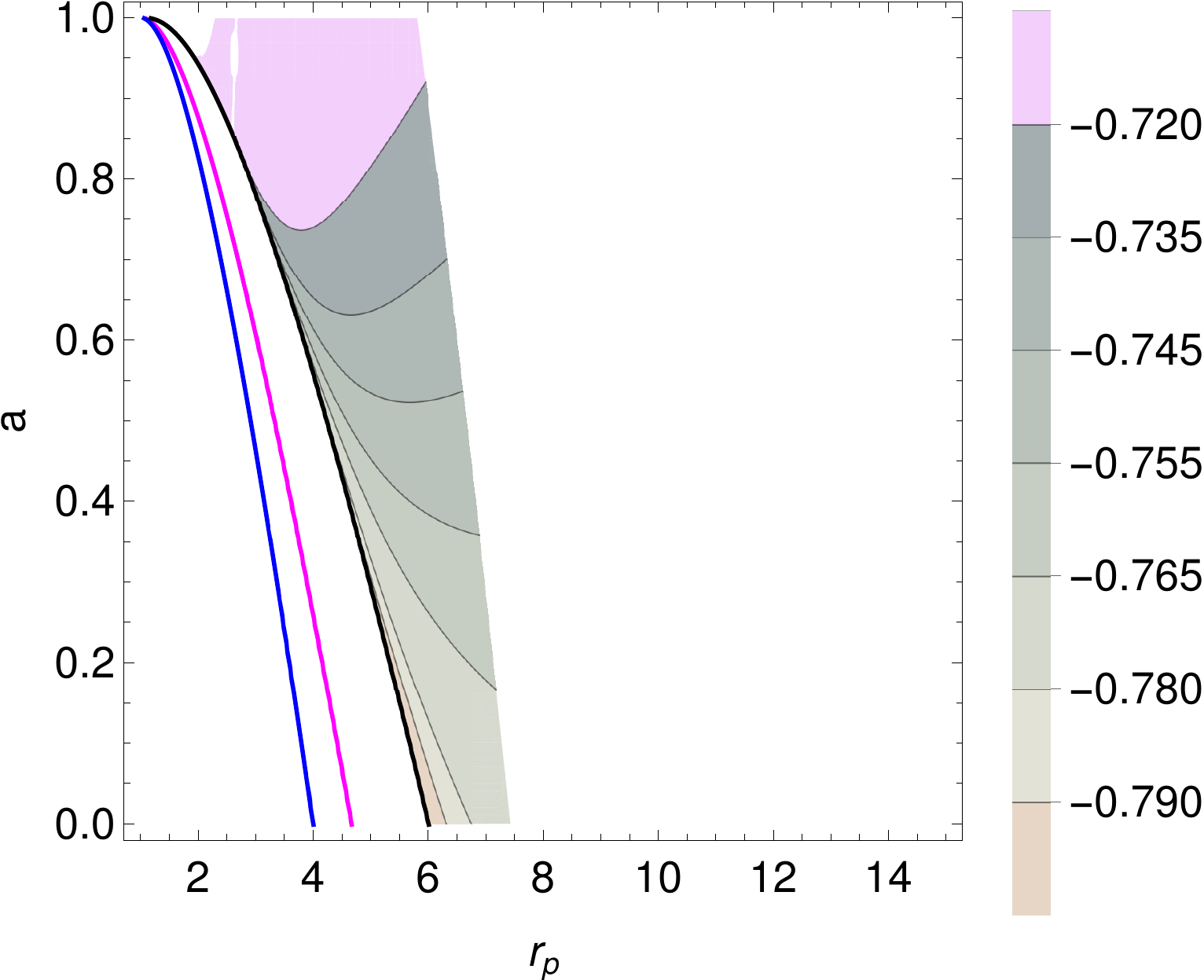}\label{nupere5Q0}}
\hspace{0.1cm}
 \subfigure[]{
\includegraphics[scale=0.4]{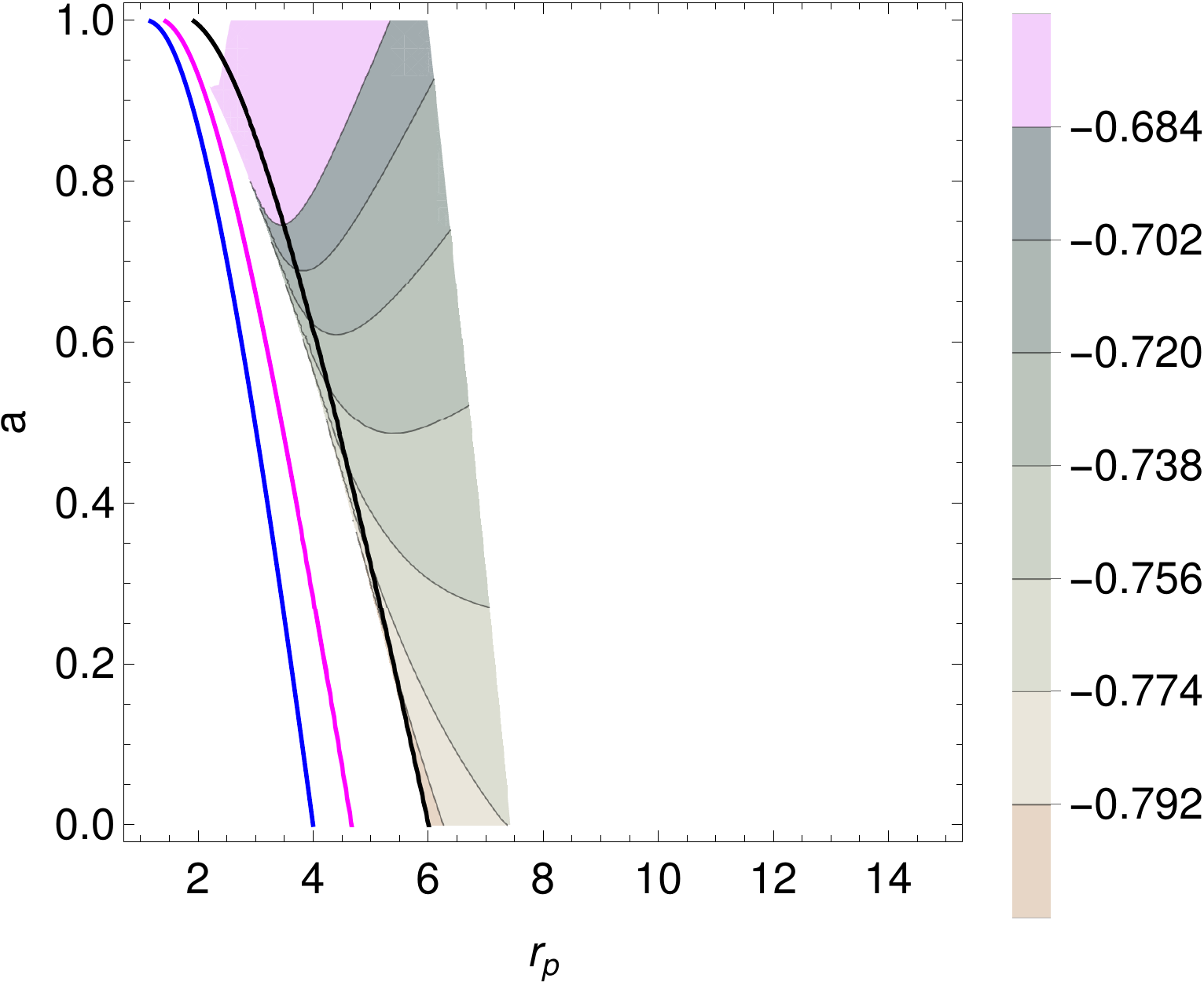}\label{nupere5Q2}}
\hspace{0.1cm}
\subfigure[]{
\includegraphics[scale=0.4]{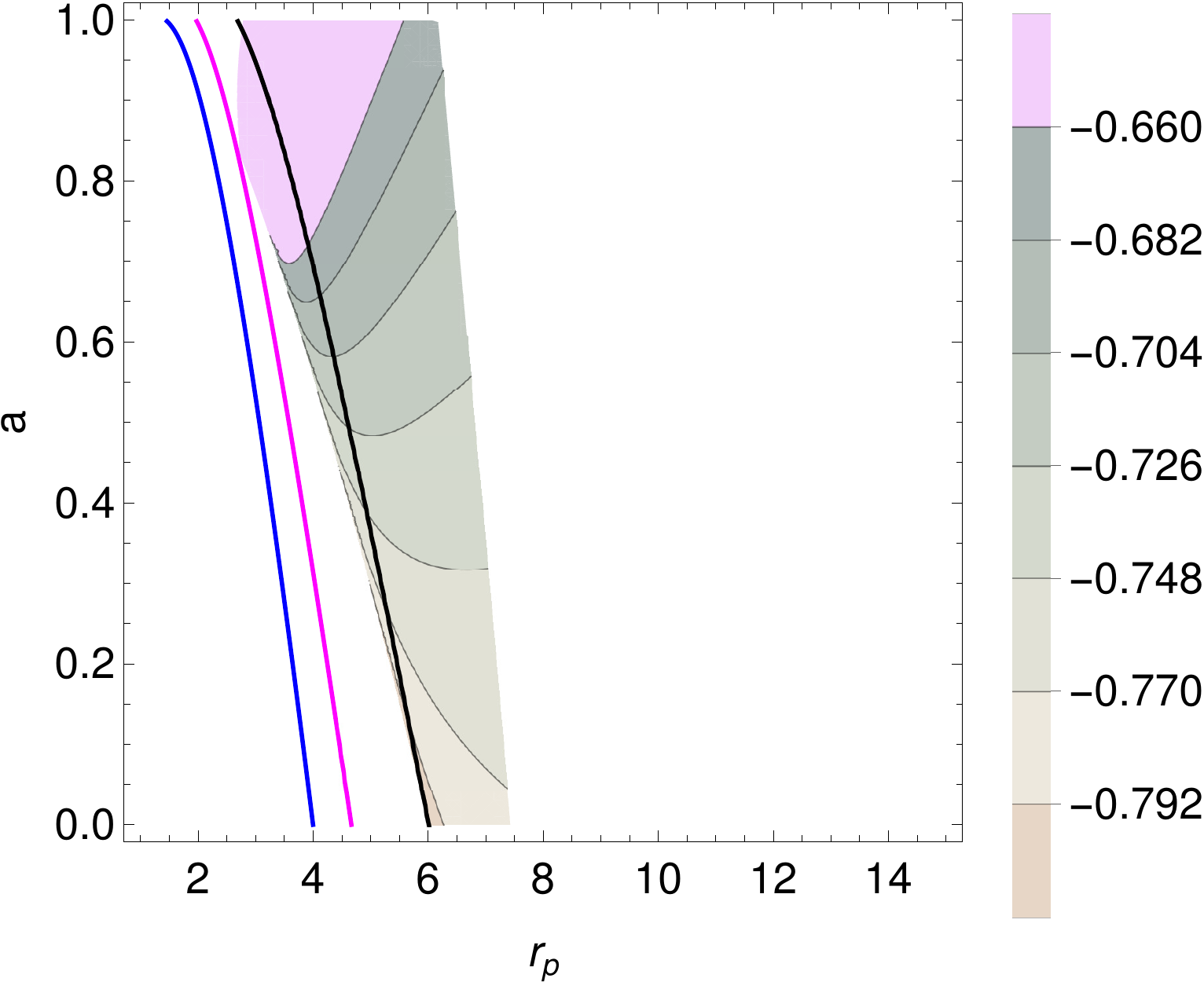}\label{nupere5Q4}}}
\caption{\label{nupereccentriccontrs} The contours of $\delta_{\rm pp}\left(e, r_p, a, Q \right)$ are shown in the ($r_p$, $a$) plane for eccentric orbits around a Kerr black hole, where the parameter combinations are (a) \{$e=0.25, Q=0$\}, (b) \{$e=0.25, Q=2$\}, (c) \{$e=0.25, Q=4$\}, (d) \{$e=0.5, Q=0$\}, (e) \{$e=0.5, Q=2$\}, and (f) \{$e=0.5, Q=4$\}.}
\end{figure}

\begin{figure}
\mbox{
 \subfigure[]{
  \hspace{-0.9cm}
\includegraphics[scale=0.4]{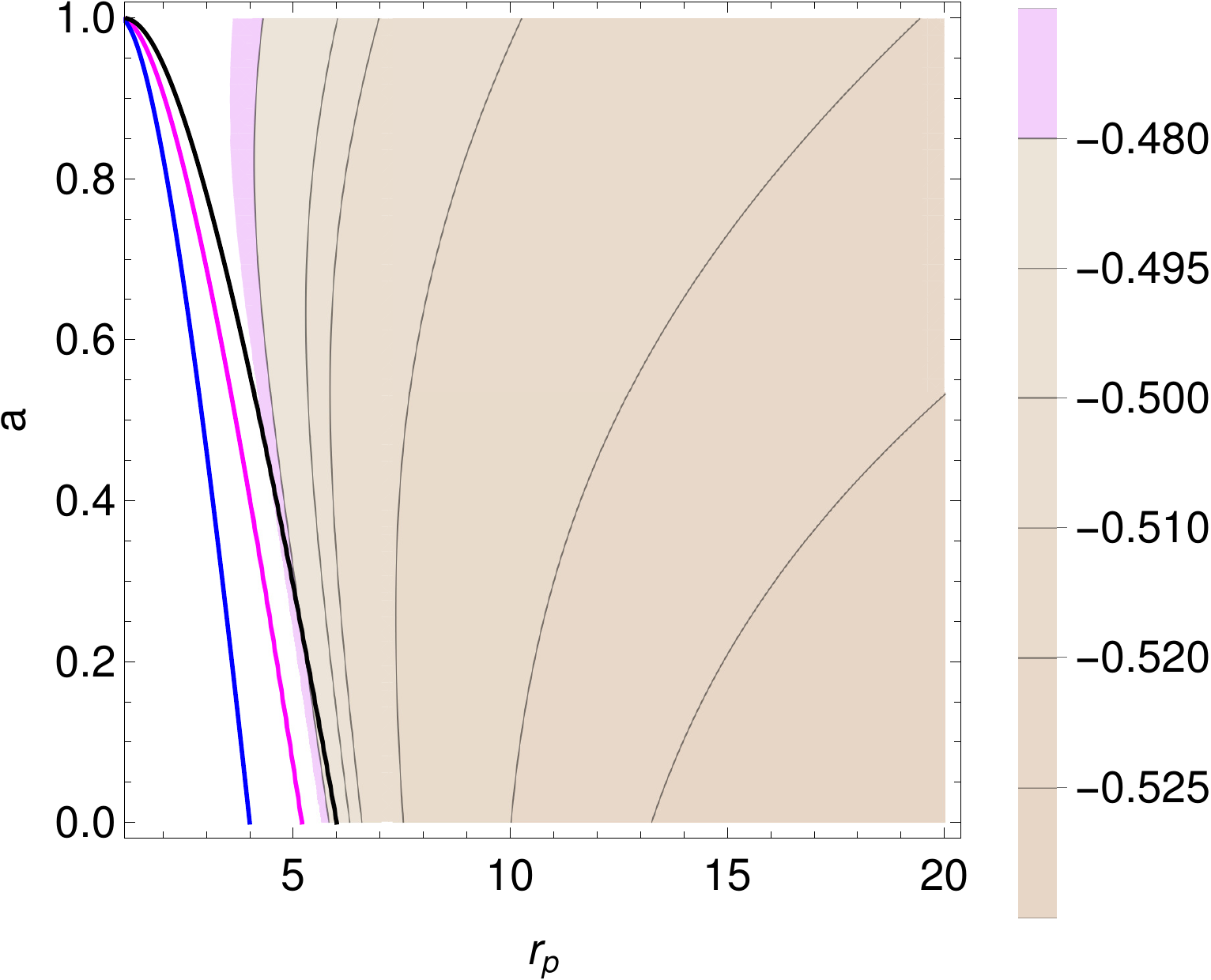}\label{nunode25Q0}}
\hspace{0.1cm}
 \subfigure[]{
\includegraphics[scale=0.4]{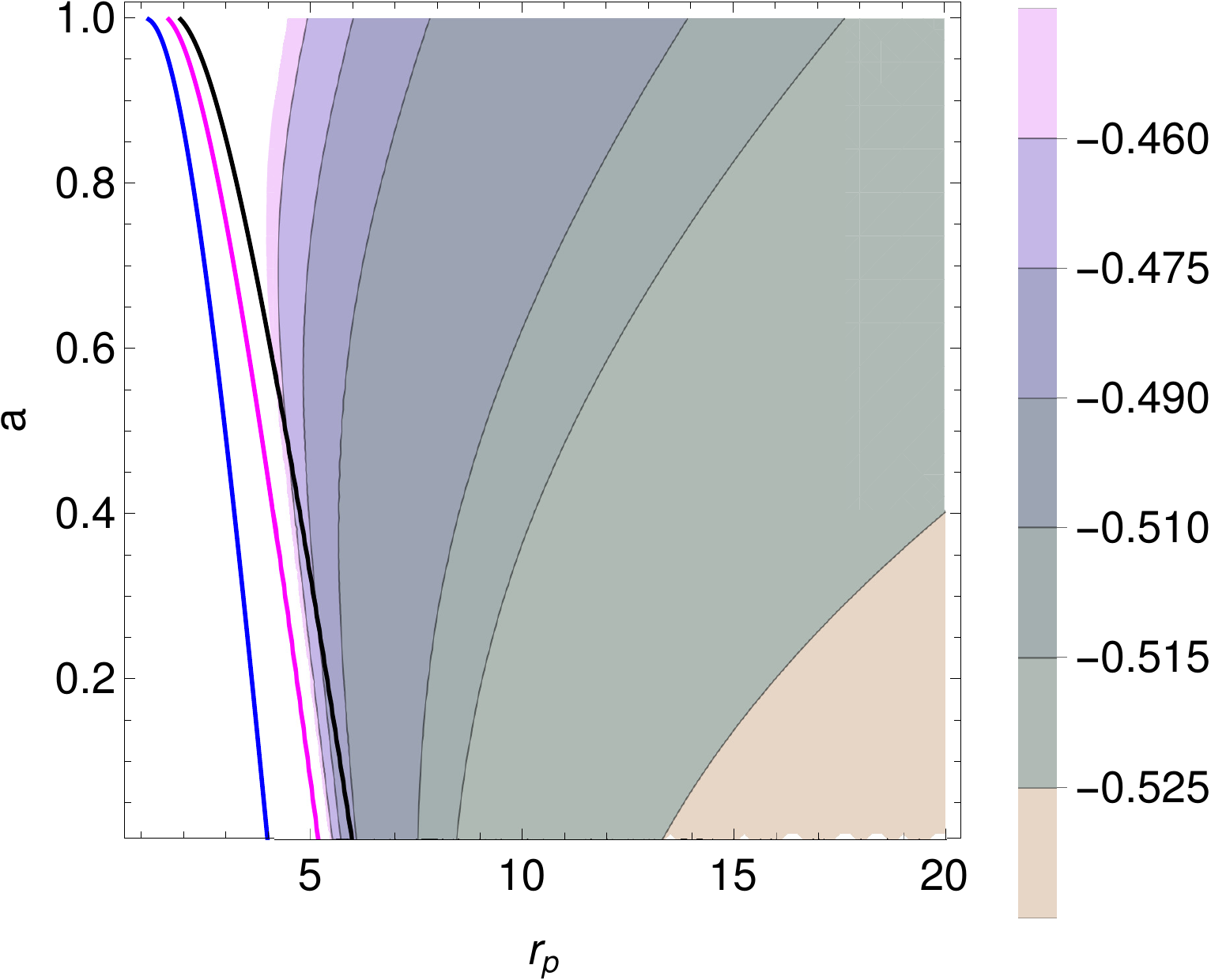}\label{nunode25Q2}}
\hspace{0.1cm}
 \subfigure[]{
\includegraphics[scale=0.4]{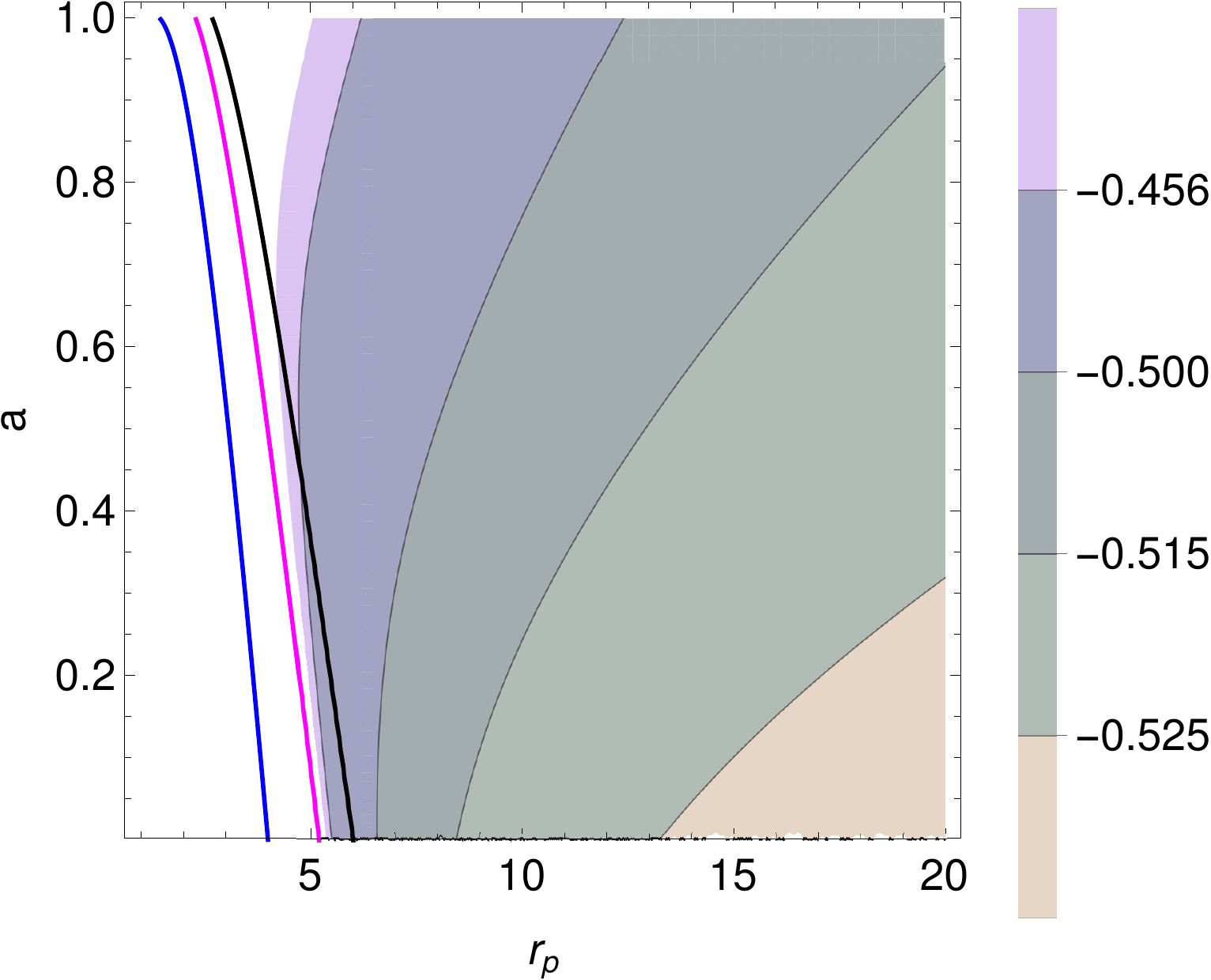}\label{nunode25Q4}}}
\mbox{
 \subfigure[]{
  \hspace{-0.9cm}
\includegraphics[scale=0.4]{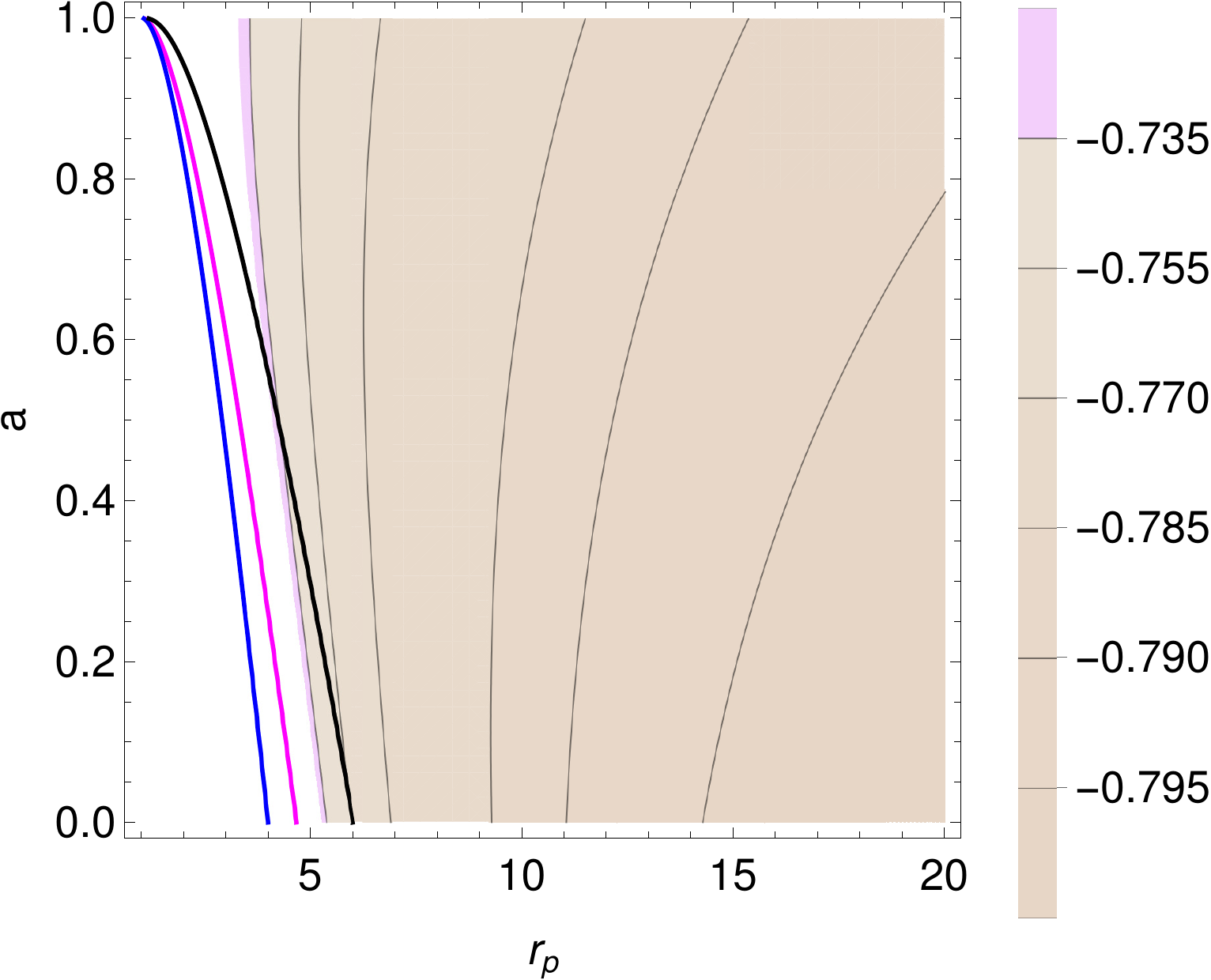}\label{nunode5Q0}}
\hspace{0.1cm}
\subfigure[]{
\includegraphics[scale=0.4]{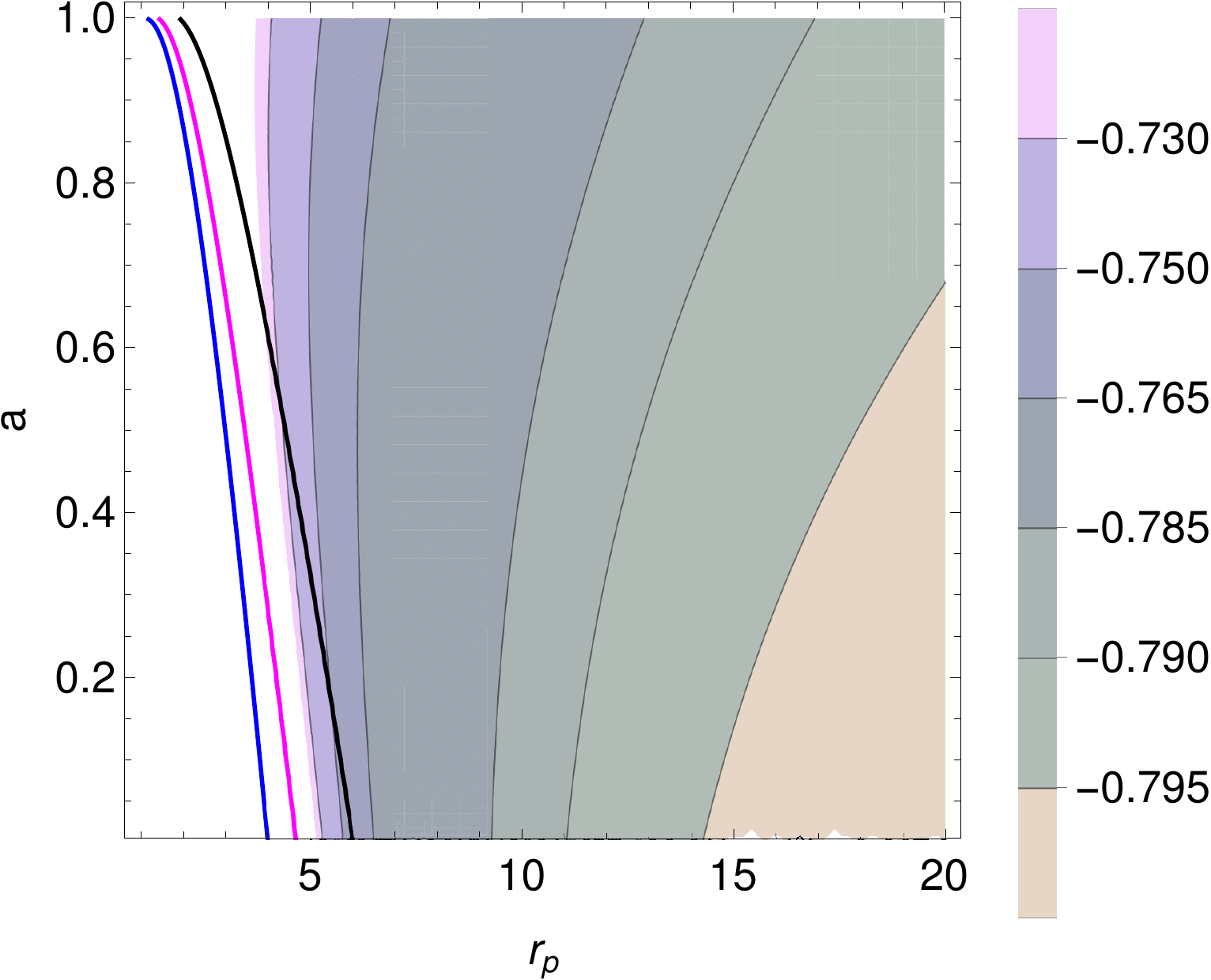}\label{nunode5Q2}}
\hspace{0.1cm}
\subfigure[]{
\includegraphics[scale=0.4]{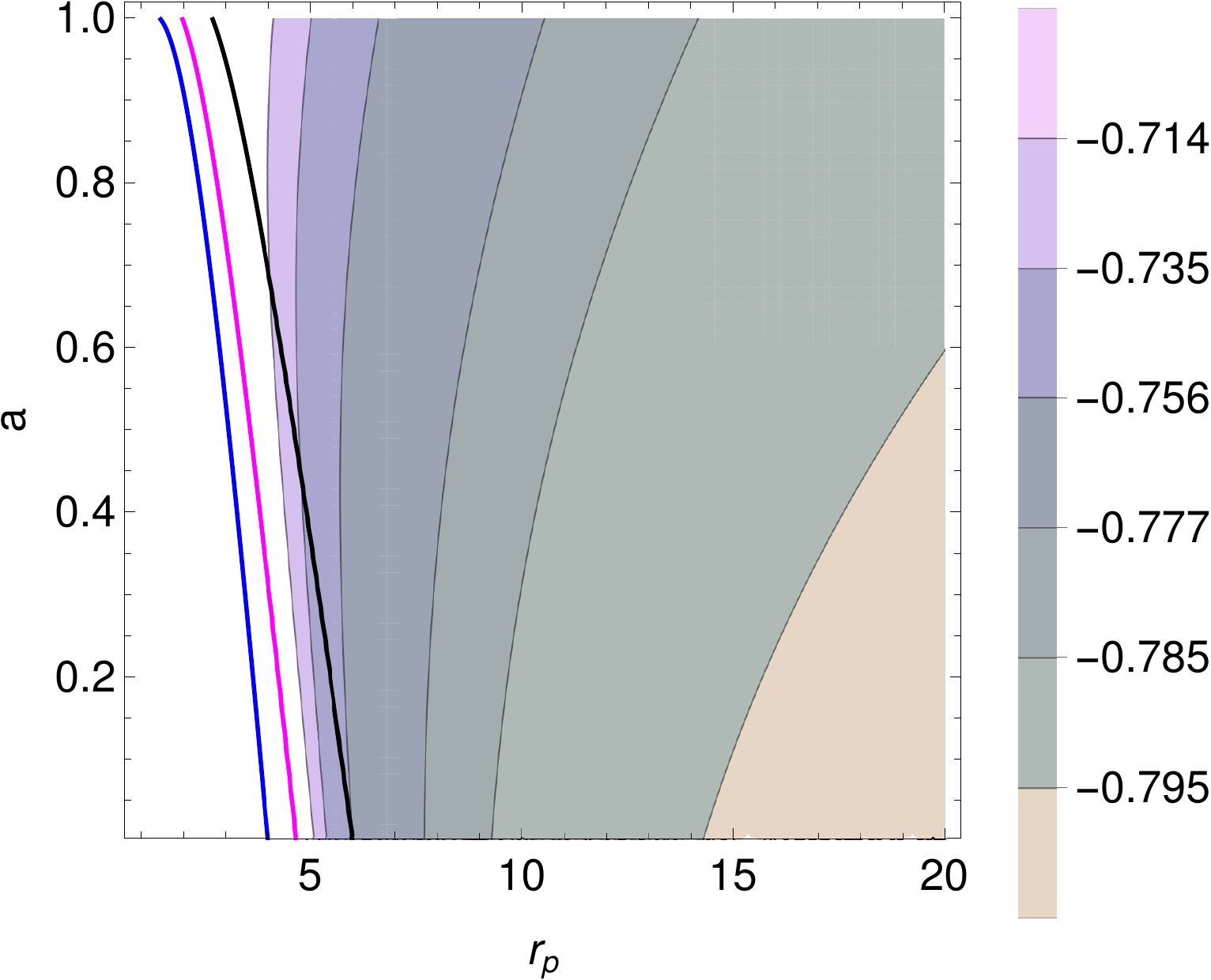}\label{nunode5Q4}}}\caption{\label{nunodeccentriccontrs} The contours of $\delta_{\rm np}\left(e, r_p, a, Q \right)$ are shown in the ($r_p$, $a$) plane for eccentric orbits around a Kerr black hole, where the parameter combinations are (a) \{$e=0.25, Q=0$\}, (b) \{$e=0.25, Q=2$\}, (c) \{$e=0.25, Q=4$\}, (d) \{$e=0.5, Q=0$\}, (e) \{$e=0.5, Q=2$\}, and (f) \{$e=0.5, Q=4$\}.} 
\end{figure}

\begin{figure}
\mbox{
 \subfigure[]{
  \hspace{-0.6cm}
\includegraphics[scale=0.38]{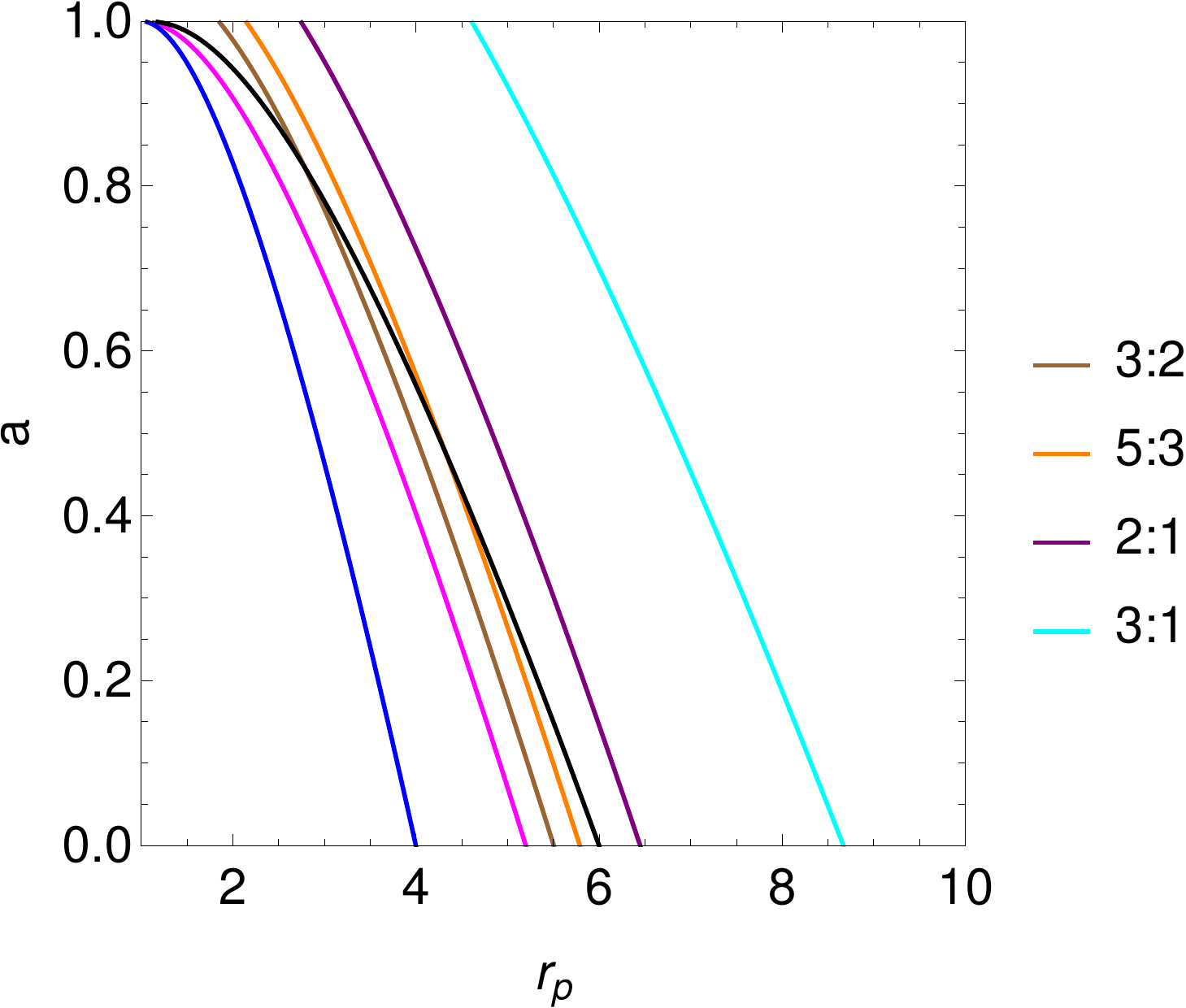}\label{ratiocntre25}}
\subfigure[]{
  \hspace{0.5cm}
\includegraphics[scale=0.52]{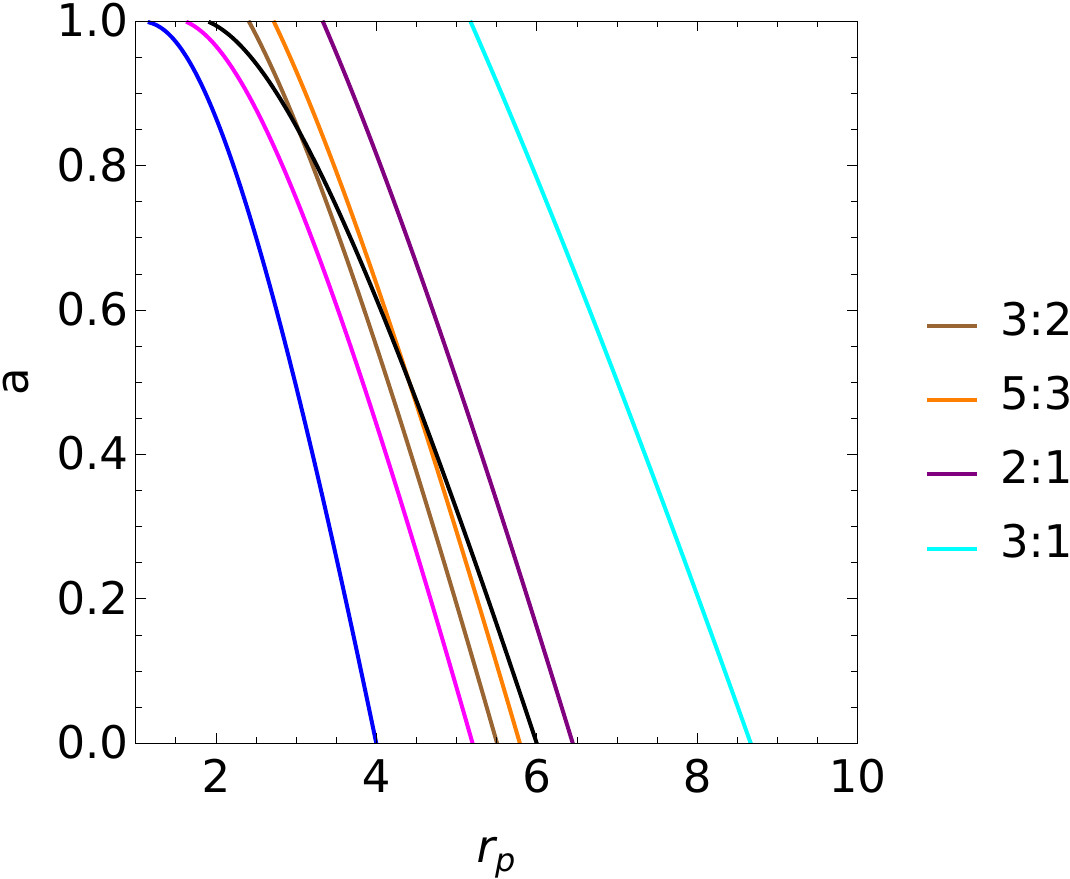}\label{ratiocntre25Q2}}
\subfigure[]{
  \hspace{0.5cm}
\includegraphics[scale=0.52]{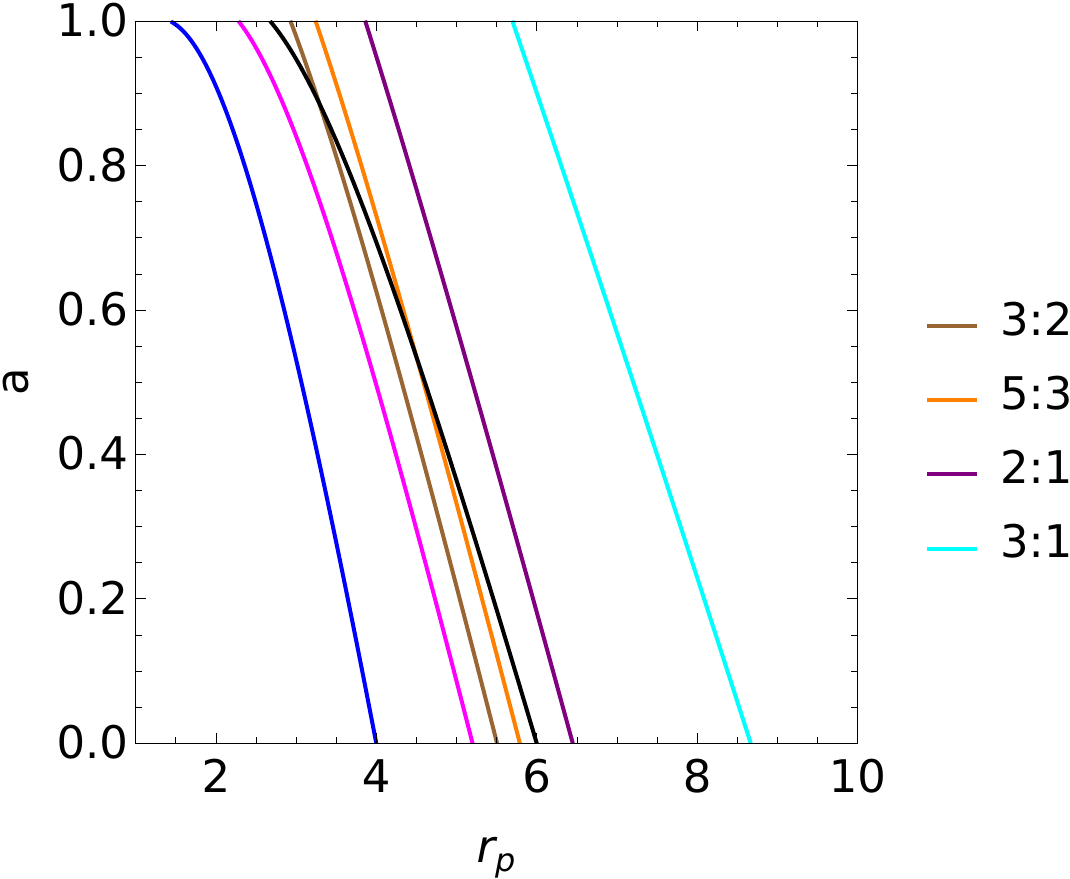}\label{ratiocntre25Q4}}}
\mbox{
 \subfigure[]{
  \hspace{-0.6cm}
\includegraphics[scale=0.38]{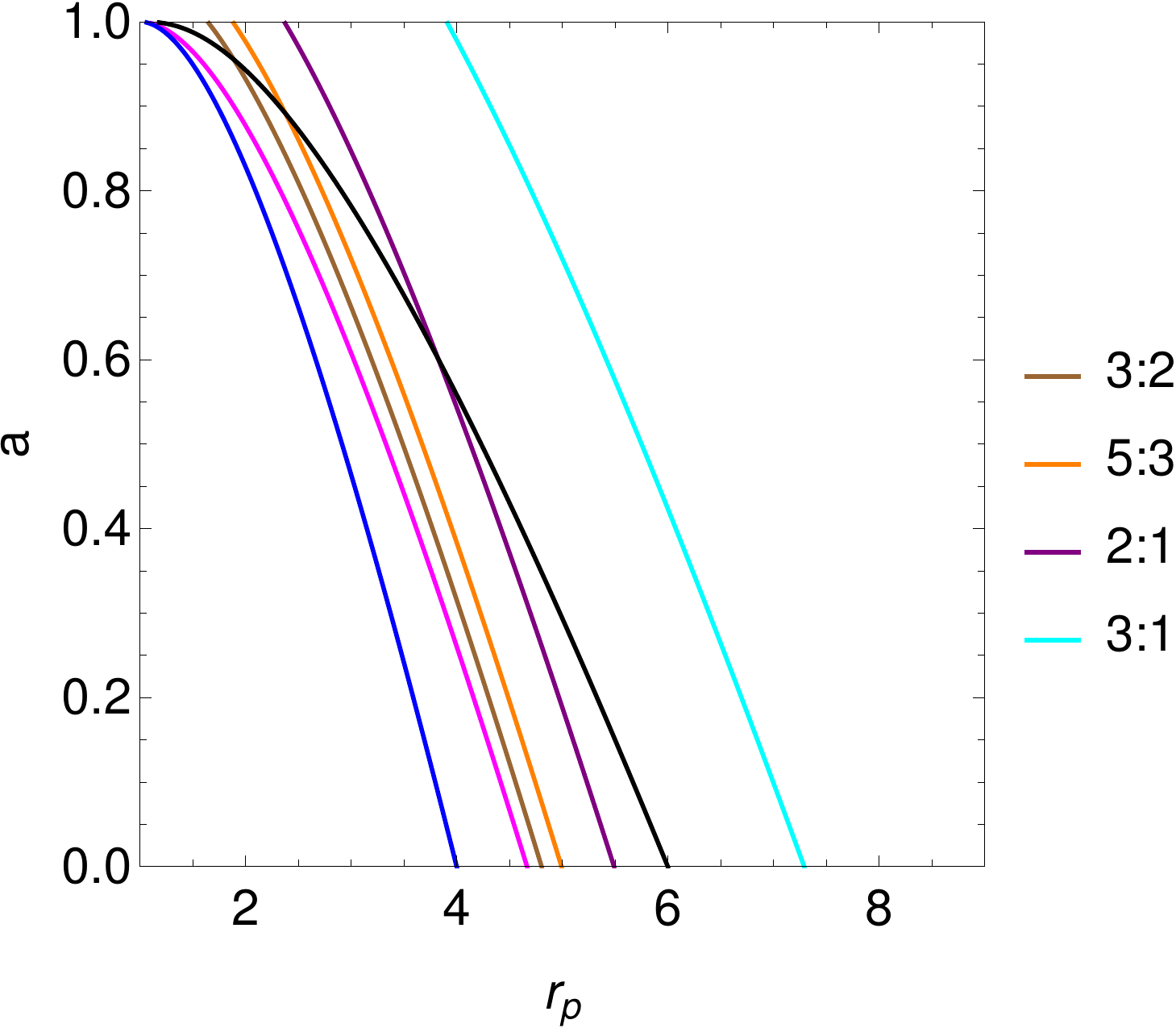}\label{ratiocntre5}}
\subfigure[]{
  \hspace{0.5cm}
\includegraphics[scale=0.52]{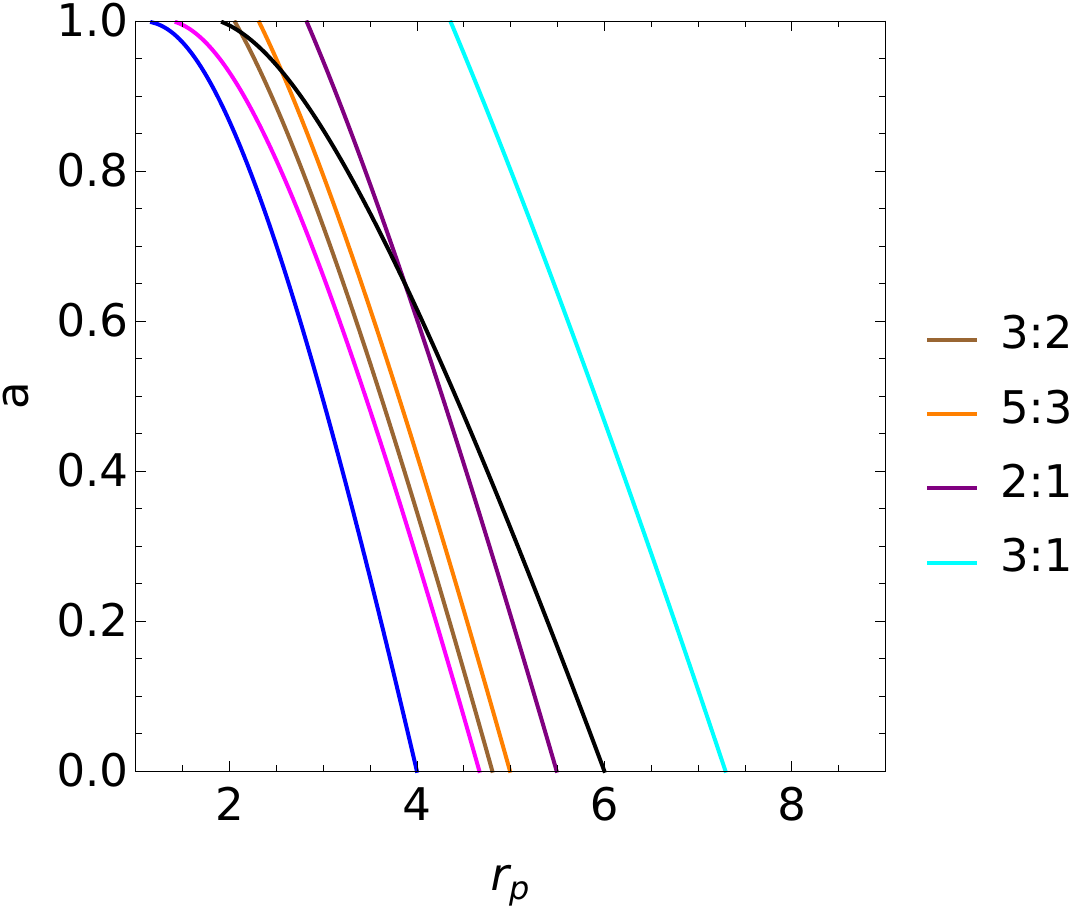}\label{ratiocntre5Q2}}
\subfigure[]{
  \hspace{0.5cm}
\includegraphics[scale=0.52]{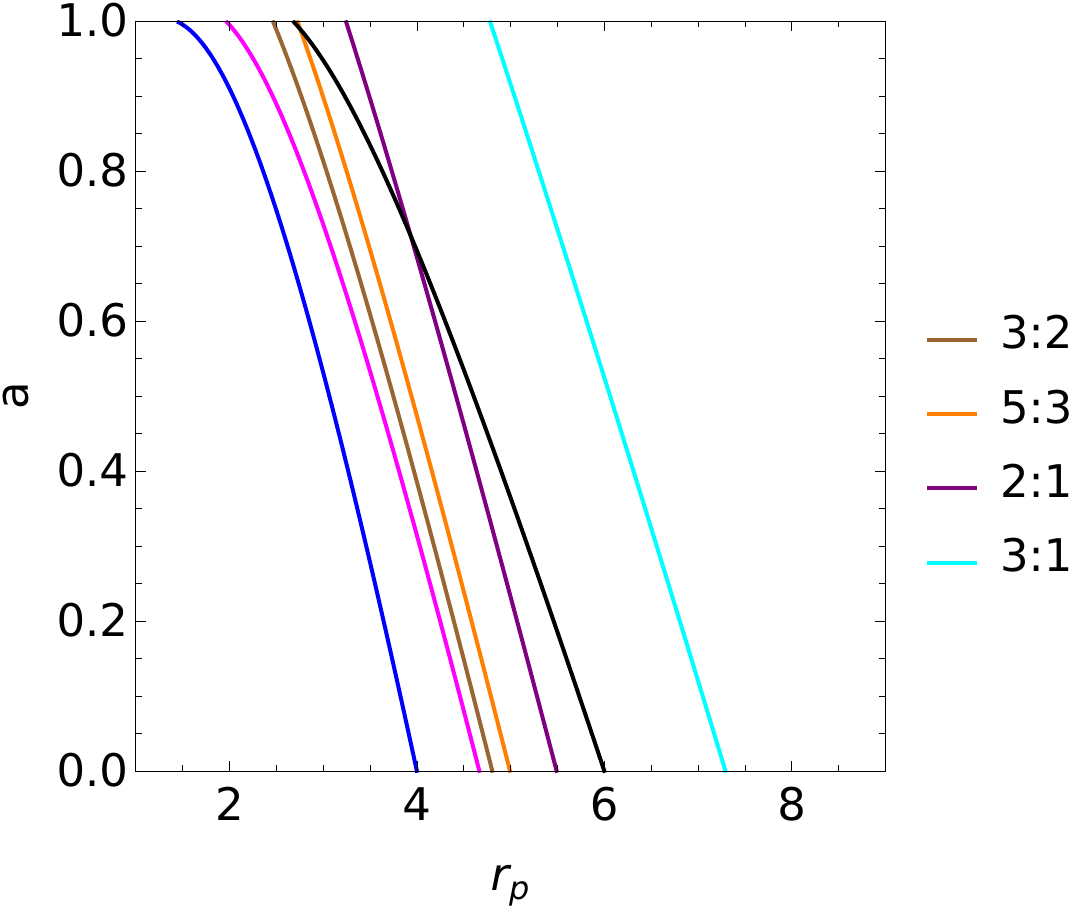}\label{ratiocntre5Q4}}}
\caption{\label{ratiocntrseccentric}The HFQPO frequency ratio, $\bar{\nu}_{\phi}/ \bar{\nu}_{\rm pp}$, contours are shown for eccentric orbits around a Kerr black hole in the ($r_p$, $a$) plane, assuming the GRPM, where the parameter combinations are (a) \{$e=0.25$, $Q=0$\}, (b) \{$e=0.25$, $Q=2$\}, (c) \{$e=0.25$, $Q=4$\}, (d) \{$e=0.5$, $Q=0$\}, (e) \{$e=0.5$, $Q=2$\}, and (f) \{$e=0.5$, $Q=4$\}.}
\end{figure}

Now, we use these frequency formulae, Equations (\ref{nuphi}$-$\ref{nutheta}), to deduce the suitable parameter range of parameters \{$e$, $r_p$, $a$, $Q$\} for $eQ$ and Equations (\ref{eqnuphi}$-$\ref{eqnutheta}) for $e0$ trajectories to find \{$e$, $r_p$, $a$\} to retrodict the observed range of QPOs in BHXRBs, which is provided in Table \ref{QPOfreqtable}. In Figures \ref{nuphieccentriccontrs}$-$\ref{nunodeccentriccontrs}, we have shown the variation of the quantities
\begin{subequations}
\begin{eqnarray}
 \delta_{\phi}\left(e, r_p, a, Q \right) &&=\dfrac{\left[\bar{\nu}_{\phi}\left(e, r_p, a, Q\right)-\bar{\nu}_{\phi}\left(e=0, r_p, a, Q=0\right)\right]}{\bar{\nu}_{\phi}\left(e=0, r_p, a, Q=0\right)}, \\
\delta_{\rm pp}\left(e,r_p, a, Q \right) &&=\dfrac{\left[\bar{\nu}_{\rm pp}\left(e, r_p, a, Q\right) -\bar{\nu}_{\rm pp}\left(e=0, r_p, a, Q=0\right)\right]}{\bar{\nu}_{\rm pp}\left(e=0, r_p, a, Q=0\right)}, \\
 \delta_{\rm np}\left(e,r_p, a, Q \right) &&=\dfrac{\left[\bar{\nu}_{\rm np}\left(e, r_p, a, Q\right)- \bar{\nu}_{\rm np}\left(e=0, r_p, a, Q=0\right)\right]}{\bar{\nu}_{\rm np}\left(e=0, r_p, a, Q=0\right)},
\end{eqnarray}
\label{deltasecc}
\end{subequations}
 in the ($r_p$, $a$) plane for combinations of $e=$\{0.25, 0.5\} and $Q=$\{0, 2, 4\}. These quantities provide a fractional deviation between frequencies of general eccentric orbits and circular orbits for the same spin and periastron radius. For this comparison, we have calculated the frequency corresponding to a circular orbit at the same radius, $r_p=r$, for a fixed value of parameter $a$. Hence, the deviation, $\delta$, between frequencies defined in this manner is dominated by the parameters $e$ and $Q$. Also, these deviations are shown only in the region where $\bar{\nu}_{\phi}\left(e, r_p, a, Q\right)$, $\bar{\nu}_{\rm pp}\left(e, r_p, a, Q\right)$, and $\bar{\nu}_{\rm np}\left(e, r_p, a, Q\right)$ are in the range of QPO frequencies allowed by the observations, as provided in Table \ref{QPOfreqtable}. Hence, these plots together give us the information of deviation of frequencies from circularity, as the $e$ and $Q$ parameters are varied, along with information on the range of $\left(e, r_p, a, Q\right)$ for general eccentric orbits allowed by the observed range of QPO frequency. The 3:2 and 5:3 ratios of the simultaneous HFQPOs, seen in a few BHXRBs, are also a remarkable phenomenon that we need to fathom; for example, 300 Hz and 450 Hz HFQPOs were seen in GROJ 1655-40 \citep{RemillardMorgan1999, Strohmayer2001a}, and 240 Hz and 160 Hz HFQPOs in H1743-322 \citep{Homan2005, Remillard2006}. Assuming the GRPM, this ratio is given by ${\nu}_{\phi}/{\nu}_{pp}=\bar{\nu}_{\phi}/\bar{\nu}_{\rm pp}$, which is a dimensionless quantity. The contours of this ratio are shown in Figure \ref{ratiocntrseccentric} for the six combinations from the set $e=\{0.2, \ 0.4\}$, $Q=\{0, \ 2, \ 4\}$. The blue contours  in Figures \ref{nuphieccentriccontrs}$-$\ref{ratiocntrseccentric} represent the ISSO radius, and black contours represent the MBSO radius as also indicated in Figure \ref{radii}, whereas the magenta color contours represent the separatrix orbits, given by the equality in Equation \eqref{boundcondition}, defining the innermost limit for $r_p$ of an eccentric orbit with a given $e$.
 
A summary of the results is given below: 
\begin{enumerate}
 \setlength\itemsep{-0.5em}
 \item A novel and reduced form of $\bar{\nu}_{\theta}\left(e, r_p, a\right)$ for $e0$ trajectories, given by Equation \eqref{eqnutheta}, is derived in Appendix \ref{nuthetaderivation}.
\item Assuming the GRPM, (non)equatorial eccentric trajectories with small to moderate eccentricities, $e\lesssim 0.5$, with $Q\sim 0 -4$ also generate the expected range of QPO frequencies, \{$\bar{\nu}_{\phi}$, $\bar{\nu}_{\rm pp}$, $\bar{\nu}_{\rm np}$\}, in BHXRBs, as shown in Table \ref{QPOfreqtable}. We have not taken very high values for the $Q$ parameter, as the particle oscillation is expected to be close to the equatorial plane in typical BHXRB scenarios. 
\item The effective $r_p$ ranges that produce the required QPO frequency ranges are $\Delta r_p\sim 2-15 $ for $\bar{\nu}_{\phi}$, $\Delta r_p\sim 2-10 $ for $\bar{\nu}_{\rm pp}$, and $\Delta r_p\sim 4-20 $ for $\bar{\nu}_{\rm np}$, where $a$ varies from 0 to 1. While these $\Delta r_p$ values are strongly dependent on $e$, they are only weakly dependent on the $Q$ parameter. The frequency $\bar{\nu}_{\rm np}$ (see Figure \ref{nunodeccentriccontrs}) increases with $a$, which implies that we expect to find high type C LFQPO values (nearly $\bar{\nu}_{\rm np}\sim 0.001$) for the black holes with high spin. 
\item As $e$ increases, the allowed region shifts close to the black hole. In other words, we expect (non)equatorial eccentric orbits close to the black hole to create the allowed frequency range, whereas circular orbits at comparatively larger radius cater to the same frequency range (see Figure \ref{nucirccontrs}). This is consistent with the finding that the GRPM favors slightly eccentric and strongly relativistic orbits. We also see that as $e$ increases, the frequencies deviate and decrease from corresponding circular orbit frequencies; for example, $\bar{\nu}_{\phi}$ decreases by 30\% for $e=0.25$ to 60\% for $e=0.5$ (see Figure \ref{nuphieccentriccontrs}), $\bar{\nu}_{\rm pp}$ decreases by 40\% for $e=0.25$ to 79\% for $e=0.5$ (see Figure \ref{nupereccentriccontrs}), and $\bar{\nu}_{\rm np}$ decreases by 40\% for $e=0.25$ to 80\% for $e=0.5$ (see Figure \ref{nunodeccentriccontrs}). 
\item The dependence of these frequencies on $Q$ is very weak. Although the change is comparatively small, we see that these frequencies increase with $Q$. For example, the maximum increase in $\bar{\nu}_{\phi}$ is $\sim$3\% (see Figure \ref{nuphieccentriccontrs}) and $\sim$10\% for $\bar{\nu}_{\rm pp}$ (see Figure \ref{nupereccentriccontrs}), whereas it is $\sim$3\% for $\bar{\nu}_{\rm np}$ (see Figure \ref{nunodeccentriccontrs}) as $Q$ changes from $0$ to $4$. Even for high $Q$ values, say $Q \sim 10$, the change in frequencies is of the same order. 
\item Expectedly, the associated frequencies increase as the $r_p$ of a trajectory decreases for a given \{$e, \ a ,\ Q$\}, where $r_p$ of an eccentric trajectory is limited by the corresponding separatrix orbit, having the same \{$e$, $a$, $Q$\} values.
\item As shown in Figure \ref{ratiocntrseccentric}, the 3:2 or 5:3 ratios of HFQPOs originate in the region very close to the separatrix orbits, which is between MBSO and ISSO radii corresponding to typically $\Delta r_p\sim 2-6$; this range is dependent on $a$ since $r_p$ decreases as $a$ increases. The frequency ratio contours shift close to the black hole as $e$ is increased, whereas these contours move toward large $r_p$ as $Q$ is increased. This indicates that nonequatorial orbits show a 3:2 or 5:3 ratio of HFQPO frequencies farther away from the black hole than the equatorial orbits, and eccentric orbits have such ratios comparatively closer to the black hole than the circular orbits. Therefore, $eQ$ and $00$ orbits close to the black hole can account for these ratios, as $e$ and $Q$ have canceling effects.
\end{enumerate}

\subsection{Spherical Orbits: $Q0$}
\label{sphericalmotivation}
\begin{figure}
\mbox{
 \subfigure[]{
  \hspace{-0.8cm}
\includegraphics[scale=0.52]{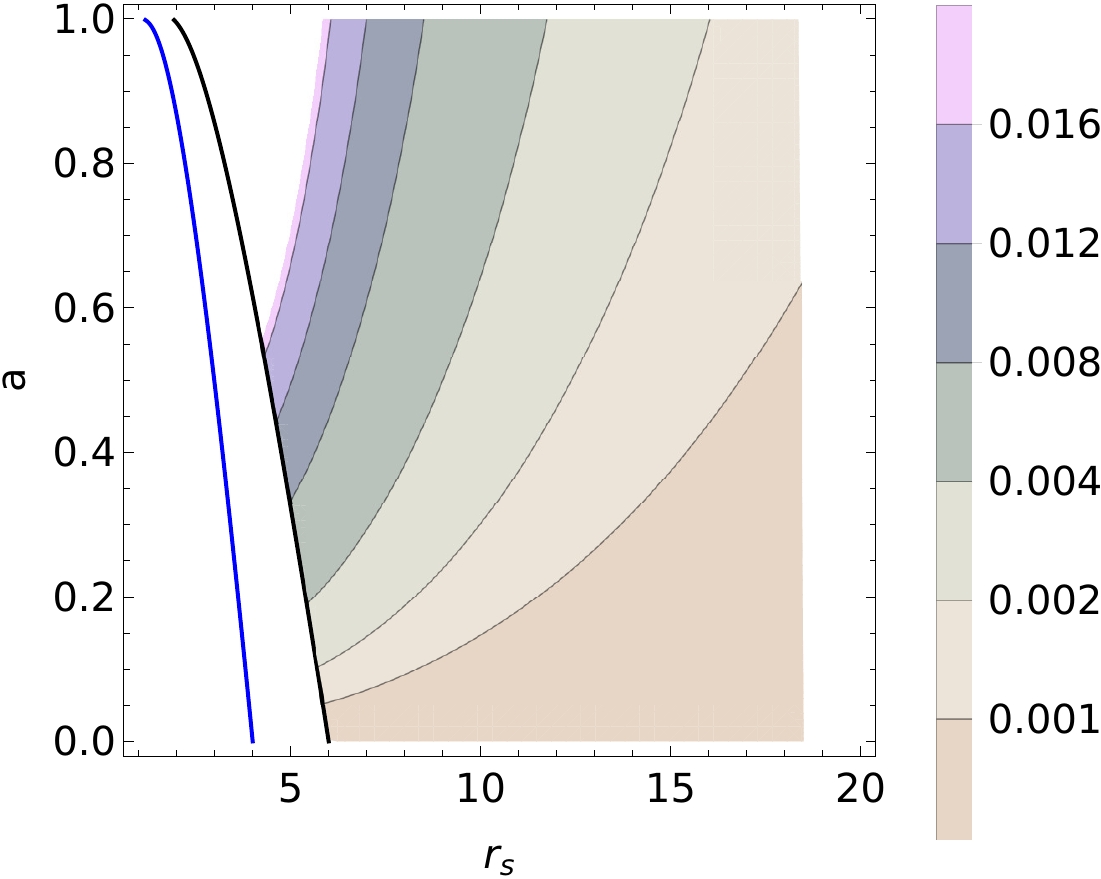}\label{nuphie0Q2}}
\hspace{0.6cm}
 \subfigure[]{
\includegraphics[scale=0.52]{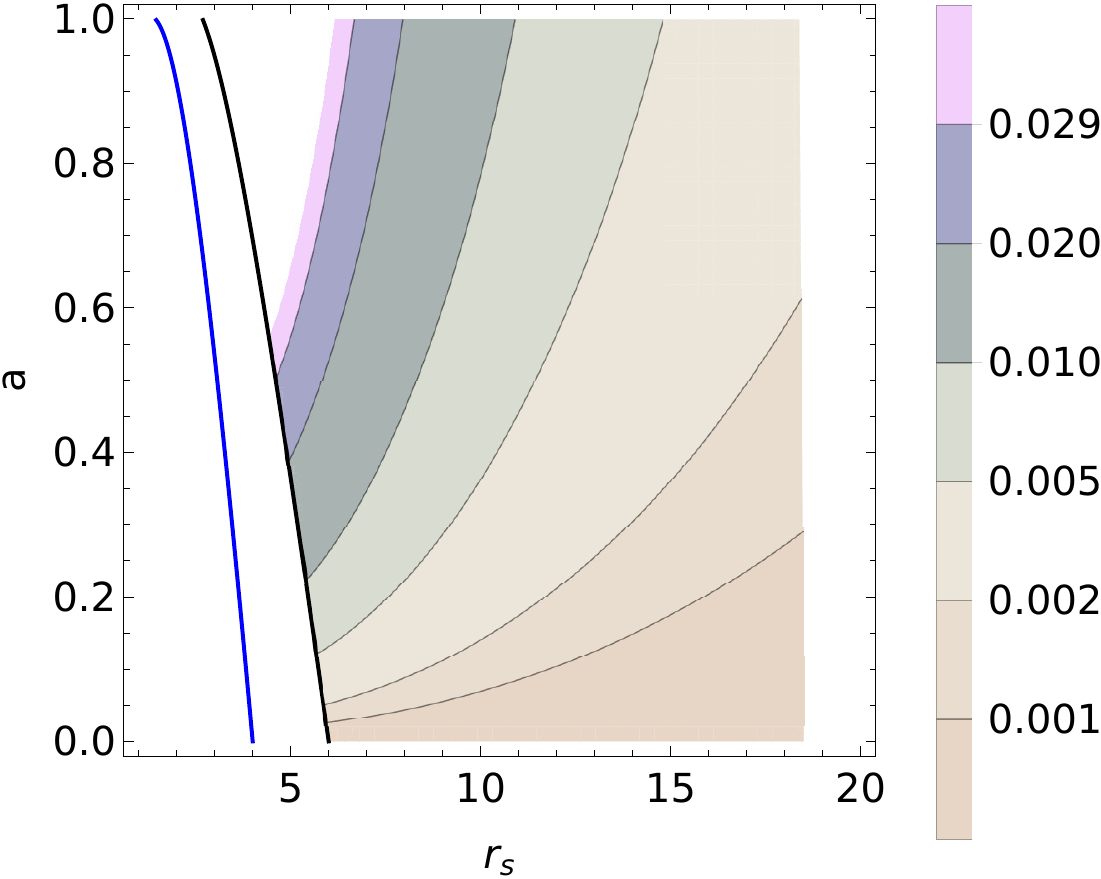}\label{nuphie0Q4}}
\hspace{0.6cm}
 \subfigure[]{
\includegraphics[scale=0.52]{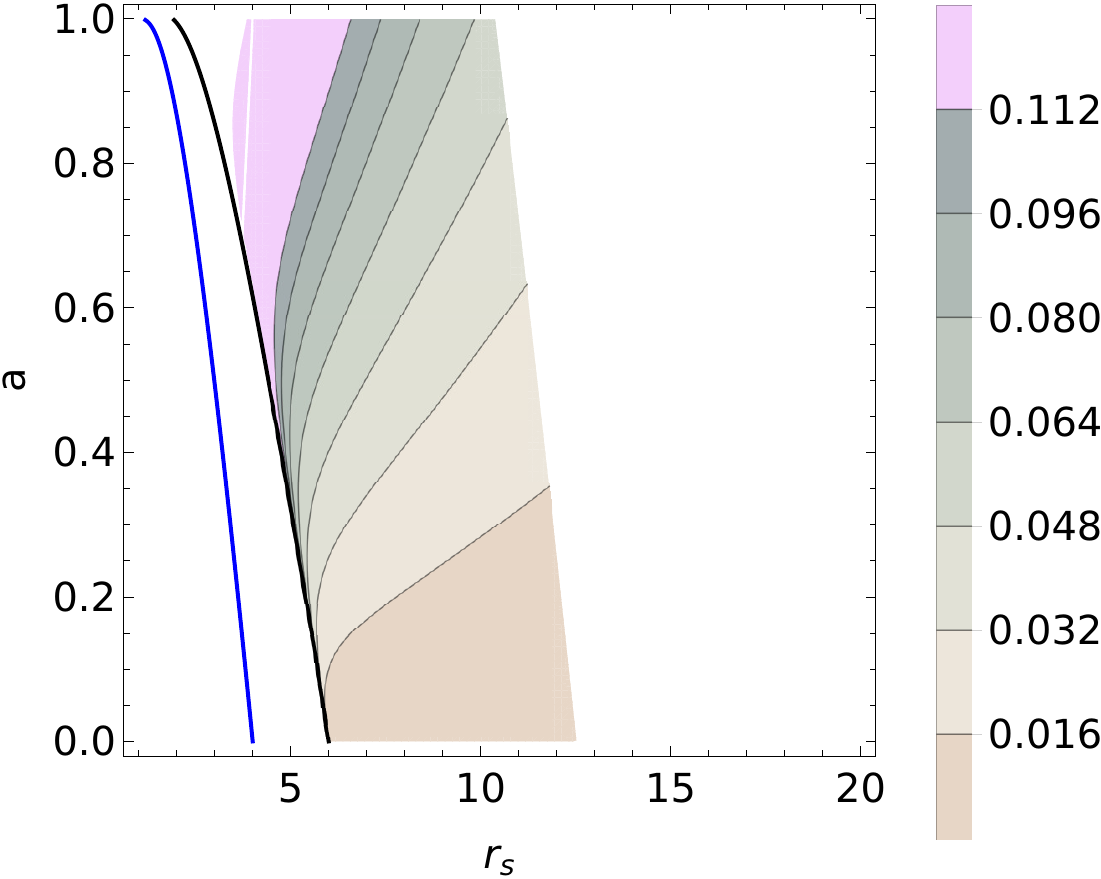}\label{nupere0Q2}}}
\mbox{
 \subfigure[]{
  \hspace{-0.8cm}
\includegraphics[scale=0.52]{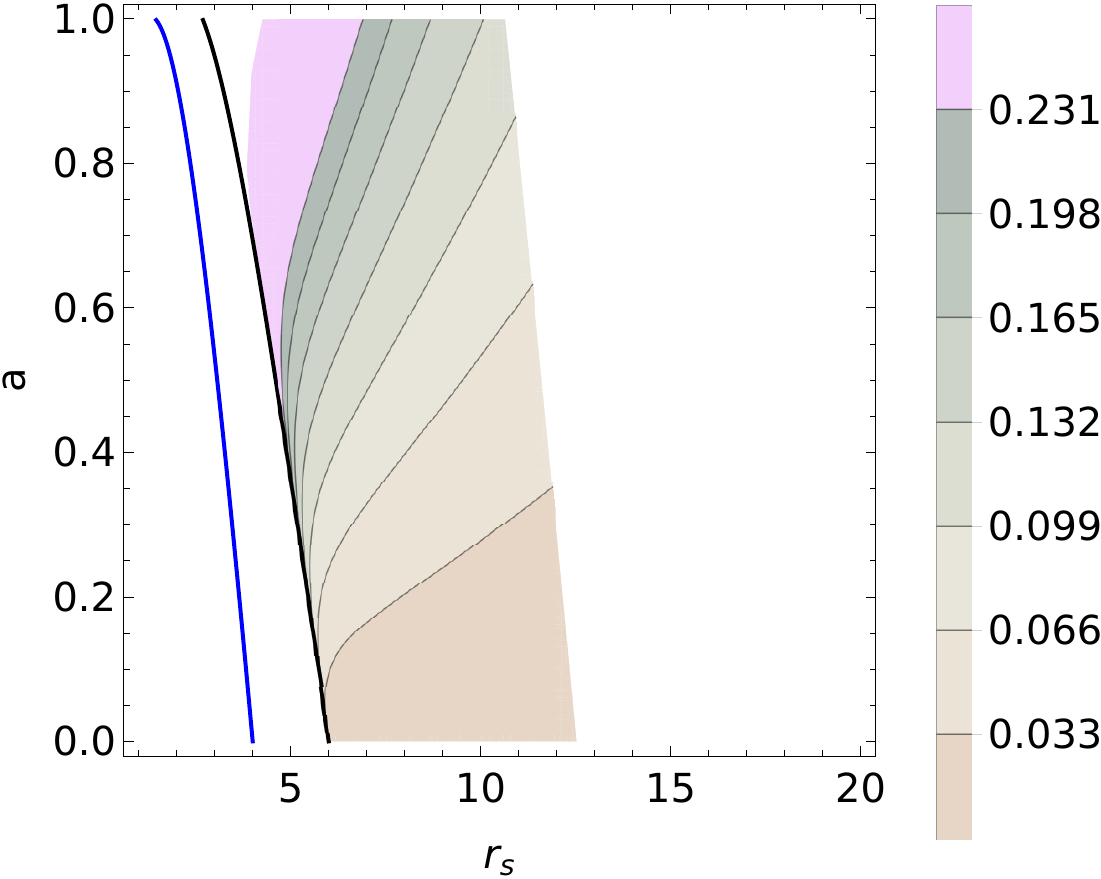}\label{nupere0Q4}}
\hspace{0.6cm}
 \subfigure[]{
\includegraphics[scale=0.52]{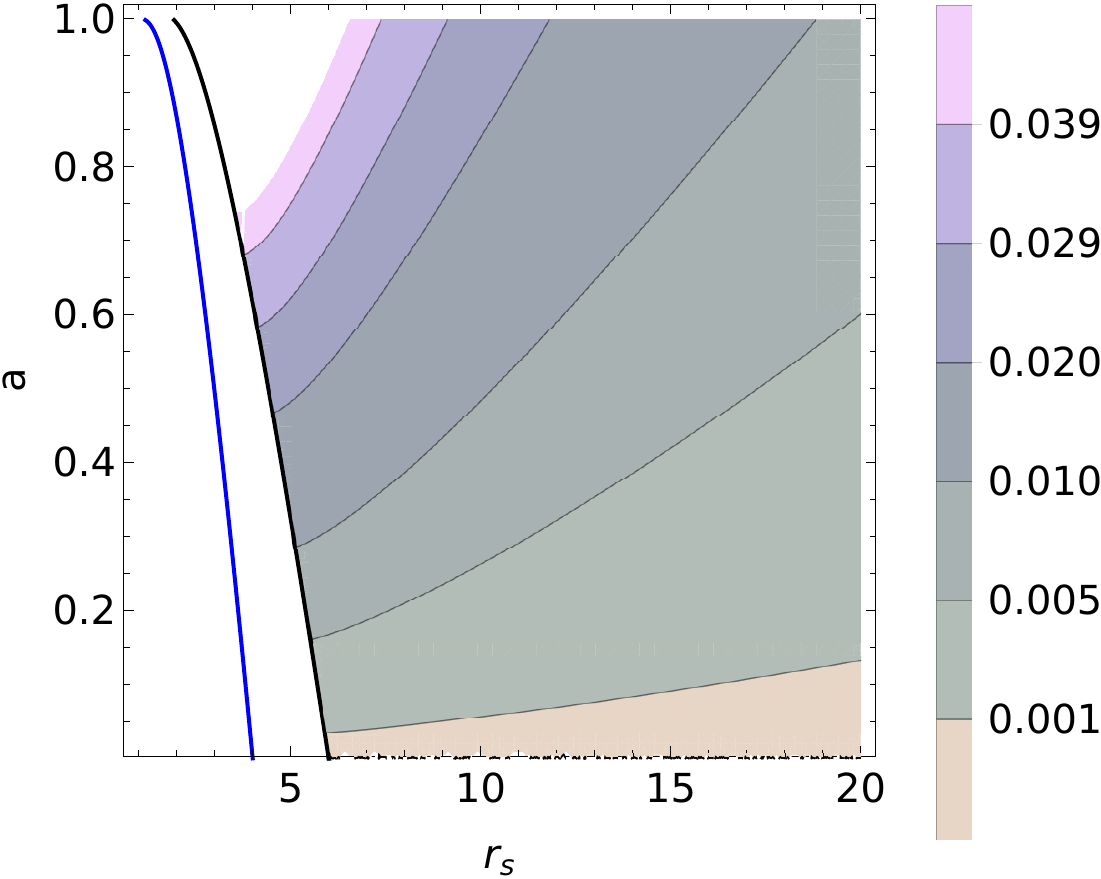}\label{nunode0Q2}}
\hspace{0.6cm}
 \subfigure[]{
\includegraphics[scale=0.52]{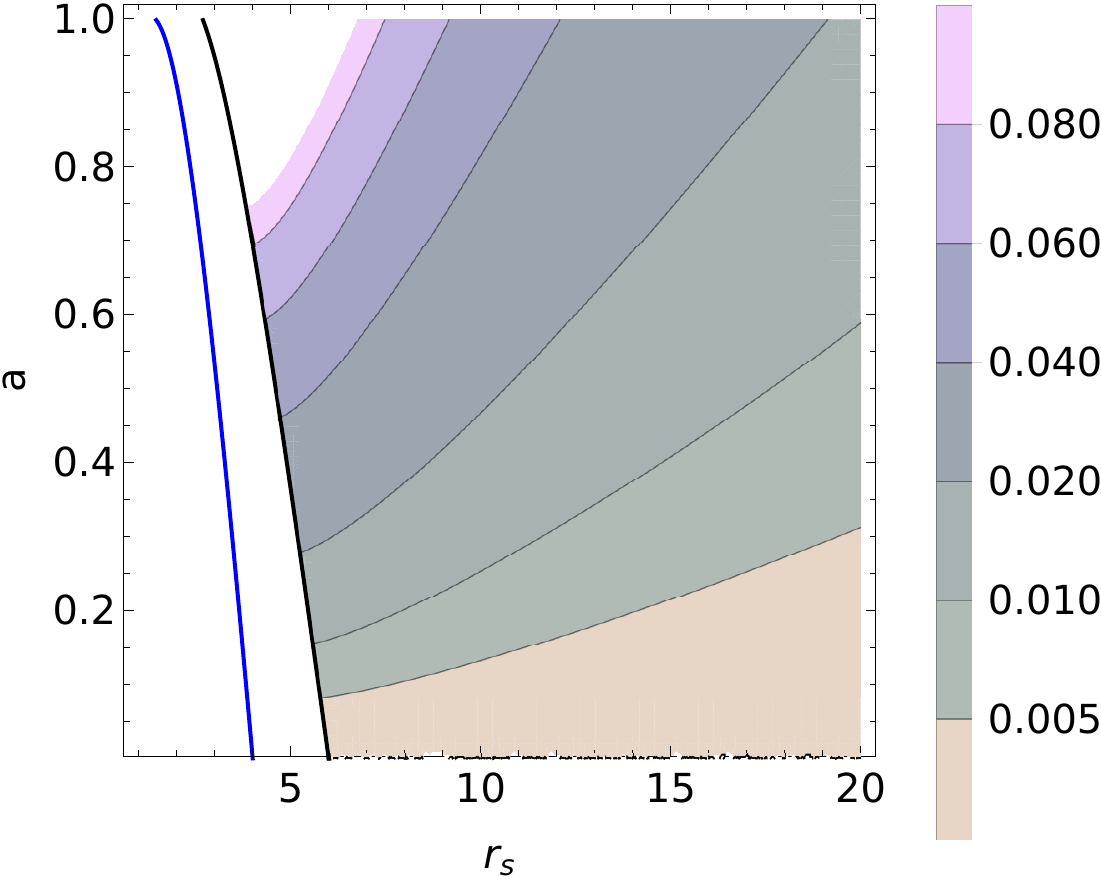}\label{nunode0Q4}}}
\caption{\label{freqcntrspherical}The contours of $\delta_{\phi}\left(r_s, a, Q \right)$ are shown for (a) $Q$=2, (b) $Q$=4; $\delta_{\rm pp}\left(r_s, a, Q \right)$ for (c) $Q$=2, (d) $Q$=4; and $\delta_{\rm np}\left(r_s, a, Q \right)$ for (e) $Q$=2, (f) $Q$=4 in the ($r_s$, $a$) plane for the spherical orbits around a Kerr black hole.}
\end{figure}

\begin{figure}
\mbox{
\hspace{1.5cm}
 \subfigure[]{
\includegraphics[scale=0.41]{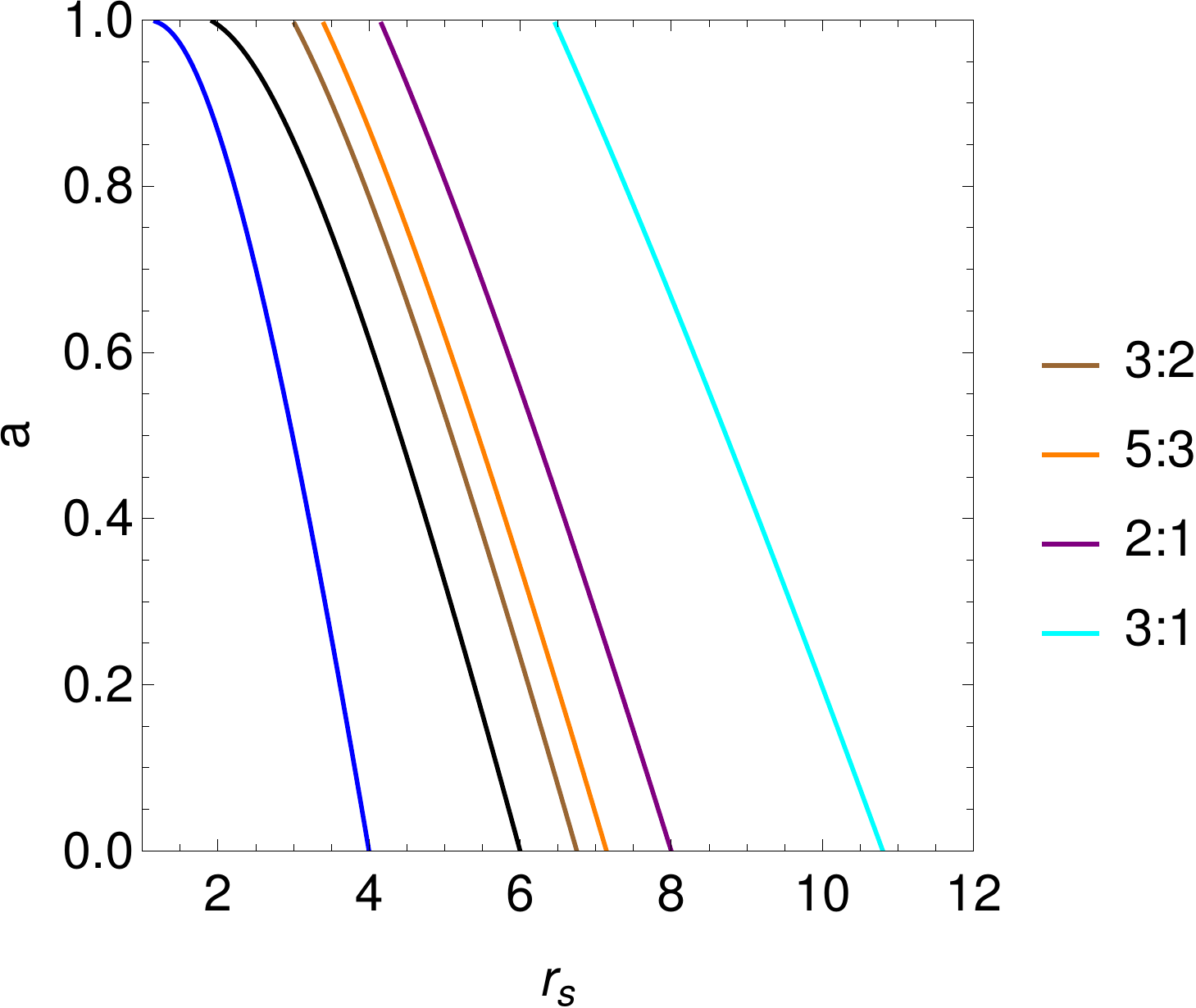}\label{ratiocntre0Q2}}
\hspace{2.0cm}
 \subfigure[]{
\includegraphics[scale=0.41]{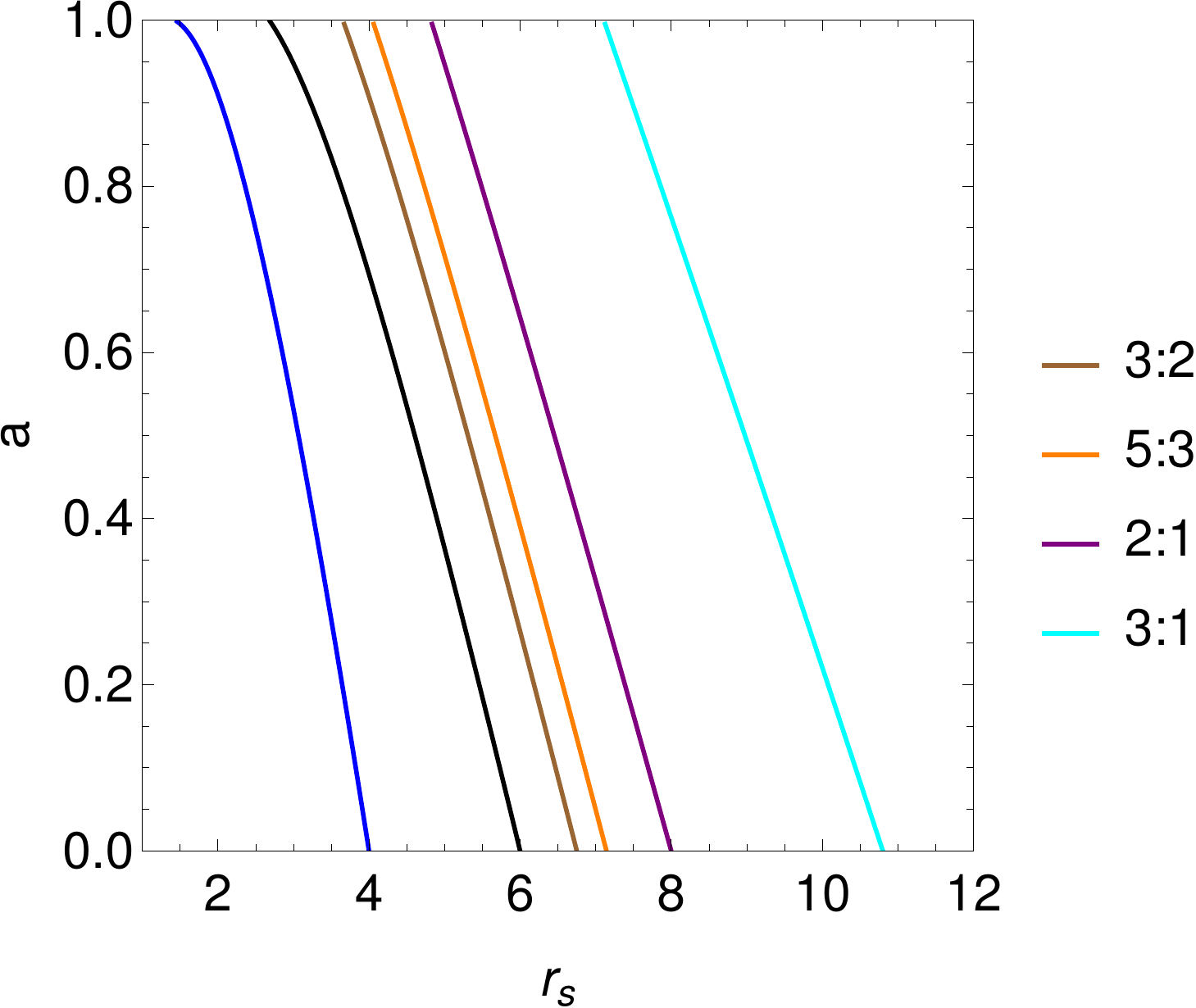}\label{ratiocntre0Q4}}}
\caption{\label{ratiocntrse0}The HFQPO frequency ratio, $\bar{\nu}_{\phi}/\bar{\nu}_{\rm pp}$, contours are shown for the spherical orbits around a Kerr black hole in the ($r_s$, $a$) plane, assuming the GRPM for (a) $Q=2$ and (b) $Q=4$.}
\end{figure}

Similar to the $eQ$ trajectories, the spherical orbits ($Q0$) are also specific to the rotating black holes. They are the orbits with a constant radius, $r_s$, where the orbital plane precesses on a sphere about the spin axis of the black hole. Similar to the ISCO and MBCO radii for circular orbits, ISSO and MBSO radii exist for the spherical orbits that are functions of the $a$ and $Q$ parameters \citep{RMCQG2019,RMarxiv2019}. We explore the ranges of parameters, \{$r_s$, $a$, $Q$\}, for spherical orbits allowed by the observed frequency range of QPOs (see Table \ref{QPOfreqtable}). The fundamental frequency formulae for the spherical orbits reduce to the form given by (see Appendix \ref{sphericalorbitsderivations} for a derivation: Equations \eqref{nuthetasph1}, \eqref{nuphisph1}, and \eqref{nursph1})
\begin{subequations}
\begin{eqnarray}
\bar{\nu}_{\phi}\left( r_s, a, Q \right)=&&\frac{\left\lbrace \left[ - \dfrac{\left( 2L_z r_s -L_z r_s^2 -2r_s aE\right) }{\Delta} -  L_z \right] F\left( \frac{\pi}{2},\frac{z_{-}^{2}}{z_{+}^{2}}\right)  + L_z \cdot \Pi\left( z_{-}^2, \frac{\pi}{2},\frac{z_{-}^{2}}{z_{+}^{2}}\right)  \right\rbrace }{2 \pi \left\lbrace\left[\dfrac{\left[ E \left( a^2 r_s^2 +r_s^4 +2a^2 r_s\right) -2 L_z a r_s \right] }{\Delta}  +  a^2 z_{+}^2 E  \right] F\left( \frac{\pi}{2},\frac{z_{-}^{2}}{z_{+}^{2}}\right)  -  a^2 z_{+}^2 E \cdot K\left( \frac{\pi}{2},\frac{z_{-}^{2}}{z_{+}^{2}}\right)  \right\rbrace }, \nonumber \\
 \label{nuphisph2} \\
\bar{\nu}_r \left(r_s, a, Q \right)=&&\frac{\sqrt{r_s^4 \left( 1- E^2\right) + \left( 3 Q a^2 -2 x^2 r_s -2 Q r_s\right) } \cdot F\left( \frac{\pi}{2},\frac{z_{-}^{2}}{z_{+}^{2}}\right) }{2 \pi r_s \left\lbrace\left[\dfrac{\left[ E \left( a^2 r_s^2 +r_s^4 +2a^2 r_s\right) -2 L_z a r_s \right] }{\Delta}  +  a^2 z_{+}^2 E  \right] F\left( \frac{\pi}{2},\frac{z_{-}^{2}}{z_{+}^{2}}\right)  -  a^2 z_{+}^2 E \cdot K\left( \frac{\pi}{2},\frac{z_{-}^{2}}{z_{+}^{2}}\right)  \right\rbrace },\nonumber \\
 \label{nursph2} \\
\bar{\nu}_{\theta} \left( r_s , a, Q \right)=&&\frac{a \sqrt{1- E^2} z_{+} }{4\left\lbrace\left[\dfrac{\left[ E \left( a^2 r_s^2 +r_s^4 +2a^2 r_s\right) -2 L_z a r_s \right] }{\Delta}  +  a^2 z_{+}^2 E  \right] F\left( \frac{\pi}{2},\frac{z_{-}^{2}}{z_{+}^{2}}\right)  -  a^2 z_{+}^2 E \cdot K\left( \frac{\pi}{2},\frac{z_{-}^{2}}{z_{+}^{2}}\right)  \right\rbrace },\nonumber \\
 \label{nuthetasph2}
\end{eqnarray}
\label{sphfreq}
\end{subequations}
where $\Delta=r_s^2 +a^2 -2r_s$, and $z_{\pm}$ are given by Equation (9d) of \cite{RMCQG2019}. In Figure \ref{freqcntrspherical}, we show the contours of the quantities
\begin{subequations}
\begin{eqnarray}
\delta_{\phi}\left(r_s, a, Q \right)&&=\dfrac{\left[\bar{\nu}_{\phi}\left(r_s, a, Q\right)-\bar{\nu}_{\phi}\left(r_s, a, Q=0\right)\right]}{\bar{\nu}_{\phi}\left(r_s, a, Q=0\right)}, \\  
\delta_{\rm pp}\left(r_s, a, Q \right)&&=\dfrac{\left[\bar{\nu}_{\rm pp}\left(r_s, a, Q\right)-\bar{\nu}_{\rm pp}\left(r_s, a, Q=0\right)\right]}{\bar{\nu}_{\rm pp}\left(r_s, a, Q=0\right)},  \\
\delta_{\rm np}\left(r_s, a, Q \right)&&=\dfrac{\left[\bar{\nu}_{\rm np}\left(r_s, a, Q\right) - \bar{\nu}_{\rm np}\left(r_s, a, Q=0\right)\right]}{\bar{\nu}_{\rm np}\left(r_s, a, Q=0\right)},
\end{eqnarray}
\label{deltassph}
\end{subequations}
for QPOs in the ($r_s$, $a$) plane for spherical orbits with $Q=\{2, \ 4\} $ assuming the GRPM, using Equations (\ref{nuphisph2}$-$\ref{nuthetasph2}). The blue contours in Figures \ref{freqcntrspherical} and \ref{ratiocntrse0} represent the ISSO radii, and the black contours represent the MBSO radii. The results for spherical orbits are enumerated below:
\begin{enumerate}
 \setlength\itemsep{-0.3em}
\item Novel and reduced forms for the equations of motion \{$\phi\left(r_s, a , Q \right)$, $t\left( r_s, a , Q \right)$\}, given by Equation \eqref{sphphitfinal}, and the fundamental frequencies \{$\bar{\nu}_{\phi}\left(r_s, a , Q \right)$, $\bar{\nu}_{r}\left(r_s, a , Q \right)$, $\bar{\nu}_{\theta}\left(r_s, a , Q \right)$\}, given by Equation \eqref{sphfreq}, for spherical trajectories are derived in Appendix \ref{sphericalorbitsderivations}.
\item Assuming the GRPM, we see that the spherical orbits with $Q\sim 0 -4$ are in the expected range of QPO frequencies for BHXRBs. The allowed range of $r_s$ to source the QPOs is typically $\sim 3-18 $ for $\bar{\nu}_{\phi}$ (see Figures \ref{nuphie0Q2} and \ref{nuphie0Q4}), $\sim 3-12$ for $\bar{\nu}_{\rm pp}$ (see Figures \ref{nupere0Q2} and \ref{nupere0Q4}), and $\sim 3-20 $ for $\bar{\nu}_{\rm np}$ (see Figures \ref{nunode0Q2} and \ref{nunode0Q4}), where $a$ varies from 0 to 1. 
\item The frequencies change weakly with $Q$. The maximum changes in frequencies are $\sim$2$-$3\% for $\bar{\nu}_{\phi}$, $\sim$11$-$23\% for $\bar{\nu}_{\rm pp}$, and $\sim$4$-$8\% for $\bar{\nu}_{\rm np}$ as $Q$ changes from 2 to 4 for the spherical orbits. The associated frequencies increase as $r_s$ decreases for a given \{$a$, $Q$\}.
\item  We see from Figure \ref{ratiocntrse0} that the 3:2 or 5:3 ratio of HFQPOs, $\bar{\nu}_{\phi}/ \bar{\nu}_{\rm pp}$, for spherical orbits should emanate in the region $r_s\sim 3- 7 $ for $Q$=2 and $r_s\sim 3.5-7.5 $ for $Q$=4. The ranges of $r_s$ are also dependent on $a$, where $r_s$ for a given ratio contour decreases as $a$ increases.
\end{enumerate}

\section{Parameter Estimation of Orbits in Black Hole Systems \\ with Observed QPOs}
\label{parameterestimation}
Next, we take up a few cases of black hole systems that are known to have shown either two or three simultaneous QPOs in their PDS, and we extract the parameter values of the nonequatorial eccentric ($eQ$), equatorial eccentric ($e0$), and the spherical orbits ($Q0$) corresponding to the observed QPO frequencies using our GRPM. The solution for a given GRPM class ($eQ$, $Q0$, $e0$) being attempted here is based on balancing the knowns (number of simultaneous frequencies, two or three) with the number of unknown parameters $\{e, r_p, a, Q\}$ (see Table \ref{modelparameters} illustrating this criterion). For the three frequency cases (M82 X-1 and GROJ 1655-40), we have to either supply $a$ from available data or deduce this using a procedure that involves minimizing $\chi^2$ in the unknown parameter volume. For the geometric study of orbits that is of importance here, we have taken the view that the best approximation to $a$ is to be determined first, and then the solution vector $\{e, r_p, Q\}$ (which is crucial for the orbital shape) for the peak probability is found. We have taken slightly different approaches for the two sources as exact solutions are found only in one of the two sources (M82 X-1), where we minimize $\chi^2$ in the $a$ dimension to isolate $a$. In the other case where no exact solution vector is found (GROJ 1655-40), and where it is computationally expensive to explore the full four-dimensional parameter volume of $\{e, r_p, Q, a\}$ in a fine-grained manner, we have only done a primary coarse-grained search to find $a$ sufficiently accurately and then proceeded to determine the unknown parameters $\{e, r_p, Q\}$ by a fine-grid search. The two QPO frequency cases (XTEJ 1550-564, 4U 1630-47, and GRS 1915+105) are searched by direct fine-grid computations assuming $a$ from available data (see Table \ref{modelparameters} and \ref{sourcelist}).

\begin{deluxetable}{c c c c}
\tablecaption{\label{modelparameters}Various GRPMs, Their Corresponding Unknown Parameters, and Underlying Assumptions for BHXRBs with Three and Two Simultaneous QPOs.}
\tablewidth{500pt}
\tabletypesize{\footnotesize}
\tablehead{
 \colhead{\textbf{BHXRBs with Three QPOs}} & & &  
}
\startdata
GRPM & Model Parameters & Number of Parameters & Number of Observed QPOs\\
\hline
$eQ$ & \{$e$, $r_p$, $a$, $Q$\} & 4 & 3$^{\rm a}$ \\
$e0$ & \{$e$, $r_p$, $a$\} & 3 & 3 \\
$Q0$ & \{$a$, $r_s$, $Q$\} & 3 & 3 \\
\hline
\colhead{\textbf{BHXRBs with Two QPOs}} & & & \\
\hline
$e0$ & \{$e$, $r_p$\} & 2 & 2$^{\rm b}$ \\
$Q0$ & \{$r_s$, $Q$\} & 2 & 2$^{\rm b}$ \\
\enddata
\tablecomments{$^{\rm a}$need to supply $a$ from the best fit of $\chi^2$; $^{\rm b} a$ is fixed from the available data (see Table \ref{sourcelist}).}
\end{deluxetable}

We describe our parameter search criteria below:

\begin{enumerate}
\item For BHXRBs with three simultaneous QPOs, that is, M82 X-1 and GROJ 1655-40 (see Table \ref{sourcelist}), since a type C LFQPO is also present, which corresponds to the nodal oscillation frequency ($\nu_{\rm np}$), we search for all $eQ$, $e0$, and $Q0$ orbit solutions. We use Equations (\ref{nuphi}$-$\ref{nutheta}) and (\ref{eqnuphi}$-$\ref{eqnutheta}) to equate the QPO frequencies to \{$\nu_{\phi}$, $\nu_{\rm pp}$, $\nu_{\rm np}$\} and find the parameters \{$e$, $r_{p}$, $a$\} of $eQ$ and $e0$ orbits for M82 X-1 and GROJ 1655-40. Next, we calculate the most probable spin of the black hole to estimate \{$e$, $r_p$, $Q$\} of the orbit. Similarly, we study the $Q0$ orbits as solutions to the QPOs using Equations (\ref{nuphisph2}$-$\ref{nuthetasph2}) and find the parameters \{$r_s$, $a$, $Q$\} for these BHXRBs. Hence, the parameters searched for these cases are
\begin{subequations}
\begin{eqnarray}
    \mathrm{3 QPOs}&&= 
\begin{cases}
    
    eQ \ \mathrm{and} \ e0,& \{\mathcal{M}=\mathrm{fixed \ from \ observations}, e, r_p, a, Q=\{0, 1, 2, 3, 4\} \},\\
    Q0,   & \{\mathcal{M}=\mathrm{fixed \ from \ observations}, e=0, r_s, a, Q \}.
\end{cases} 
\label{3QPOs}
\end{eqnarray}

\item For BHXRBs with two simultaneous QPOs, that is, XTEJ 1550-564, 4U 1630-47, and GRS 1915+105 (see Table \ref{sourcelist}), we expect that the solutions are likely to be equatorial as the LFQPO, or $\nu_{\rm np}$ oscillation, is absent (this is consistent with no large-amplitude nodal oscillations and strictly equatorial orbits). Hence, we search for $e0$ solutions using Equations (\ref{eqnuphi}$-$\ref{eqnur}) for \{$\nu_{\phi}$, $\nu_{\rm pp}$\} to find \{$e$, $r_p$\} of the orbit. However, we also check for the $Q0$ orbital solution in these systems and estimate the parameters \{$r_s$, $Q$\} using \{$\nu_{\phi}$, $\nu_{\rm pp}$\}. Hence, the parameters searched for in these cases are
\begin{eqnarray}
  \mathrm{2 QPOs}&&= 
\begin{cases}
    
     \ e0,& \{\left(\mathcal{M}, a\right)=\mathrm{fixed \ from \ observations}, e, r_p, Q=0 \},\\
    Q0,   & \{\left(\mathcal{M}, a\right)=\mathrm{fixed \ from \ observations}, e=0, r_s, Q \}.
\end{cases}
\label{2QPOs}
\end{eqnarray}
\end{subequations}
\end{enumerate}

We have summarized the history of black hole systems considered here in Appendix \ref{sourcehistory}. In \S \ref{samplesel}, we summarize the observations related to QPO detection, mass, and spin estimation and the parameters we estimated for each source. In \S \ref{method}, we explain the method used to estimate the parameters of these orbits and corresponding errors and then present the results for the (non)equatorial eccentric orbits in \S \ref{resultseccentric}, and spherical orbits in \S \ref{resultspherical}. 
\subsection{Source Selection}
\label{samplesel}
Here we summarize the QPO observations of the black hole systems that we have selected to implement the GRPM for the general eccentric and spherical trajectories. We have chosen cases where either two or three simultaneous QPOs have been detected, which are as follows:
\begin{enumerate}
\item M82 X-1: We use the HF-analog QPOs of M82 X-1 along with the other detected LFQPOs \citep{Pasham2013ApJb} to estimate the parameters \{$e$, $r_p$, $a$\} of the $eQ$ and $e0$ trajectories, where the QPOs are created, by varying $Q$ in the range $0-4$ using the GRPM. Next, using these results, we calculate the most probable value of $a$ to estimate the remaining parameters \{$e$, $r_p$, $Q$\}, using three simultaneous QPO frequencies, in \S \ref{resultseccentric}. In our analysis, we have assumed the mass of the black hole to be $\mathcal{M}=428$ \citep{Pasham2014}. We also search for the $Q0$ orbit solution and estimate the corresponding parameters \{$r_s$, $a$, $Q$\} assuming the GRPM in \S \ref{resultspherical}. In this paper, we have assumed that the LFQPOs are simultaneous with 3.32$\pm$0.06 Hz and 5.07$\pm$0.06 Hz QPOs, because these HF-analog QPOs were found to be stable over a few years \citep{Pasham2014}, and during this period LFQPOs were also detected; see Table \ref{sourcelist}. Hence, we explore the parameter space of \{$\mathcal{M}=428$, $e$, $r_p$, $a$, $Q$\} (see Equation \eqref{3QPOs}).

\item GROJ 1655-40: We use three simultaneous frequencies detected, 441$\pm$2 Hz, 298$\pm$4 Hz, and 17.3$\pm$0.1 Hz \citep{Motta2014a}, to associate them with the general $eQ$ and $e0$ trajectories assuming the GRPM in \S \ref{resultseccentric}. We also explore a $Q0$ trajectory solution. For this BHXRB, we fixed the mass of the black hole to the previously known value, $\mathcal{M}=5.4$ \citep{Beer2002}. We did not find any $Q0$ orbit solution for this BHXRB. Hence, we explore the parameter space of \{$\mathcal{M}=5.4$, $e$, $r_p$, $a$, $Q$\} (see Equation \eqref{3QPOs}).

\item XTEJ 1550-564: We use the simultaneous frequencies, 268$\pm$3 Hz and 188$\pm$3 Hz \citep{Miller2001}, in our GRPM and calculate \{$e$, $r_p$\} of the orbit assuming the $e0$ orbit in \S \ref{resultseccentric}. We also estimate the parameters \{$r_s$, $Q$\} of the $Q0$ orbit using these QPO frequencies in \S \ref{resultspherical}. We assumed that the mass of the black hole is $\mathcal{M}=9.1$, as estimated using the optical spectro-photometric observations \citep{Orosz2011}, and that the spin of the black hole is $a=0.34$ \citep{Miller2015}, estimated from the disk continuum spectrum. Hence, we explore the parameter space of \{$\mathcal{M}=9.1$, $a=0.34$, $e$, $r_p$\} for $e0$ orbits and \{$\mathcal{M}=9.1$, $a=0.34$, $r_s$, $Q$\} for $Q0$ orbits (see Equation \eqref{2QPOs}).

\item 4U 1630-47: We use the twin HFQPOs at 179.3$\pm$5.7 Hz and 38.06$\pm$7.3 Hz \citep{KleinWolt2004} and associate them with the fundamental frequencies of the $e0$ orbits to find the parameters \{$e$, $r_p$\} in \S \ref{resultseccentric}. We assumed the mass of the black hole to be $\mathcal{M}=10$, calculated from the scaling of the photon index of the Comptonized spectral component with the LFQPOs \citep{Seifina2014}. We fixed the spin of the black hole to $a=0.985$, as previously estimated from the fit to the reflection spectrum using NuSTAR observations \citep{King2014}. We did not find a $Q0$ orbit solution for this BHXRB. Hence, we explore the solution space of \{$\mathcal{M}=10$, $a=0.985$, $e$, $r_p$\} for the $e0$ orbit (see Equation \eqref{2QPOs}).

\item GRS 1915+105: We take simultaneous HFQPOs at 69.2$\pm$0.15 Hz and 41.5$\pm$0.4 Hz \citep{Strohmayer2001b} to study the $e0$ orbits using the GRPM and calculate the corresponding parameters \{$e$, $r_p$\} in \S \ref{resultseccentric}. We fixed the mass of the black hole to $\mathcal{M}=10.1$, estimated using the near-infrared spectroscopic observations \citep{Steeghs2013}. We assumed the spin of the black hole to be $a=0.98$, calculated by fitting to the disk reflection spectrum using NuSTAR observations \citep{Miller2013}. We did not find a $Q0$ orbit solution for this BHXRB. Hence, we explore the solution space of \{$\mathcal{M}=10.1$, $a=0.98$, $e$, $r_p$\} for the $e0$ orbit (see Equation \eqref{2QPOs}).
\end{enumerate}
\begin{deluxetable}{c l c c c c c c}
\tablecaption{\label{sourcelist}Summary of Existing BHXRBs That Exhibit Either Three or Two Simultaneous QPOs.}
\tablewidth{500pt}
\tabletypesize{\footnotesize}
\tablehead{
\colhead{\textbf{S.No.}} & \colhead{\textbf{BHXRB}}  & \colhead{\textbf{$\nu_{1}$ (Hz)}} & \colhead{\textbf{$\nu_{2}$ (Hz)}} & \colhead{\textbf{$\nu_{3}$ (Hz)}} & \colhead{\textbf{$\mathcal{M}$}} & \colhead{\textbf{$a$}}  & \colhead{\textbf{Model}} \\
& & & & & & & \colhead{\textbf{Classes}} \\
}
\startdata
1. & M82 X-1 & 5.07$\pm$0.06$^{\rm (a)}$ & 3.32$\pm$0.06$^{\rm (a)}$ & $\left(204.8\pm6.3\times10^{-3}\right)^{ \rm (b)}$ & 428$\pm$105$^{\rm (a)}$ & -  & $eQ$, $e0$, $Q0$\\
2. & GROJ 1655-40 & 441$\pm$2$^{\rm (c)}$ & 298$\pm$4$^{\rm (c)}$ & 17.3$\pm$0.1$^{\rm (c)}$ & 5.4$\pm$0.3$^{\rm (d)}$ & -  & $eQ$, $e0$, $Q0$\\
& &  & &  & & & \\
\hline
3. & XTE J1550-564 & 268$\pm$3$^{\rm (e)}$  & 188$\pm$3$^{\rm (e)}$ & - & 9.1$\pm$0.61$^{\rm (f)}$ & 0.34$^{+0.37}_{-0.45}$ $^{\rm (g)}$   & $e0$, $Q0$\\
4. & 4U 1630-47  & 179.3$\pm$5.7$^{\rm (h)}$ & 38.06$\pm$7.3$^{\rm (h)}$ & - & 10$\pm$0.1$^{\rm (i)}$ & 0.985$^{+0.005}_{-0.014}$ $^{\rm (j)}$  & $e0$, $Q0$\\
5. & GRS 1915+105 & 69.2$\pm$0.15$^{\rm (k)}$ & 41.5$\pm$0.4$^{\rm (k)}$ & - & 10.1$\pm$0.6$^{\rm (l)}$ & 0.98$\pm$0.01$^{\rm (m)}$  & $e0$, $Q0$\\
& &  &  & & & & \\
\enddata
\tablecomments{The first two rows represent the cases having twin HFQPOs with simultaneous type-C QPO. The remaining rows show the cases of BHXRB having only twin HFQPOs. The columns show the source name, QPO frequencies, and previously measured mass through optical, infra-red or X-ray observations, previously known spin of the black hole measured by fit to the Fe K$\alpha$ line or to the continuum spectrum (for 1 and 2 we calculate the parameter $a$ from our method), and the class of GRPM applied to estimate the parameters. The references are indicated by lower case letters (a-m).}
\tablerefs{(a) \cite{Pasham2014}, (b) \cite{Pasham2013ApJb}, (c) \cite{Motta2014a}, (d) \cite{Beer2002}, (e) \cite{Miller2001}, (f) \cite{Orosz2011}, (g) \cite{Miller2015}, (h) \cite{KleinWolt2004}, (i) \cite{Seifina2014}, (j) \cite{King2014}, (k) \cite{Strohmayer2001b}, (l) \cite{Steeghs2013}, (m) \cite{Miller2013}.}
\end{deluxetable}

We have summarized the BHXRB data in the Table \ref{sourcelist} along with the frequencies of detected QPOs, and previously known values of mass and spin of the black hole, along with their references.

\subsection{Method Used and Results}
\label{method}
\begin{figure}
\begin{center}
\scalebox{0.8}{
\tikzstyle{decision} = [diamond, draw,  
    text width=6em, text badly centered, node distance=3cm, inner sep=0pt]
\tikzstyle{block} = [rectangle, draw, 
 minimum width=4.0cm, minimum height=1cm,  text width=35em, text centered, rounded corners]
\tikzstyle{line} = [draw, -latex']
\tikzstyle{process} = [rectangle, draw, minimum width=3cm, minimum height=1cm,   text width=10em, text centered, rounded corners ]
\tikzstyle{cloud} = [draw, ellipse,node distance=3cm,
    minimum height=2em]
\begin{tikzpicture}[node distance = 3cm, auto]
   \small
\node [block, rectangle split,rectangle split parts=2,rectangle split part fill={pink!20,blue!20}] (para1) {{Joint Probability density} \nodepart{second} $P\left(\nu \right) = \prod_{i}  P_{i}\left( \nu_{i}\right)$,  $i$=1 to $l$,\\
 where $l$=3 for M82 X-1 and GROJ 1655-40, \\
and $l$=2 for XTEJ 1550-564, 4U 1630-47, and GRS 1915+105.};
\node [block, below of=para1, node distance=3.7cm, rectangle split,rectangle split parts=2,rectangle split part fill={pink!20,blue!20}] (para3) {{Jacobian}\nodepart{second} Find $\mathcal{J}_l$ using Equation \eqref{jacobian}, where $ \ i, j$=1 to $l$. \\
 Eccentric orbits - $x_i$'s$=$\{$e$, $r_p$, $a$\} for $l$=3, and $x_i$'s$=$\{$e$, $r_p$\} for $l$=2.\\
 Spherical orbits - $x_i$'s$=$\{$r_s$, $a$, $Q$\} for $l$=3, and $x_i$'s$=$\{$r_s$, $Q$\} for $l$=2.};
\node [block, below of=para3, node distance=3.7cm, rectangle split,rectangle split parts=2,rectangle split part fill={pink!20,blue!20}] (para4) {{Exact solutions}\nodepart{second}Find the exact solutions for $x_j$'s using the frequency formulae, Equations \eqref{genfreq}, \eqref{eqfreq} and \eqref{sphfreq}. Fix $\mathcal{M}$ or \{$\mathcal{M}$, $a$\} to the previously known values for $l$=3 and $l$=2 respectively, see Table \ref{sourcelist}.};
\node [block, below of=para4, node distance=3.9cm, rectangle split,rectangle split parts=2,rectangle split part fill={pink!20,blue!20}] (para5) {{Probability density and normalization factor}\nodepart{second}Choose appropriate range for $x_j$'s near exact solution and resolutions $\Delta x_j$'s to find $P\left( \left[ x\right]\right)$ and $\mathcal{N}$ using Equations \eqref{Pera} and \eqref{normN}. For the case with no exact solution, for example GROJ 1655-40, we choose complete range of parameters to calculate $P\left( \left[ x\right]\right)$ and $\mathcal{N}$. }; 
\node [block, below of=para5, node distance=3.7cm, rectangle split,rectangle split parts=2,rectangle split part fill={pink!20,blue!20}] (para6) {{Normalized probability density}\nodepart{second}Find $\mathcal{P}\left( [x ]\right)$ using Equation \eqref{normPera}, which is $\mathcal{P}\left( x_1 , x_2, x_3\right)$ for $l$=3 and $\mathcal{P}\left( x_1 , x_2\right)$ for $l$=2. };
\node [block, below of=para6, node distance=3.7cm, rectangle split,rectangle split parts=2,rectangle split part fill={pink!20,blue!20}] (para7) {{ Integrated profiles} \nodepart{second} Integrate $\mathcal{P}\left( [x ]\right)$ to obtain the profile in each dimension: \\
Eccentric orbits - \{$\mathcal{P}_1 \left( e\right)$, $\mathcal{P}_{1}\left( r_p\right)$, $\mathcal{P}_1 \left(a\right)$\} for $l$=3, \{$\mathcal{P}_1 \left( e\right)$, $\mathcal{P}_{1}\left( r_p\right)$\} for $l$=2. \\
Spherical orbits - \{$\mathcal{P}_{1} \left( r_s\right)$, $\mathcal{P}_{1}\left( Q\right)$, $\mathcal{P}_1 \left(a\right)$\} for $l$=3, \{$\mathcal{P}_{1} \left( r_s\right)$, $\mathcal{P}_{1}\left( Q\right)$\} for $l$=2.};
\node [block, below of=para7, node distance=4.5cm, rectangle split,rectangle split parts=2,rectangle split part fill={pink!20,blue!20}] (para8) {{Three simultaneous QPOs }\nodepart{second} Multiple trajectory solutions were estimated for M82 X-1 and GROJ 1655-40 with varying spin. We choose the most probable value of $a$ and estimate the exact solutions \{$e_0$, $r_{p0}$, $Q_0$\}, probability density profiles \{$\mathcal{P}_1 \left( e\right)$, $\mathcal{P}_1 \left( r_p\right)$, $\mathcal{P}_1 \left( Q\right)$\}, and the corresponding errors using the same procedure described above. };

\draw[->,thick] (para1)--(para3);
\draw[->,thick] (para3)--(para4);
\draw[->,thick] (para4)--(para5);
\draw[->,thick] (para5)--(para6);
\draw[->,thick] (para6)--(para7);
\draw[->,thick] (para7)--(para8);

 \end{tikzpicture}
}
\end{center}
\caption{Flowchart of the method used to estimate the orbital solutions for QPOs and corresponding errors.}
\label{methodflowchart}
\end{figure}
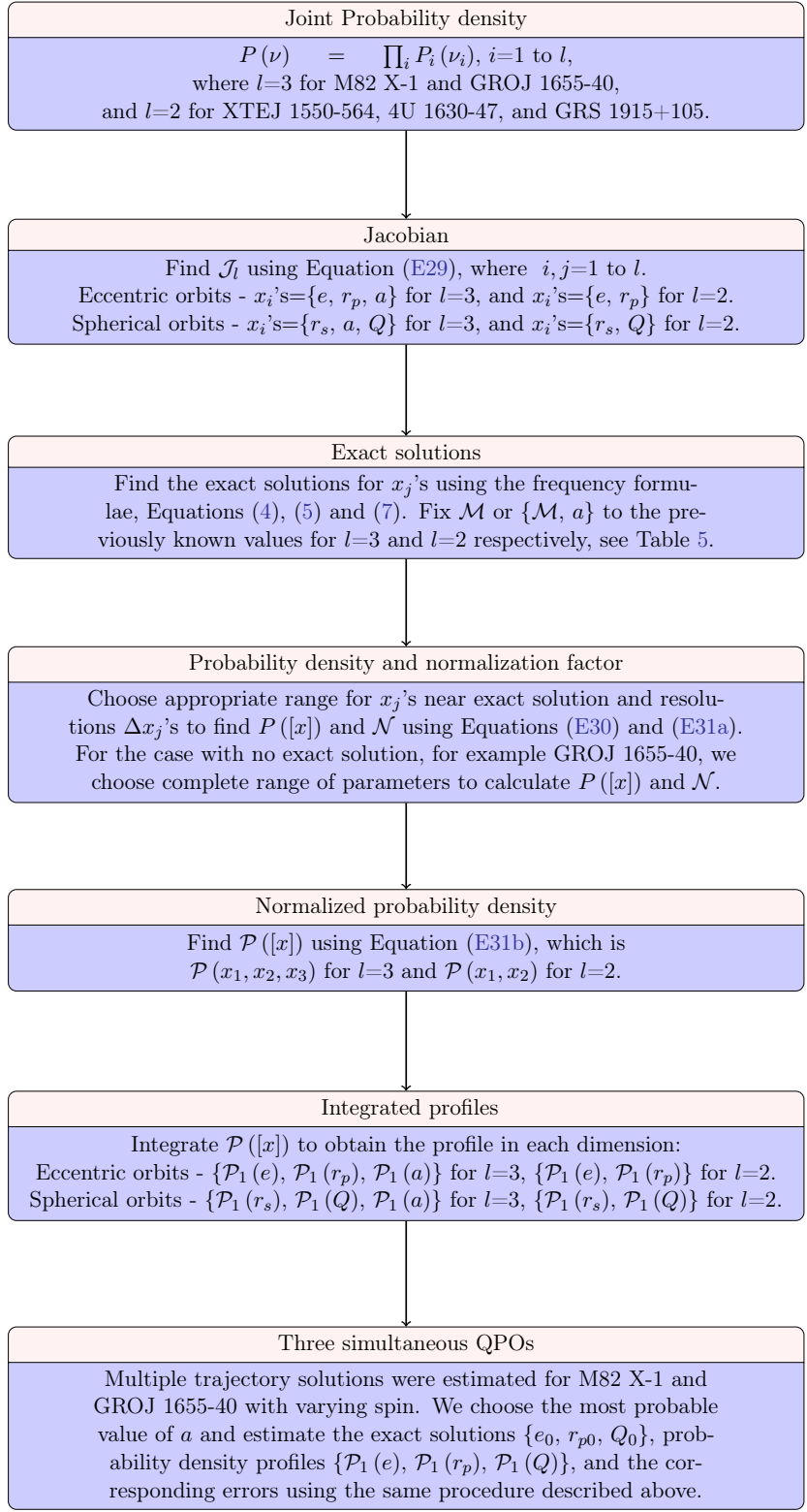

We apply the GRPM to associate the fundamental frequencies of $eQ$, $e0$, and $Q0$ orbits with QPOs. In Appendix \ref{methodsection}, we describe a generic procedure that we have used to estimate errors in the orbital parameters. A flowchart of this method is provided in Figure \ref{methodflowchart}. Next, we summarize the results corresponding to the $eQ$ and $e0$ models in \S \ref{resultseccentric} and the $Q0$ model in \S \ref{resultspherical}.

\subsubsection{Nonequatorial and Equatorial Eccentric Orbits ($eQ$ and $e0$)}
\label{resultseccentric}

We have taken the cases of five BHXRBs, known to have either three or two simultaneous detections of QPOs in their observations, to study the eccentric and nonequatorial trajectories as solutions to the QPOs assuming the GRPM. Here we summarize the results for the cases of three and two simultaneous QPOs separately, as discussed below:
\begin{enumerate}
\item \textit{Three simultaneous QPOs}: In our analysis, varying the dimensionless parameter $Q=\{0, 1, 2, 3, 4\} \propto \left( L^2 -L_{z}^2\right)$ gives us various trajectory solutions having different \{$e$, $r_p$, $a$, $Q$\} combinations. We first find the exact solutions for the parameters \{$e$, $r_p$, $a$\}, given in the Table \ref{3QPOresults}, by equating the centroid frequencies of three simultaneous QPOs (see Table \ref{sourcelist}) to \{$\nu_{\phi}$, $\nu_{\rm pp}$, $\nu_{\rm np}$\} for each value of $Q=\{0, 1, 2, 3, 4\}$ using our analytic formulae (Equations (\ref{nuphi}$-$\ref{nutheta})). We estimate errors for the parameters \{$e$, $r_p$, $a$\} using the method discussed in Appendix \ref{methodsection} (see Figure \ref{methodflowchart}) for each value of $Q$. The results of fits to the integrated profiles \{$\mathcal{P}_1 \left( e\right)$, $\mathcal{P}_1 \left( r_p\right)$, $\mathcal{P}_1 \left( a\right)$\} are summarized in the Table \ref{3QPOresults}. Since the spin of the black hole is not expected to vary, we estimate the most probable spin value for these black holes and then estimate the orbital parameters \{$e$, $r_p$, $Q$\} and their corresponding errors again using the same method discussed in Appendix \ref{methodsection} (see Figure \ref{methodflowchart}). The results for each case are as follows:
\begin{figure}
\mbox{
\subfigure[]{
\hspace{1.5cm}
\includegraphics[scale=0.29]{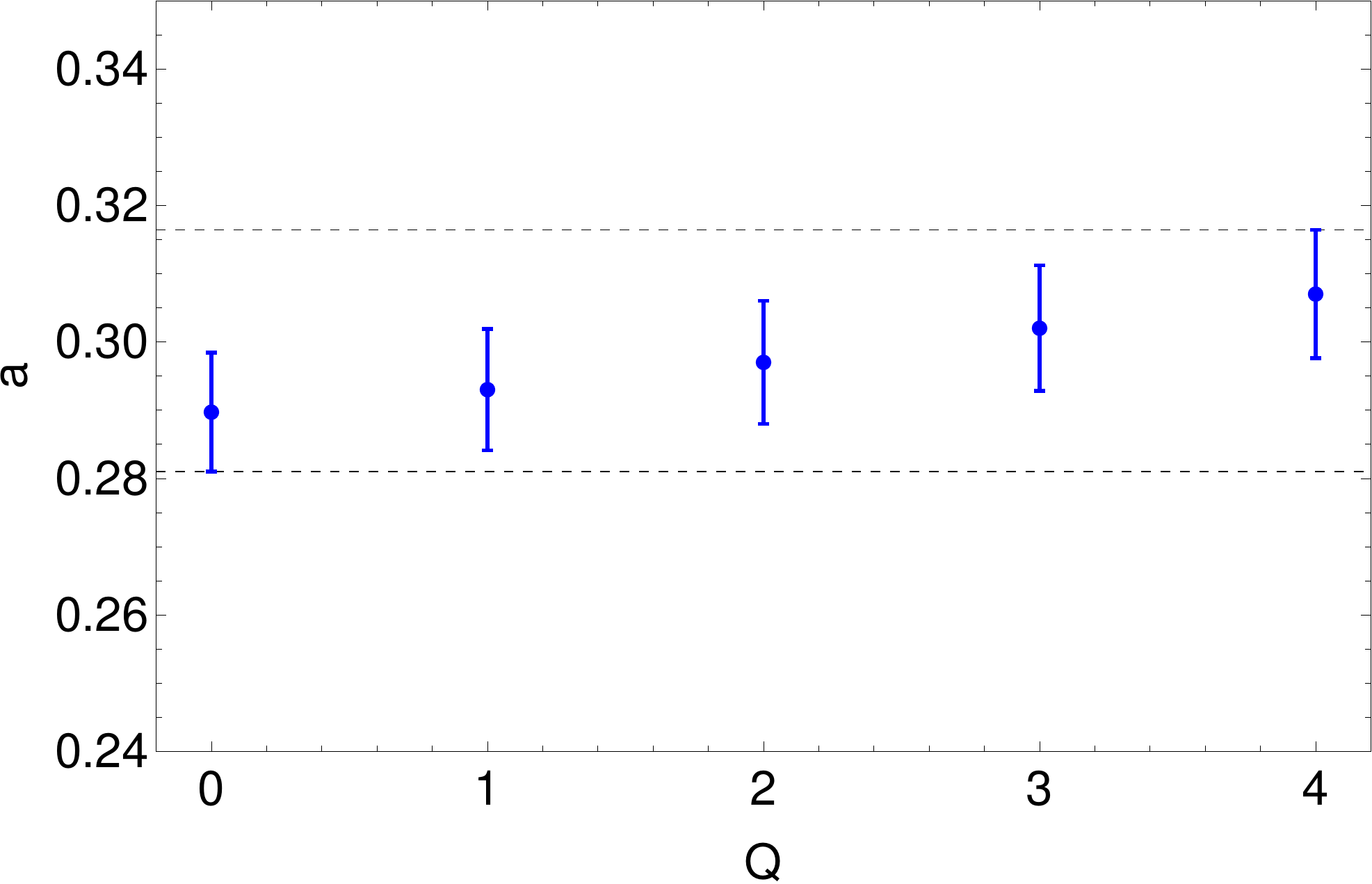}\label{ErroraQM82}}
\hspace{1.9cm}
\subfigure[]{
\includegraphics[scale=0.29]{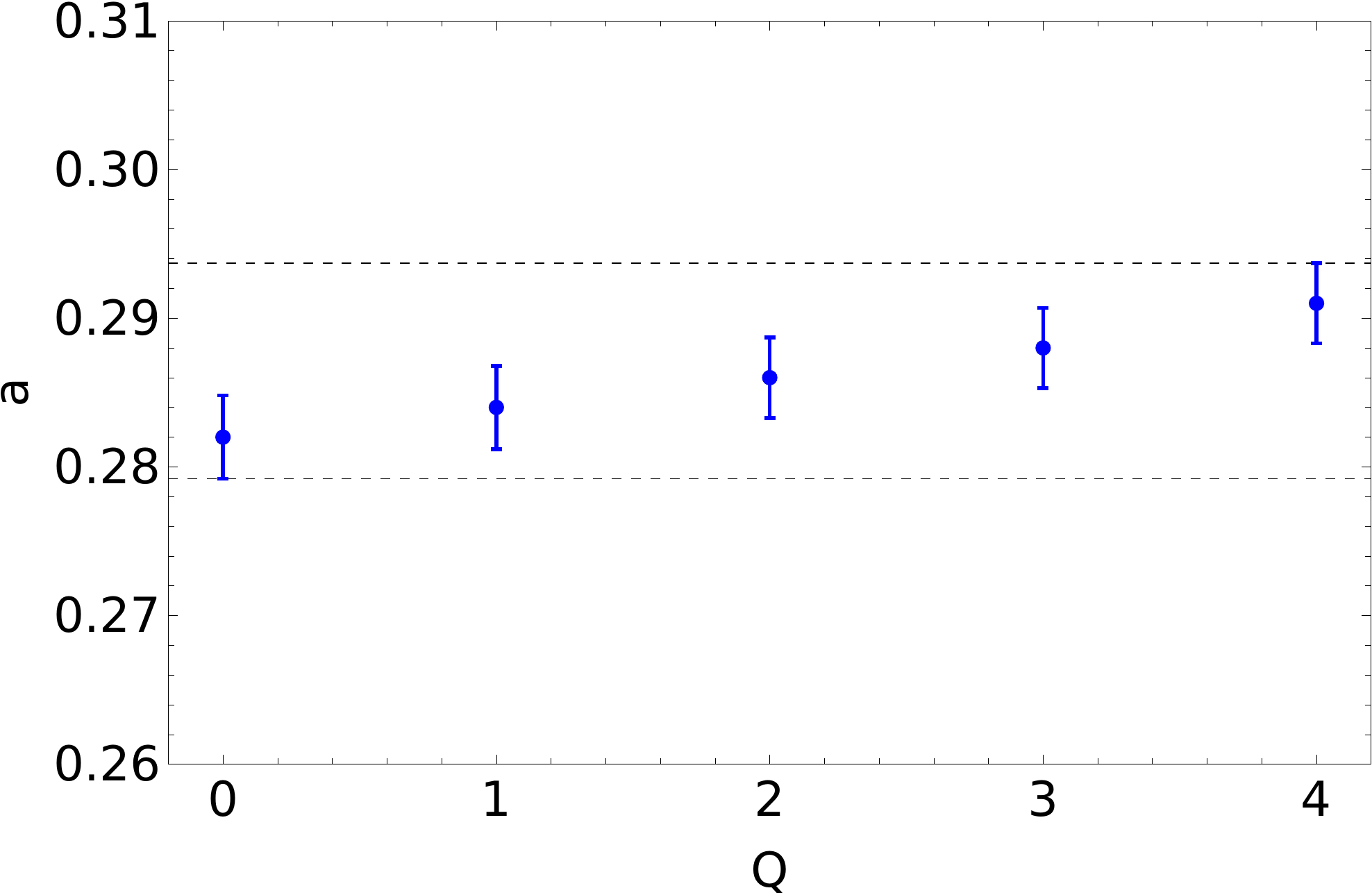}\label{ErroraQGROJ}}}
\caption{\label{ErroraQ}The figures show $1 \sigma$ errors in the spin parameters for various $Q$ values for exact solutions of (a) M82 X-1, and corresponding to the peak of the probability distributions for (b) GROJ 1655-40, as given in the Table \ref{3QPOresults}. The upper and lower dashed curves correspond to the limits of the calculated errors. Although each $Q$ value corresponds to a different spin of the black hole, the calculated values, and corresponding errors are within a narrow band which puts a sharp and reasonable constraint on the spin of the black hole.}
\end{figure}
\begin{itemize}
\item \textit{M82 X-1}: In this case, we find that the (non)equatorial trajectories with small to moderate eccentricities $e\sim$0.18$-$0.28 with $r_p=$4.6$-$5.07 and $a=$0.28$-$0.31 (see Table \ref{3QPOresults}) are possible exact solutions for the observed QPO frequencies in M82 X-1, for $Q$ between 0 and 4. Starting with these exact solutions, the most probable value of the spin is found first. In Fig \ref{ErroraQM82}, we show the spin variation in the parameter solutions for QPOs as a function of $Q$. Next, to estimate the most probable value of the spin, we minimize the function
\begin{subequations}
\begin{equation}
\chi^2_a= \sum_i \dfrac{\left( a- a_i\right)^2}{\sigma_{a_i}^2},
\end{equation}
which gives 
\begin{equation}
a=\dfrac{\sum_i \left( a_i / \sigma_{a_i}^2 \right)}{\sum_i \left( 1/  \sigma_{a_i}^2 \right) }, \label{mprobablea}
\end{equation}
\end{subequations}
where $i=1-6$ corresponds to six probable solutions for $a$, and the $\sigma_i$ values are the corresponding 1 $\sigma$ errors, where five of these are given in Table \ref{3QPOresults}, and the remaining one corresponds to the spherical orbit solution found for M82 X-1 given in Table \ref{sphresults}. By including these six solutions, we have spanned the complete ($e$, $Q$) parameter space, which is bounded by $e0$ and $Q0$ solutions. This gives us the most probable spin value of $a=0.2994$. Hence, we fix the spin of the black hole to this most probable estimate and then calculate the remaining parameters \{$e$, $r_p$, $Q$\} and corresponding errors using the method given in Appendix \ref{methodsection} and Figure \ref{methodflowchart}. We find the exact solution for QPOs at \{$e=0.2302$, $r_p=4.834$, $Q=2.362$\} calculated by equating centroid QPO frequencies to \{$\nu_{\phi}$, $\nu_{\rm pp}$, $\nu_{\rm np}$\} while fixing $a=0.2994$. The probability density distribution profiles \{$\mathcal{P}_1 \left( e\right)$, $\mathcal{P}_1 \left( r_p\right)$, $\mathcal{P}_1 \left(Q\right)$\}, along with their model fit, and the probability contours in the parameter plane \{$e$, $Q$\}, \{$r_p$, $e$\}, and \{$Q$, $r_p$\} are shown in Figure \ref{M82X1all}. The results of the model fit to the integrated profiles are summarized in the Table \ref{erQresults}. The corresponding errors are quoted with respect to the exact solution of the parameters, which slightly differ from the peak of the integrated profiles \{$\mathcal{P}_1 \left( e\right)$, $\mathcal{P}_1 \left( r_p\right)$, $\mathcal{P}_1 \left(Q\right)$\}, as expected (see Figure \ref{M82X1all}).

  \begin{figure}
  \mbox{
\subfigure[]{
\includegraphics[scale=0.45]{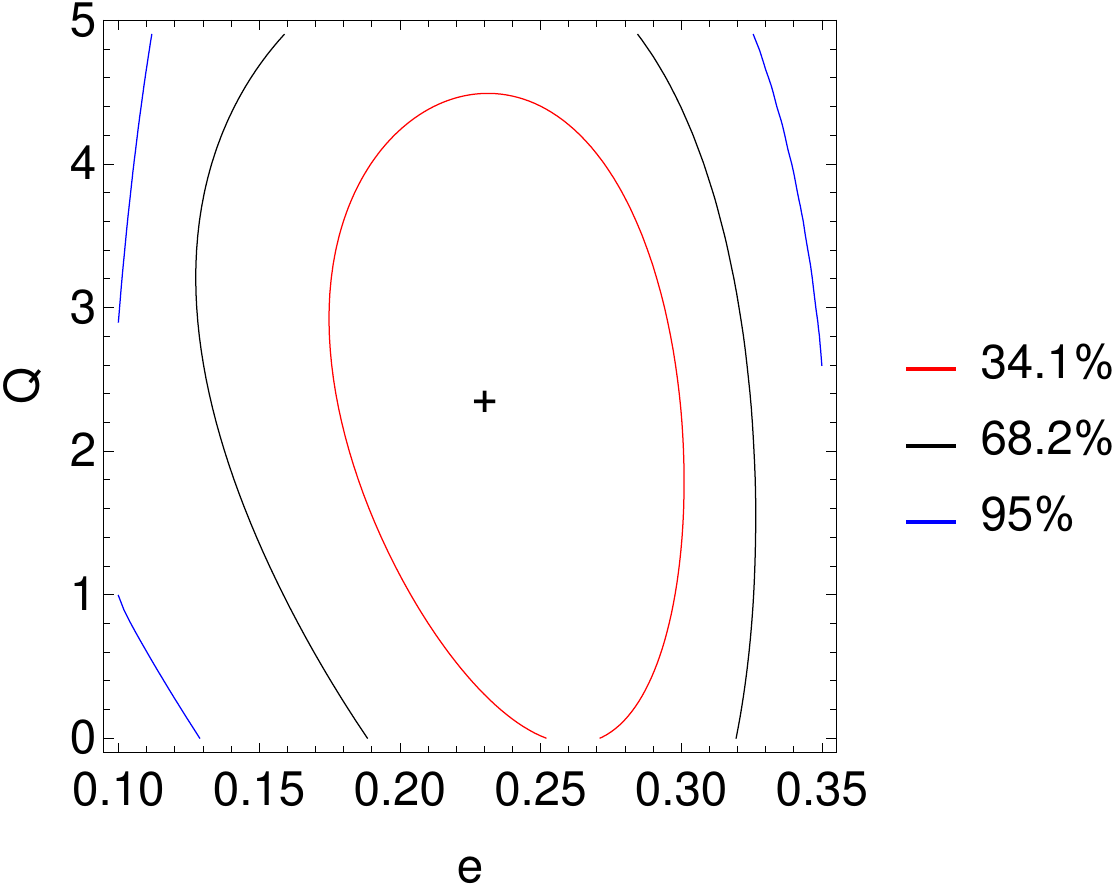}\label{M82a}}
\hspace{0.8cm}
\subfigure[]{
\includegraphics[scale=0.45]{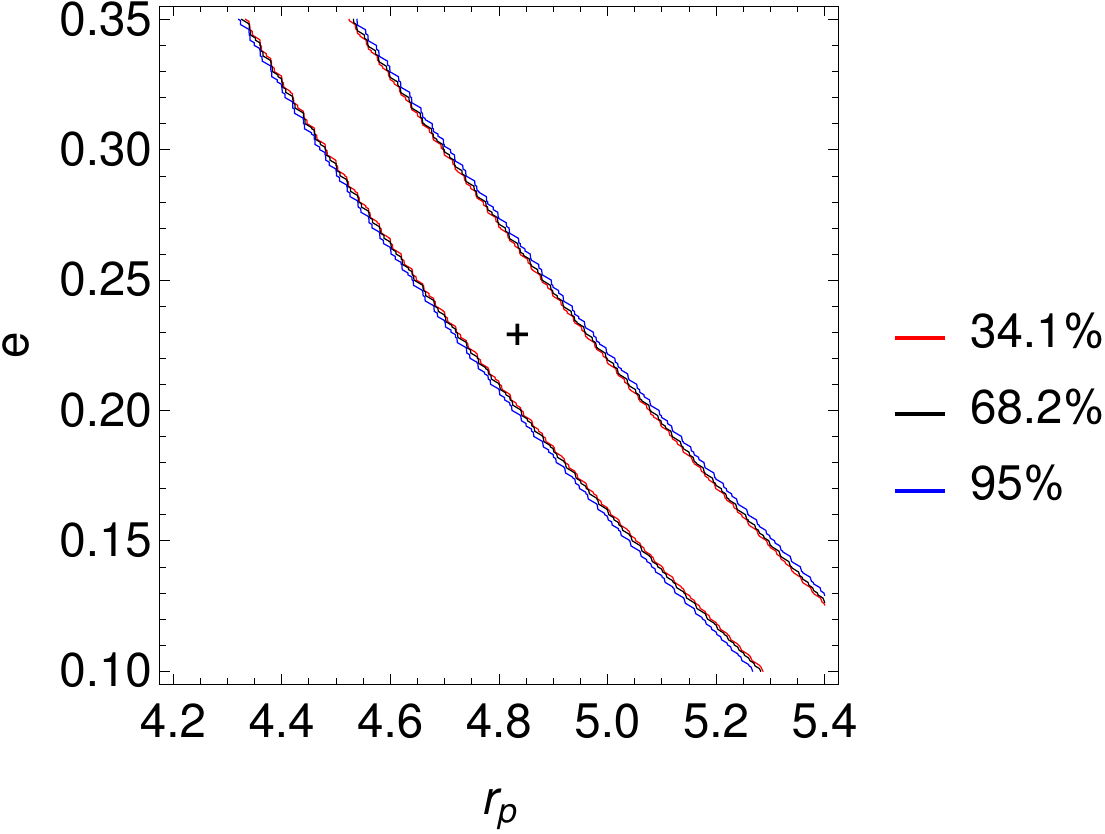}\label{M82b}}
\hspace{0.8cm}
\subfigure[]{
\includegraphics[scale=0.45]{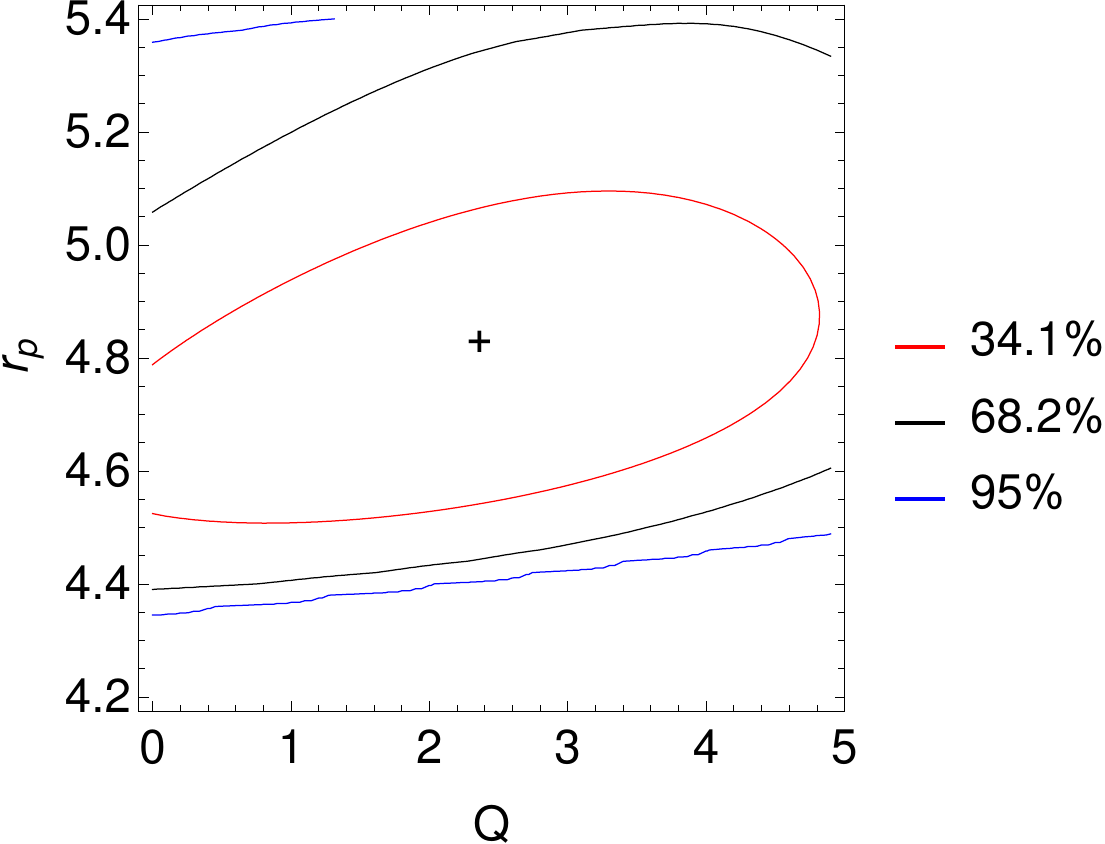}\label{M82c}}}
  \mbox{
\subfigure[]{
\includegraphics[scale=0.26]{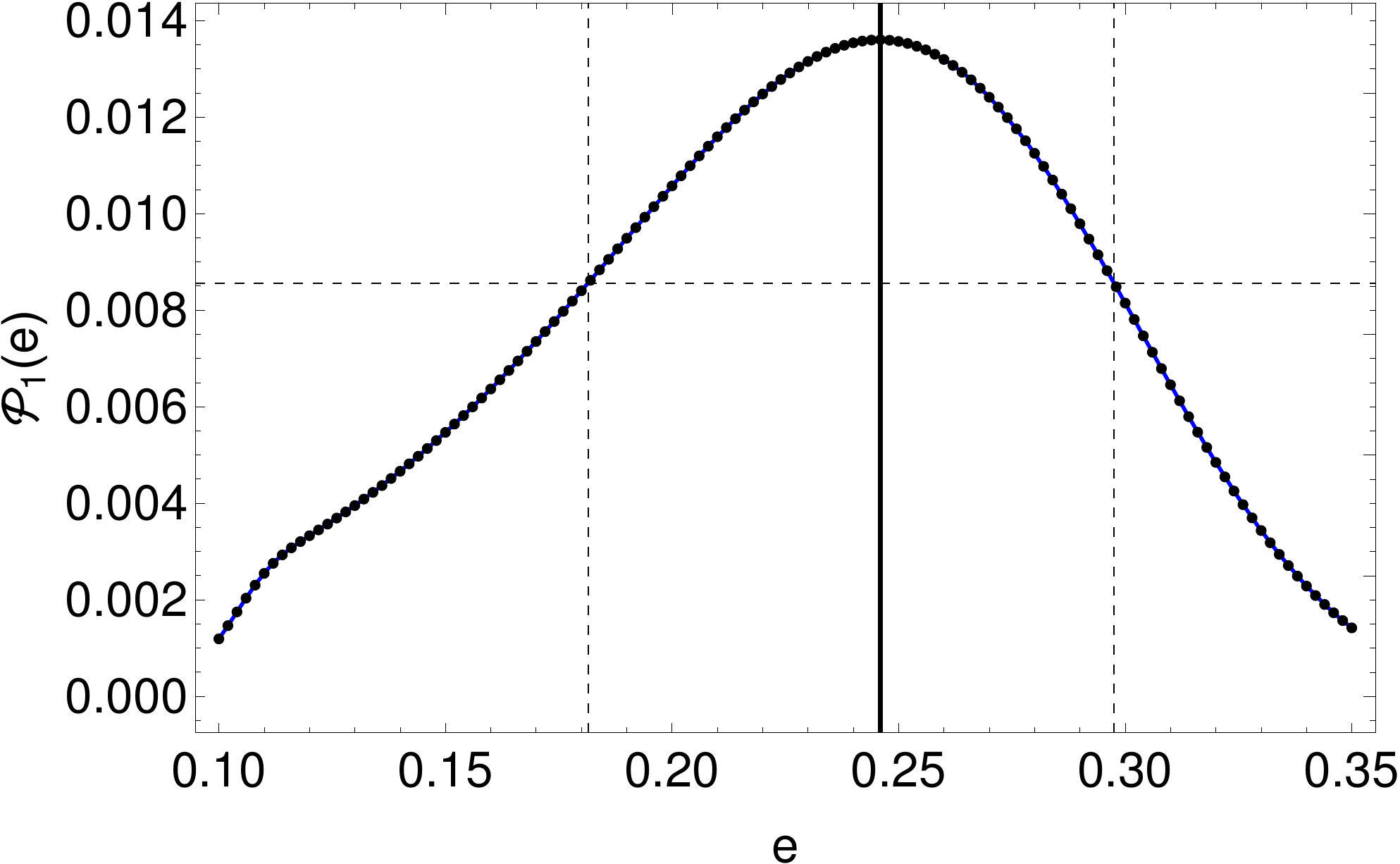}\label{M82d}}
\hspace{0.3cm}
\subfigure[]{
\includegraphics[scale=0.26]{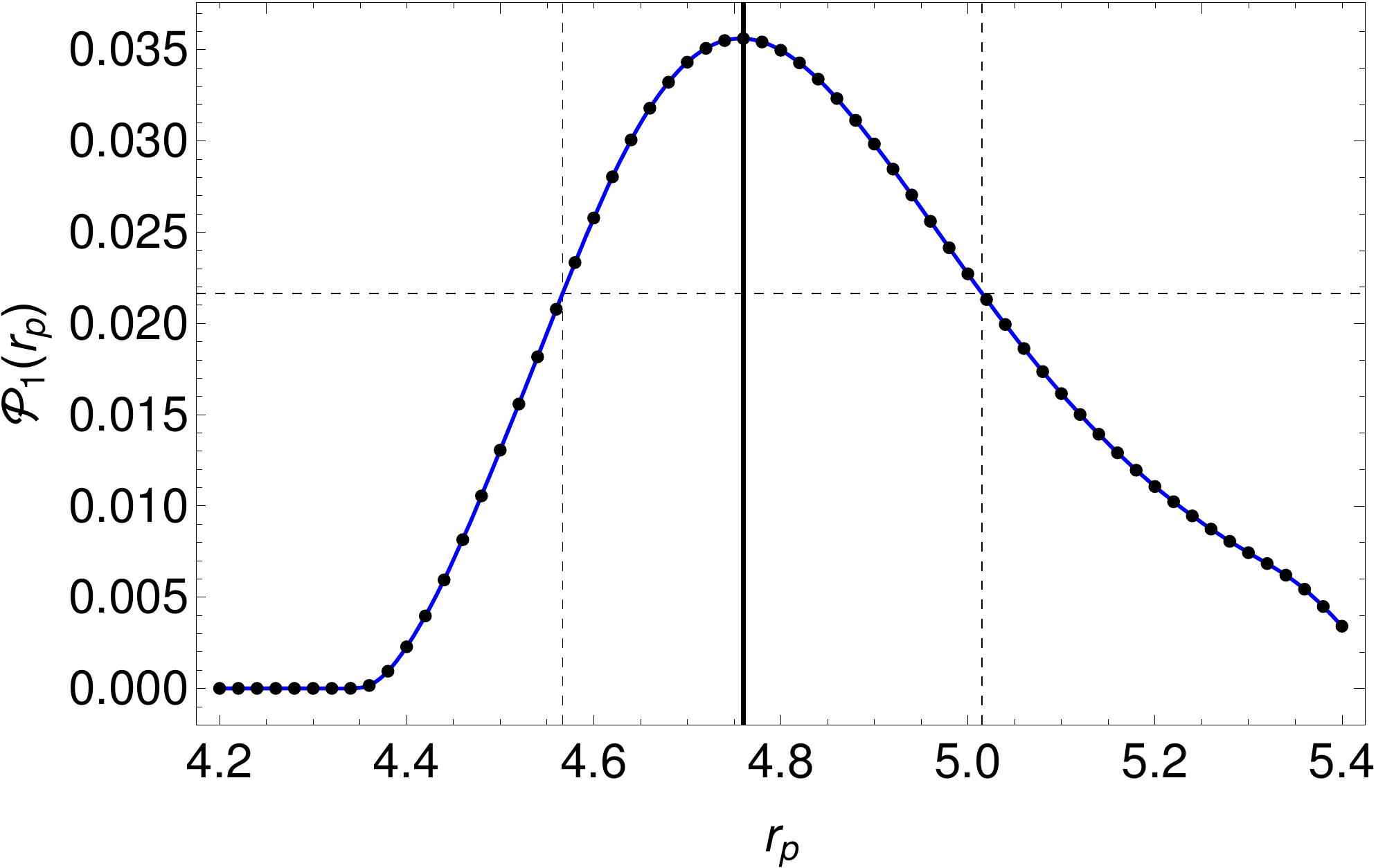}\label{M82e}}
\hspace{0.3cm}
\subfigure[]{
\includegraphics[scale=0.26]{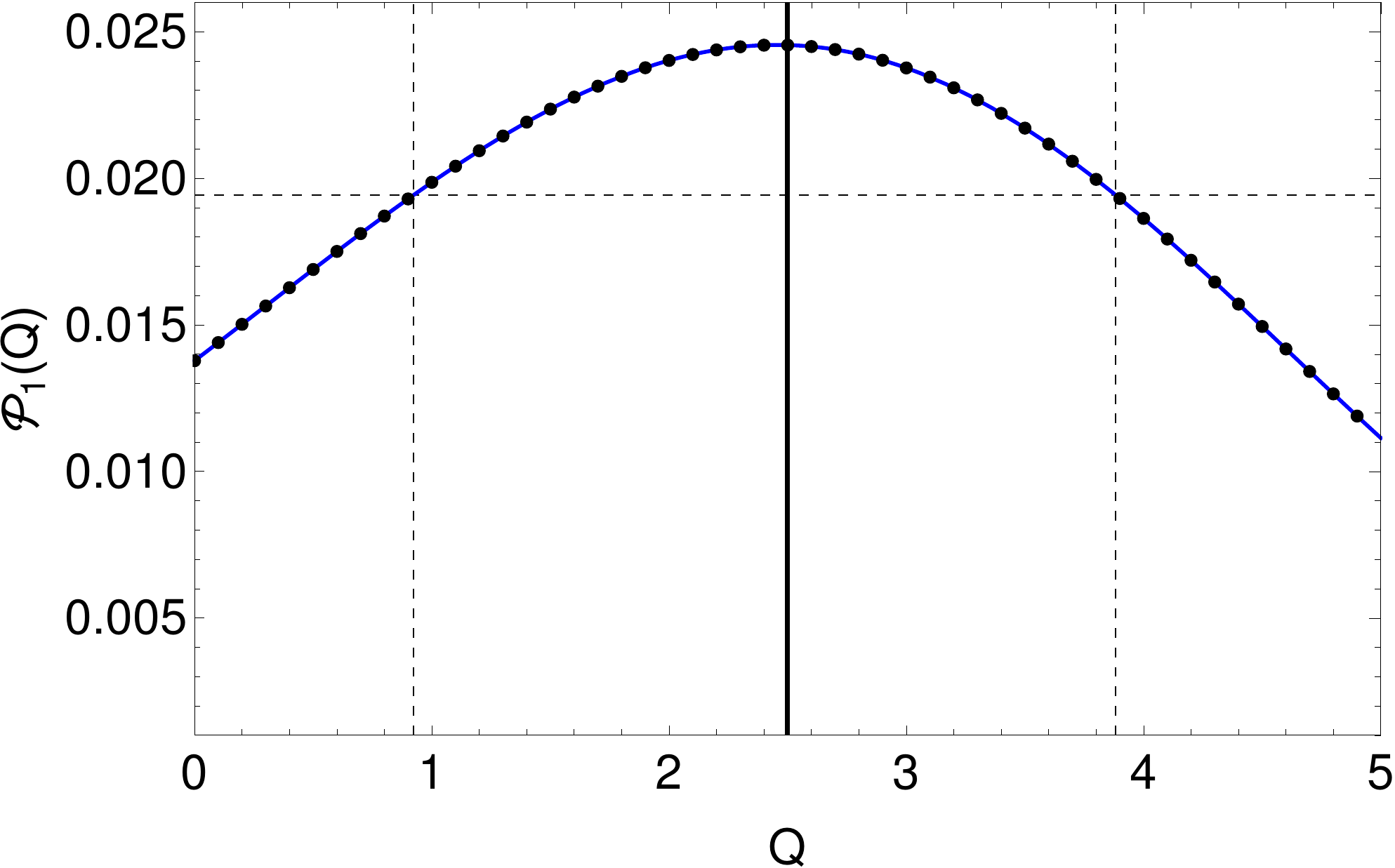}\label{M82f}}}
  \caption{\label{M82X1all} The probability contours in the parameter planes are shown in (a) \{$e$, $Q$\}, (b) \{$r_p$, $e$\}, and (c) \{$Q$, $r_p$\}, where the $+$ sign marks the exact solution for the parameters for QPOs in M82 X-1 with $a=0.2994$. The probability density profiles are shown in (d) $\mathcal{P}_1 \left( e\right)$, (e) $\mathcal{P}_1 \left( r_p\right)$, and (f) $\mathcal{P}_1 \left(Q\right)$, where the black points represent integrated probability densities and the blue curves are their model fit. The dashed vertical lines enclose a region with 68.2\% probability, and the solid vertical line marks the peak of the profiles. }
  \end{figure}

\item \textit{GROJ 1655-40}: For this case, we did not find the exact solution for the parameters \{$e$, $r_p$, $a$\} when the centroid frequencies of QPOs, Table \ref{sourcelist}, are equated to \{$\nu_{\phi}$, $\nu_{\rm pp}$, $\nu_{\rm np}$\}. However, we generate the probability density profiles $\mathcal{P}_{1} \left( e\right)$, $\mathcal{P}_{1} \left( r_p\right)$, and $\mathcal{P}_{1} \left( a\right)$ for each value of $Q$ between 0 and 4. The results of fits for these profiles are summarized in Table \ref{3QPOresults}. We found that the probability density peaks near very small eccentricities $e\sim0.05-0.07$ for various values of $Q$, whereas $r_p$ ranges between $5.24$ and $5.43$ and $a$ ranges between $0.282$ and $0.291$; see Table \ref{3QPOresults}. The change in the value of the spin of the black hole as a function of $Q$ is shown in Figure \ref{ErroraQGROJ} for GROJ 1655-40. Next, we find the most probable value of the spin for this BHXRB. Since we did not find any exact solution for the parameters by equating centroid frequencies of QPOs to the frequency formulae, we calculated the $\chi^2$ function given by 
\begin{equation}
\chi^2=\dfrac{\left(\nu_{\phi}-\nu_{10}\right)^2}{\sigma_{1}^2}+\dfrac{\left(\nu_{\rm pp}-\nu_{20}\right)^2}{\sigma_{2}^2}+\dfrac{\left(\nu_{\rm np}-\nu_{30}\right)^2}{\sigma_{3}^2}, \label{chisqr}
\end{equation}
in the four-dimensional parameter space \{$e$, $r_p$, $a$, $Q$\} using Equations \eqref{nuphi}$-$\eqref{nutheta} for \{$\nu_{\phi}$, $\nu_{\rm pp}$, $\nu_{\rm np}$\}, and we numerically found the minimum $\chi^2=2.814$ for the parameter combination \{$e=0.021$, $r_p=5.51$, $a=0.283$, $Q=0$\}. This is a primary coarse-grained search to find a viable solution of $a$. Next, we assume $a=0.283$ corresponding to the minimum $\chi^2$ to calculate the final solution for the parameters \{$e$, $r_p$, $Q$\}, which are the key parameters for the geometric study, using the more accurate fine-grid method described in Appendix \ref{methodsection} and Figure \ref{methodflowchart}. We find that the probability density peaks near \{$e=0.071$, $r_p=5.25$, $Q=0$\}. The results of fitting to the \{$\mathcal{P}_1 \left( e\right)$, $\mathcal{P}_1 \left( r_p\right)$, $\mathcal{P}_1 \left(Q\right)$\} profiles are summarized in the Table \ref{erQresults}, whereas these profiles with their model fit and the probability contours in the parameter plane \{$e$, $Q$\}, \{$r_p$, $e$\}, and \{$Q$, $r_p$\} are shown in Figure \ref{GROJall}.
   \begin{figure}
  \mbox{
\subfigure[]{
\includegraphics[scale=0.45]{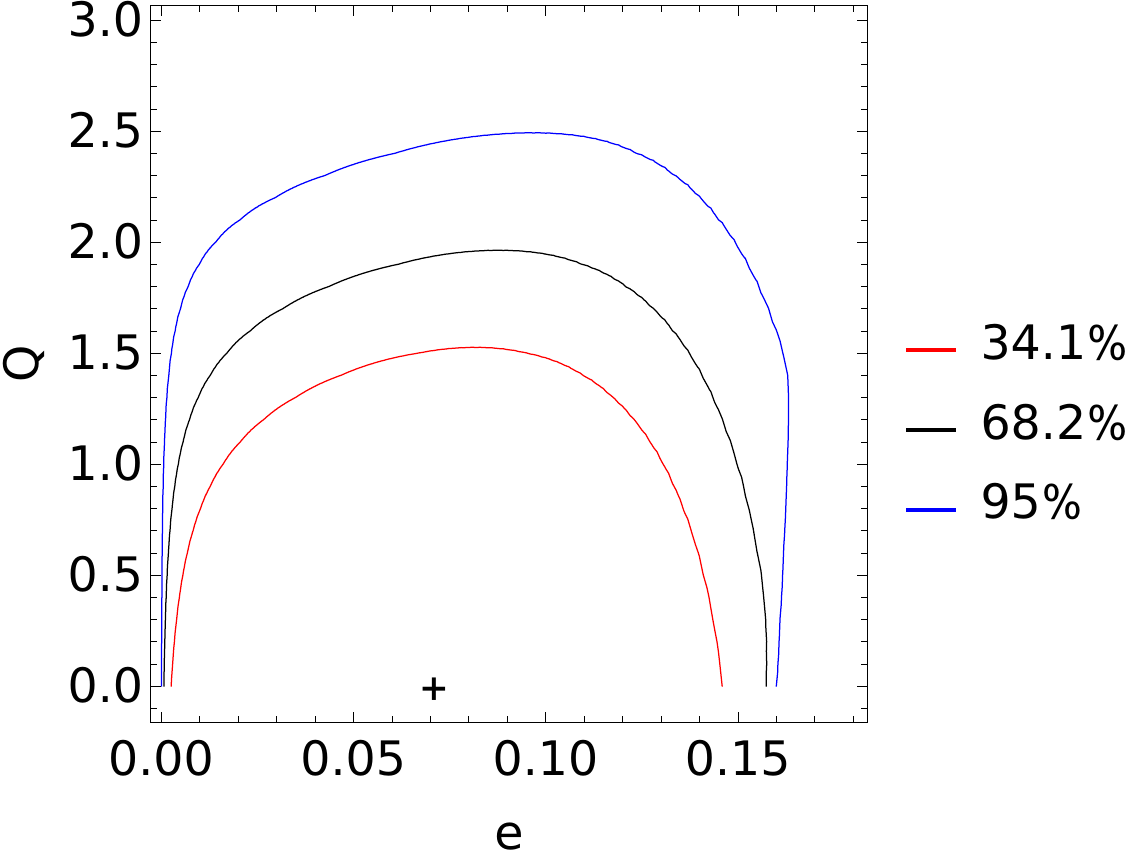}\label{GROJa}}
\hspace{0.8cm}
\subfigure[]{
\includegraphics[scale=0.43]{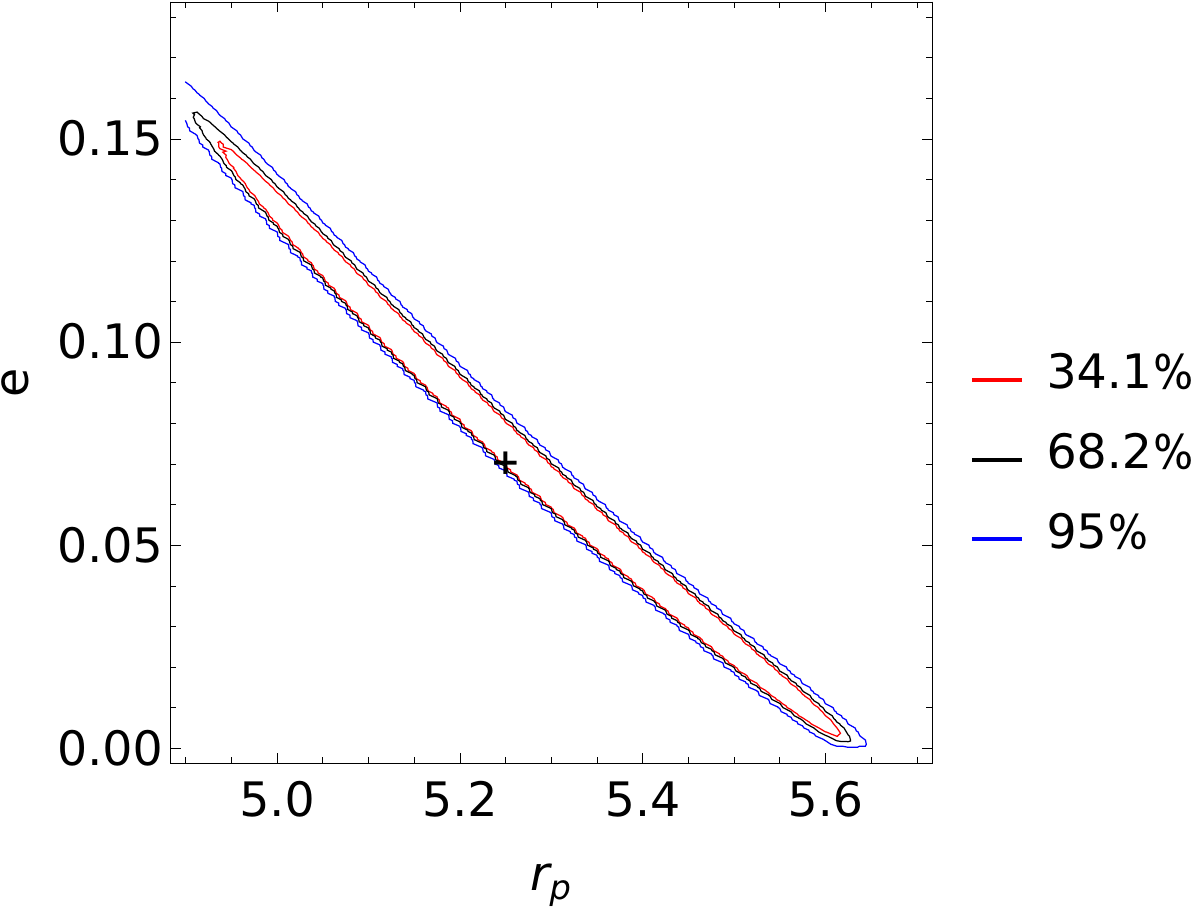}\label{GROJb}}
\hspace{0.8cm}
\subfigure[]{
\includegraphics[scale=0.45]{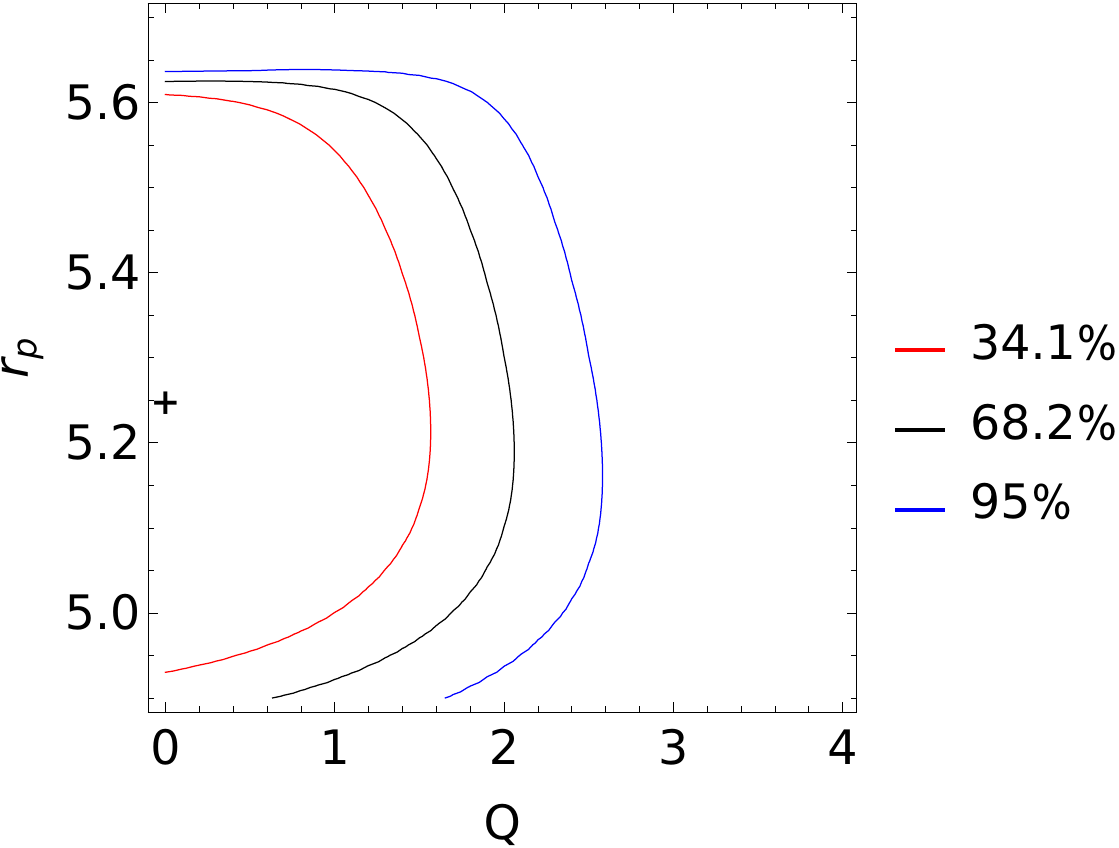}\label{GROJc}}}
  \mbox{
\subfigure[]{
\includegraphics[scale=0.26]{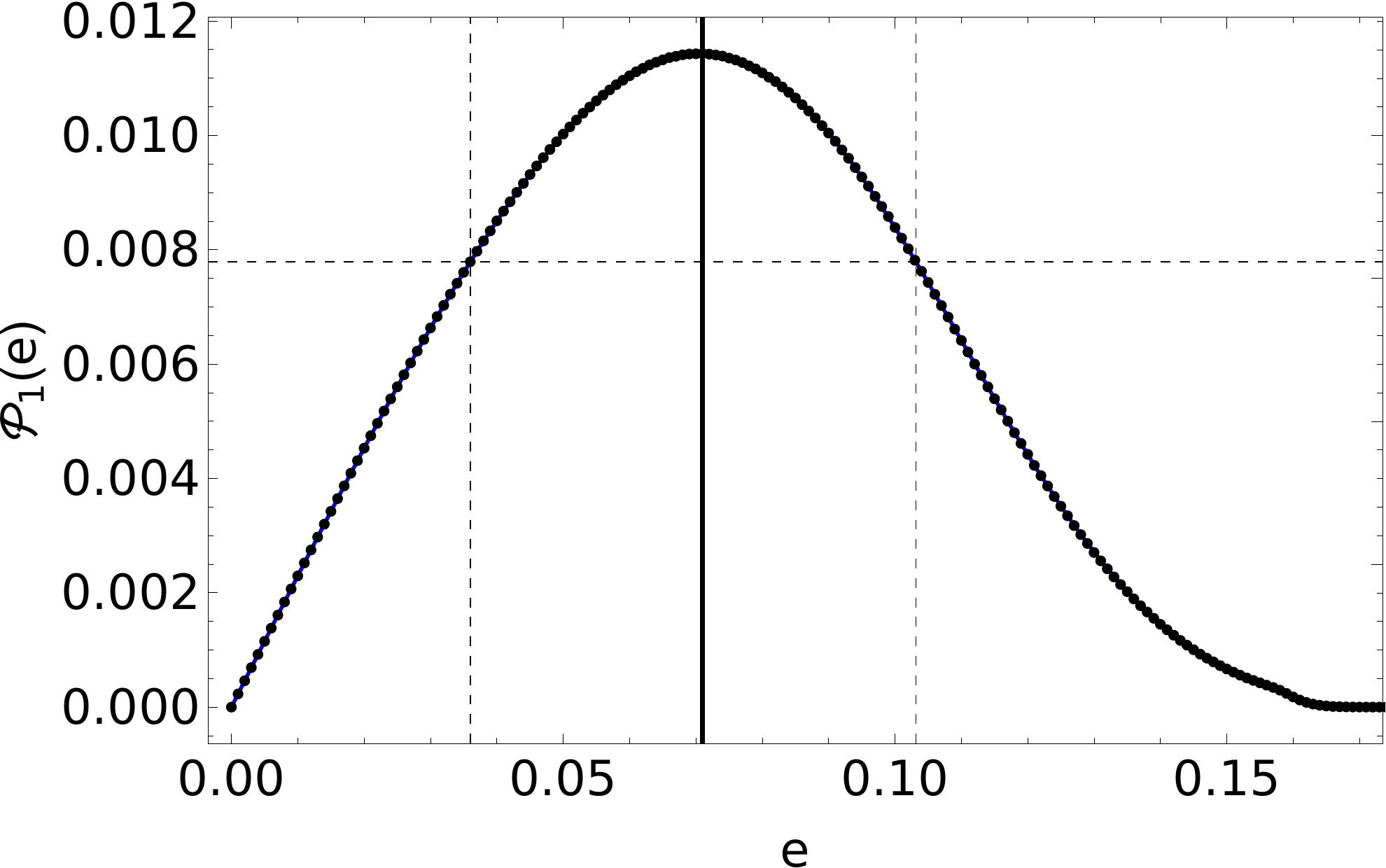}\label{GROJd}}
\hspace{0.3cm}
\subfigure[]{
\includegraphics[scale=0.26]{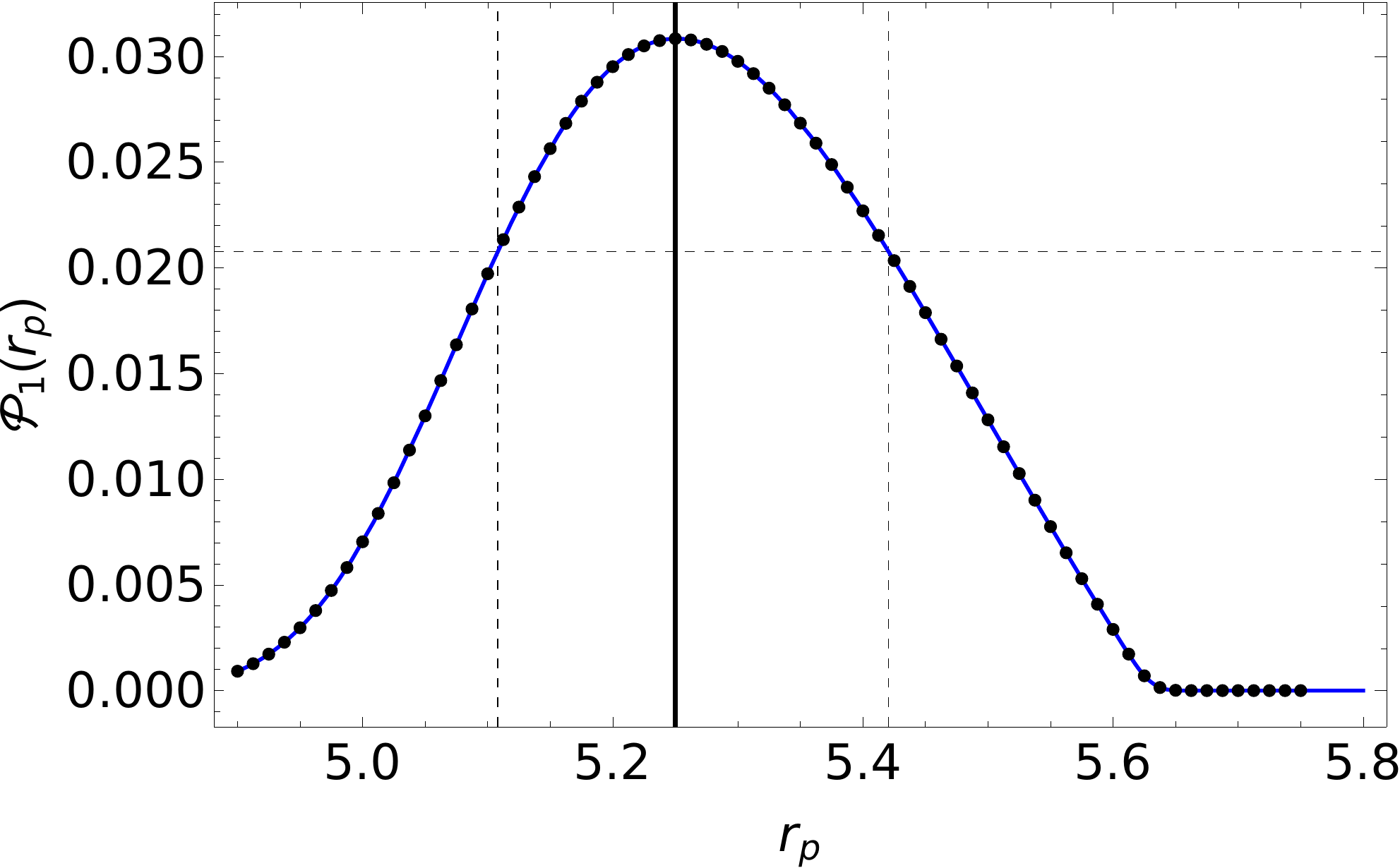}\label{GROJe}}
\hspace{0.3cm}
\subfigure[]{
\includegraphics[scale=0.26]{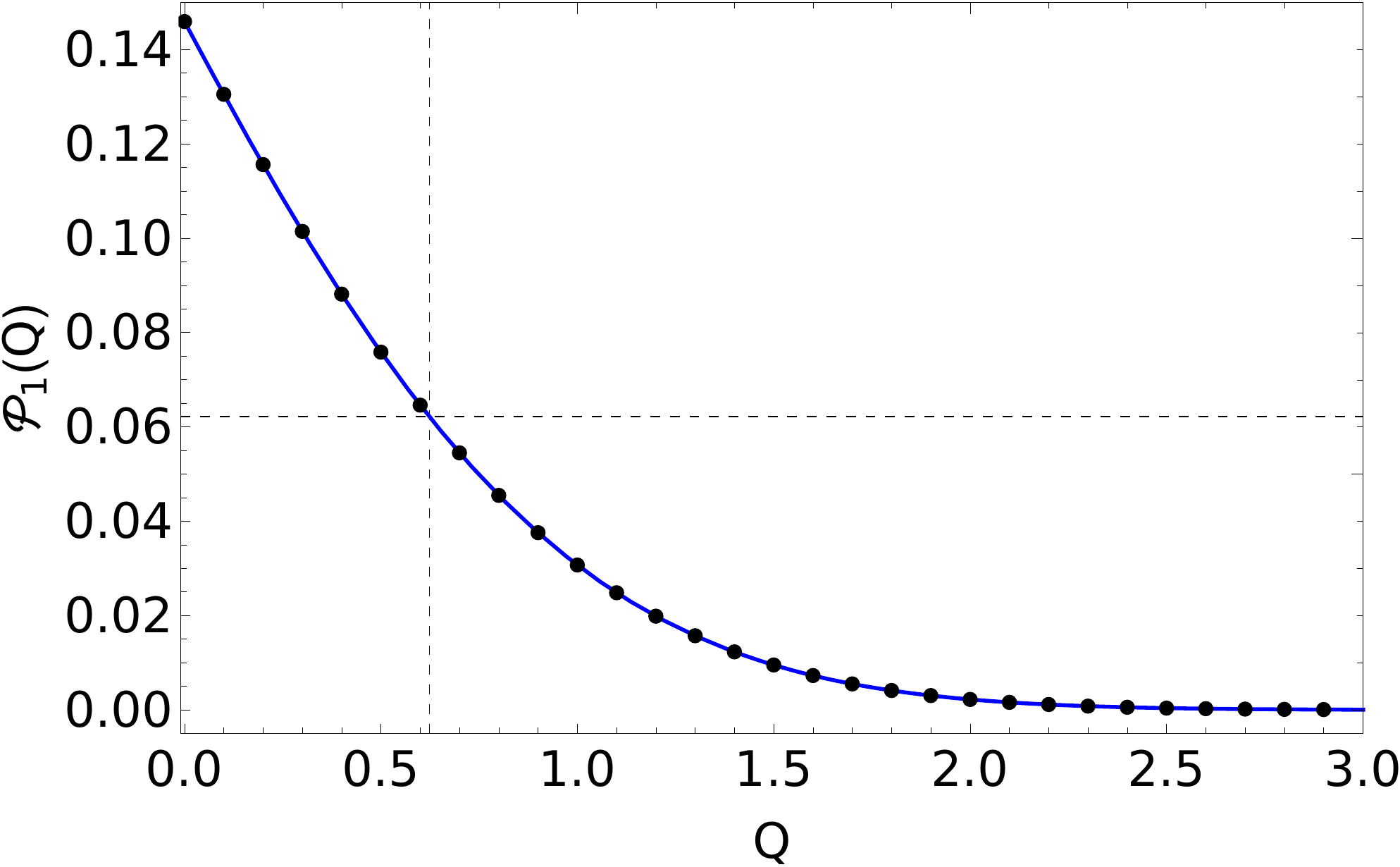}\label{GROJf}}}
  \caption{\label{GROJall} The probability contours in the parameter planes are shown in (a) \{$e$, $Q$\}, (b) \{$r_p$, $e$\}, and (c) \{$Q$, $r_p$\}, where the $+$ sign marks the peak of the probability density for GROJ 1655-40 with $a=0.283$. The probability density profiles are shown in (d) $\mathcal{P}_1 \left( e\right)$, (e) $\mathcal{P}_1 \left( r_p\right)$, and (f) $\mathcal{P}_1 \left(Q\right)$ , where the dashed vertical lines enclose a region with 68.2\% probability and the solid vertical line marks the peak of the profiles. }
  \end{figure}
\end{itemize}

  Hence, we conclude for both M82 X-1 and GROJ 1655-40 that (non)equatorial trajectories (both $eQ$ and $e0$) with small or moderate eccentricities in the region very close to the black hole are the solutions for the observed QPOs assuming our GRPM. A self-emitting blob of matter close to a Kerr black hole can have enough energy and angular momentum to attain an eccentric and nonequatorial trajectory. These results are also consistent with the conclusions made in \S \ref{eccentricmotivation} that the trajectories having small to moderate eccentricities with $Q=0-4$ are also possible solutions for the observed range of QPO frequencies in BHXRBs. 
    
  The errors in QPO frequencies cause to a distribution in the solution space \{$e$, $r_p$, $Q$\} as solutions using our GRPM, as shown in Figures \ref{M82X1all} and \ref{GROJall}. We take various combinations of these parameters within the range of 1$\sigma$ errors, as summarized in Table \ref{erQresults}, as any such parameter combination is a probable solution for the frequencies within the width of QPOs observed in the power spectrum. In Figure \ref{trajectories}, we have plotted together the trajectories for these parameter combinations for both BHXRBs M82 X-1 and GROJ 1655-40. Each trajectory has different parameter values \{$e$, $r_p$, $Q$\} and is indicated by a different color, where we fixed the spin of the black hole to $a=0.2994$ for M82 X-1 and $a=0.283$ for GROJ 1655-40. Hence, these trajectories, having fundamental frequencies very close to each other and within the width of the QPO, together simulate the strong rms of the observed QPOs. The trajectories together span a torus in the region $4.7-9.08$ for M82 X-1 and $5.11-6.67$ for GROJ 1655-40, which should be the emission region for QPOs, where we expect precession frequencies of both the $eQ$ and $e0$ trajectories. The ISCO radius is $\sim5$ for both the cases of BHXRB. We suggest that the simultaneous HFQPO and LFQPO emission should be from a region that is close to the inner edge of the accretion disk ($r_{\rm in}$), where both $eQ$ and $e0$ trajectories span a torus; the disk edge could be a source of blobs that are generating QPOs, as we will argue later in \S \ref{gasflowmodel}. In contrast, a rigid body precession model is invoked by some authors \citep{Ingram2009,Ingram2011,Ingram2012}, where Lense$-$Thirring precession of a rigid torus is suggested as the origin of the type C QPOs. Here, instead of the rigid precession of a solid torus, we propose that a collective precession of various trajectories, spanning a torus region, explains the origin of HFQPOs and LFQPOs simultaneously. We argue that HFQPOs originate when $r_{\rm in}$ comes in very close to the black hole at some point during the outburst (the soft state). In the hard state, $r_{\rm in}$ is farther out, and the type C QPO is more frequent and it is more prone to the vertical oscillations ($\nu_{\rm np}$). This scenario explains the increase in the frequency of type C QPOs with a decrease in $r_{\rm in}$, while the spectrum transits from the hard to soft state. 
\begin{figure}
\mbox{
\subfigure[]{
\hspace{0.5cm}
\includegraphics[scale=0.5]{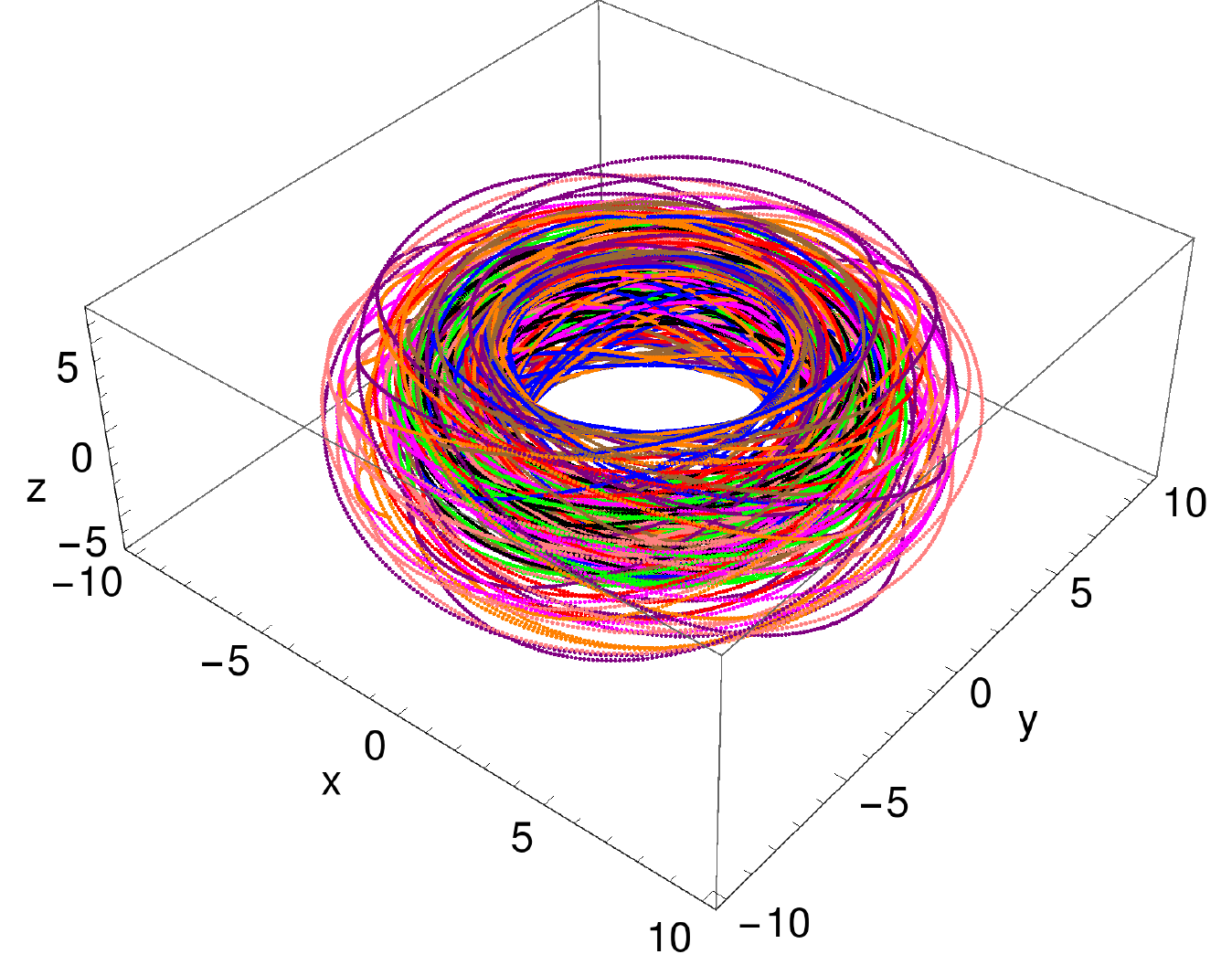}\label{trajectoriesM82}}
\hspace{1.7cm}
\subfigure[]{
\includegraphics[scale=0.5]{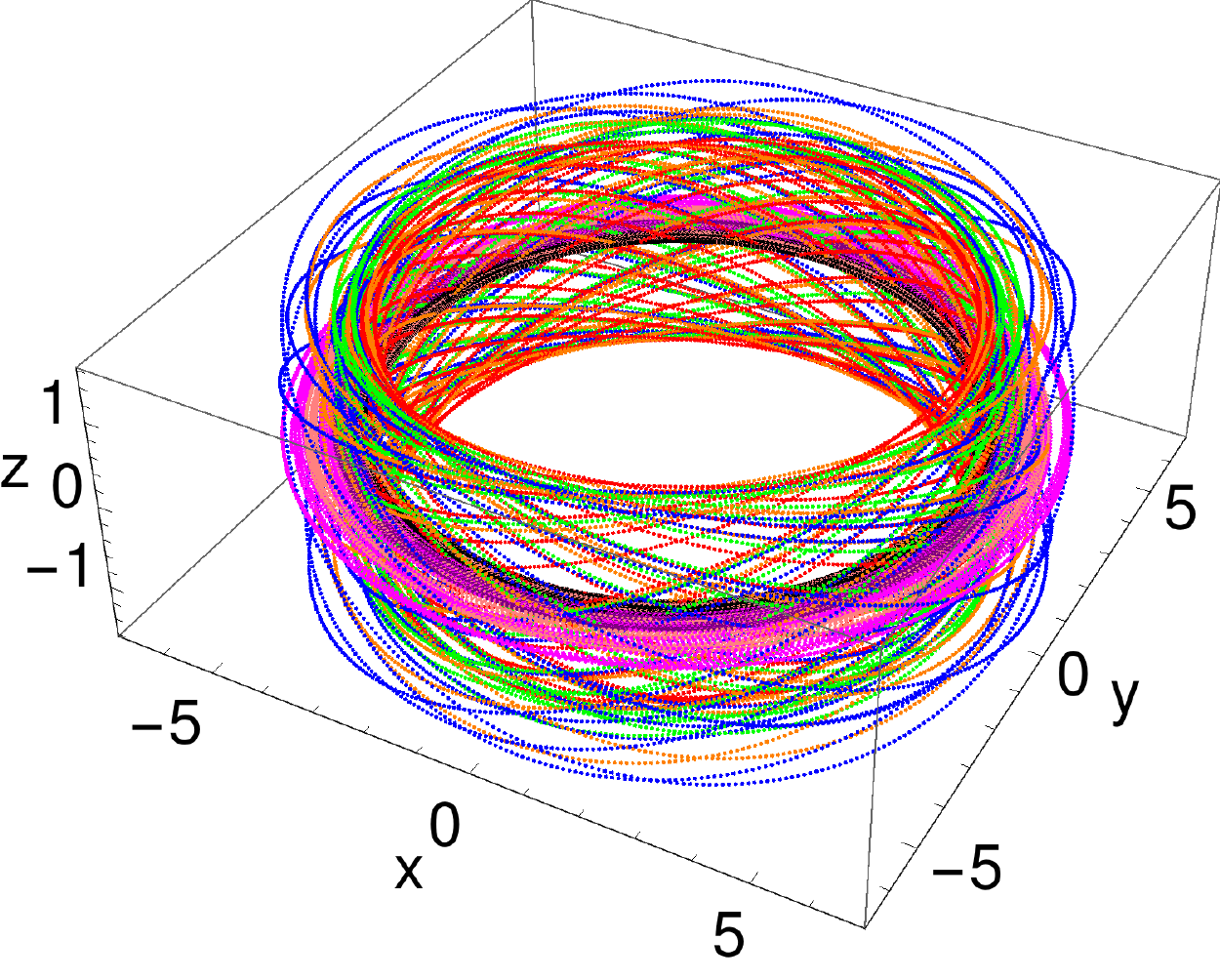}\label{trajectoriesGROJ}}}
\caption{\label{trajectories}The figures show various trajectories together having parameter combinations \{$e$, $r_p$, $Q$\} within the estimated range of 1$\sigma$ errors, as tabulated in the Table \ref{erQresults}, for (a) M82 X-1 and (b) GROJ 1655-40. The spin of the black hole is fixed to the most probable estimates, which are $a=0.2994$ for M82 X-1 and $a=0.283$ for GROJ 1655-40. Each color corresponds to a different parameter combination, where \{$e=0.18-0.29$; $r_p=4.7-5$; $Q=1-4$\} for M82 X-1 and \{$e=0.035-0.103$; $r_p=5.11-5.42$; $Q=0-0.6234$\} for GROJ 1655-40.}
\end{figure}
\item \textit{Two simultaneous QPOs}: We have considered only equatorial eccentric trajectories, $Q$=0, for these BHXRBs, as we can estimate only two parameters of the orbit corresponding to two simultaneous QPOs. First, we find the exact solutions for the parameters \{$e$, $r_p$\}, summarized in Table \ref{2QPOresults}, by equating the centroid frequencies of two simultaneous QPOs (see Table \ref{sourcelist}) to \{$\nu_{\phi}$, $\nu_{\rm pp}$\} using our analytic formulae for $Q$=0, Equations \eqref{eqnuphi} and \eqref{eqnur}. Then we calculate the errors in the parameters \{$e$, $r_p$\} using the method discussed in Appendix \ref{methodsection} (see Figure \ref{methodflowchart}). The results are summarized in Table \ref{2QPOresults}. These results are described below:
 \begin{itemize}
\item \textit{XTEJ 1550-564}: We find that an equatorial trajectory with eccentricity $e=0.262$ with $r_p=4.365$ (see Table \ref{2QPOresults}) as a solution for the observed QPO frequencies in XTEJ 1550-564. The calculated probability density profiles in $e$ and $r_p$ dimensions, $\mathcal{P}_{1} \left( e\right)$ and $\mathcal{P}_{1} \left( r_p\right)$, were found to be skew symmetric and were fit by an interpolating function. The corresponding errors were obtained by taking the integrated probability of 68.2\% about the peak value of the probability density distributions. The quoted errors are calculated with respect to the exact solution of the parameters, which slightly deviates from the peak of the integrated profiles \{$\mathcal{P}_1 \left( e\right)$ and $\mathcal{P}_1 \left( r_p\right)$\}; see Figure \ref{plotXTEJ} and Table \ref{2QPOresults}. These profiles, corresponding model fit, and the probability contours in the ($e$, $r_p$) plane are shown in Figure \ref{plotXTEJ}.  
\begin{longrotatetable}
\hspace{-7.0cm}
\begin{deluxetable}{l c c c c c c c c c c c c c}
\hspace{-7.0cm}
\tablecaption{\label{3QPOresults}Summary of Results Corresponding to (Non)equatorial Eccentric Solutions ($eQ$ and $e0$) for BHXRBs M82 X-1 and GROJ 1655-40.}
\tablewidth{300pt}
\tabletypesize{\tiny}
\tablehead{
\colhead{\textbf{BHXRB}} & \colhead{\textbf{ $Q$}}  & \colhead{\textbf{$e$ Range}} & \colhead{\textbf{Resolution}} & \colhead{\textbf{Exact}} & \colhead{\textbf{Model Fit}} & \colhead{\textbf{$r_p$ Range}}  & \colhead{\textbf{Resolution}} & \textbf{Exact} &  \textbf{\textbf{Model Fit}} & \textbf{$a$ Range}& \textbf{Resolution} & \textbf{Exact} & \textbf{\textbf{Model Fit}}\\
&  &  &\colhead{\textbf{$\Delta e$}} & \colhead{\textbf{Solution }}&  \colhead{\textbf{to $\mathcal{P}_1 \left( e \right)$}}  & & \colhead{\textbf{$\Delta r_p$}} & \colhead{ \textbf{Solution }} & \colhead{\textbf{ to $\mathcal{P}_{1} \left( r_p \right)$}} & & \colhead{\textbf{$\Delta a$}} & \colhead{\textbf{ Solution}} & \colhead{\textbf{ to $\mathcal{P}_{1} \left( a \right)$}}\\
&  &  &  & \colhead{\textbf{ $e_{0}$}} &  &  & & \colhead{\textbf{$r_{p0}$}} &  &  & & \colhead{\textbf{$a_0$}} &  
}
\startdata
M82 X-1 & 0 & $0.23-0.32$ & 0.001 & 0.277 &  0.277$^{+0.066}_{-0.045}$  & $4.4-4.85$ & 0.005 & 4.616  & 4.616$^{+0.069}_{-0.126}$ & $0.26-0.32$ & 0.001 & 0.290 &  $0.290 \pm 0.009$\\
&  &  &  & &  &  & &  &  &  & & &  \\
& 1 & $0.21-0.31$ & 0.001 & 0.259 &  0.259$^{+0.072}_{-0.045}$ & $4.3-5$ & 0.01 & 4.698 &  4.698$^{+0.068}_{-0.154}$ & $0.265-0.315$ & 0.001 & 0.294 &  $0.294 \pm 0.009$\\
&  &  &  & &  &  & &  &  &  & & &  \\
& 2 & $0.19-0.29$ & 0.001 & 0.239 &  0.239$^{+0.080}_{-0.046}$  & $4.45-5.1$ & 0.01 & 4.795 &  4.795$^{+0.066}_{-0.166}$ & $0.27-0.32$ & 0.001 & 0.298 &  $0.298 \pm 0.009$ \\
&  &  &  & &  &  & &  &  &  & & &  \\
& 3 & $0.16-0.26$ & 0.001 & 0.214 &  0.214$^{+0.090}_{-0.045}$ & $4.55-5.25$ & 0.01 & 4.913 &  4.913$^{+0.081}_{-0.163}$ & $0.28-0.33$ & 0.001 & 0.302 &  0.302$\pm$0.009 \\
&  &  &  & &  &  & &  &  &  & & &  \\
& 4 & $0.12-0.24$ & 0.001 & 0.187 &  0.187$^{+0.113}_{-0.047}$ & $4.65-5.35$ & 0.01 & 5.067 &  5.067$^{+0.076}_{-0.221}$ & $0.285-0.335$ & 0.001 & 0.308 &  0.308$\pm$0.009 \\
&  &  &  & &  &  & &  &  &  & & &  \\
&  &  &  & &  &  & &  &  &  & & &  \\
GROJ 1655-40& 0  & $0-0.22$ & 0.002 & - &  0.07$^{+0.042}_{-0.038}$ & $4.6-5.7$ & 0.01 & - & 5.24$^{+0.191}_{-0.186}$ & $0.265-0.3$ & 0.001 & - &  0.282$\pm$0.003\\
&  &  &  & &  &  & &  &  &  & & &  \\
& 1 & $0-0.22$ & 0.002 & - &  0.062$^{+0.040}_{-0.034}$ & $4.6-5.8$ & 0.015 & - &  5.305$^{+0.170}_{-0.185}$ & $0.24-0.32$ & 0.002 & - &  0.284$\pm$0.003 \\
&  &  &  & &  &  & &  &  &  & & &  \\
& 2 & $0-0.2$ & 0.002 & - &  0.056$^{+0.038}_{-0.031}$ & $4.7-5.85$ & 0.015 & - &  5.345$^{+0.167}_{-0.169}$ & $0.27-0.31$ & 0.001 & - &  0.286$\pm$0.003\\
&  &  &  & &  &  & &  &  &  & & & \\
& 3 & $0-0.2$ & 0.002 & - &  0.052$^{+0.036}_{-0.029}$ & $4.75-5.9$ & 0.015 & - &  5.395$^{+0.151}_{-0.170}$ & $0.275-0.32$ & 0.001 & - &  0.288$\pm$0.003\\
&  &  &  & &  &  & &  &  &  & & &  \\
& 4 & $0-0.2$ & 0.002 & - &  0.05$^{+0.034}_{-0.028}$ & $4.8-5.95$ & 0.015 & - &  5.43$^{+0.147}_{-0.162}$ & $0.278-0.31$ & 0.001 & - &  0.291$\pm$0.003\\
&  &  &  & &  &  & &  &  &  & & &  \\
\enddata
\tablecomments{ The columns describe the range of parameter volume considered for \{$e$, $r_p$, $a$\} with a chosen resolution to calculate the normalized probability density at each point inside the parameter volume using Equation \eqref{normPera}, the exact solutions for \{$e$, $r_p$, $a$\} calculated using Equations \eqref{nuphi}$-$\eqref{nutheta}, and the results of the model fit to $\mathcal{P}_{1} \left( e \right)$, $\mathcal{P}_{1} \left( r_p \right)$, and $\mathcal{P}_{1} \left( a \right)$, for each value of $Q$ between 0 and 4.}
\end{deluxetable}
\end{longrotatetable}
\begin{longrotatetable}
\begin{deluxetable}{l c c c c c c c c c c c c}
\tablecaption{\label{erQresults}\scriptsize{Summary of Results for \{$e$, $r_p$, $Q$\} Parameter Solutions and Corresponding Errors for QPOs in BHXRBs M82 X-1 and GROJ 1655-40.}}
\tablewidth{300pt}
\tabletypesize{\tiny}
\tablehead{
&   &  & &  &  & &  &  &  & & &  \\
\colhead{\textbf{BHXRB}} & \colhead{\textbf{$e$ Range}} & \colhead{\textbf{Resolution}} & \colhead{\textbf{Exact}} &  \colhead{ \textbf{Model Fit}}  & \colhead{\textbf{$r_p$ Range}} & \colhead{\textbf{Resolution}} & \colhead{\textbf{Exact}} &  \colhead{ \textbf{Model Fit}}  & \colhead{\textbf{$Q$ Range}}& \colhead{\textbf{Resolution}} & \colhead{\textbf{Exact}} &  \colhead{ \textbf{Model Fit}} \\
&   &\colhead{\textbf{$\Delta e$}} & \colhead{\textbf{ Solution}} & \colhead{\textbf{to $\mathcal{P}_{1} \left( e \right)$}}  & & \colhead{\textbf{$\Delta r_p$}} & \colhead{ \textbf{Solution}} & \colhead{\textbf{to $\mathcal{P}_{1} \left( r_p \right)$} }& & \colhead{\textbf{$\Delta Q$}} & \colhead{\textbf{ Solution}} & \colhead{\textbf{to $\mathcal{P}_{1} \left( Q \right)$}} \\
&  &   & \colhead{\textbf{$e_0$}}&  &  & & \colhead{\textbf{$r_{p0}$}} &  &  & & \colhead{\textbf{$Q_0$}} &   
}
\startdata
M82 X-1  & $0.1-0.35$ & 0.002 & 0.230 & $0.230_{-0.049}^{+0.067}$  & $4.2-5.4$ & 0.02 & 4.834 & $4.834_{-0.268}^{+0.181}$  & $0-5$  & 0.1  & 2.362 & $2.362_{-1.439}^{+1.519}$ \\
&   &  & &  &  & &  &  &  & & &  \\
GROJ 1655-40 & $0-0.18$ & 0.001 & - & $0.071_{-0.035}^{+0.031}$  & $4.9-5.75$ & 0.0125 & - & $5.25_{-0.142}^{+0.171}$  & $0-3$  & 0.1  & - & $0^{+0.623}$ \\
&   &  & &  &  & &  &  &  & & &  \\ 
\enddata
\tablecomments{ The columns describe the range of parameter volume taken for \{$e$, $r_p$, $Q$\}, and the chosen resolution to calculate the normalized probability density at each point inside the parameter volume, the exact solutions, and the results of the model fit to the integrated profiles. The spin of the black hole is fixed to the most probable estimates, which are $a=0.2994$ for M82 X-1 and $a=0.283$ for GROJ 1655-40.}
\end{deluxetable}
\end{longrotatetable}
\item \textit{4U 1630-47}: We found an exact solution at \{$e=0.734$, $r_p=2.249$\} (see Table \ref{2QPOresults}) by equating \{$\nu_{\phi}$, $\nu_{\rm np}$\} instead of \{$\nu_{\phi}$, $\nu_{\rm pp}$\} to the centroid QPO frequencies. This might be because the QPO with a lower frequency of $\sim 38$Hz (see Table \ref{sourcelist}) is too small to be an HFQPO. The calculated probability density profiles in the $e$ and $r_p$ dimensions, the corresponding model fit, and the probability contours in the ($e$, $r_p$) plane are shown in Figure \ref{plot4U}. In this case, too, we see that $\mathcal{P}_{1} \left( e\right)$ and $\mathcal{P}_{1} \left( r_p\right)$ profiles are skew, such that the integrated probability is 68.2\% about the peak value of the probability density distributions, and the errors are quoted with respect to the exact solution of the parameters, which slightly deviates from the peak of the integrated profiles $\mathcal{P}_1 \left( e\right)$ and $\mathcal{P}_1 \left( r_p\right)$ (see Figure \ref{plot4U} and Table \ref{2QPOresults}). We see that a highly eccentric orbit is found as the most probable solution.
\begin{deluxetable}{l c c c c c c c c}
\tablecaption{\label{2QPOresults}Summary of Results Corresponding to the Equatorial Eccentric Orbit Solutions for BHXRBs XTEJ 1550-564, 4U 1630-47, and GRS 1915+105.}
\tablewidth{250pt}
\tabletypesize{\footnotesize}
\tablehead{
\colhead{\textbf{BHXRB}} & \colhead{\textbf{$e$ Range}} & \colhead{\textbf{Resolution}} & \colhead{\textbf{Exact}} & \colhead{\textbf{Model Fit}}  & \colhead{\textbf{$r_p$ Range}} & \colhead{\textbf{Resolution}} & \colhead{\textbf{Exact}} & \colhead{\textbf{Model Fit}} \\
&   &\colhead{\textbf{$\Delta e$}} & \colhead{\textbf{ Solution}} &  \colhead{\textbf{to $\mathcal{P}_1 \left( e \right)$}}  & & \colhead{\textbf{$\Delta r_p$}} &  \colhead{\textbf{Solution}} & \colhead{\textbf{ to $\mathcal{P}_{1} \left( r_p \right)$}} \\
&   &  & \colhead{\textbf{$e_{0}$}} &  &  &  & \colhead{\textbf{$r_{p0}$}} & 
}
\startdata
 XTEJ 1550-564 & $0.01-0.7$ & 0.0005 & 0.262 &  0.262$^{+0.090}_{-0.062}$  & $2.5-6.5$ & 0.005 & 4.365 &  4.365$^{+0.169}_{-0.279}$  \\
&   &  & &  &  & & & \\
4U 1630-47 & $0.4-0.99$ & 0.0005 & 0.734 &  0.734$^{+0.066}_{-0.048}$ & $1-4.5$ & 0.005 & 2.249 &  2.249$^{+0.249}_{-0.353}$ \\
&   &  & &  &  & & &  \\
GRS 1915+105 & $0.6-0.999$ & 0.0005 & 0.918 &  0.918$\pm$0.002 & $0.3-4$ & 0.005 & 1.744 & 1.744$^{+0.025}_{-0.011}$  \\
&   &  & &  &  & & &\\
\enddata
\tablecomments{ The columns describe the parameter range considered for \{$e$, $r_p$\}, its resolution, the exact solutions for \{$e$, $r_p$\} calculated using \{$\nu_{\phi}$, $\nu_{\rm pp}$\} for XTEJ 1550-564 and GRS 1915+105, and using \{$\nu_{\phi}$, $\nu_{\rm np}$\} for 4U 1630-47 using Equations \eqref{eqnuphi}, \eqref{eqnur}, and \eqref{eqnutheta}, and results of the model fit to $\mathcal{P}_{1} \left( e \right)$ and $\mathcal{P}_{1} \left( r_p \right)$.}
\end{deluxetable}
    \begin{figure}
  \mbox{
  \hspace{1.3cm}
\subfigure[]{
\includegraphics[scale=0.5]{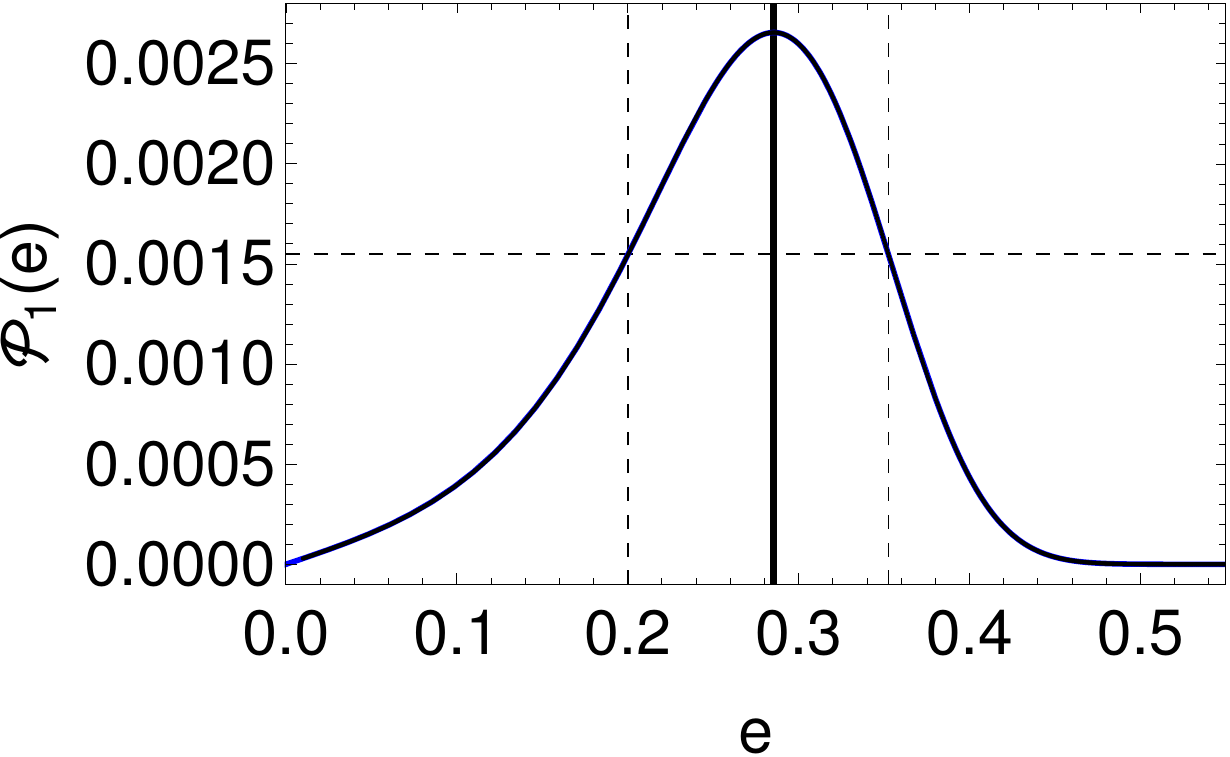}\label{XTEJa}}
\hspace{1.8cm}
\subfigure[]{
\includegraphics[scale=0.53]{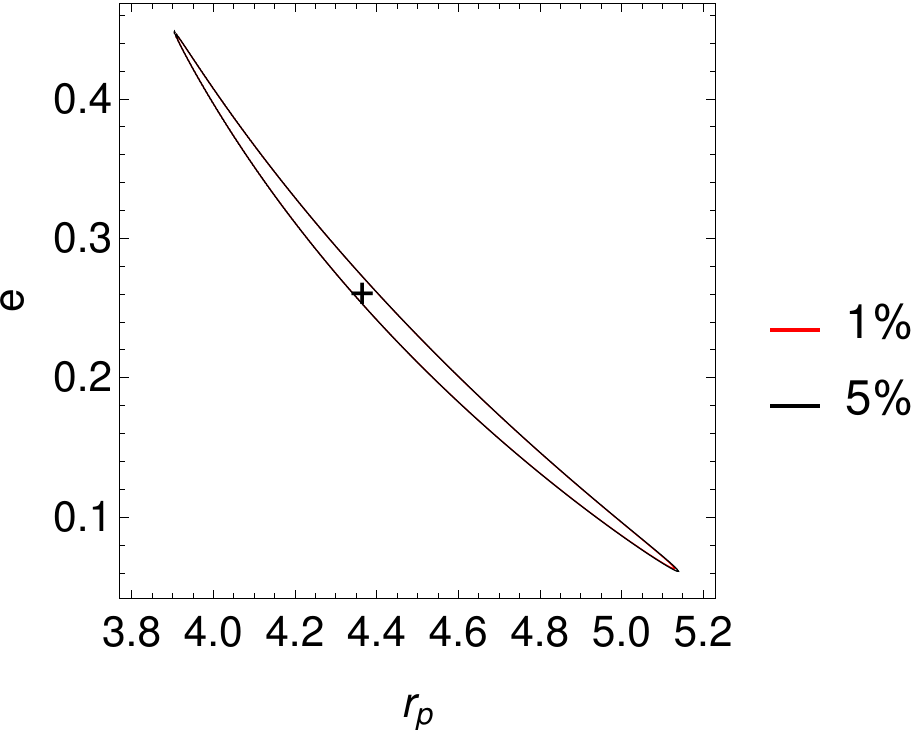}\label{XTEJb}}}
  \mbox{
  \hspace{1.8cm}
\subfigure[]{
\includegraphics[scale=0.53]{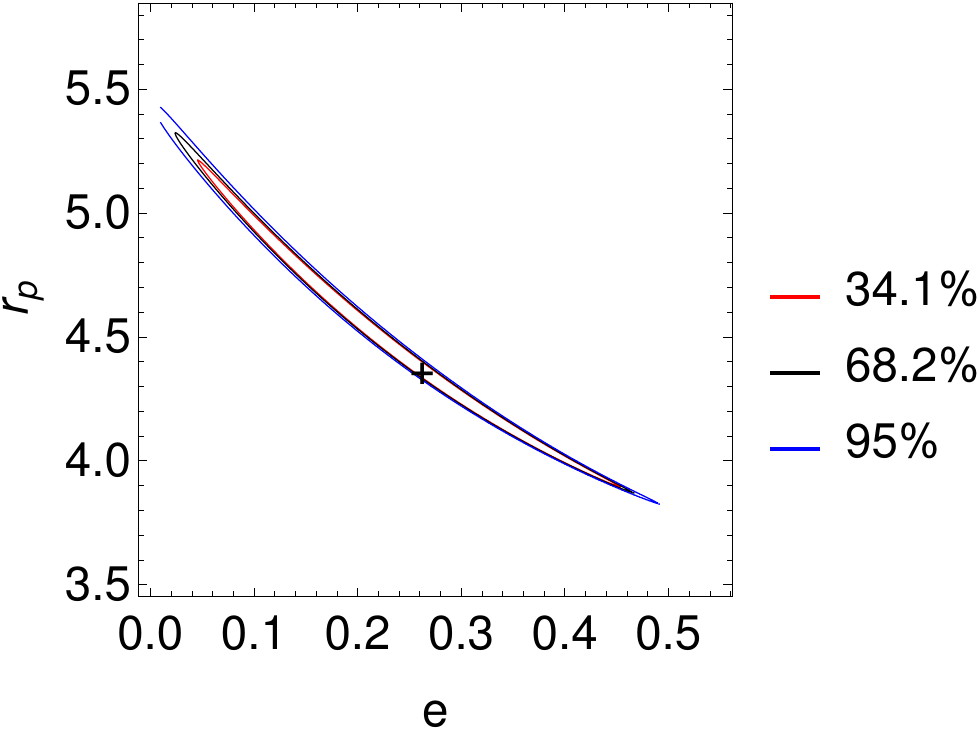}\label{XTEJd}}
\hspace{1.8cm}
\subfigure[]{
\includegraphics[scale=0.5]{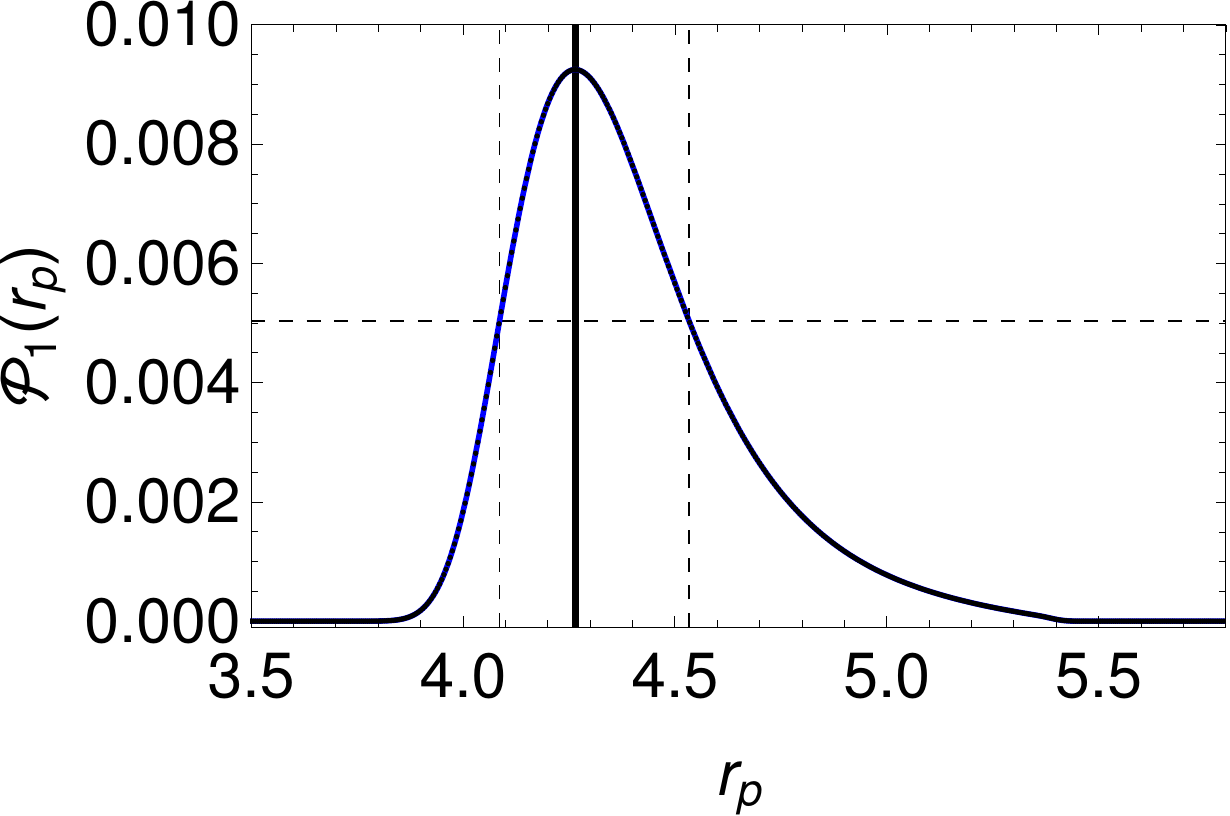}\label{XTEJe}}}
  \caption{\label{plotXTEJ} The integrated density profiles of BHXRB XTEJ 1550-564 are shown in (a) $\mathcal{P}_1 \left( e\right)$ and (d) $\mathcal{P}_1 \left(r_p\right)$ , where the dashed vertical lines enclose a region with 68.2\% probability, and the solid vertical line corresponds to the peak of the profiles. The probability contours of the parameter solution are shown in the (b) ($r_p$, $e$) and (c) ($e$, $r_p$) planes, where the $+$ sign marks the exact solution.}
  \end{figure}
 
  \begin{figure}
  \mbox{
  \hspace{1.3cm}
\subfigure[]{
\includegraphics[scale=0.5]{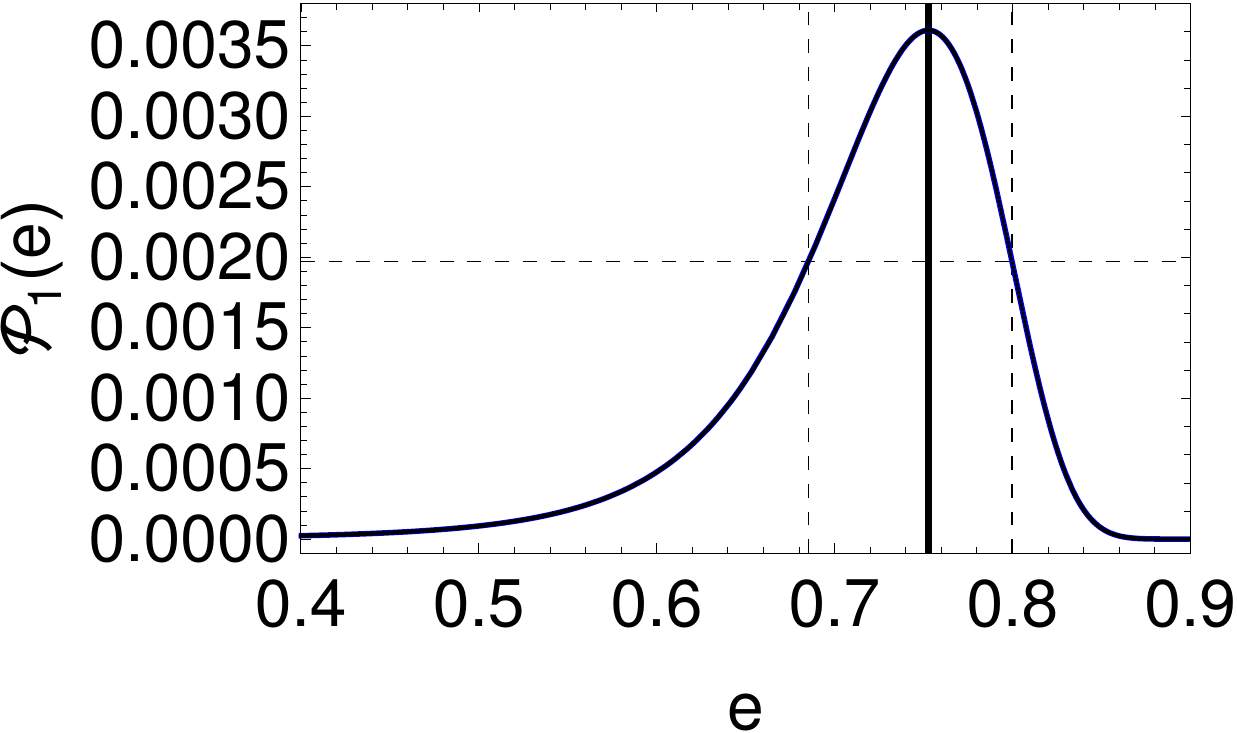}\label{4Ua}}
\hspace{1.8cm}
\subfigure[]{
\includegraphics[scale=0.53]{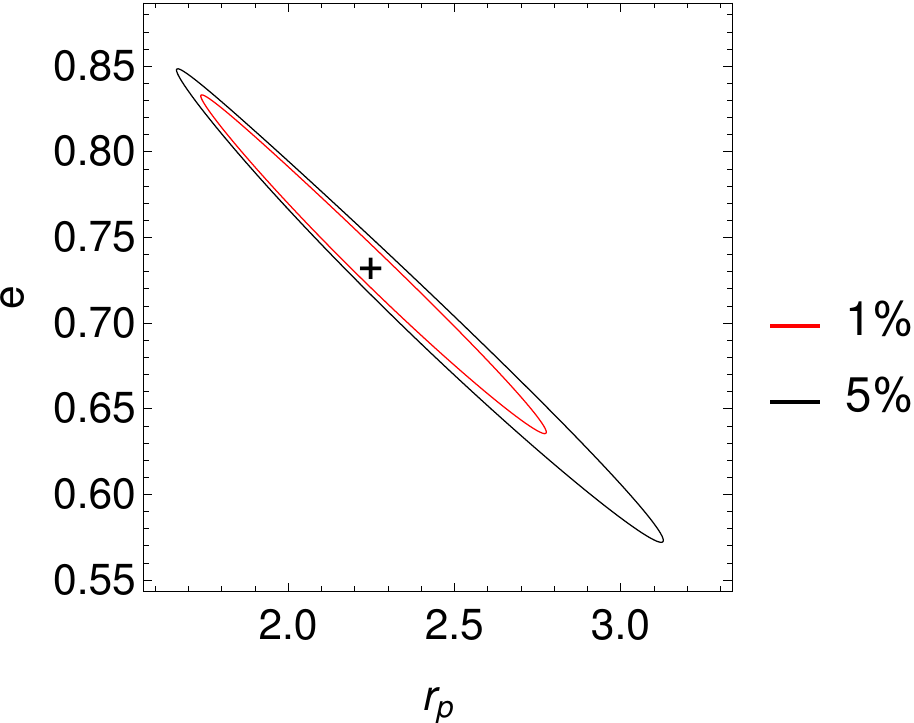}\label{4Ub}}}
  \mbox{
  \hspace{1.8cm}
\subfigure[]{
\includegraphics[scale=0.53]{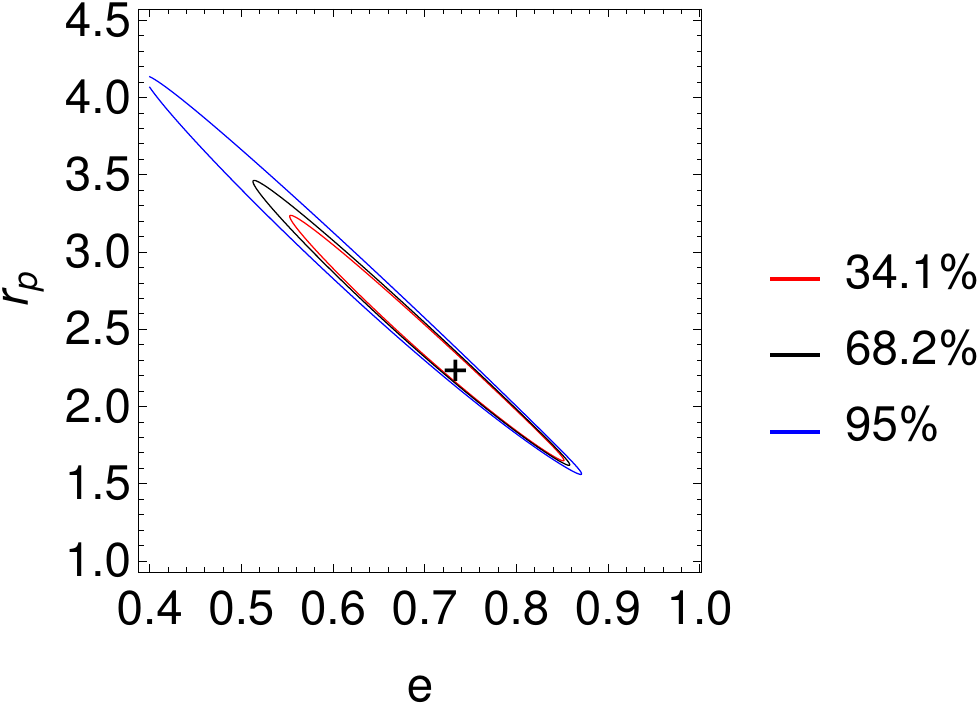}\label{4Ud}}
\hspace{1.8cm}
\subfigure[]{
\includegraphics[scale=0.5]{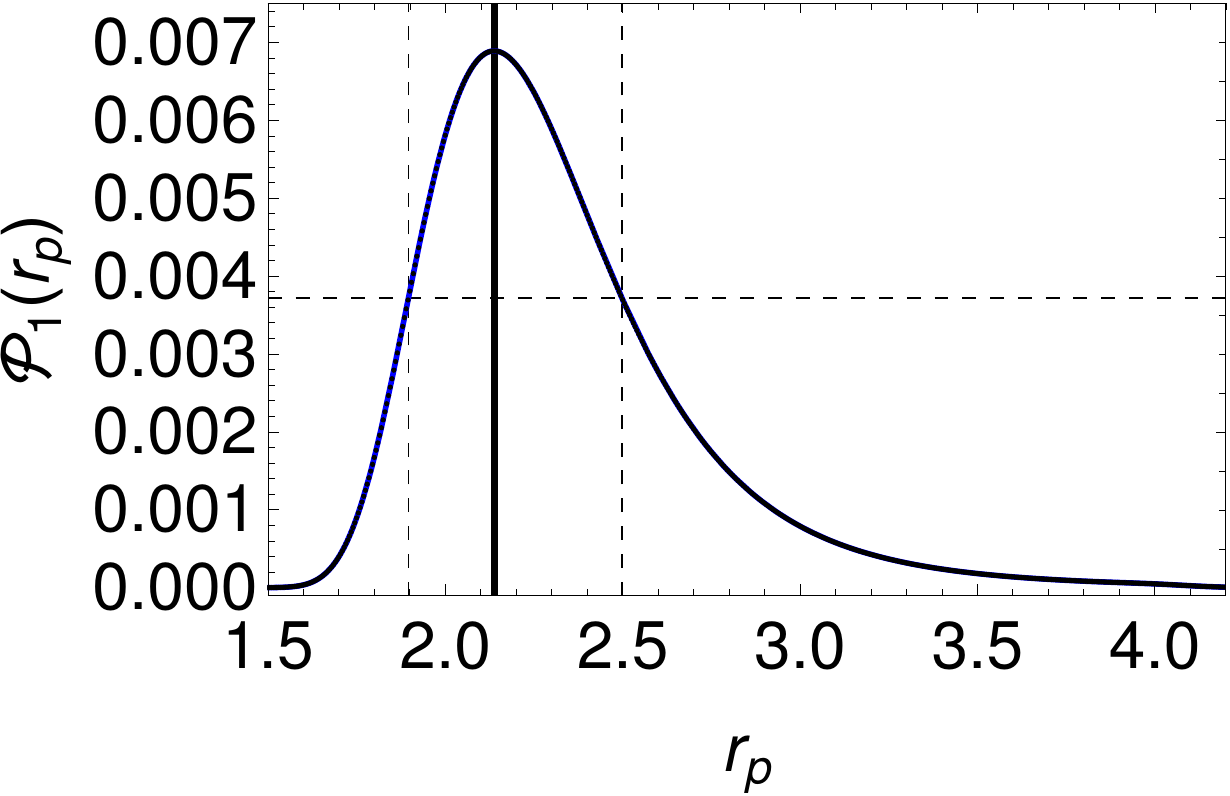}\label{4Ue}}}
  \caption{\label{plot4U} The integrated density profiles are shown in (a) $\mathcal{P}_1 \left( e\right)$ and (d) $\mathcal{P}_1 \left(r_p\right)$ for BHXRB 4U 1630-47, where the dashed vertical lines enclose a region with 68.2\% probability, and the solid vertical line corresponds to the peak of the profiles. The probability contours of the parameter solution are shown in the (b) ($r_p$, $e$) and (c) ($e$, $r_p$) planes, where the $+$ sign marks the exact solution.}
  \end{figure}
\item \textit{GRS 1915+105}: We found an exact solution at \{$e=0.918$, $r_p=1.744$\}; see Table \ref{2QPOresults}. We find a highly eccentric equatorial trajectory as the most probable solution that can give the observed QPO frequencies in GRS 1915+105. This result is similar to the case of 4U 1630-47, which leads us to observe that a black hole with a high spin value prefers a highly eccentric orbit solution to simultaneous QPOs. The calculated probability density profiles $\mathcal{P}_{1} \left( e\right)$ and $\mathcal{P}_{1} \left( r_p\right)$ are well fit by the Gaussian. The corresponding model fit and the probability contours in the ($e$, $r_p$) plane are shown in Figure \ref{plotGRS}. 
\end{itemize} 
 \begin{figure}
  \mbox{
  \hspace{1.3cm}
\subfigure[]{
\includegraphics[scale=0.3]{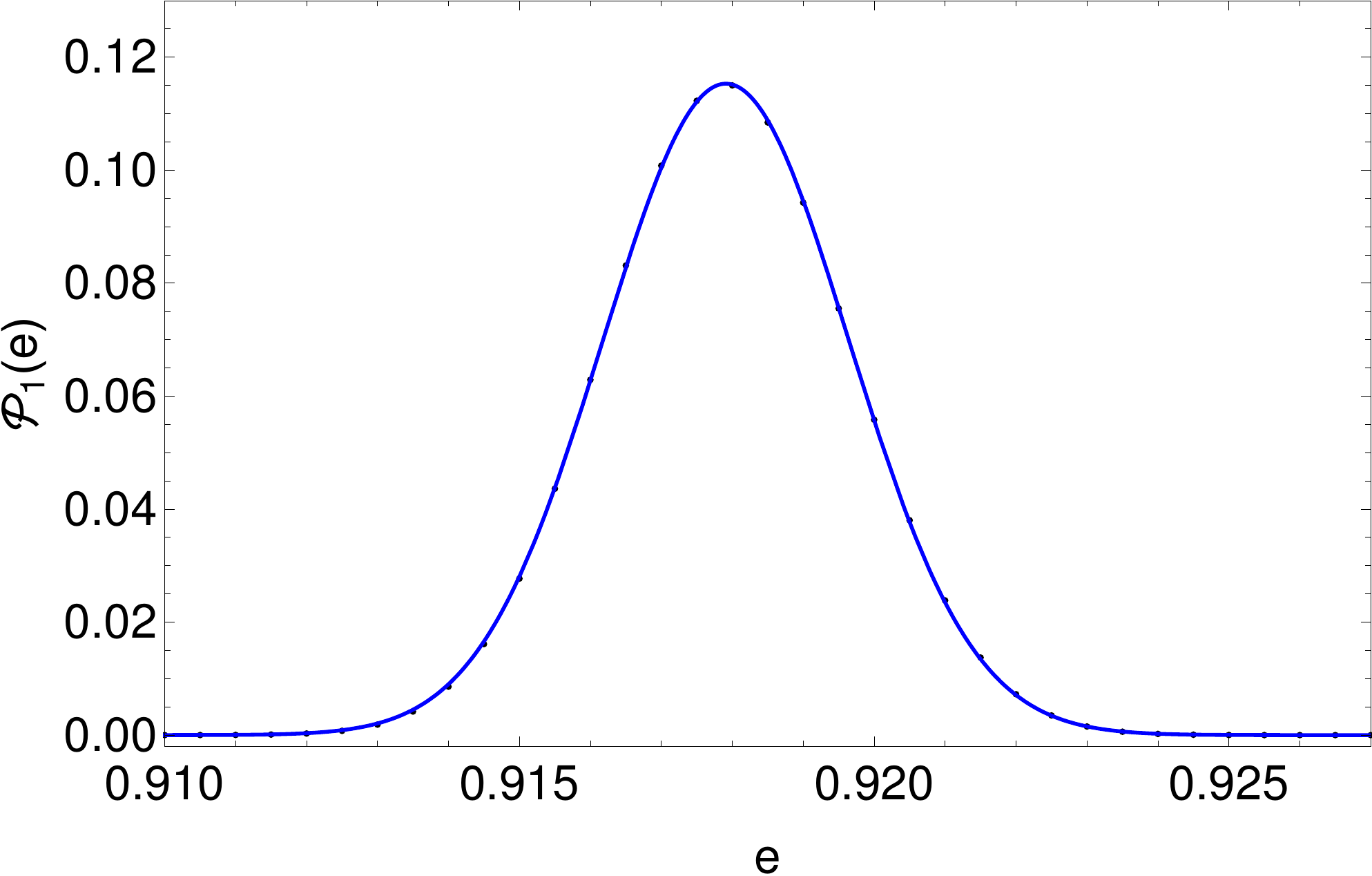}\label{GRSa}}
\hspace{1.8cm}
\subfigure[]{
\includegraphics[scale=0.35]{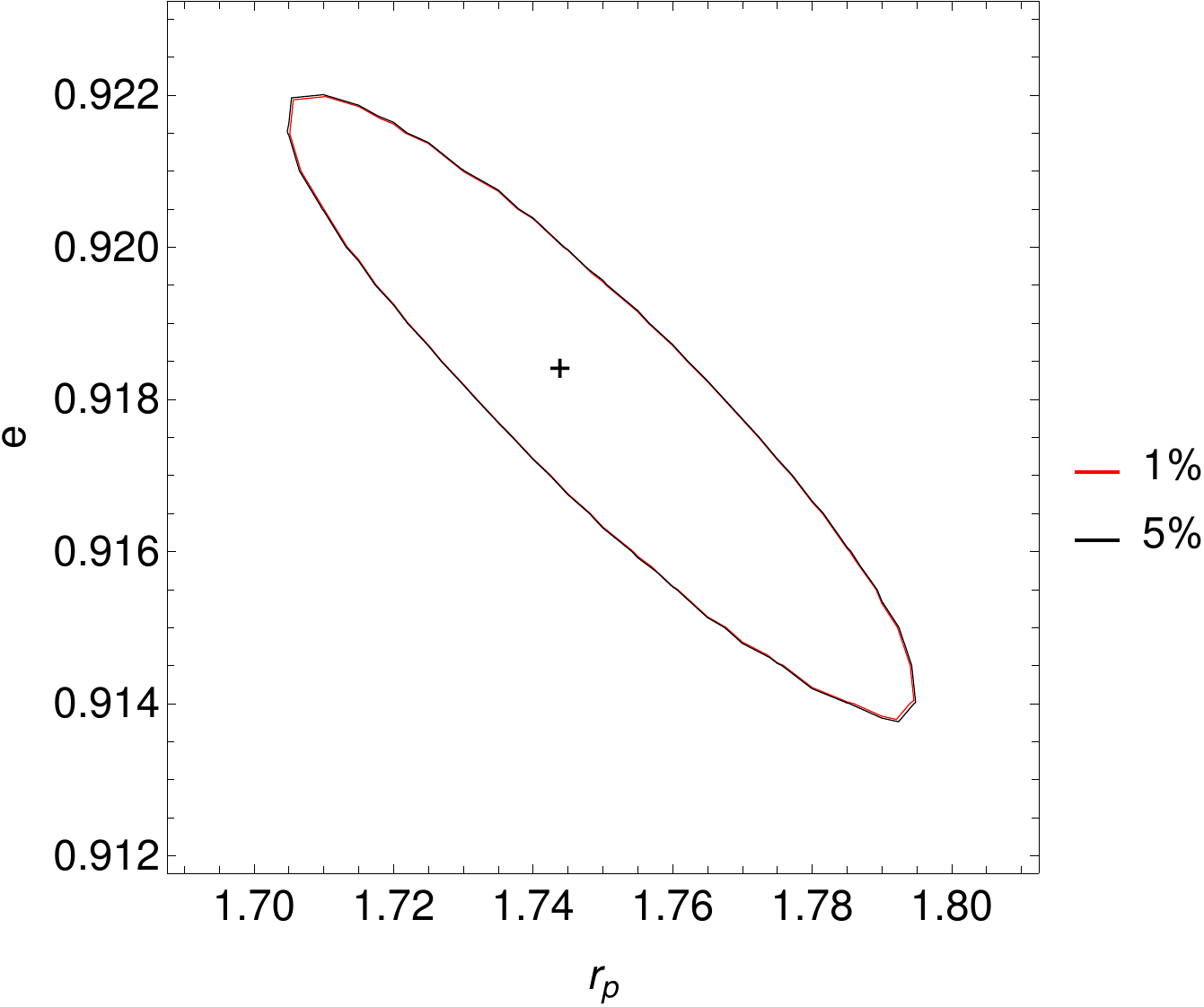}\label{GRSb}}}
  \mbox{
  \hspace{1.8cm}
\subfigure[]{
\includegraphics[scale=0.36]{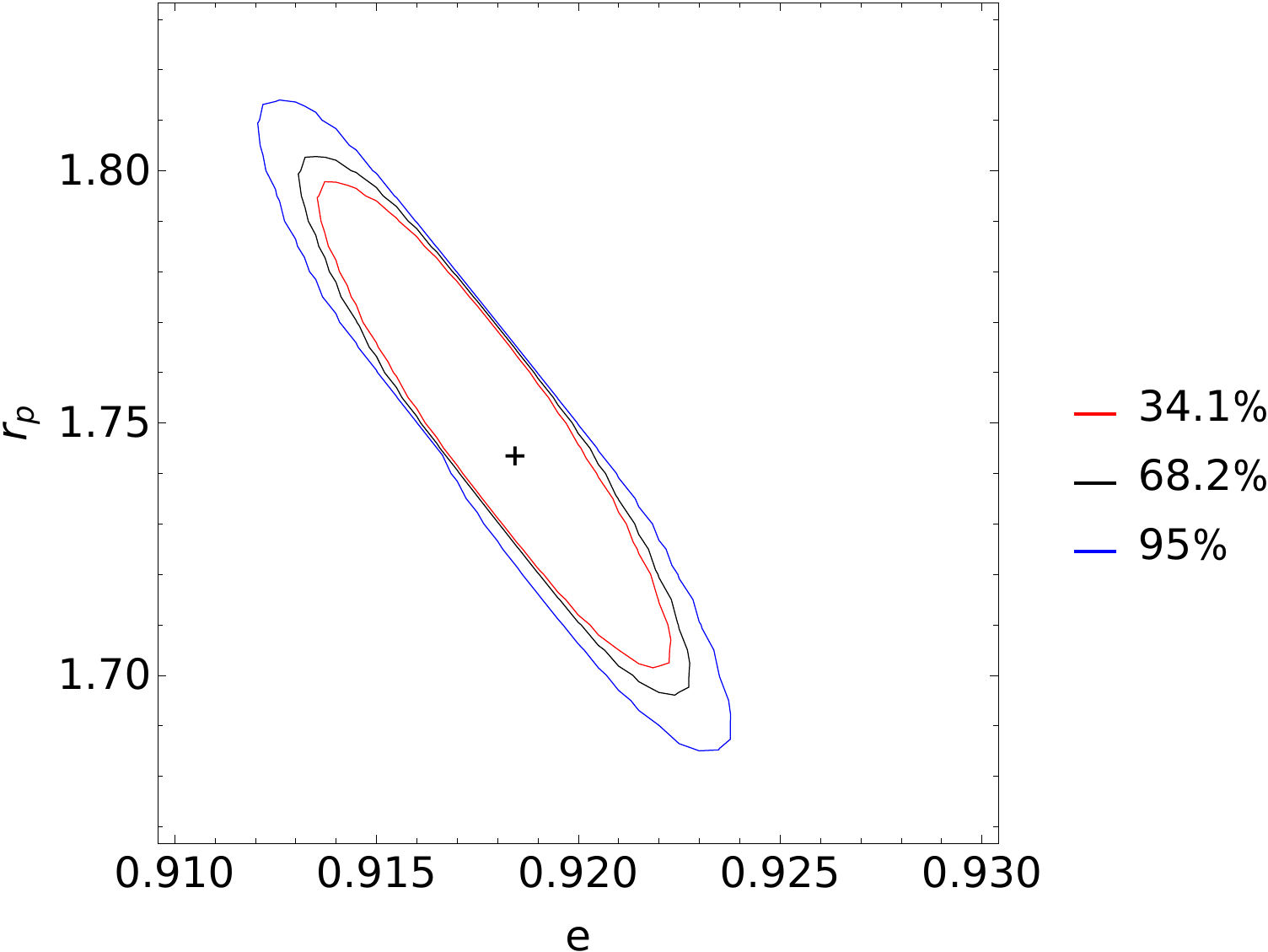}\label{GRSd}}
\hspace{1.8cm}
\subfigure[]{
\includegraphics[scale=0.3]{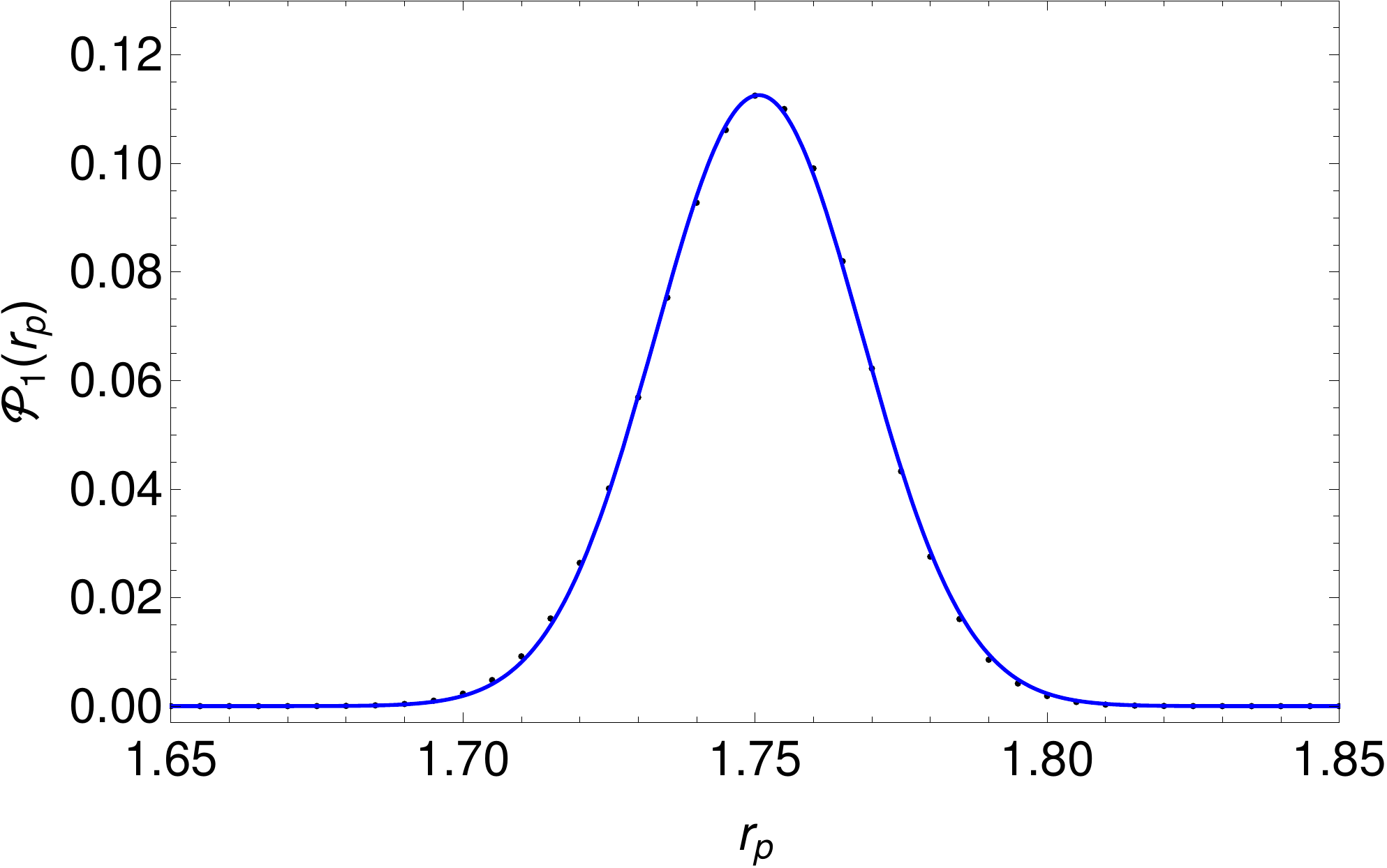}\label{GRSe}}}
  \caption{\label{plotGRS} The integrated density profiles are shown in (a) $\mathcal{P}_1 \left( e\right)$ and (d) $\mathcal{P}_1 \left(r_p\right)$ for BHXRB GRS 1915+105, where the dashed vertical lines enclose a region with 68.2\% probability, and the solid vertical line corresponds to the peak of the profiles. The probability contours of the parameter solution are shown in the (b) ($r_p$, $e$) and (c) ($e$, $r_p$) planes, where the $+$ sign marks the exact solution.}
  \end{figure}
 Hence, we conclude that for XTEJ 1550-564, 4U 1630-47, and GRS 1915+105, the $e0$ model in the region $r_p=1.74-4.36$ are the probable cause of the observed QPOs in the power spectrum. We found high eccentricity values for the orbits as solutions for QPOs in the cases of BHXRB 4U 1630-47 and GRS 1915+105, and this indicates that black holes with very high spin values prefer highly eccentric orbits in the QPO solutions.
\end{enumerate}  
\begin{figure}
\mbox{
 \subfigure[]{
 \hspace{1.0cm}
\includegraphics[scale=0.28]{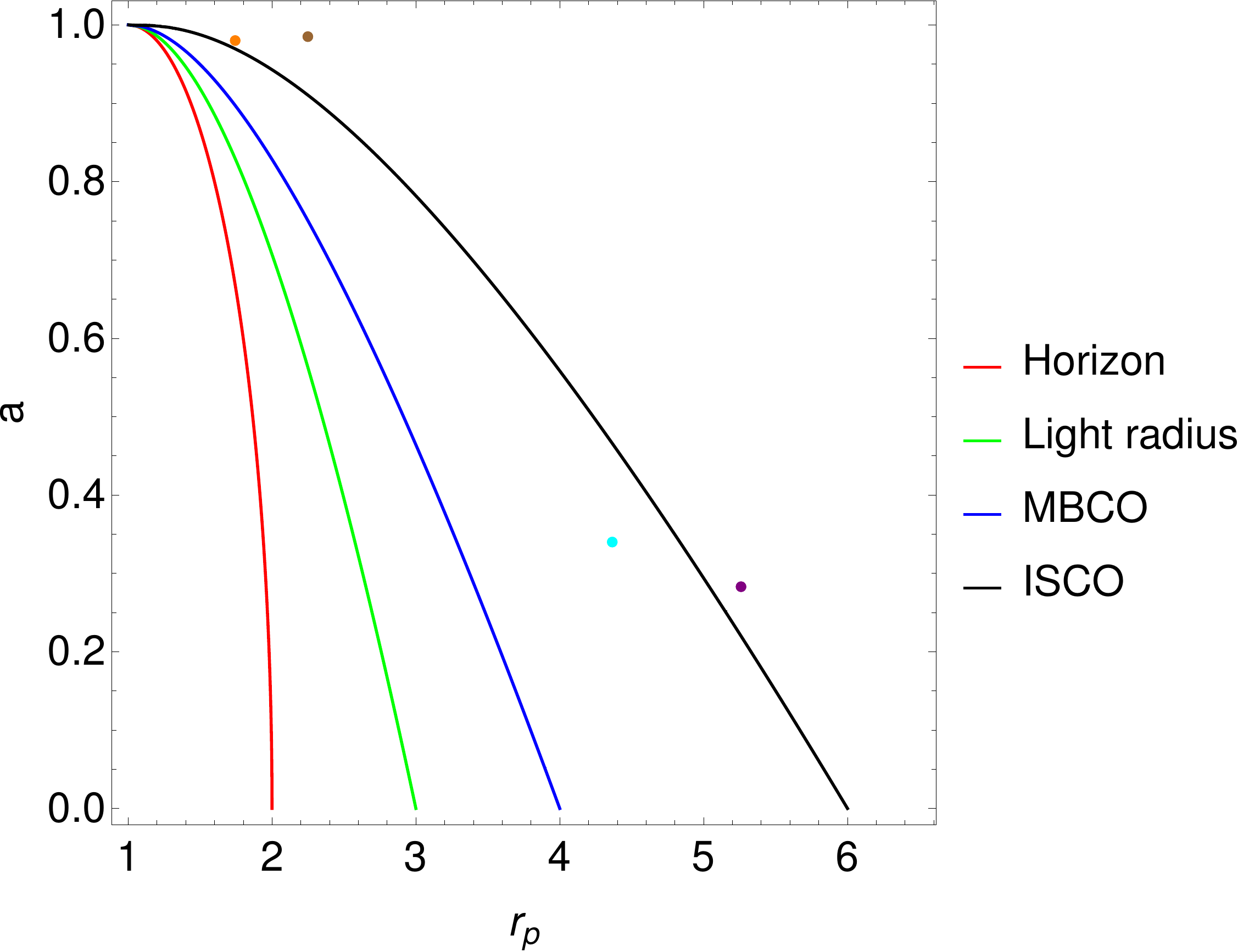}\label{plotrpaplaneQ0}}
\hspace{1.5cm}
 \subfigure[]{
\includegraphics[scale=0.4]{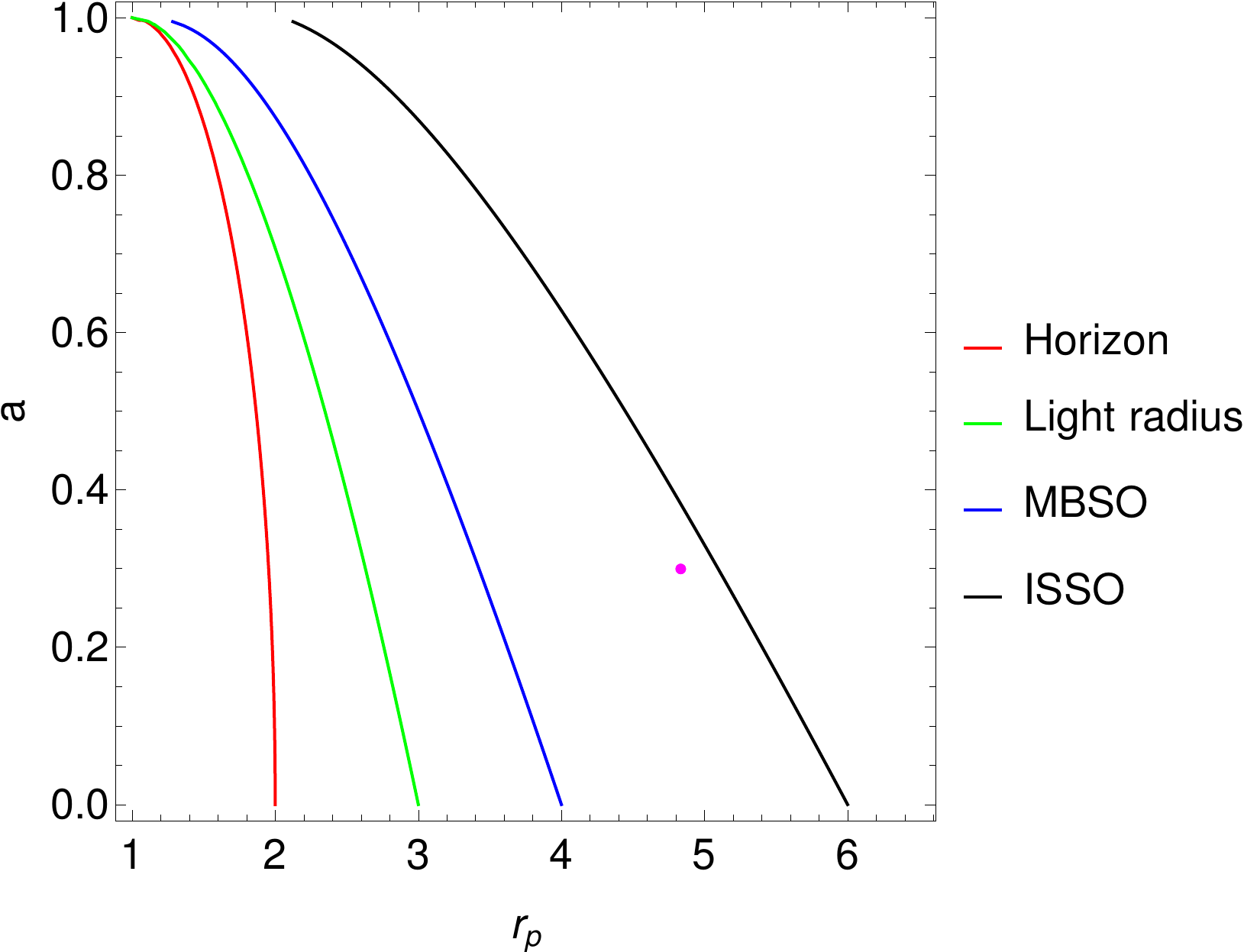}\label{plotrpaplaneQ2}}}
\caption{\label{plotrpaplane}Equatorial eccentric orbit solutions for QPOs observed in BHXRBs GROJ 1655-40 (purple), XTEJ 1550-564 (cyan), 4U 1630-47 (brown), and GRS 1915+105 (orange) for (a) $Q=0$; and (b) the nonequatorial eccentric orbit solution for BHXRB M82 X-1 (magenta) for $Q=2.362$.  }
\end{figure}

 We show all of the eccentric trajectory solutions together for both $Q=0$ and $Q\neq0$ in Figure \ref{plotrpaplane} in the ($r_p$, $a$) plane along with the radii ISCO (ISSO), MBCO (MBSO), light radius, and the horizon. We see that the calculated eccentric orbit solutions are found in region 1 of the ($r_p$, $a$) plane (as defined in Figure \ref{radii}) and near ISCO for $Q=0$ in the cases of BHXRB 4U 1630-47, GROJ 1655-40, and GRS 1915+105. The trajectory solutions are found in region 2 near ISCO for XTEJ 1550-564 ($Q=0$) and near ISSO for M82 X-1 ($Q=2.362$; as defined in Figure \ref{radii}). These results are also consistent with the results discussed in \S \ref{eccentricmotivation}, except that very high $e$ values are found for trajectories in BHXRB 4U 1630-47 and GRS 1915+105. Hence, we conclude that the eccentric trajectory solutions with $Q=0$ and $Q\neq0$ for the observed QPOs in BHXRBs are found either in the region 1 or region 2 of the ($r_p$, $a$) plane but close to the ISCO (ISSO) curve; we call this radius as $R_0$. As all these orbit solutions are distributed near $R_0$, it is expected that this radius is very close to the inner edge radius, $r_{\rm in}$, of the circular accretion disk, which could also be a source of blobs that are generating these QPOs. The torus region, shown in Figure \ref{trajectories}, spans a part of regions 1 and 2 near $R_0$, which can be represented as $\left({R_0}^{+\Delta_1}_{-\Delta_2}\right)$, where $\Delta_i$ represents a small deviation from $R_0$ (which need not be the center point of the torus in this scenario). This means that the orbits near $R_0$ are induced by the instabilities in the inner flow to be (non)equatorial and eccentric.
 
\subsubsection{Spherical Orbits}
\label{resultspherical}
Here we summarize the results of associating the spherical orbits around a Kerr black hole with QPOs in BHXRBs. We limited this study to the cases of BHXRBs M82 X-1 and XTEJ 1550-564, as we found the exact solutions for the parameters \{$r_s$, $a$, $Q$\} or \{$r_s$, $a$\} for only these two BHXRBs when we solved \{$\nu_{\phi}=\nu_{10}$, $\nu_{\rm pp}=\nu_{20}$, $\nu_{\rm np}=\nu_{30}$\} for M82 X-1 and \{$\nu_{\phi}=\nu_{10}$, $\nu_{\rm pp}=\nu_{20}$\} for XTEJ 1550-564 using Equations (\ref{nuphisph2}$-$\ref{nuthetasph2}). We calculated errors for the parameters using the method discussed in Appendix \ref{methodsection} (also see Figure \ref{methodflowchart}); these results are summarized in the Table \ref{sphresults} and are presented below:
  \begin{figure}[hbt!]
\mbox{
\subfigure[]{
\hspace{1.5cm}
\includegraphics[scale=0.4]{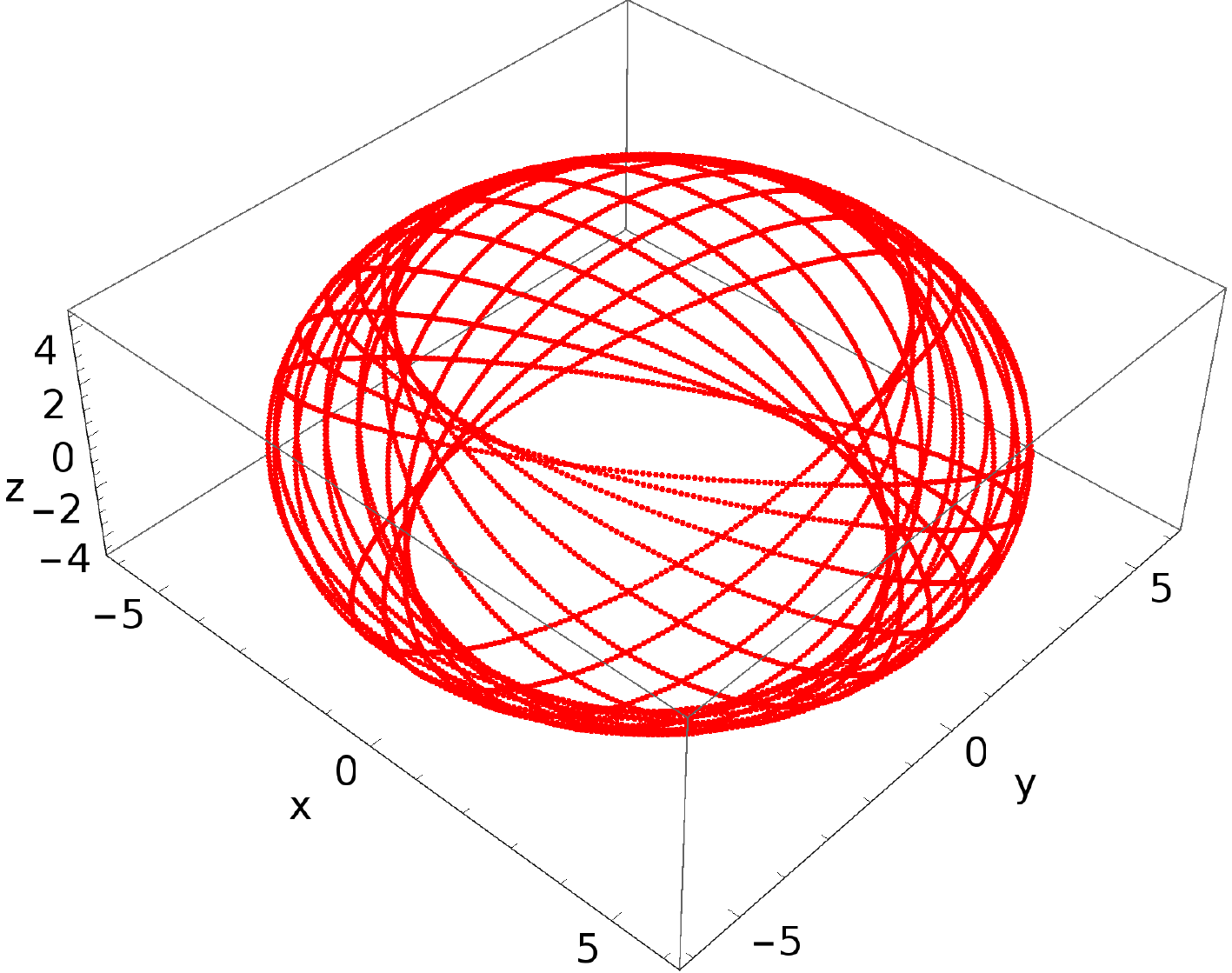}\label{M82sphtrajec}}
\hspace{1.7cm}
\subfigure[]{
\includegraphics[scale=0.4]{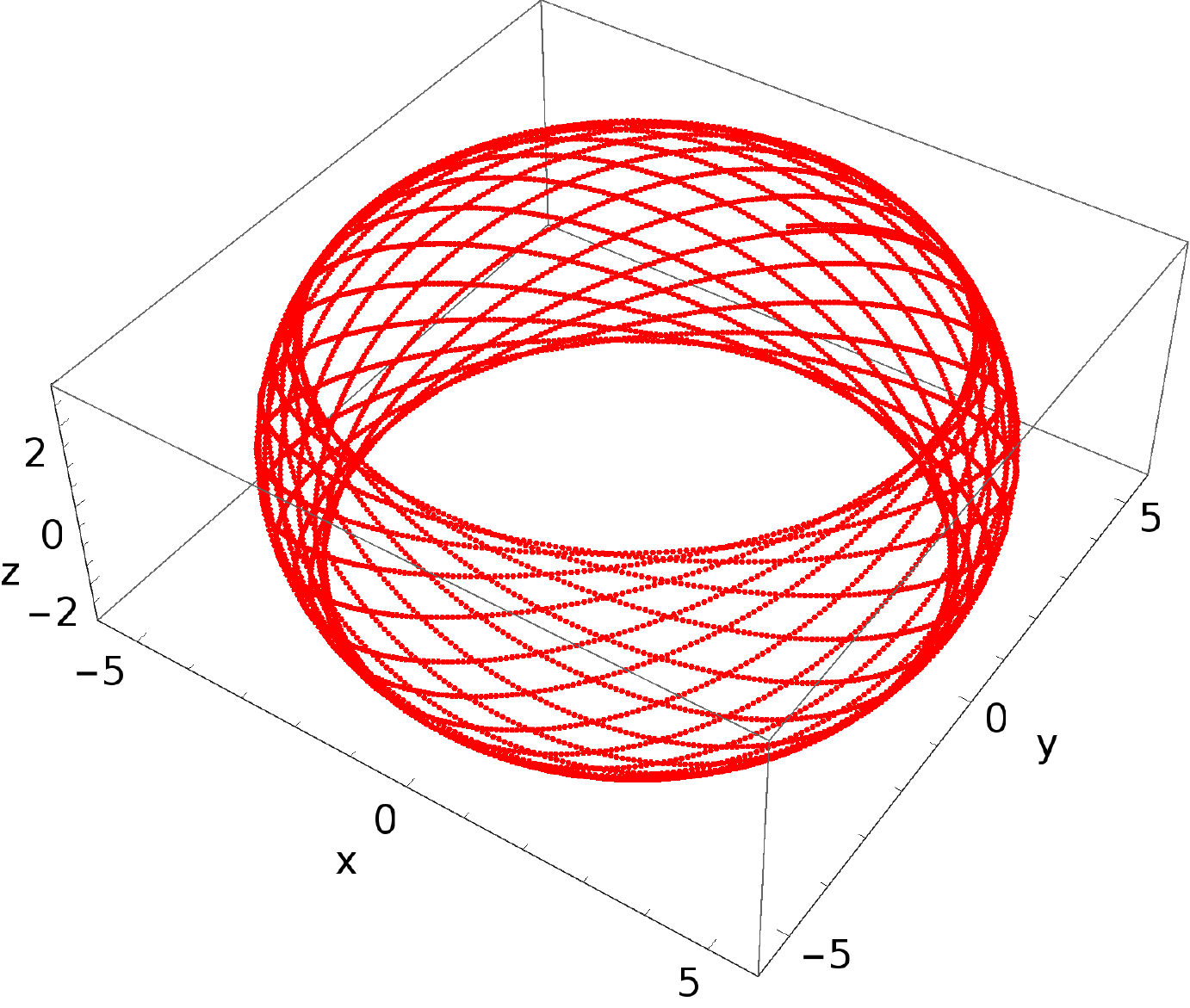}\label{XTEJsphtrajec}}}
\caption{\label{sphericaltrajectories}Spherical trajectories corresponding to the exact solutions calculated for (a) M82 X-1 at \{$r_s=6.044$, $a=0.321$, $Q=6.113$\} and for (b) XTEJ 1550-564 at \{$r_s=5.538$, $a=0.34$, $Q=2.697$\}, as also provided in Table \ref{sphresults}. }
\end{figure}
\begin{figure}[hbt!]
\mbox{
\hspace{-0.7cm}
 \subfigure[]{
\includegraphics[scale=0.29]{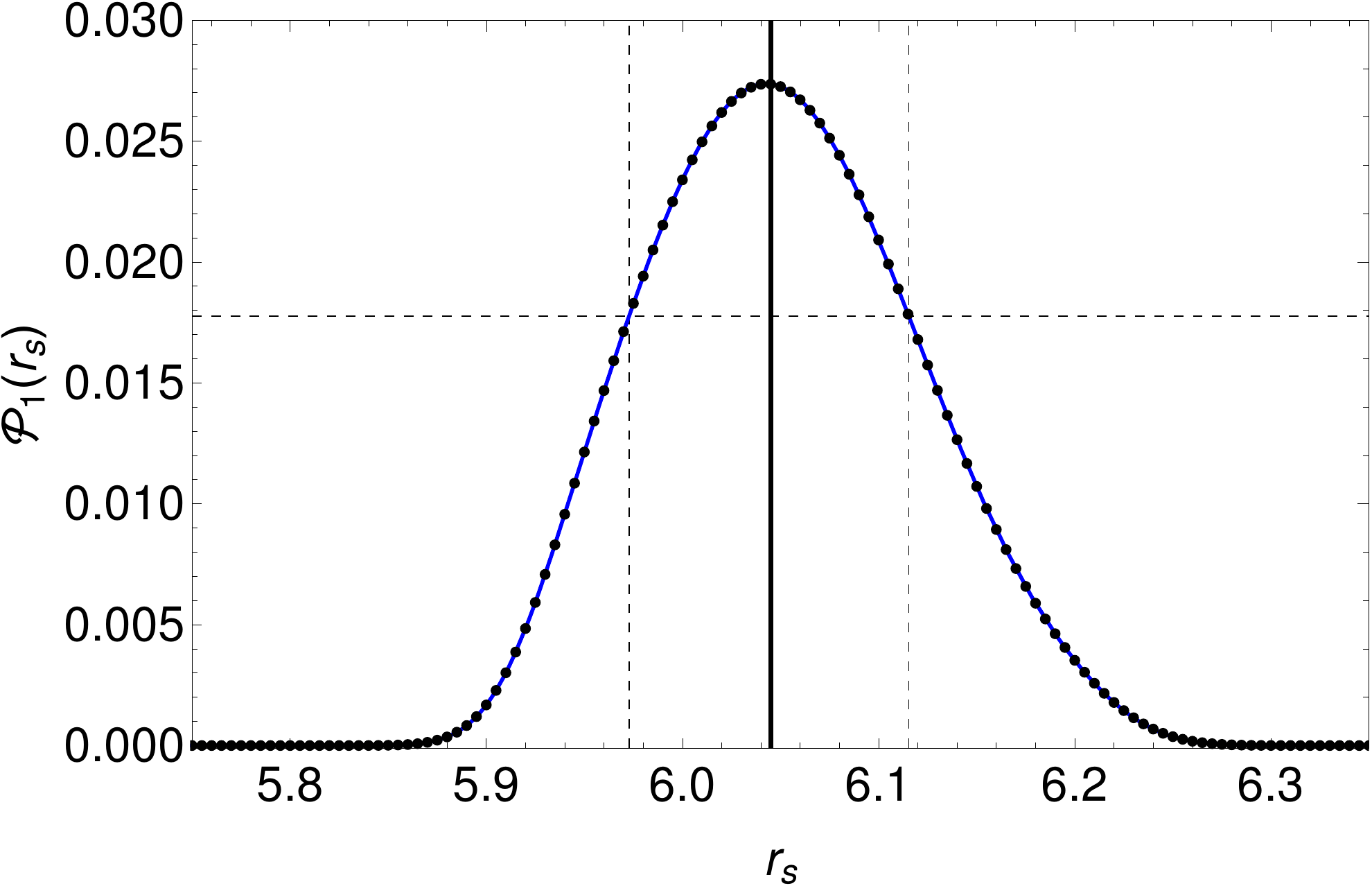}\label{plotrsM82sph}}
\hspace{0.1cm}
 \subfigure[]{
\includegraphics[scale=0.3]{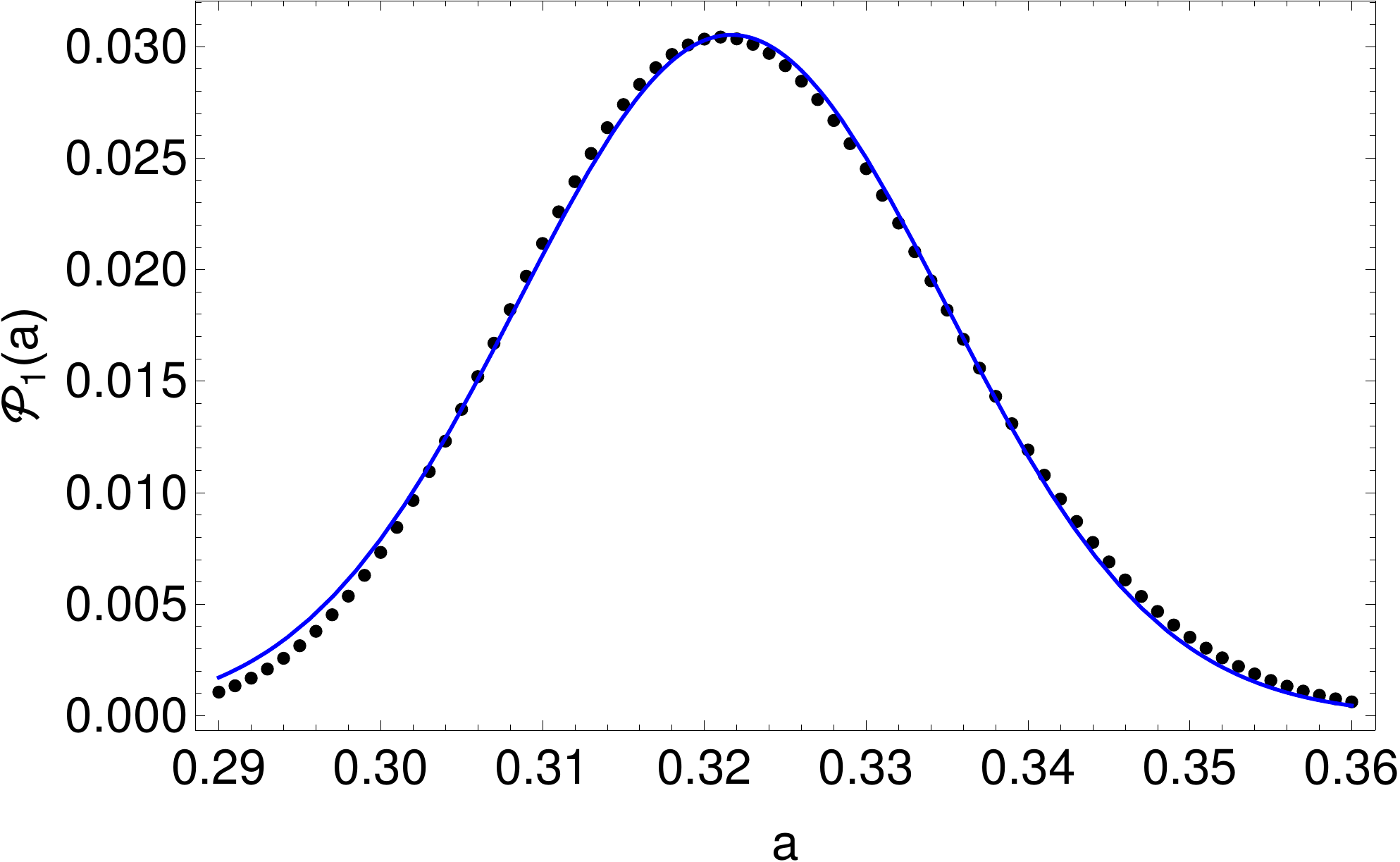}\label{plotaM82sph}}
\hspace{0.1cm}
\subfigure[]{
\includegraphics[scale=0.3]{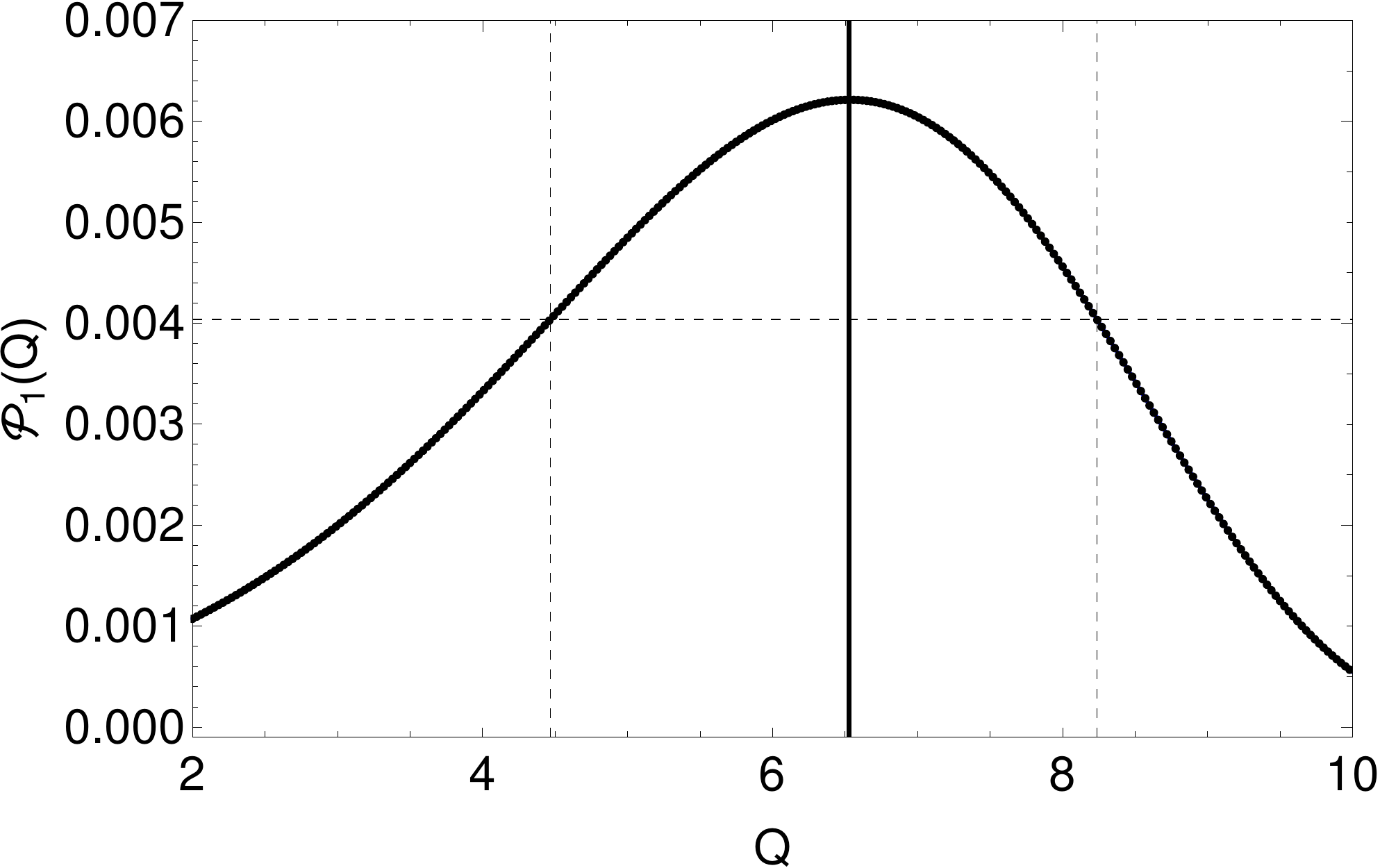}\label{plotQM82sph}}}
\caption{\label{plotM82sph}Probability density profiles in \{$r_s$, $a$, $Q$\} dimensions for M82 X-1: (a) $\mathcal{P}_{1} \left( r_s\right)$, (b) $\mathcal{P}_{1} \left( a\right)$, and (c) $\mathcal{P}_{1} \left( Q\right)$. The black points represent normalized probability density profiles generated using the method described in \S \ref{method}, and the blue curves are the model fit, and the results are summarized in Table \ref{sphresults}. The errors for the $\mathcal{P}_{1} \left( r_s\right)$ and $\mathcal{P}_{1} \left( Q\right)$ profiles are obtained such that the integrated probability between the vertical dashed curves is 68.2\%, whereas the vertical thick curves correspond to the peak value of the reduced probability density distributions. }
\end{figure} 
\begin{itemize}
\item \textit{M82 X-1}: We found the exact solution for a spherical orbit at \{$r_s=6.044$, $a=0.321$, $Q=6.113$\} for M82 X-1. The spherical trajectory with these parameter values is shown in Figure \ref{M82sphtrajec}. The calculated probability density profiles and the model fit are shown in Figure \ref{plotM82sph}. The $\mathcal{P}_{1} \left( r_s\right)$ and $\mathcal{P}_{1} \left( Q\right)$ profiles were found to be skew symmetric, and the integrated probability is 68.2\% about the peak of the probability density distribution between the error bars, while $\mathcal{P}_{1} \left( a\right)$ is well fit by a Gaussian. We see that the spin of the black hole is also found very close to the spin solutions estimated in \S \ref{resultseccentric}. We conclude that along with the $eQ$ trajectories having moderate eccentricities, as discussed in \S \ref{resultseccentric}, a spherical trajectory ($Q0$) at $r_s=6.044$ with $Q=6.113$ is also a viable solution that can produce the observed QPO frequencies in M82 X-1. The corresponding spin estimate $a=0.321\pm0.0132$ was utilized in \S \ref{resultseccentric} using Equation \eqref{mprobablea} to calculate the most probable value of the spin for M82 X-1. 
 \item \textit{XTEJ 1550-564}: A spherical trajectory solution was found at $r_s=5.538$ and $Q=2.697$ for BHXRB XTEJ 1550-564 that is shown in Figure \ref{XTEJsphtrajec}, and the calculated probability density profiles, the Gaussian model fit, and the probability contours in the \{$r_s$, $Q$\} plane are shown in Figure \ref{plotXTEJsph}. So, along with an $e0$ trajectory, as discussed in \S \ref{resultseccentric}, a $Q0$ orbit is also a viable candidate for the observed QPOs in the temporal power spectrum of XTEJ 1550-564.
\begin{longrotatetable}
\begin{deluxetable}{l c c c c c c c c c c c c}
\tablecaption{\label{sphresults}Summary of Results Corresponding to the Spherical Orbit Solutions for BHXRBs M82 X-1 and XTEJ 1550-564.}
\tablewidth{300pt}
\tabletypesize{\tiny}
\tablehead{
&   &  & &  &  & &  &  &  & & &  \\
\colhead{\textbf{BHXRB}} & \colhead{\textbf{$r_s$ Range}} & \colhead{\textbf{Resolution}} & \colhead{\textbf{Exact}} &  \colhead{ \textbf{Model Fit}} & \colhead{\textbf{$Q$ Range}} & \colhead{\textbf{Resolution}} & \colhead{\textbf{Exact}} &  \colhead{ \textbf{Model Fit}}  & \colhead{\textbf{$a$ Range}}& \colhead{\textbf{Resolution}} & \colhead{\textbf{Exact}} &  \colhead{ \textbf{Model Fit}}\\
&   &\colhead{$\Delta r_s$} & \colhead{\textbf{Solution}} & \colhead{\textbf{to $\mathcal{P}_{1} \left( r_s \right)$}}  & & \colhead{$\Delta Q$} & \colhead{ \textbf{Solution}} & \colhead{\textbf{to $\mathcal{P}_{1} \left( Q \right)$}}& & \colhead{$\Delta a$} & \colhead{\textbf{Solution}} &  \colhead{\textbf{to $\mathcal{P}_{1} \left( a \right)$}}\\
&  &   & \colhead{$r_{s0}$}&  &  & & \colhead{$Q_{0}$} &  &  & & \colhead{$a_0$} & 
}
\startdata
M82 X-1  & $5.75-6.35$ & 0.005 & 6.044 &  6.044$^{+0.071}_{-0.072}$  & $2-10$ & 0.03 & 6.113 & 6.113$^{+2.124}_{-1.645}$ & $0.29-0.36$ & 0.001 & 0.321 & 0.321$\pm$0.013 \\
&   &  & &  &  & &  &  &  & & &  \\
XTEJ 1550-564 & $3-8$ & 0.005 & 5.538 & 5.538$\pm$0.054 & $0.01-5$ & 0.01 & 2.697 &  2.697$^{+1.738}_{-1.627}$ & - & - & -  & - \\
&   &  & &  &  & &  &  &  & & &  \\
\enddata
\tablecomments{ The columns describe the range of parameter volume considered for \{$r_s$, $a$, $Q$\} and its resolution to calculate the normalized probability density using Equation \eqref{normPera}, the exact solutions for \{$r_s$, $a$, $Q$\} calculated using Equations \eqref{nuphisph2}$-$\eqref{nuthetasph2}, the value of parameters corresponding to the peak of the integrated profiles in 
\{$r_s$, $a$, $Q$\}, and results of the model fit to $\mathcal{P}_{1} \left( r_s \right)$, $\mathcal{P}_{1} \left( Q \right)$, and $\mathcal{P}_{1} \left( a \right)$.}
\end{deluxetable}
\end{longrotatetable}
 
\begin{figure}
  \mbox{
  \hspace{1.3cm}
\subfigure[]{
\includegraphics[scale=0.45]{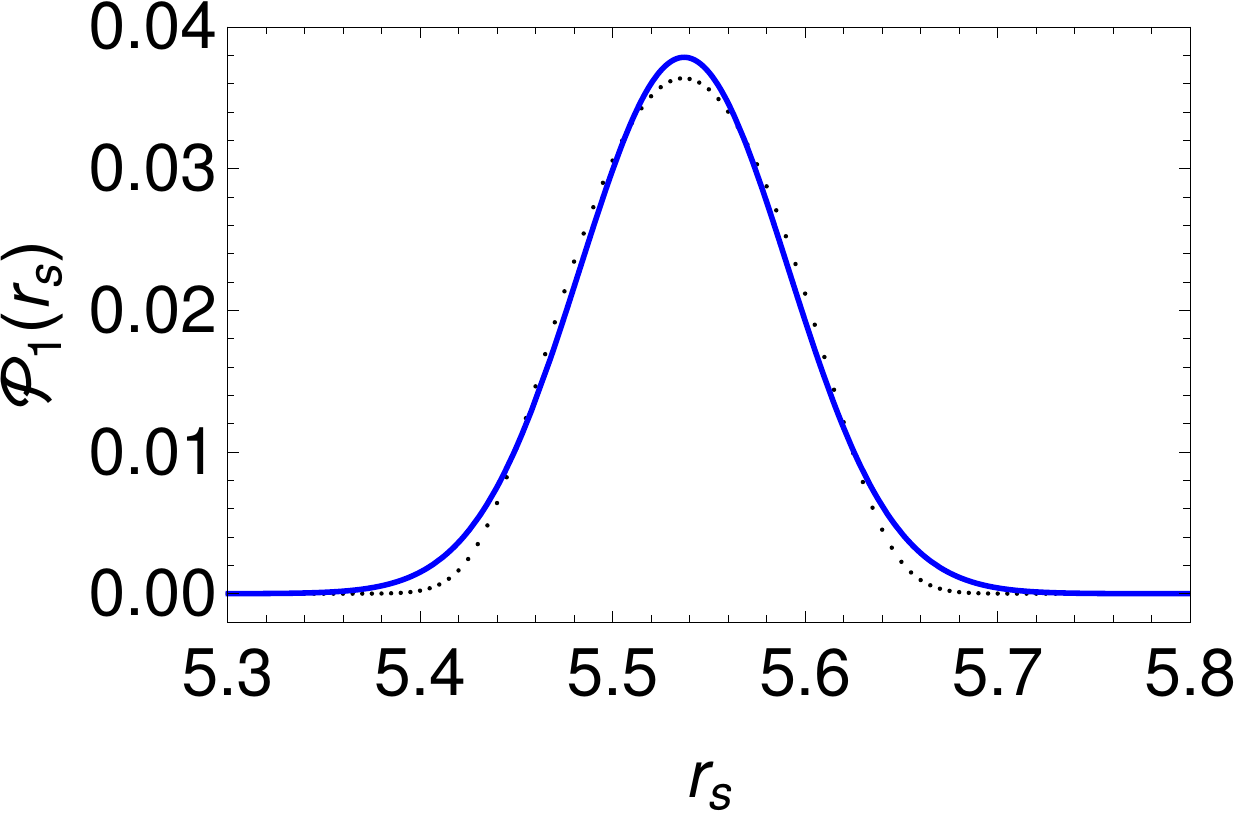}}
\hspace{1.8cm}
\subfigure[]{
\includegraphics[scale=0.5]{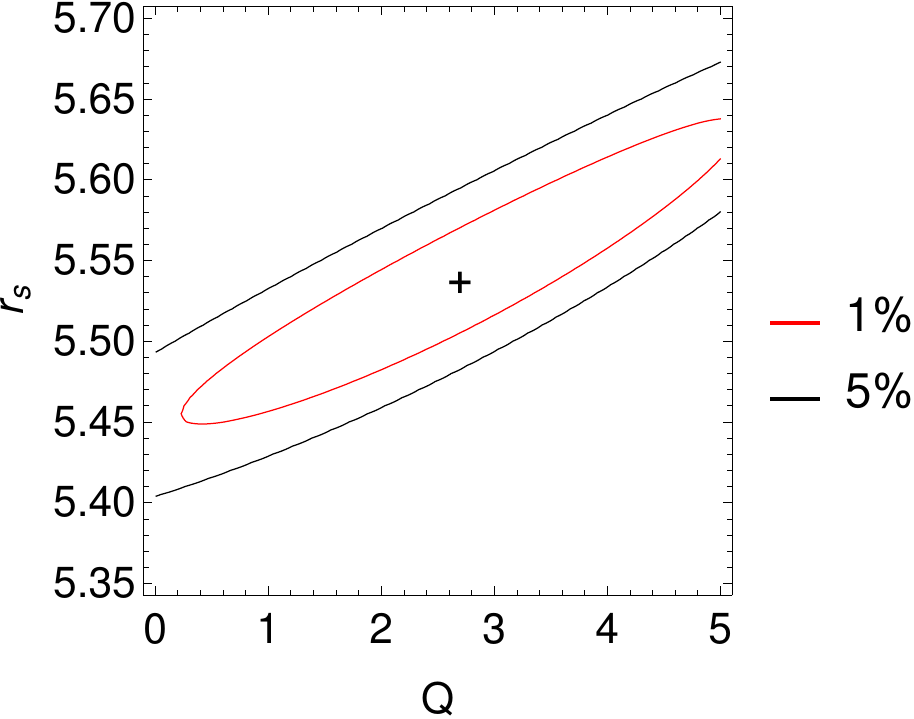}}}
  \mbox{
  \hspace{1.8cm}
\subfigure[]{
\includegraphics[scale=0.5]{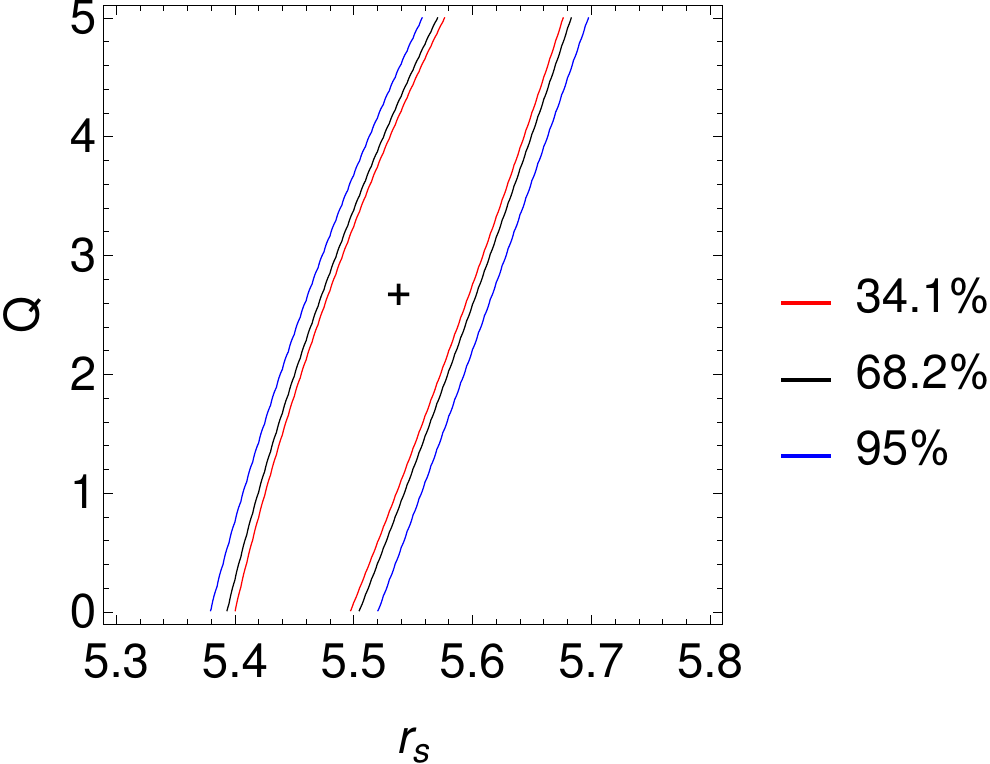}}
\hspace{1.8cm}
\subfigure[]{
\includegraphics[scale=0.45]{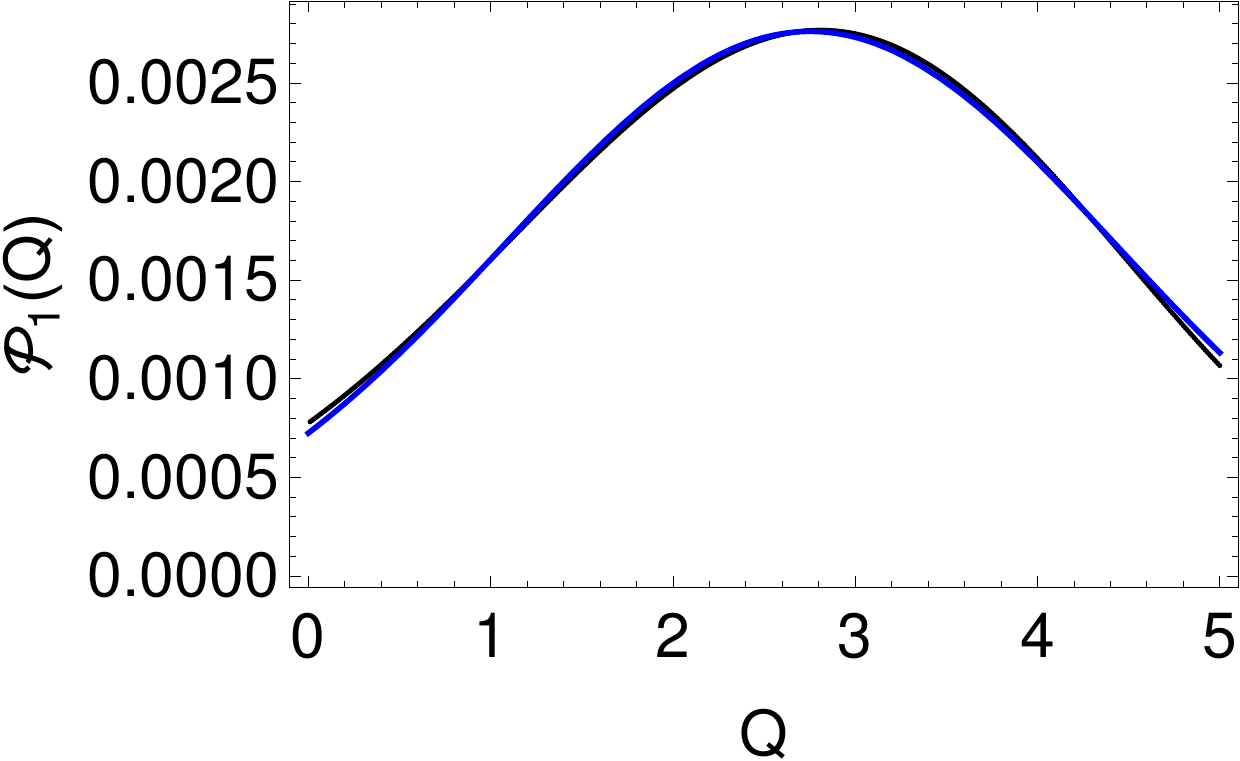}}}
  \caption{\label{plotXTEJsph} The integrated density profiles are shown in (a) $\mathcal{P}_1 \left( r_s\right)$ and (d) $\mathcal{P}_1 \left(Q\right)$ for the spherical orbit solution of BHXRB XTEJ 1550-564, where the dashed vertical lines enclose a region with 68.2\% probability, and the solid vertical line corresponds to the peak of the profiles. The probability contours of the parameter solution are shown in the (b) ($Q$, $r_s$) and (c) ($r_s$, $Q$) planes, where the $+$ sign marks the exact solution.}
\end{figure}


\end{itemize}
We found that the spherical trajectories are also possible solutions for QPOs in BHXRBs M82 X-1 ($a=0.321$, $Q=6.113$, $r_s=6.044$, $r_I=5.258$) and XTEJ 1550-564 ($a=0.34$, $Q=2.697$, $r_s=5.538$, $r_I=4.988$). This indicates that the spherical trajectory solutions are in region 1 of the ($r$, $a$) plane, as defined in Figure \ref{radii}; for both BHXRBs, and they are very close to the ISSO radius, $r_I$. These results are also consistent with the results discussed in \S \ref{sphericalmotivation}, where the QPO-generating region is close to the ISSO curve in the ($r$, $a$) plane. For the case of M82 X-1, the spherical trajectory solution has a different value of spin compared to the ones estimated in \S \ref{resultseccentric}, but it is very close to the other estimates given in Table \ref{3QPOresults}. This value of spin, together with other results in Table \ref{3QPOresults}, is used to estimate the most probable value of spin of the black hole for M82 X-1, which is $a=0.2994$. We also see that a low eccentric trajectory prefers a high $Q$ value and vice versa, as seen from the results shown in Table \ref{3QPOresults}. As the $Q$ value of the orbit is increased, the eccentricity of the trajectory solution decreases for both BHXRBs M82 X-1 and GROJ 1655-40. This trend is also followed here: for the spherical orbit ($e=0$), $Q\sim 6$ is found as a solution for M82 X-1 and $Q\sim2.7$ for XTEJ 1550-564, whereas a moderately eccentric trajectory solution was found with $Q=0$ for XTEJ 1550-564; see Table \ref{2QPOresults}.  

We conclude that various kinds of Kerr orbits, for example, spherical \{$e=0$, $Q\neq0$\}, equatorial eccentric \{$e\neq0$, $Q=0$\}, and nonequatorial eccentric \{$e\neq0$, $Q\neq0$\}, are also viable solutions for QPOs in BHXRBs. Hence, such trajectories with similar fundamental frequencies can together give a strong QPO signal in the temporal power spectrum.
\section{The PBK Correlation}
\label{PBK}
A tight correlation between the frequencies of two components in the PDS of various sources, including black hole and neutron star X-ray binaries, was discovered \citep{PBK1999}. Such a correlation among various variability components of the PDS in both types of sources suggests a common and important emission mechanism for these signals. This correlation is either between two QPOs, an LFQPO and either of the two HFQPOs, or it is between an LFQPO and high-frequency broadband noise components. We adopt the definition of \cite{Belloni2002ApJ} for these variability components: $L_{\rm LF}$ for LFQPO, and $L_l$ and $L_u$ for lower and upper HFQPOs or broad noise components. A systematic study of 571 RXTE observations was carried out for BHXRB GROJ 1655-40 between 1996 March and 2005 October \citep{Motta2014a}, and they also found such correlation between the type C QPOs and high-frequency QPOs and broadband components (either $L_l$ or $L_u$; see Tables 1 and 2 and Figure 5 of \cite{Motta2014a}). In this study, they calculated mass, spin of the black hole, and the radius at which QPOs originated \{$\mathcal{M}=5.31$, $a=0.29$, $r=5.68$\} \citep{Motta2014a} using \{$L_u=\nu_{\phi}$, $L_l=\nu_{\rm pp}$, $L_{\rm LF}=\nu_{\rm np}$\}, assuming that circular equatorial orbits are the origin of three simultaneous QPOs in the RPM ($00$ model as defined in Figure \ref{orbitflowchart}). Using the estimated values of $\mathcal{M}$ and $a$, they fit the PBK correlation of variability components in GROJ 1655-40 by varying $r$. 
 \begin{figure}
\mbox{
\hspace{0.5cm}
\subfigure[]{
\includegraphics[scale=0.3]{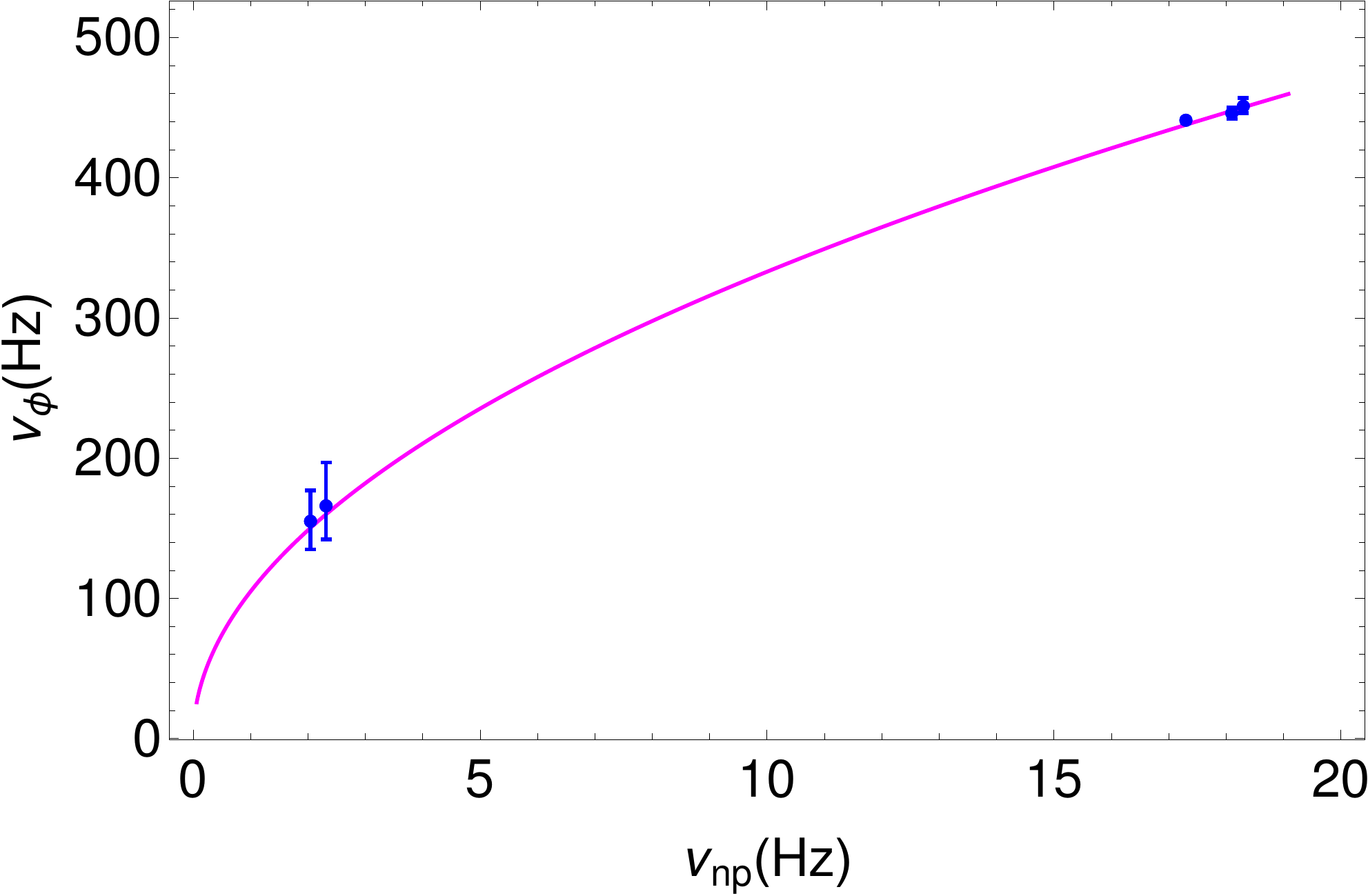}\label{plotphiGROJ}}
\hspace{1.5cm}
\subfigure[]{
\includegraphics[scale=0.3]{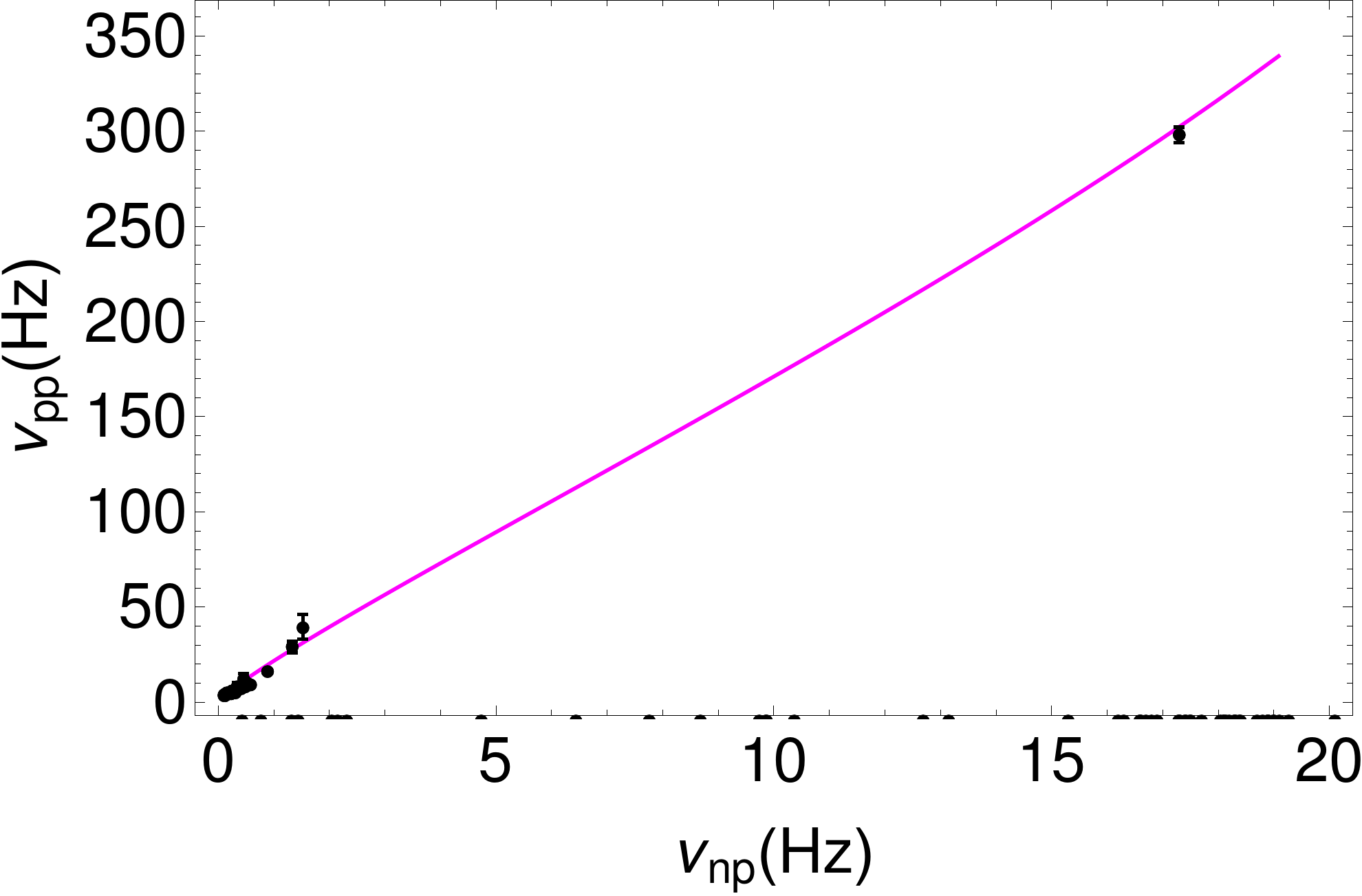}\label{plotperGROJ1}}}
\mbox{
\hspace{0.5cm}
\subfigure[]{
\includegraphics[scale=0.3]{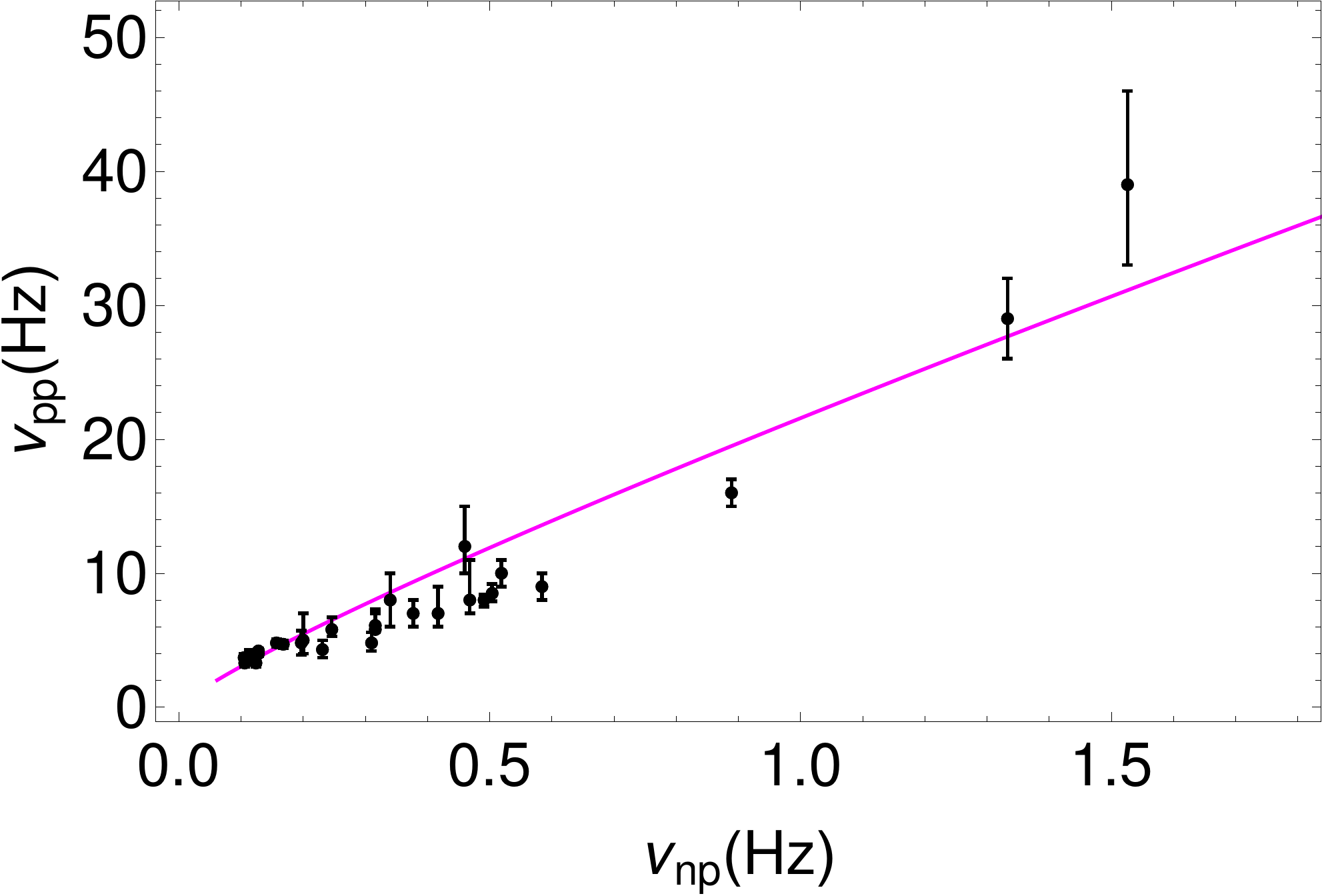}\label{plotperGROJ}}
\hspace{1.5cm}
\subfigure[]{
\includegraphics[scale=0.3]{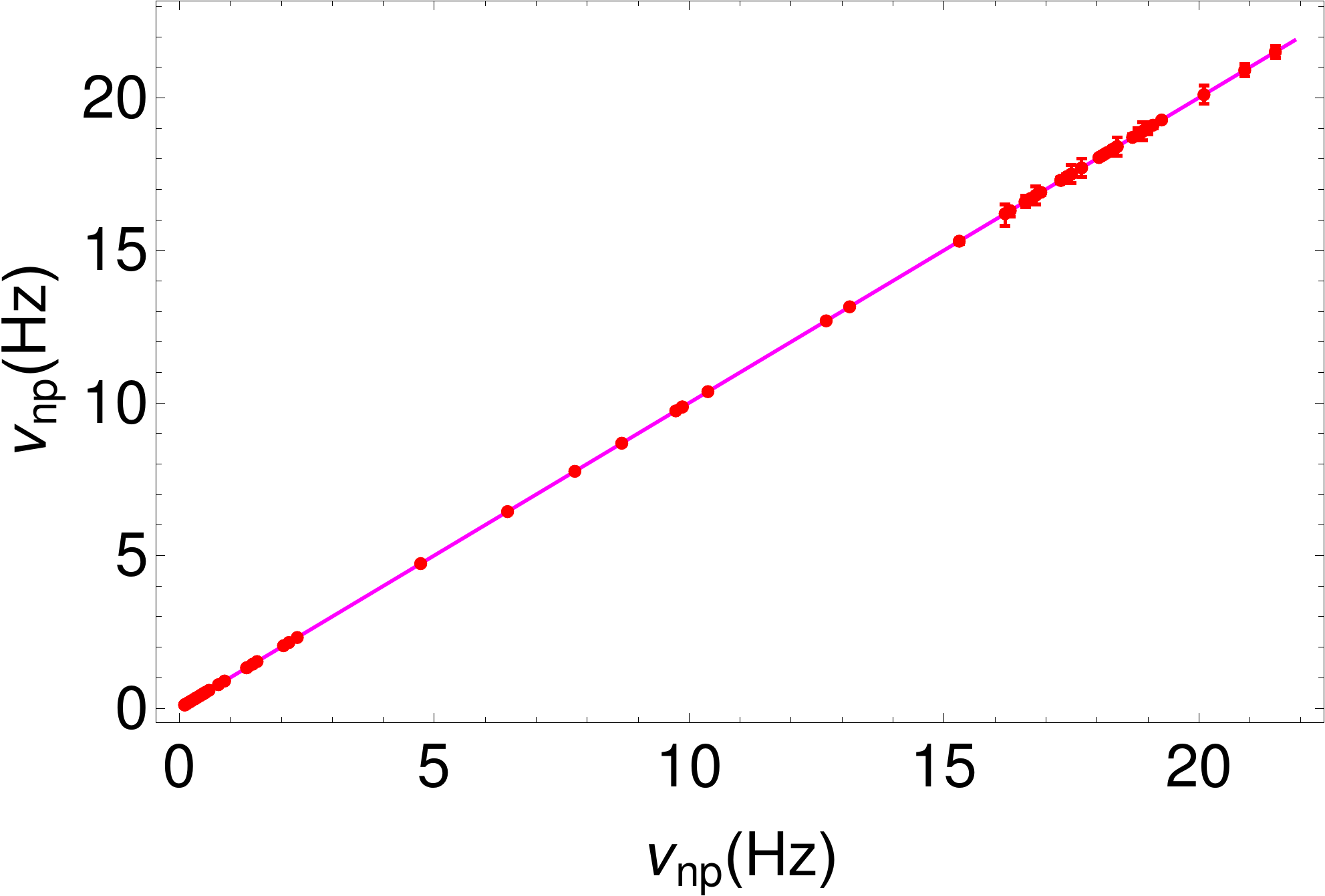}\label{plotnodGROJ}}}
\caption{\label{PBKGROJ}The PBK correlation is shown for BHXRB GROJ 1655-40 as previously observed [data points are from \cite{Motta2014a}]. The observed correlation is in good agreement with the frequencies of the $e0$ solution estimated, where \{$e=0.071$, $a=0.283$, $Q=0$, $\mathcal{M}=5.4$\}, for GROJ 1655-40 in \S \ref{resultseccentric}, where (a) $\nu_{\phi}$, (b) $\nu_{\rm pp}$ in the low-frequency range, (c) $\nu_{\rm pp}$ in the high-frequency range, and (d) $\nu_{\rm np}$ are shown. The blue, black, and red data points represent the $L_u$, $L_l$, and $L_{\rm LF}$ components of the PDS, respectively. The magenta curves show the theoretical values of frequencies.}
\end{figure}

 \begin{figure}
\mbox{
\hspace{0.8cm}
\subfigure[]{
\includegraphics[scale=0.43]{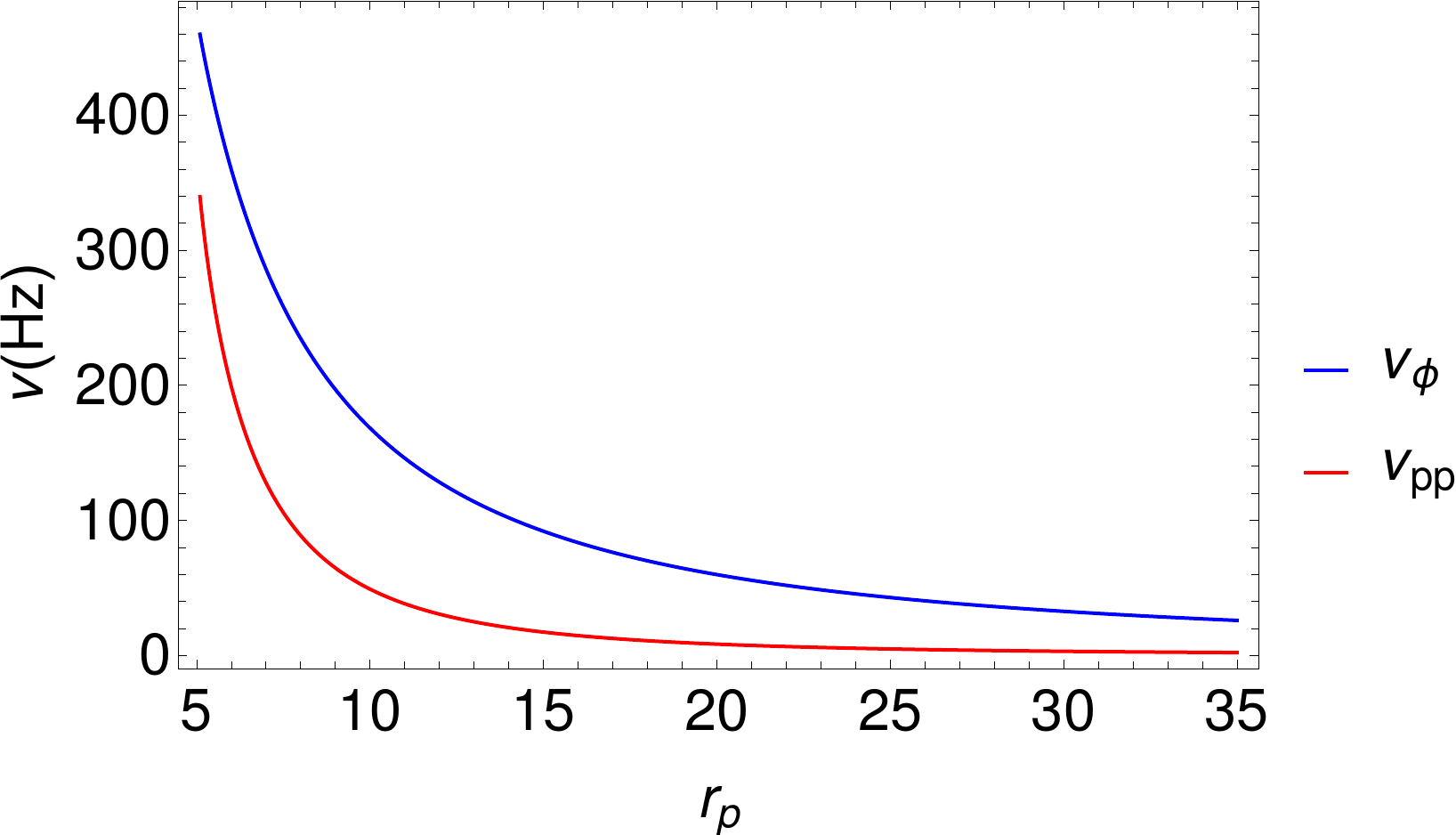}\label{plotphirp}}
\hspace{1.5cm}
\subfigure[]{
\includegraphics[scale=0.3]{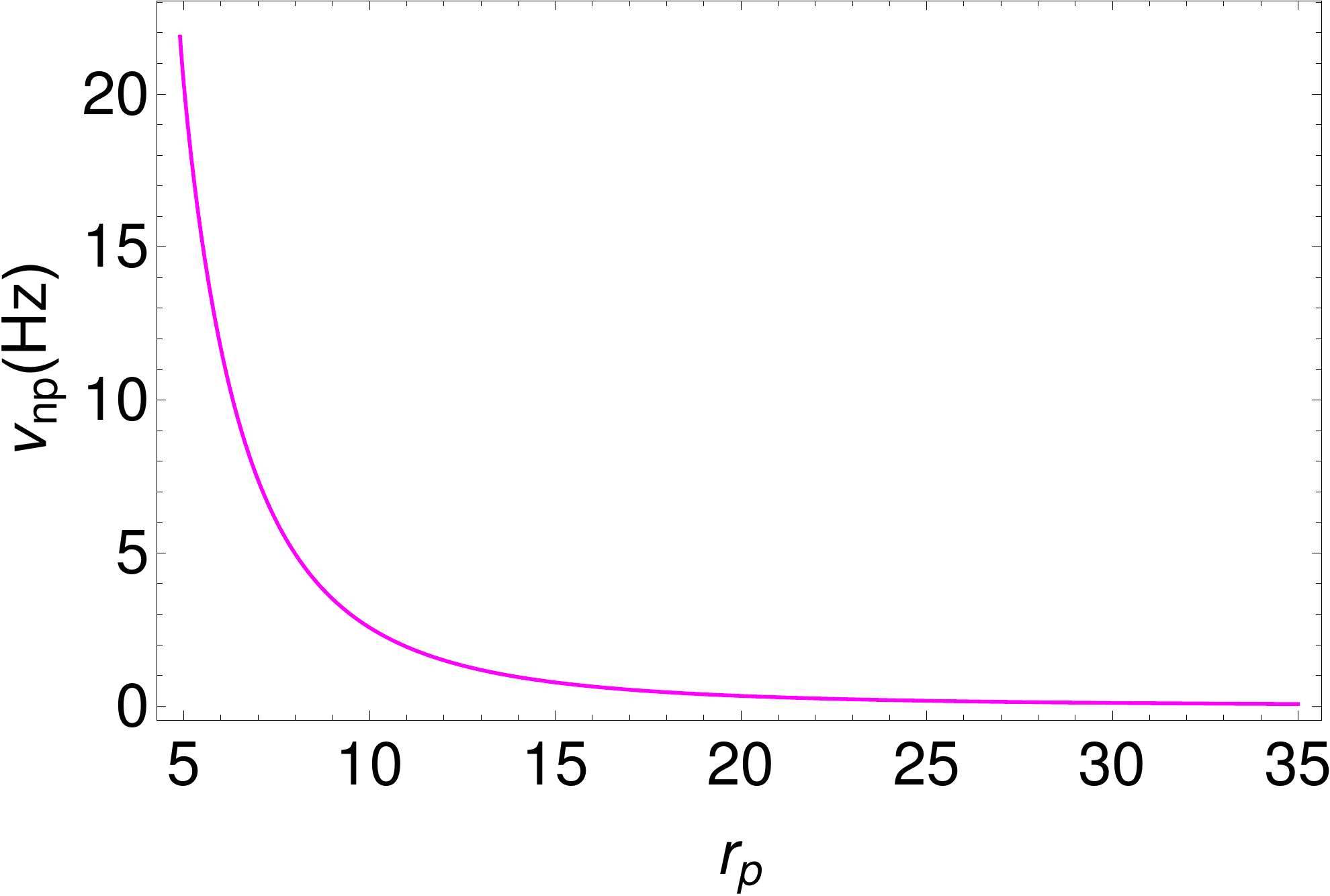}\label{plotnodrp}}}
\caption{\label{PBKrp}The frequencies (a) $\nu_{\phi}$ and $\nu_{\rm pp}$, (b) $\nu_{\rm np}$ are shown as function of $r_p$, for the $e0$ solution vector \{$e=0.071$, $a=0.283$, $Q=0$, $\mathcal{M}=5.4$\}.}
\end{figure}

Here we apply the $e0$ model solution calculated in \S \ref{resultseccentric} assuming \{$L_u=\nu_{\phi}$, $L_l=\nu_{\rm pp}$, $L_{\rm LF}=\nu_{\rm np}$\}, using the observation ID having three simultaneous QPOs detected in GROJ 1655-40 (shown in Table \ref{sourcelist}), to fit the PBK correlation. We fix the mass of the black hole to $\mathcal{M}=5.4$ \citep{Beer2002} and the spin of the black hole to the most probable value, $a=0.283$, estimated by minimizing the $\chi^2$ function, given by Equation \eqref{chisqr}. We fix $e$ and $Q$ to the values estimated by the fine-grid method \{$e=0.071$, $Q=0$\} and vary $r_p$ to calculate the frequencies. In Figure \ref{PBKGROJ}, we show the correlations of the frequencies corresponding to the parameters \{$e=0.071$, $a=0.283$, $Q=0$\}, which are in good agreement with the PBK correlation. In Figure \ref{PBKrp}, these frequencies are shown as functions of $r_p$. We see that the data points for $L_u$ components fit very well (see Figure \ref{plotphiGROJ}), whereas the $L_l$ components show a good fit in the high-frequency range [see Figures \ref{plotperGROJ1},\ref{plotperGROJ}]. The $L_{\rm LF}$ components also show good agreement with the eccentric orbit solution (see Figure \ref{plotnodGROJ}).

\begin{deluxetable}{c c c c c}
\tablecaption{\label{eQsolutionsGROJ}Nonequatorial Eccentric Orbit ($eQ$) Solutions for $L_l$ and $L_{\rm LF}$ Components Detected in RXTE Observations of GROJ 1655-40 \citep{Motta2014a}, Where the First Row Corresponds to the Observation ID with Three Simultaneous QPOs.}
\tablewidth{300pt}
\tabletypesize{\scriptsize}
\tablehead{
\colhead{ \textbf{$L_{\rm LF}$}}& \colhead{ $\ \ \ \ \ \ $ \textbf{$L_l$} } $\ \ \ \ \ \ \ \ \ $ & \colhead{ \textbf{$r_p$}} & \colhead{ $\ \ \ \ \ \ \ $ \textbf{$e$}} $\ \ \ \ \ \ \ \ \ \ \ \ \ $ &  \colhead{ \textbf{$Q$}} \\
(Hz) & (Hz) & & & 
}
\startdata
17.3 & 298 & 5.25 & 0.071 & 0 \\
\hline
0.106 & 3.3    & 29.179 & 0.077 & 24.423\\
0.117 & 3.9    & 28.228 & 0.083 & 33.903\\
0.123 & 4     & 27.758 & 0.083 & 33.392\\
0.128 & 4   & 27.389 & 0.083 & 32.642\\
0.11 & 3.5   & 28.818 & 0.082 & 33.622\\
0.115 & 3.7    & 28.392 & 0.083 & 34.028\\
0.128 & 4.2    & 27.389 & 0.083 & 33.010\\
0.157 & 4.8    & 25.576 & 0.083 & 30.964\\
1.333 & 29   & 12.464 & 0.079 & 10.921\\
0.46 & 12   & 17.826 & 0.085 & 22.343\\
\enddata
\tablecomments{ The mass of the black hole was fixed to $\mathcal{M}=5.4$ and spin was fixed to $a=0.283$.}
\end{deluxetable}

Thirty-four $L_l$ and $L_{\rm LF}$ components which were detected simultaneously in the same observation ID [see Table 1 of \cite{Motta2014a}]. To calculate $r_p$, we first solve for $L_{\rm LF}=\nu_{\rm np}$ for the solution vector \{$e=0.071$, $Q=0$, $a=0.283$, $\mathcal{M}=5.4$\}; this locates the $r_p$, where oscillations are present, to a good approximation. Using these $r_p$ values, we simultaneously solve \{$\nu_{\rm pp}=L_{l}$, $\nu_{\rm np}=L_{\rm LF}$\} using the centroid frequencies of these components and estimate the exact solutions for parameters \{$e$, $Q$\} with \{$a=0.283$, $\mathcal{M}=5.4$\}. In 10 out of 34 cases, we found low-eccentricity $eQ$ solutions for these PDS components, where the calculated parameters are shown in Table \ref{eQsolutionsGROJ}. We find orbits with high $Q$ values at large $r_p$ (this is expected as $Q\propto L^2-L_{z}^2$) as solutions for these PDS components. This exercise confirms the existence of $eQ$ in addition to $e0$ solutions for QPOs.
\section{Gas Flow near ISSO (ISCO)}
\label{gasflowmodel}
In this section, we discuss our torus picture of eccentric trajectories, and we examine the model of fluid flow in the general-relativistic thin disk around a Kerr black hole \citep{2012MNRAS.420..684P,2014ApJ...791...74M} with the aim of finding a source of the $e0$, $eQ$, and $Q0$ trajectories. In this model, the region around the rotating black hole was divided into various regimes: (1) the plunge region between the ISCO radius and black hole horizon dominated by gas pressure and electron scattering based opacity, (2) the edge region at and very near to the ISCO radius dominated by gas pressure and electron scattering based opacity, (3) the inner region outside the edge region with small radii comparable to ISCO dominated by radiation pressure and electron scattering based opacity, (4) the middle region outside the inner region where gas pressure again dominates over the radiation pressure and electron scattering based opacity, (5) the outer region far from the black hole horizon and outside the middle region dominated by gas pressure and electron scattering based opacity. The analytic forms for the important quantities like flux of radiant energy, $F$, temperature, $T$, and radial velocity in the locally nonrotating frame, $\beta_r$, were given for these different regions (as functions of $r$, $a$, viscosity, $\alpha$, accretion rate, $\dot{m}=\dot{M_\bullet}/\dot{M}_{Edd}$, and $M_{\bullet}$) where nonzero stresses were incorporated at the inner edge of the disk in this model \citep{2012MNRAS.420..684P}. Also, the expression for quality factor $Q_{\phi}\left(r, a, \beta_r \right)$ was derived for $\nu_{\phi}$ QPO frequencies in the equatorial plane, which is given by [\cite{2014ApJ...791...74M}, typo fixed in Equation (10)]
\begin{eqnarray}
Q_{\phi}\left(r, a, \beta_r \right)=\dfrac{-\sqrt{A}}{3 \pi \beta_r \Delta r^{1/2}}\left[ 1- \dfrac{\left(A \Omega -2ar \right)^2 }{\Sigma^2 \Delta}\right]^{-1/2}, \label{Qualityf}
\end{eqnarray}
where $A=\left(r^2 + a^2 \right)^2-a^2 \Delta \sin^2 \theta$, $\Delta=r^2 + a^2-2r$, $\Sigma= r^2 + a^2 \cos^2 \theta$, and $\Omega=1/ \left( r^{3/2} +a \right)$, and where $\theta=\pi/2$ is assumed in Equation \eqref{Qualityf}. Using this formula, one can obtain the quality factor of the QPO in various regions close to the black hole by substituting the $\beta_r$ of the corresponding region as defined above. The expressions for $\beta_r$ in the edge and inner regions are given by (Equations (12), (13) of \cite{2014ApJ...791...74M})
\begin{subequations}
\begin{eqnarray}
\beta_{r,edge}&&=-7.1 \times 10^{-5} \alpha^{4/5} m_1^{-1/5} \dot{m}^{2/5} r^{-2/5} \mathcal{B}^{4/5} \mathcal{C}^{-1/2} \mathcal{D}^{3/10} \Phi^{-3/5}, 
\end{eqnarray}
\begin{eqnarray}
\beta_{r,inner}&&=-124.416 \ \alpha \ \dot{m}^2 r^{-5/2} \mathcal{A}^2 \mathcal{B}^{-3} \mathcal{C}^{-1/2} \mathcal{D}^{-1/2} \mathcal{S}^{-1} \Phi,
\end{eqnarray}
\end{subequations}
where $m_1=M_{\bullet}/10 M_{\odot}$, $\mathcal{C}=1-3 r^{-1} +2 a r^{-3/2}$ [there is a typo in the expression of $\mathcal{C}$, Equation (A4c), in \cite{2012MNRAS.420..684P}]; and $\mathcal{A}$, $\mathcal{B}$, $\mathcal{D}$, $\mathcal{S}$, and $\Phi$ are given in \cite{2012MNRAS.420..684P} (Equations A4(a), (b), (d), (o) and (3.6)).

\begin{figure}[h!]
\mbox{
\hspace{-0.7cm}
\subfigure[]{
\includegraphics[scale=0.5]{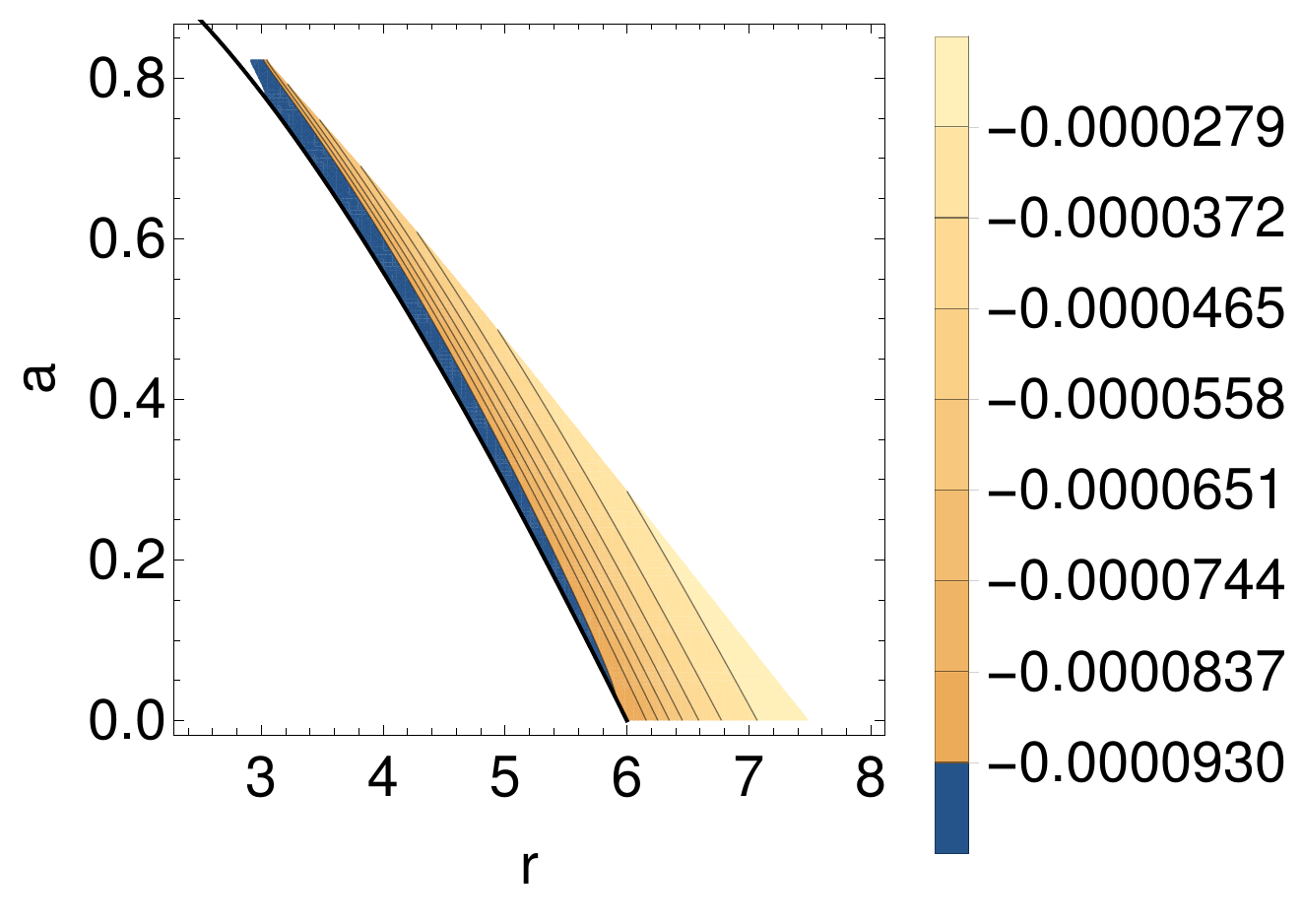}\label{betar}}
\subfigure[]{
\includegraphics[scale=0.5]{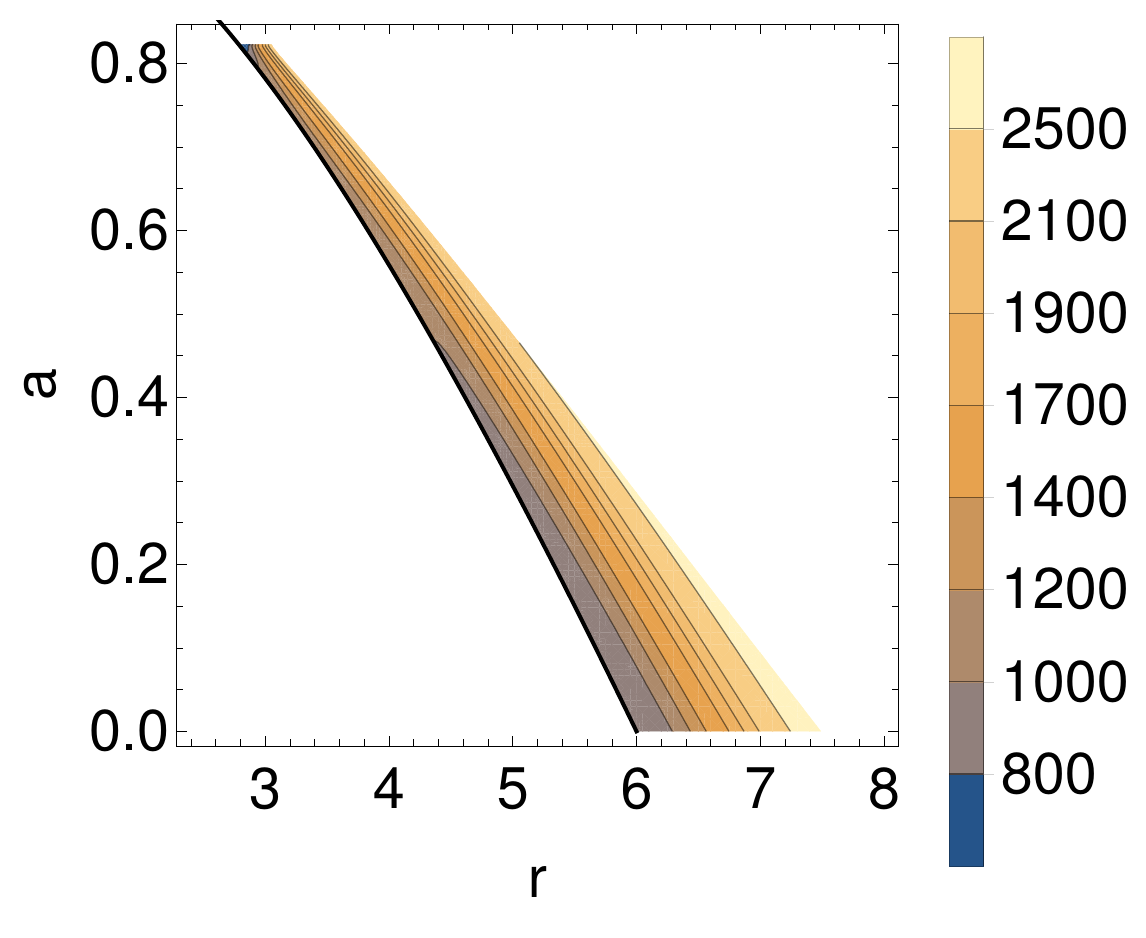}\label{Qf}}
\subfigure[]{
\includegraphics[scale=0.32]{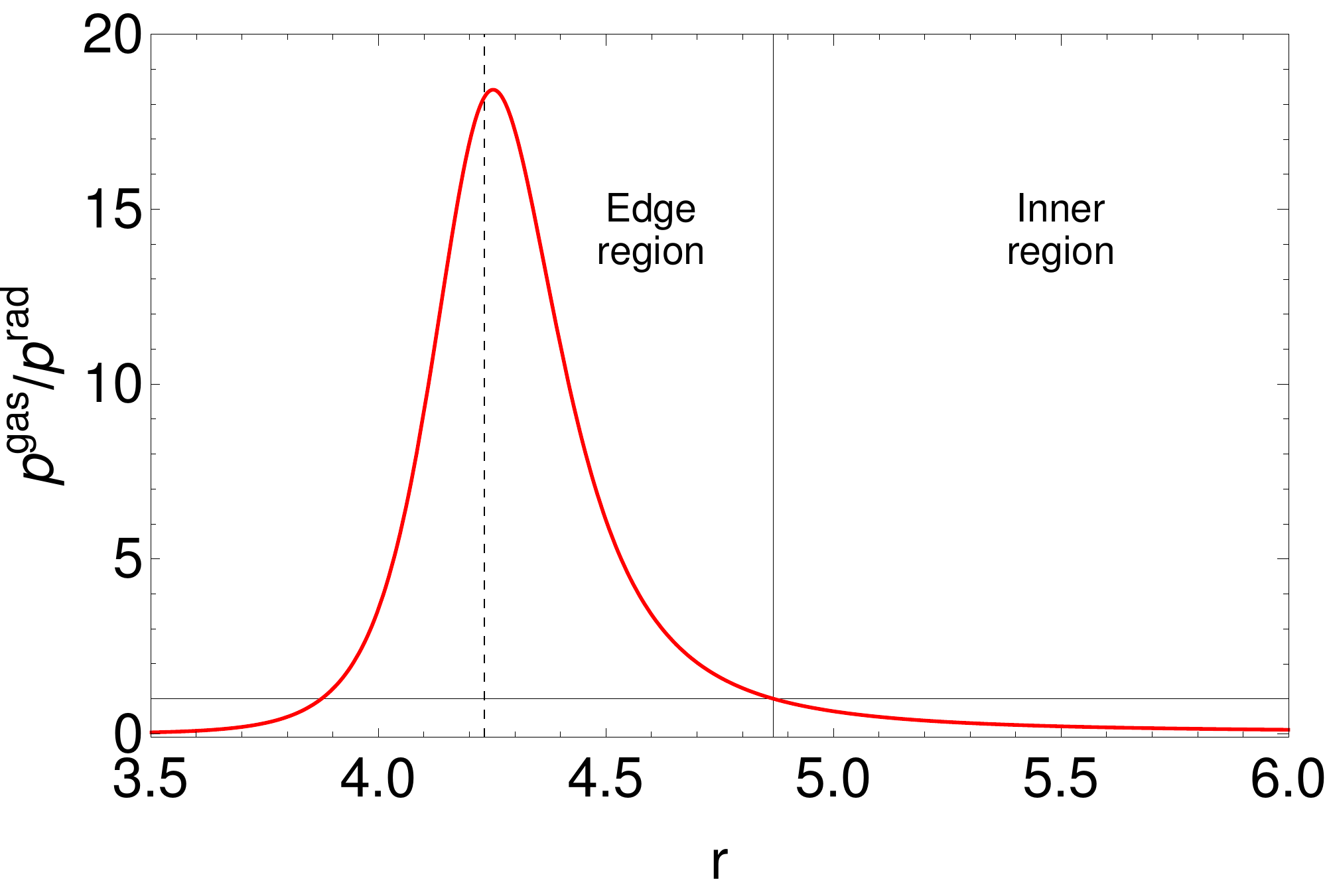}\label{pgasprad}}}
\caption{\label{edgeregion}Contours of (a) $\beta_r$ and (b) $Q_{\phi}$ in the ($r$, $a$) plane  in the edge region of the general-relativistic thin disk, and (c) $p^{\rm gas}/p^{\rm rad}$ as a function of $r$ with $a=0.5$ (where the dotted vertical curve corresponds to ISCO and the solid vertical curve corresponds to $r$ when $p^{\rm gas}/p^{\rm rad}=1$). We have fixed \{$\alpha=0.1$, $m_1=1$, $\dot{m}=0.1$\}. }
\end{figure}

In Figure \ref{betar} and \ref{Qf}, we have shown the contours for $\beta_r$ and $Q_{\phi}$ for the edge region in the ($r$, $a$) plane, and the $p^{\rm gas}/p^{\rm rad}$ ratio as a function of $r$ in Figure \ref{pgasprad}. One can discern the transition from the inner to edge region by the sudden increase of the $p^{\rm gas}/p^{\rm rad}$ ratio, as seen in Figure \ref{pgasprad}, which is given by [\cite{2012MNRAS.420..684P}, Equation (3.7g)]
\begin{equation}
\dfrac{p^{\rm gas}}{p^{\rm rad}}= 1.983 \times 10^{-8} m_1^{-1/4} \alpha^{-1/4} \dot{m}^{-2} r^{21/8} \mathcal{A}^{-5/2} \mathcal{B}^{9/2} \mathcal{D} \mathcal{S}^{5/4} \Phi^{-2}.
\end{equation}

\begin{deluxetable}{c c c c c}[h!]
\tablecaption{\label{regionstab}Ranges of $r$, Pressure Ratio, $p^{\rm gas}/p^{\rm rad}$, Quality Factor, $Q_{\phi}$, and Radial Velocity, $\beta_r$, in the Edge and Inner Regions of Fluid Flow in the Relativistic Thin Accretion Disk around a Kerr Black Hole \citep{2012MNRAS.420..684P,2014ApJ...791...74M}, Where We Have Fixed \{$m_1=1$, $\alpha=0.1$\} for BHXRBs.}
\tablewidth{300pt}
\tabletypesize{\footnotesize}
\tablehead{
 Region &\colhead{\textbf{$\left(a=0.3,\dot{m}=0.1\right)$}} & \colhead{\textbf{$\left(a=0.5,\dot{m}=0.1\right)$}} & \colhead{\textbf{$\left(a=0.3,\dot{m}=0.3\right)$}} & \colhead{\textbf{$\left(a=0.5,\dot{m}=0.3\right)$}}  \\
 & \colhead{\textbf{$\left(r, p^{\rm gas}/p^{\rm rad},\beta_{r}, Q_{\phi} \right)$}} & \colhead{\textbf{$\left(r,  p^{\rm gas}/p^{\rm rad}, \beta_{r}, Q_{\phi} \right)$}} &  \colhead{\textbf{$\left(r, p^{\rm gas}/p^{\rm rad}, \beta_{r}, Q_{\phi} \right)$}} &  \colhead{\textbf{$\left(r,  p^{\rm gas}/p^{\rm rad}, \beta_{r}, Q_{\phi} \right)$}}
}
\startdata
Edge & $4.98-5.93$& $4.23-4.87$ & $4.98-5.25$ & $4.23-4.35$ \\
& $ 1.002  - 29.84$ & $ 1.003 - 18.41$ & $ 1.026  - 1.921$ & $ 1.003  - 1.186$ \\
& -$ \left(2.84-10.2 \right) \times 10^{-5}$ & -$\left(3.81-11.37\right)\times 10^{-5}$ & -$\left(1.03-1.34 \right)\times 10^{-4}$ & -$\left(1.36-1.49 \right) \times 10^{-4}$ \\
& $914.46-2624.12$ & $1019.24-2473.07$ & $694.29-844.59$ & $773.62-814.27$ \\
\hline
Inner & $5.93-85.22$ & $4.87-87.81$ & $5.25-226.2$ & $4.35-229.45$ \\
& $0.0589-0.998$ & $0.0373-0.999$ & $0.0065-0.998$& $0.0041-0.998$ \\
& -$\left(1.2052-69.28\right) \times 10^{-6}$ & -$\left(1.1601-110.71\right)\times 10^{-6}$ & -$\left( 1.127-626.95\right) \times 10^{-6}$& -$\left( 1.107-1001.39\right) \times 10^{-6}$ \\
& $688.09-9830.85$ & $501.112-10046.3$ & $76.21-6333.26$ & $55.52-6397.35$ \\
\enddata
\end{deluxetable}

In Table \ref{regionstab}, we give the range of \{$r$, $Q_{\phi}$, $\beta_r$, $p^{\rm gas}/p^{\rm rad}$\} for the edge and inner regions for different combinations of $a$ and $\dot{m}$, fixing \{$m_1=1$, $\alpha=0.1$\} for BHXRBs, with a low accretion rate ($\dot{m}\simeq0.1$) corresponding to the hard spectral state and a high accretion rate ($\dot{m}\simeq0.3$) corresponding to the soft spectral state of BHXRBs. We see a sharp rise in $p^{gas}/p^{rad}$ values in the edge region in Figure \ref{pgasprad}. The ranges of $Q_{\phi}$ in both the edge and inner regions are very high compared to those observed in BHXRBs ($Q_{\phi}=5-40$). We, therefore, suggest that the QPOs are coming from a region very close to and inside ISCO; we identify this with the torus region, consisting of geodesics \citep{2012MNRAS.420..684P}, and hence $Q_{\phi}$ is different. This is also supported by the observation that the edge-flow-sourced geodesics populate the torus region obtained here for M82 X-1 ($r=4.7-9.08$) and GROJ 1655-40 ($r=5.1-6.67$); see Figure \ref{trajectories}. Specifically, the sharp pressure ratio gradient suggests that the edge region can be a launchpad for the instabilities that then oscillate with fundamental frequencies, causing geodesic flows in the torus region inside ISCO ($r<r_{\rm ISCO}$), where the fluid motion is close to Hamiltonian flow. A further understanding of this proposal (or conjecture) can be gained by carrying out a detailed model or simulation of the GRMHD flow in the edge region.

\section{Discussion, Caveats, and Conclusions}
\label{discconcl}
\begin{figure}[hbt!]
 \hspace{1.9cm}
\includegraphics[scale=0.45]{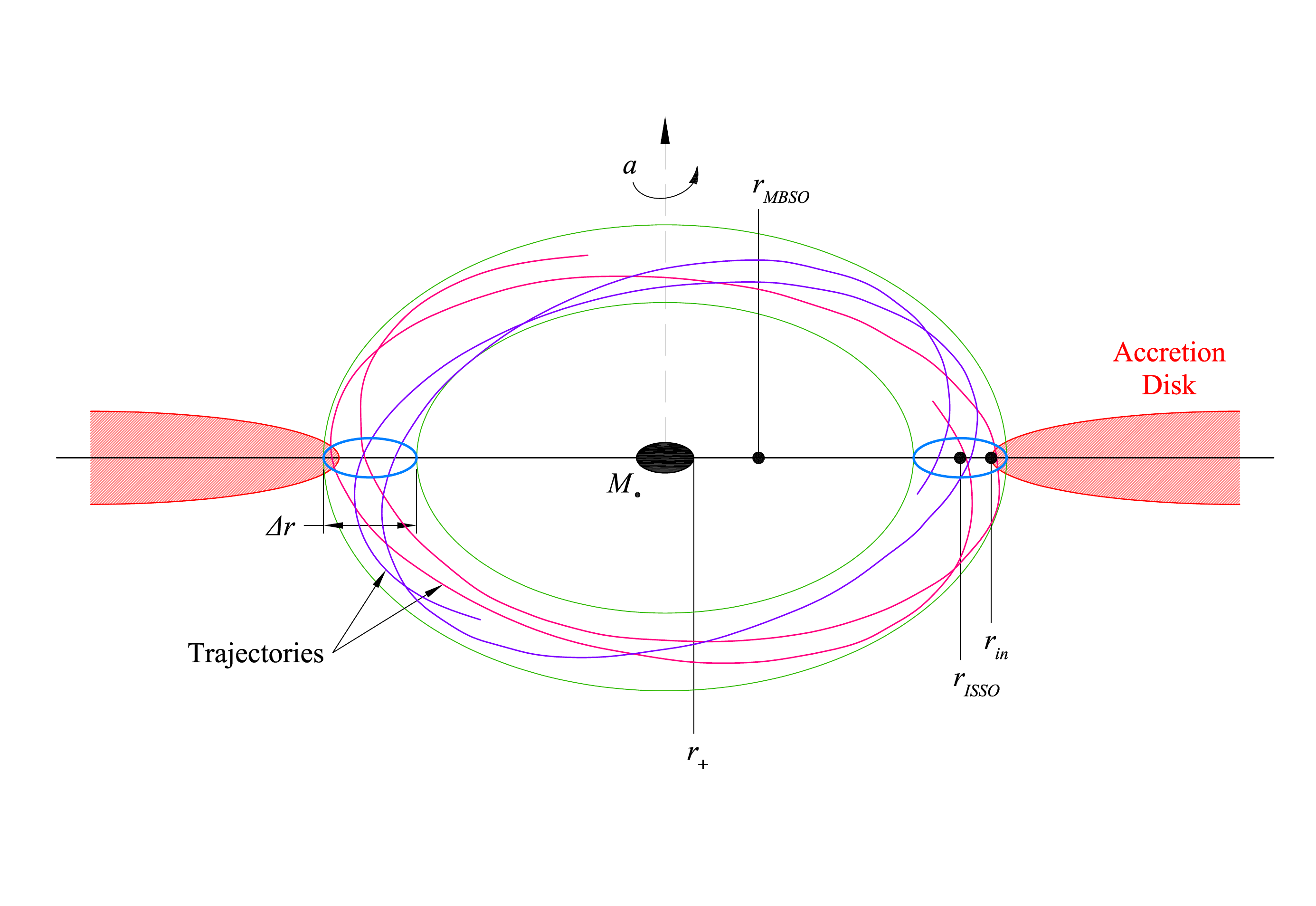} 
 \caption{\label{model}The cartoon shows a geometric model explaining the region of origin of QPOs assuming the more general nonequatorial eccentric trajectories in the GRPM, where the torus extent is ${R_0}^{+\Delta_1}_{-\Delta_2}$ (and torus width $\Delta r=\Delta_1 + \Delta_2$).}
\end{figure}

The QPOs in BHXRBs have been an important probe for comprehending the inner accretion flow close to the rotating black hole. Many theoretical models have been proposed in the past to explain its origin and in particular LFQPOs and HFQPOs \citep{Kato2004b,Torok2005,Tagger2006,Germana2009,Ingram2009, Ingram2011, Ingram2012}. These various models have been able to explain different properties of QPOs. For example, one of these models attributes the HFQPOs to the Rossby instability under the general relativistic regime \citep{Tagger2006}, whereas another model attributes type C QPOs to the Lense$-$Thirring precession of a rigid torus of matter around a Kerr black hole \citep{Ingram2009, Ingram2011, Ingram2012}. Although these models can explain either LFQPOs or HFQPOs, they do not explain the simultaneity of these QPOs, as previously observed in BHXRB GROJ 1655-40 \citep{Motta2014a}. The RPM, which is based on the geometric phenomenon of the relativistic precession of particle trajectories, explains these simultaneous QPOs as \{$\nu_{\phi}$, $\left(\nu_{\phi}-\nu_{r}\right)$, $\left(\nu_{\phi}-\nu_{\theta}\right)$\} of a self-emitting blob of matter (or instability) in a bound orbit near a Kerr black hole. We have extended the RPM for QPOs in BHXRBs to study and associate the fundamental frequencies of the bound particle trajectories near a Kerr black hole, which are $eQ$, $e0$, and $Q0$ solutions with the frequencies of QPOs. We call this as the generalized RPM (GRPM). Recently, novel and compact analytic forms for the trajectories around a Kerr black hole and their fundamental frequencies were derived \citep{RMCQG2019,RMarxiv2019}. We applied these formulae to the GRPM to extract the QPO frequencies. Graphical examples of these trajectories around a Kerr black hole are shown in Figures \ref{trajectories}, \ref{sphericaltrajectories}, and \ref{model}. A summary of these results is given in Table \ref{radiisummary}.

\begin{deluxetable}{l c c c c c c c c c c}
\tablecaption{\label{radiisummary}Summary of Orbital Solutions Found for QPOs Observed in Five BHXRBs Using the GRPM in This Article, and the Corresponding Region of the ($r_p$, $a$) Plane Where QPOs Originate.}
\tablewidth{300pt}
\tabletypesize{\scriptsize}
\tablehead{
& & & &  & & & & &\\
\colhead{\textbf{BHXRB}} & \colhead{\textbf{Number}} & \colhead{\textbf{Model}} & \colhead{\textbf{$e$}} & \colhead{\textbf{$r_p$}} & \colhead{\textbf{$a$}} & \colhead{\textbf{$Q$}} & \colhead{\textbf{MBSO}} & \colhead{\textbf{ISCO}} & \colhead{\textbf{ISSO}} & \colhead{\textbf{Region in} } \\
 & \colhead{\textbf{of QPOs}} & \colhead{\textbf{Class}} & & &  & & & & & \colhead{\textbf{($r_p$, $a$) Plane}}
}
\startdata
M82 X-1 & 3 & $eQ$ &$0.230_{-0.049}^{+0.067}$ & $4.834_{-0.268}^{+0.181}$ & $0.299$ & $2.362_{-1.439}^{+1.519}$ & 3.424 & 4.981 & 5.096 & 2\\
& & $Q0$& 0 &  6.044$^{+0.071}_{-0.072}$ & 0.321$\pm$0.013 & 6.113$^{+2.124}_{-1.645}$ & 3.475 & 4.903 &5.258 & 1\\
& & & &  & & & & & &\\
GROJ 1655-40 & 3 & $eQ$ &$0.071^{+0.031}_{-0.035}$ & $5.25_{-0.142}^{+0.171}$ & $0.283$ & $0^{+0.623}$  & - & 5.039 & - & 1 \\
& & & & &  & & & & & \\
XTE J1550-564 & 2 & $e0$ &0.262$^{+0.090}_{-0.062}$ & 4.365$^{+0.169}_{-0.279}$ & 0.34 & 0 &  - & 4.835 & -& 2 \\
& & $Q0$ &0 &5.538$\pm$0.054 & 0.34 & 2.697$^{+1.738}_{-1.627}$ & 3.35 & 4.835 & 4.988 & 1\\
& & & & &  & & & & &\\
4U 1630-47 & 2 & $e0$ & 0.734$^{+0.066}_{-0.048}$ & 2.249$^{+0.249}_{-0.353}$ & 0.985 & 0 & - & 1.541 & - & 1\\
& & & & &  & & & & & 1\\
GRS 1915+105 & 2 & $e0$ & 0.918$\pm$0.002 & 1.744$^{+0.025}_{-0.011}$ & 0.98 & 0 & - & 1.614 & - & 1 \\
\enddata
\end{deluxetable}
We add the following caveats and conclusions:
\begin{enumerate}
\setlength\itemsep{-0.05em}
\item \textit{Novel and useful formulae}: We have derived novel forms for the spherical trajectory solutions \{$\phi\left(r_s, a , Q \right)$, $t\left( r_s, a , Q \right)$\}, given by Equation \eqref{sphphitfinal}, and their fundamental frequencies \{$\bar{\nu}_{\phi}\left(r_s, a , Q \right)$, $\bar{\nu}_{r}\left(r_s, a , Q \right)$, $\bar{\nu}_{\theta}\left(r_s, a , Q \right)$\}, given by Equation \eqref{sphfreq}. A reduced form of the vertical oscillation frequency, $\bar{\nu}_{\theta}\left(e, r_p, a \right)$  given by Equation \eqref{eqnutheta}, for equatorial eccentric orbits is also derived in Appendix \ref{nuthetaderivation}. These new and compact formulae are useful for various theoretical studies of Kerr orbits, besides other astrophysical applications (e.g., \cite{RMgalaxies2020}).
\item \textit{Orbital solutions}: The fundamental frequencies of the $eQ$, $e0$, and $Q0$ trajectories are in the range of QPO signals observed in BHXRBs, so these are viable solutions for explaining the observed QPOs in BHXRBs M82 X-1, GROJ 1655-40, XTEJ 1550-564, 4U 1630-47, and GRS 1915+105 in the GRPM paradigm. We see that these trajectory solutions are found in either region 1 or 2 of the ($r$, $a$) plane, as defined in Figure \ref{radii}, and shown in Figure \ref{plotrpaplane}. The values of the black hole spin for BHXRBs M82 X-1 and GROJ 1655-40 were fixed to their most probable values calculated in \S \ref{resultseccentric}, and to the previously observed values for the other BHXRBs for eccentric orbit solutions. For BHXRBs with two QPOs, fixing the spin to previously known values increases the uncertainty in the estimated orbital parameters, because the spin values assumed have uncertainties associated with the X-ray spectroscopic methods that are influenced by systematics, with the general finding that the solution lies near ISCO. However, our exercise still supports the GRPM. A spin value was also calculated for M82 X-1 for a $Q0$ solution. A summary of these parameter solutions and corresponding MBSO, ISCO, and ISSO radii for all BHXRBs is given in Table \ref{radiisummary}.

\item \textit{Trajectories in the torus}: We found trajectories, having different parameter combinations within the estimated range of errors in the orbital parameters and having fundamental frequencies within the width of the observed QPOs, as solutions for QPOs in BHXRBs M82 X-1 and GROJ 1655-40. We also found that the distinct parameter solutions found for these cases follow a trend that, as the eccentricity of the orbit decreases, the $Q$ value increases for a given QPO frequency pair. This behavior can also be understood from Figures \ref{nuphieccentriccontrs}$-$\ref{nunodeccentriccontrs}, where the frequencies increase as $Q$ increases, but decrease as $e$ increases for a given $r_p$. This implies that to obtain the degenerate parameter solutions for the same set of frequencies, a low eccentricity$ \ \displaystyle{\Longleftrightarrow} \ $high $Q$ trend is expected. We also found that these trajectories span a torus region near the Kerr black hole, as shown in Figure \ref{trajectories}, which together give rise to the same peaks in the power spectrum. This should also explain the strong rms seen for the HFQPOs and type C LFQPOs. Another possibility of a rigidly precessing torus was suggested \citep{Ingram2009,Ingram2011,Ingram2012}; our proposal consists of a nonprecessing torus, which includes all viable solutions of the GRPM: $eQ$, $e0$, and $Q0$ trajectories.

\item \textit{Torus region}: The emission of simultaneous QPOs is expected from a region where different trajectories having similar fundamental frequencies span a torus, as shown in Figure \ref{trajectories} and they can together show a strong peak in the power spectrum. The inner radius of the circular accretion disk is expected to be close to this torus region in such a scenario. In Figure \ref{model}, we depict this geometric model where the emission region of the simultaneous QPOs is shown as a torus region close to the inner edge of the accretion disk. This torus region is expected to be outside the MBSO radius, and the ISSO radius is expected to be in between the torus region for the eccentric orbit solutions, as observed in the case of M82 X-1. The torus region can be represented as ${R_0}^{+\Delta_1}_{-\Delta_2}$, where $R_0$ is an $e=0$ orbit (ISCO or ISSO) and $\Delta_i$ represents the region very close to $R_0$. The width of the torus region in this model is given by $\Delta r=\left( \Delta_1 + \Delta_2 \right)$. All of the orbit solutions are found to be distributed near $R_0$; hence, it is expected that this radius corresponds to the inner edge radius, $r_{\rm in}$, of the circular accretion disk. This torus region exists in region 1 and(or) 2 near the $R_0$ radius. Due to the instabilities in the inner flow, we argue that the nearly $e0$ orbits near the $R_0$ radius transcend to $eQ$ orbits. Based on the geometry of the orbits and the emission region, we plan to build a detailed GRMHD-based model to expand on the GRPM paradigm. More cases of BHXRBs with three simultaneous QPOs, if detected in the future, will help us test our models.

\item \textit{Highly eccentric solutions}: For QPOs in BHXRBs 4U 1630-47 and GRS 1915+105, we found highly eccentric $e0$ solutions. This indicates that black holes with high spin values prefer highly eccentric trajectories as solutions to the QPOs. This behavior can also be understood from Figures \ref{nuphieccentriccontrs}$-$\ref{nunodeccentriccontrs}, where we see that for black holes with very high spins, the QPOs originate very close to the black hole, and the solution contours move close to the black hole as $e$ increases. This implies that more eccentric orbits are preferred for a given frequency pair of QPOs for a black hole with very high spin. We do not find any spherical orbit solution for QPOs in these two BHXRBs, which confirms that the orbital solution is purely equatorial, but such highly eccentric solutions are unlikely. We expect more and better estimates of the orbital solutions in the future if a more precise estimate of the spin is available, or if three simultaneous QPOs are discovered in BHXRBs 4U 1630-47 and GRS 1915+105. For the case that we studied in this paper of 4U 1630-47, the lower frequency of the QPO pair probably has a different origin than the high-frequency feature suggested by \cite{KleinWolt2004}. However, even in such a scenario, the frequency range of this QPO still implies an origin near the torus region in our model. There was also another pair of QPOs observed in 4U 1630-47 \citep{KleinWolt2004}, for which there was no exact solution found in the orbital parameter space.

\item \textit{Nonequatorial solutions}: In the case of BHXRBs M82 X-1 and XTE 1550-564, we found both $eQ$ ($e0$ for XTE 1550-564) and $Q0$ solutions, and the spin determinations are slightly different for the two different types of trajectory solutions. These solutions were found close to and outside their corresponding ISSO radii. The mass of the black hole in case of M82 X-1 was fixed to the intermediate-mass black hole (IMBH) range, $\mathcal{M}=428$, because the QPOs observed in the low-frequency range (3$-$5 Hz) were found to be very stable, unlike LFQPOs, implying that they are HFQPO counterparts of BHXRBs, and hence indicating an IMBH \citep{Pasham2014}. Although this mass estimation stems from the mass-scaling relation of QPOs, which is not very reliable, a more accurate estimate of $\mathcal{M}$, if found in the IMBH range, will not significantly change the result. However, if, in the future, a more reliable and precise  estimate places it in the stellar-mass range, then the outcome from the GRPM will be dramatically different. The QPOs observed in XTE 1550-564 by \cite{Miller2001} were later shown to be the result of the data averaging by \cite{Motta2014b}, where the same QPO moved up in the frequency, appearing as a distinct QPO. As in the case of 4U 1630-47, the range of this QPO frequency still implies an origin near the torus region.

\item \textit{Spectral states}: We suggest that HFQPOs originate when $r_{\rm in}$ comes very close (near ISCO/ISSO) to the black hole during the soft spectral state of the outburst. When $r_{\rm in}$ is farther out as in the hard state, the resulting type C QPO frequency is of the order of millihertz. As a type C QPO occurs more frequently and is prone to the vertical oscillations, the increase in its frequency is explained as an increase in $\nu_{\rm np}$ when $r_{\rm in}$ decreases, with the spectral transition from the hard to soft state.

\item \textit{Circularity}: The RPM was previously applied to understand the QPOs observed in BHXRBs GROJ 1655-40 and XTEJ 1550-564 \citep{Motta2014a,Motta2014b} using the fundamental frequencies of $00$ orbits. We have found an $eQ$ solution for GROJ 1655-40 very close to an equatorial orbit having a very small eccentricity $e\sim0.071$ (see Table \ref{radiisummary}), which is in a very close agreement with the solution found by \cite{Motta2014a}, where their estimated mass of the black hole, $\mathcal{M}=5.307$, is also very close to our assumption, $\mathcal{M}=5.4$ (see Table \ref{sourcelist}). Our most probable spin estimated for GROJ 1655-40, $a=0.283$, is almost the same as found by \cite{Motta2014a}, $a\sim0.286$, but our solution provides a more precise estimation of $e$ and $Q$ values while confirming a near $00$ orbit solution as assumed by \cite{Motta2014a}. For the case of XTEJ 1550-564, the mass of the black hole was assumed to be $\mathcal{M}=9.1$ by \cite{Motta2014b} as also in our model. Our assumption for the spin, $a=0.34^{+0.37}_{-0.45}$ \citep{Orosz2011}, is also nearly the same as the value estimated by \cite{Motta2014b}; but our model gives the $e0$ and $Q0$ solutions for XTEJ 1550-564, having moderate $e=0.262^{+0.090}_{-0.062}$ and $Q=2.697^{+1.738}_{-1.627}$ values, respectively (see Table \ref{radiisummary}). This indicates that the assumption of circularity is not always valid.

\item \textit{Solution degeneracy}: To study the impact of the GRPM (with nonzero $e$ and $Q$), we have explored the behavior of \{$\delta_{\phi}$, $\delta_{\rm pp}$, $\delta_{\rm np}$\}($e$, $r_p$, $a$, $Q$) as defined in Equation \eqref{deltasecc} as deviations from the $00$ behavior (circularity). We find that the frequencies are strongly dependent on $e$ but not so much on $Q$ (see Figures \ref{nuphieccentriccontrs}$-$\ref{nunodeccentriccontrs}). This is elaborated upon in points 3 and 4 in \S \ref{eccentricmotivation}, and in points 2 and 3 in \S \ref{sphericalmotivation} for spherical orbits. The GRPM has a built-in degeneracy in the parameter space \{$e$, $r_p$, $a$, $Q$\}, called the isofrequency pairs, for a given combination of QPO frequencies. This degeneracy is a known behavior of trajectories around a Kerr black hole \citep{Warburton2013}, where different combinations of \{$E$, $L_z$, $Q$\} can have the same set \{$\nu_{\phi}$, $\nu_{r}$, $\nu_{\theta}$\} for a fixed $a$. An evidence of this degeneracy is also seen in Figures \ref{nuphieccentriccontrs}$-$\ref{nunodeccentriccontrs}, where the contours of \{$\delta_{\phi}$, $\delta_{\rm pp}$, $\delta_{\rm np}$\}($e$, $r_p$, $a$, $Q$) have multiple solutions; that is, for a given $\delta$ value, there are different combinations of \{$e$, $Q$\} that have distinct contours on the ($r_p$, $a$) plane. Unlike RPM, the mass of the black hole is always assumed from the previous estimates in the GRPM, which is a valid assumption because the underlying physics or behavior of the Kerr orbits is independent of $M_{\bullet}$. The GRPM, along with the statistical method (Figure \ref{methodflowchart}, Appendix \ref{methodsection}) that is applied, provides a more precise estimation of the spin of the black hole.

\item \textit{Frequency ratio}: The 3:2 and 5:3 ratios of the simultaneous HFQPOs are a phenomenon observed in a few cases of BHXRBs: 300 and 450 Hz HFQPOs in GROJ 1655-40 \citep{RemillardMorgan1999, Strohmayer2001a}, 240 and 160 Hz HFQPOs in H1743-322 \citep{Homan2005, Remillard2006}. Such claims, other than the case of GROJ 1655-40, are probably not real \citep{Belloni2012}. Hence, the possibility of such ratios is still causes skepticism. However, if true, the GRPM suggests that the origin of these ratios is very close to the torus region and $r_{\rm in}$.

\item \textit{The PBK correlation}: In \S \ref{PBK}, we show that the $e0$ solution \{$e=0.071$, $a=0.283$, $Q=0$, $\mathcal{M}=5.4$\}, estimated using a fine-grid method in \S \ref{resultseccentric}, fits the PBK correlation that was previously observed in BHXRB GROJ 1655-40 \citep{Motta2014a}. This fit is shown in Figure \ref{PBKGROJ}. We also found that 10 observation IDs, where $L_l$ and $L_{\rm LF}$ (broad frequency components) were detected simultaneously \citep{Motta2014a}, show low-eccentricity $eQ$ solutions, where the calculated parameters are shown in Table \ref{eQsolutionsGROJ}. The calculated $Q$ values are consistent with large $r_p$ and small $e$ values. This exercise suggests that $eQ$ solutions for QPOs are viable.

\item \textit{Probing the disk edge with a GR fluid model}: We study a model of fluid flow in the general-relativistic thin accretion disk \citep{2012MNRAS.420..684P,2014ApJ...791...74M}. We find that the disk edge flows into a torus region containing Hamiltonian geodesics that was obtained for M82 X-1 ($r=4.7-9.08$) and GROJ 1655-40 ($r=5.11-6.67$). Specifically, the sharp gradient of the $p^{\rm gas}/p^{\rm rad}$ pressure ratio, seen in Figure \ref{pgasprad}, suggests that the edge region is a launch pad for the instabilities that orbit with fundamental frequencies of the geodesics in the edge and geodesic regions, which then follow the geodesics inside the torus region and also close to the edge region, where Hamiltonian dynamics is applicable, that is built into the GRPM. The range of \{$r$, $Q_{\phi}$, $\beta_r$, $p^{\rm gas}/p^{\rm rad}$\} for the edge and inner regions for different combinations of $a$ and $\dot{m}$, fixing \{$m_1=1$, $\alpha=0.1$\}, is given in Table \ref{regionstab}, and the contours of $\beta_r$ and $Q_{\phi}$ in the ($r$, $a$) plane for the edge region are shown in Figures \ref{betar} and \ref{Qf}. The ranges of $Q_{\phi}$ (tuned to $\Delta \nu$, the width of the observed QPO), which is defined by orbits in the torus which was provided by observed frequency centroids, in both the edge and inner regions are very high compared to those observed in BHXRBs ($Q_{\phi}=5-40$). We are suggesting that the QPOs originate in the geodesic region. We also see that the edge is adjacent to the torus region (consisting of geodesics) found for M82 X-1 and GROJ 1655-40, implying that the QPOs are originating from geodesics close to the edge region. Hence, the particle and gas dynamics models together justify the scenario sketched in Figure \ref{model}, of a unified fluid-particle picture that is the following: the source of the particles in the torus are dynamical instabilities of plasma blobs ejected from the edge region. These blobs have zero $\alpha$ and therefore obey the Hamiltonian dynamics. The clue that the torus region physically overlaps with the edge and geodesic regions is a subject of future detailed GRMHD models (and simulations).

\item \textit{Isofrequency combinations}: In the cases with three simultaneous QPOs, once $a$ is fixed (to the most probable value or the previously estimated value), it is easy to predict the remaining parameters \{$e$, $r_p$, $Q$\} using three QPO frequencies. In the case of M82 X-1 and GROJ 1655-40, when $a$ was fixed to the most probable value (Table \ref{erQresults}), we obtained a single solution for \{$e$, $r_p$, $Q$\} and their errors \{$\Delta e$, $\Delta r_p$, $\Delta Q$\}, where this range of parameters spans the torus region based on the GRPM. However, there is a finite possibility \citep{Warburton2013} that distinct solutions for the \{$e$, $r_p$, $Q$\} triad are obtained for the same triple QPO frequency set, subject to the bound orbit conditions: $0\leq e <1$, $Q\geq 0$, and Equation \eqref{boundcondition}. This completely depends on the values of the QPO frequency set that are further subject to the constraints of bound orbit conditions. In the cases where only two simultaneous QPOs exist, it is difficult to predict whether an $e=0$ orbit will be preferred over an $e>0$ orbit, or a $Q=0$ orbit will be preferred over a $Q>0$ orbit, or vice versa. This will be clear when more cases of three simultaneous QPOs are found and whether they yield distinct solution sets for \{$e$, $r_p$, $a$, $Q$\}, thereby indicating if the torus region at the disk edge is indeed the geometric origin of QPOs. From our numerical experiment, we find a distinct exact solution for \{$e$, $r_p$, $Q$\} for the three simultaneous QPOs case, where $a$ was fixed to the most probable value. The RPM restricts the search to \{$e=0$, $Q=0$\} orbital solutions, while the GRPM expands it to  more general but astrophysically possible \{$e\neq0, \ Q\neq0$\} solutions and thereby subsumes the RPM within its framework. Hence, the GRPM provides more realistic orbit solutions around a Kerr black hole that are outside the scope of the RPM, thus giving more impetus to probes of physical models of the origin of QPOs.
\item \textit{Caveats}: The results predicted by the GRPM are subject to the veracity of the observed data that are inputs to our model. For example, in the case of 4U 1630-47 and GRS 1915+105, very highly eccentric orbit solutions obtained by the GRPM are unlikely; this implies that very high spin values in these cases are probably unreliable. Similarly, if M82 X-1 does not host an IMBH but a stellar-mass black hole or a neutron star, then the results predicted by the GRPM will change drastically. Also, for 4U 1630-47 and XTEJ 1550-564, where the input frequencies of QPOs are not very reliable \citep{KleinWolt2004,Motta2014b}, as discussed before, the results obtained by the GRPM might not be physically meaningful. As most of the measured frequencies do exist in a similar range, then their geometric origin in the torus region (as predicted by the GRPM) is valid.
\item \textit{Future work}: In the near future, we expect suitable observational results from the currently operative Indian X-ray satellite, AstroSat, and from future missions, such as eXTP, which is proposed to have instruments with much higher sensitivity for fast variations and X-ray timing. If simultaneous QPO signals are observed from these missions, we expect to test our GRPM further.
\end{enumerate}

\vspace{0.5cm}
We would like to thank the anonymous referee for detailed and insightful suggestions that have improved our paper significantly. We acknowledge the DST SERB grant No. CRG 2018/003415 for financial support. We would like to thank Dr. Prashanth Mohan for his useful suggestions and comments. We would like to thank Saikat Das for helping us with Figure \ref{precessionplots} and \ref{model}. We acknowledge the use and support of the IIA-HPC facility.

\appendix
 \section{Vertical Oscillation Frequency for Eccentric Orbits About \\
 Equatorial Plane with $Q=0$}
 \label{nuthetaderivation}
 Here, we derive the $\theta$ oscillation frequency for the equatorial eccentric orbits about the equatorial plane. Using Equations \eqref{nur} and \eqref{nutheta}, we can write
 \begin{subequations}
 \begin{equation}
\frac{\bar{\nu}_{\theta}}{\bar{\nu}_{r}} = \frac{a \sqrt{1- E^2} z_{+} I_8 \left(  e, \mu, a, Q \right)}{2 F\left( \frac{\pi}{2},\frac{z_{-}^{2}}{z_{+}^{2}}\right) },
 \end{equation}
 where the substitution of $I_8 \left(  e, \mu, a, Q \right)$ from Equation (6h) of \cite{RMCQG2019} into the above equation yields
 \begin{equation}
 \frac{\bar{\nu}_{\theta}}{\bar{\nu}_{r}} = \frac{\mu \left( 1- e^2\right) a \sqrt{1- E^2} z_{+} F\left( \dfrac{\pi}{2}, k^2\right)}{\sqrt{C-A+\sqrt{{B}^2-4 AC}} F\left( \frac{\pi}{2},\frac{z_{-}^{2}}{z_{+}^{2}}\right) }. \label{eqnuthetanur}
 \end{equation}
 \end{subequations}
 By the substitution of $A$, $B$, and $C$ using Equations (7f$-$7h) of \cite{RMCQG2019}, and using $Q=0$ for the equatorial orbits, we find
 \begin{subequations}
 \begin{equation}
 \sqrt{C-A+\sqrt{{B}^2-4 AC}}= \mu^{1/2} \left( 1-e^2\right) \left[ 1- \mu^2 x^2 \left( 3- e^2 -2e\right) \right]^{1/2}. \label{eqexp}
\end{equation}
Also, from Equation (9d) of \cite{RMCQG2019}, we see that 
\begin{equation}
z_{-}=0, \ \ \ \ z_{+}=\frac{\sqrt{L_z^2+ a^2 \left(1- E^2\right) }}{a\sqrt{\left(1- E^2\right)}}=\frac{\sqrt{x^2+a^2+ 2a E x }}{a\sqrt{\left(1- E^2\right)}}, \label{eqzpm}
\end{equation} 
 for $Q=0$, which implies that 
\begin{equation}
F\left( \frac{\pi}{2},\frac{z_{-}^{2}}{z_{+}^{2}}\right)= \frac{\pi}{2}. \label{eqellipticF}
\end{equation}
\end{subequations}
Hence, Equations \eqref{eqnuthetanur}$-$\eqref{eqellipticF} together reduce $\bar{\nu}_{\theta}/ \bar{\nu}_{r}$  for equatorial orbits to
\begin{equation}
\frac{\bar{\nu}_{\theta}}{\bar{\nu}_{r}}=\frac{2 \mu^{1/2} \sqrt{x^2+a^2+ 2a E x } \cdot F\left( \dfrac{\pi}{2}, k^2\right)}{\pi \left[ 1- \mu^2 x^2 \left( 3- e^2 -2e\right) \right]^{1/2} }.
\end{equation}
We see from Equations (7f$-$7j) of \cite{RMCQG2019} that $k^2=\left( n^2-m^2 \right)/ \left(1-m^2 \right)$ can be written in terms of $A$, $B$, and $C$ as
\begin{equation}
k^2 =\frac{2\sqrt{B^2 -2 A C}}{\left( -A+C+\sqrt{B^2 -2 A C}\right) }, \label{ksqrABC}
\end{equation}
where the substitution of $A$, $B$, and $C$ for $Q=0$ gives
\begin{equation}
k^2=m^2=\frac{4e x^2 \mu^2}{\left[ 1- \mu^2 x^2 \left( 3-e^2 -2e\right)\right]}. \label{ksqreq}
\end{equation}
Hence, we can write $\bar{\nu}_{\theta}$ for the equatorial orbits as
\begin{equation}
\bar{\nu}_{\theta}\left( e, \mu, a\right)=\frac{2 \bar{\nu}_{r}\left( e, \mu, a\right) \mu^{1/2} \sqrt{\left( x^2+a^2+ 2a E x \right) } \cdot F\left( \dfrac{\pi}{2}, k^2\right)}{\pi \left[ 1- \mu^2 x^2 \left( 3- e^2 -2e\right) \right]^{1/2} }, 
\end{equation}
where $\bar{\nu}_{r}\left( e, \mu, a\right)$ is given by Equation \eqref{eqnur} and $k^2$ is given by Equation \eqref{ksqreq}.

\section{Trajectory and Frequency Formulae for Spherical Orbits}
\label{sphericalorbitsderivations}
\begin{enumerate}
\item Azimuthal angle and coordinate time:
The integrals of motion for a general nonequatorial trajectory of a particle with rest mass $m_{0}$ around a Kerr black hole have been derived using the Hamilton$-$Jacobi method, in terms of the Boyer$-$Lindquist coordinates ($r$, $\phi$, $\theta$, $t$) \citep{Carter1968,Schmidt2002}:
\begin{subequations}
\begin{eqnarray}
\phi - \phi_{0}=&& -\frac{1}{2}\int_{r_{0}}^{r} \frac{1}{\Delta \sqrt{R}}\frac{\partial R}{\partial L_z} {\rm d} r^{'} -\frac{1}{2}  \int_{\theta_{0}}^{\theta} \frac{1}{\sqrt{\Theta}}\frac{\partial \Theta}{\partial L_z}{\rm d}\theta^{'} = -\frac{1}{2} I_{1} -\frac{1}{2}  H_{1}, \label{phiintmtn}\\
t - t_{0}=&& \frac{1}{2}\int_{r_{0}}^{r} \frac{1}{\Delta \sqrt{R}}\frac{\partial R}{\partial E} {\rm d} r^{'}  + \frac{1}{2}\int_{\theta_{0}}^{\theta} \frac{1}{\sqrt{\Theta}}\frac{\partial \Theta}{\partial E}{\rm d}\theta^{'} = \frac{1}{2} I_2 + \frac{1}{2} H_2, \label{tintmtn} \\
\int_{r_{0}}^{r} \frac{{\rm d} r^{'}}{\sqrt{R}}=&& \int_{\theta_{0}}^{\theta} \frac{{\rm d}\theta^{'}}{\sqrt{\Theta}}\Rightarrow I_8=H_3, \label{Rthetaint}
\end{eqnarray}
where $R$ and $\Theta$ are given by
\begin{eqnarray}
R=&&\left[\left( {r^{'}}^2 +a^2 \right)E -a L_z \right] ^{2}- \Delta \left[ {r^{'}}^2+\left(L_z-a E \right)^{2}+Q  \right], \label{Rdefn} \\
\Theta =&& Q - \left[\left(1 - E^2 \right)a^2+\frac{L_{z}^2}{\sin^2 \theta^{'}}  \right] \cos^{2}\theta^{'}.
\end{eqnarray}
\end{subequations}
We have from Equation \eqref{Rthetaint} that 
\begin{equation}
\frac{{\rm d} r^{'}}{\sqrt{R}}= \frac{{\rm d}\theta^{'}}{\sqrt{\Theta}};
\end{equation}
the substitution of the above equation into Equations (\ref{phiintmtn}, \ref{tintmtn}) for the spherical orbits  reduces the expressions of the azimuthal angle and coordinate time to
\begin{eqnarray}
\phi - \phi_{0}=&& - \dfrac{1}{2} \left[\frac{1}{\Delta }\frac{\partial R}{\partial L_z} H_3 + H_1 \right], \ \ \ \ \
 t - t_{0}= \dfrac{1}{2} \left[ \frac{1}{\Delta }\frac{\partial R}{\partial E} H_3 + H_2  \right]. \label{phitsph}
\end{eqnarray}
Since $r=r_s$ is constant for the spherical orbits, the expressions of $\frac{1}{\Delta }\frac{\partial R}{\partial L_z}$ and $\frac{1}{\Delta }\frac{\partial R}{\partial E}$ can be written as
\begin{eqnarray}
\frac{1}{\Delta }\frac{\partial R}{\partial L_z}=&&\dfrac{2\left( 2L_z r_s -L_z r_{s}^2 -2r_s aE\right) }{\Delta}, \ \  \ \ \
 \frac{1}{\Delta }\frac{\partial R}{\partial E}= \dfrac{2\left[ E \left( a^2 r_{s}^2 +r_{s}^4 +2a^2 r_s\right) -2 L_z a r_s \right] }{\Delta}, \label{constantexp}
\end{eqnarray}
and the integrals $H_1$, $H_2$, and $H_3$ have been previously derived to be \citep{Fujita2009,RMCQG2019}
\begin{subequations}
\begin{equation}
H_1\left( \theta , \theta_{0}, e, \mu , a, Q \right)=  \frac{2 L_z}{ z_{+} a \sqrt{1-E^2}} \left\lbrace F\left( \arcsin \left( \frac{\cos \theta_{0}}{z_{-}}\right) , \frac{z_{-}^2}{z_{+}^2}\right)-F\left( \arcsin \left( \frac{\cos \theta}{z_{-}}\right), \frac{z_{-}^2}{z_{+}^2}\right)  + \right. \nonumber
\end{equation}
\begin{equation}
\left. \Pi\left( z_{-}^2, \arcsin \left( \frac{\cos \theta}{z_{-}}\right) , \frac{z_{-}^2}{z_{+}^2}\right) -\Pi \left( z_{-}^2, \arcsin \left( \frac{\cos \theta_{0}}{z_{-}}\right) , \frac{z_{-}^2}{z_{+}^2}\right)    \right\rbrace , \label{H1}
\end{equation}
\begin{equation}
H_2\left(  \theta , \theta_{0}, e, \mu , a, Q \right)= \frac{2E a z_{+}}{\sqrt{1-E^2}}  \left\lbrace   K\left( \arcsin \left( \frac{\cos \theta}{z_{-}}\right)  , \frac{z_{-}^2}{z_{+}^2} \right) - F\left( \arcsin \left( \frac{\cos \theta}{z_{-}}\right)  , \frac{z_{-}^2}{z_{+}^2} \right) -  \right. \nonumber
\end{equation}
\begin{equation}
 \left. K\left(  \arcsin \left( \frac{\cos \theta_{0}}{z_{-}}\right), \frac{z_{-}^2}{z_{+}^2} \right)  +F\left( \arcsin \left( \frac{\cos \theta_{0}}{z_{-}}\right)  , \frac{z_{-}^2}{z_{+}^2} \right)   \right\rbrace , \label{H2}
\end{equation}
\begin{equation}
H_3\left(  \theta , \theta_{0}, e, \mu , a, Q \right) = \frac{1}{a\sqrt{1-E^2}z_{+}}\left\lbrace F\left( \arcsin \left( \frac{\cos \theta_{0}}{z_{-}}\right) ,\frac{z_{-}^{2}}{z_{+}^{2}}\right) -F\left( \arcsin \left( \frac{\cos \theta}{z_{-}}\right),\frac{z_{-}^{2}}{z_{+}^{2}}\right)  \right\rbrace,  \label{integral2}
\end{equation}
\label{Hequations}
\end{subequations} 
where $z_{\pm}$ are given by Equation (9d) of \cite{RMCQG2019}. Hence, the substitution of Equations (\ref{constantexp}) and (\ref{Hequations}) into Equation \eqref{phitsph} yields the expressions of ($\phi-\phi_{0}$, $t-t_{0}$) for the spherical orbits, given by
\begin{subequations}
\begin{eqnarray}
\phi-\phi_{0}=&& \dfrac{1}{a \sqrt{1-E^2}z_{+}} \left\lbrace \dfrac{\left(a^2 L_z -2 a E r_s \right) }{\Delta} \left[ F\left( \arcsin \left( \frac{\cos \theta}{z_{-}}\right) , \frac{z_{-}^2}{z_{+}^2} \right) - F\left( \arcsin \left( \frac{\cos \theta_{0}}{z_{-}}\right) , \frac{z_{-}^2}{z_{+}^2} \right) \right]  \right. \nonumber  \\
&& \left. - L_z \left[ \Pi\left( z_{-}^2, \arcsin \left( \frac{\cos \theta}{z_{-}}\right) , \frac{z_{-}^2}{z_{+}^2}\right) -\Pi \left( z_{-}^2, \arcsin \left( \frac{\cos \theta_{0}}{z_{-}}\right) , \frac{z_{-}^2}{z_{+}^2}\right) \right] \right\rbrace , \\
t-t_{0}=&& \dfrac{1}{a \sqrt{1-E^2}z_{+}} \left\lbrace E a^2 z_{+}^{2} \left[ K\left( \arcsin \left( \frac{\cos \theta}{z_{-}}\right) , \frac{z_{-}^2}{z_{+}^2} \right) - K \left( \arcsin \left( \frac{\cos \theta_0}{z_{-}}\right) , \frac{z_{-}^2}{z_{+}^2} \right)\right]   \right. \nonumber \\
&& \left.  + \left[F\left( \arcsin \left( \frac{\cos \theta_0}{z_{-}}\right) , \frac{z_{-}^2}{z_{+}^2} \right) \right.  - F\left( \arcsin \left( \frac{\cos \theta}{z_{-}}\right) , \frac{z_{-}^2}{z_{+}^2} \right) \right] \cdot \nonumber \\
&& \left. \left[E a^2 z_{+}^{2}  + \dfrac{E \left( a^2 r_{s}^2 + r_{s}^4 + 2 a^2 r_s \right) -2 L_z a r_s }{\Delta} \right]   \right\rbrace .
\end{eqnarray}
\label{sphphitfinal}
\end{subequations}
\item Fundamental frequencies:
The closed forms for fundamental frequencies associated with the nonequatorial eccentric bound trajectories have been previously derived \citep{Schmidt2002,RMCQG2019} and are given by Equations \eqref{nuphi}$-$\eqref{nutheta}. We first reduce the common denominator of these expressions to the case of spherical orbits. If we take $I_8\left( e, \mu, a, Q \right)$ common from the denominator, it gives
\begin{eqnarray}
&&\left[ \left( I_2  + 2 a^2 z_{+}^2 E I_8 \right) F\left( \frac{\pi}{2},\frac{z_{-}^{2}}{z_{+}^{2}}\right)- 2 a^2 z_{+}^2 E I_8 K\left( \frac{\pi}{2},\frac{z_{-}^{2}}{z_{+}^{2}}\right)  \right]  = I_8 \left[   \left( \frac{I_2}{I_8}  + 2 a^2 z_{+}^2 E \right) F\left( \frac{\pi}{2},\frac{z_{-}^{2}}{z_{+}^{2}}\right) \right. \nonumber \\
&& \left. - 2 a^2 z_{+}^2 E K\left( \frac{\pi}{2},\frac{z_{-}^{2}}{z_{+}^{2}}\right) \right], \label{redsph}
\end{eqnarray}
where by definition $I_2/I_8=\frac{1}{\Delta}\frac{\partial R}{\partial E}$ for spherical orbits, which is given by Equation \eqref{constantexp}. Hence, Equations \eqref{redsph}, \eqref{constantexp}, and \eqref{nutheta} combine to give the vertical oscillation frequency for the spherical orbits:
\begin{equation}
\bar{\nu_{\theta}} \left( r_s , a, Q \right) =  \frac{a \sqrt{1- E^2} z_{+} }{4\left\lbrace\left[\dfrac{\left[ E \left( a^2 r_{s}^2 +r_{s}^4 +2a^2 r_s\right) -2 L_z a r_s \right] }{\Delta}  +  a^2 z_{+}^2 E  \right] F\left( \frac{\pi}{2},\frac{z_{-}^{2}}{z_{+}^{2}}\right)  -  a^2 z_{+}^2 E  K\left( \frac{\pi}{2},\frac{z_{-}^{2}}{z_{+}^{2}}\right)  \right\rbrace }. \label{nuthetasph1}
\end{equation}
Next, using Equation \eqref{redsph}, the azimuthal frequency, Equation \eqref{nuphi}, can be written as
\begin{subequations}
\begin{eqnarray}
\bar{\nu_{\phi}}\left( r_s, a, Q \right)=\frac{\left\lbrace \left[ - \dfrac{I_1}{I_8} - 2 L_z \right] F\left( \frac{\pi}{2},\frac{z_{-}^{2}}{z_{+}^{2}}\right)  + 2 L_z \cdot \Pi\left( z_{-}^2, \frac{\pi}{2},\frac{z_{-}^{2}}{z_{+}^{2}}\right)  \right\rbrace }{4 \pi \left\lbrace\left[\dfrac{\left[ E \left( a^2 r_{s}^2 +r_{s}^4 +2a^2 r_s\right) -2 L_z a r_s \right] }{\Delta}  +  a^2 z_{+}^2 E  \right] F\left( \frac{\pi}{2},\frac{z_{-}^{2}}{z_{+}^{2}}\right)  -  a^2 z_{+}^2 E  K\left( \frac{\pi}{2},\frac{z_{-}^{2}}{z_{+}^{2}}\right)  \right\rbrace }, \nonumber \\
\end{eqnarray}
where $I_1/I_8=\frac{1}{\Delta}\frac{\partial R}{\partial L_z}$, which is given by Equation \eqref{constantexp}. Hence, the azimuthal frequency for the spherical orbits is given by
\begin{eqnarray}
\bar{\nu_{\phi}}\left( r_s, a, Q \right)=\frac{\left\lbrace \left[ - \dfrac{\left( 2L_z r_s -L_z r_{s}^2 -2r_saE\right) }{\Delta} -  L_z \right] F\left( \frac{\pi}{2},\frac{z_{-}^{2}}{z_{+}^{2}}\right)  + L_z \cdot \Pi\left( z_{-}^2, \frac{\pi}{2},\frac{z_{-}^{2}}{z_{+}^{2}}\right)  \right\rbrace }{2 \pi \left\lbrace\left[\dfrac{\left[ E \left( a^2 r_{s}^2 +r_{s}^4 +2a^2 r_s\right) -2 L_z a r_s \right] }{\Delta}  +  a^2 z_{+}^2 E  \right] F\left( \frac{\pi}{2},\frac{z_{-}^{2}}{z_{+}^{2}}\right)  -  a^2 z_{+}^2 E  K\left( \frac{\pi}{2},\frac{z_{-}^{2}}{z_{+}^{2}}\right)  \right\rbrace }. \nonumber \\
\label{nuphisph1}
\end{eqnarray}
\end{subequations}
Similarly, the radial oscillation frequency, Equation \eqref{nur}, can be written for the spherical orbits by using Equation \eqref{redsph} as
\begin{subequations}
\begin{equation}
\bar{\nu_r} \left(r_s, a, Q \right) = \frac{F\left( \frac{\pi}{2},\frac{z_{-}^{2}}{z_{+}^{2}}\right) }{2 I_8 \left\lbrace\left[\dfrac{\left[ E \left( a^2 r_{s}^2 +r_{s}^4 +2a^2 r_s\right) -2 L_z a r_s \right] }{\Delta}  +  a^2 z_{+}^2 E  \right] F\left( \frac{\pi}{2},\frac{z_{-}^{2}}{z_{+}^{2}}\right)  -  a^2 z_{+}^2 E  K\left( \frac{\pi}{2},\frac{z_{-}^{2}}{z_{+}^{2}}\right)  \right\rbrace }, 
\end{equation}
where, for spherical orbits the integral $I_8$ reduces to a constant, as shown below.

We see that the expression for $k^2$, Equation \eqref{ksqrABC}, reduces to zero because $A=B=0$ (Equations (7f), (g) of \cite{RMCQG2019}) for spherical orbits ($e=0$). Hence, $I_8\left( e=0, \mu, a , Q\right)$ (Equation (6h) of \cite{RMCQG2019}) reduces to
\begin{equation}
I_8=\frac{2\mu}{\sqrt{C}}F \left(\dfrac{\pi}{2}, k^2=0\right)=\dfrac{\pi r_s}{\sqrt{r_{s}^4 \left( 1- E^2\right) + \left( 3 Q a^2 -2 x^2 r_s -2 Q r_s \right) }}.
\end{equation}
Hence, the radial oscillation frequency for spherical orbits reduces to
\begin{equation}
\bar{\nu_r} \left(r_s, a, Q \right) = \frac{\sqrt{r_{s}^4 \left( 1- E^2\right) + \left( 3 Q a^2 -2 x^2 r_s -2 Q r_s\right) } \cdot F\left( \frac{\pi}{2},\frac{z_{-}^{2}}{z_{+}^{2}}\right) }{2 \pi r_s \left\lbrace\left[\dfrac{\left[ E \left( a^2 r_{s}^2 +r_{s}^4 +2a^2 r_s\right) -2 L_z a r_s \right] }{\Delta}  +  a^2 z_{+}^2 E  \right] F\left( \frac{\pi}{2},\frac{z_{-}^{2}}{z_{+}^{2}}\right)  -  a^2 z_{+}^2 E  K\left( \frac{\pi}{2},\frac{z_{-}^{2}}{z_{+}^{2}}\right)  \right\rbrace }. \label{nursph1}
\end{equation}
\end{subequations}
\end{enumerate}

\section{Reduction of Frequency Formulae to the Equatorial Circular Case} 
\label{reducecircular}
Here, we reduce the fundamental frequency formulae to the known case of equatorial circular orbits ($00$). We show this reduction from both the equatorial eccentric ($e0$) and the spherical ($Q0$) orbits below:
\begin{enumerate}
\item \textit{Reduction from $e0$ orbits}: We see that for circular orbits  ($e=0$), the expressions of $m^2$, ${p_1}^2$, ${p_2}^2$, and ${p_3}^2$ (Equations (7i), (k) of \cite{RMCQG2019}) reduce to
\begin{equation}
m^2={p_1}^2={p_2}^2={p_3}^2=0.
\end{equation} 
We first make the subtitution $m^2=0$ in Equations (\ref{eqnuphi})$-$(\ref{eqnutheta}), which gives
\begin{subequations}
\begin{equation}
\bar{\nu}_{\phi} =\frac{  a_1  \Pi  \left( -p_{2}^2, \frac{\pi}{2}, 0\right) + b_1 \Pi\left( -p_{3}^2, \frac{\pi}{2}, 0\right) }{ 2 \pi \left\lbrace  \Pi\left( -p_{1}^2 , \frac{\pi}{2} , 0\right)  \left[ a_2  \frac{\left( p_{1}^2 +2 \right) }{2\left( 1+p_{1}^2\right) } +b_2\right]  + c_2 \Pi\left( -p_{2}^2 , \frac{\pi}{2} ,0\right) + d_2 \Pi\left( -p_{3}^2 , \frac{\pi}{2} , 0\right)   \right\rbrace},  
\end{equation}
\begin{equation}
\bar{\nu}_{r}=\frac{1}{2 \left\lbrace  \Pi\left( -p_{1}^2 , \frac{\pi}{2} , 0\right)  \left[ a_2  \frac{\left( p_{1}^2 +2 \right) }{2\left( 1+p_{1}^2\right) } +b_2\right]  + c_2 \Pi\left( -p_{2}^2 , \frac{\pi}{2} ,0\right) + d_2 \Pi\left( -p_{3}^2 , \frac{\pi}{2} , 0\right)   \right\rbrace },
\end{equation}
\begin{equation}
\bar{\nu}_{\theta}=\frac{ \bar{\nu}_{r} \mu^{1/2} \sqrt{\left( x^2+a^2+ 2a E x \right) } }{ \sqrt{ 1- 3\mu^2 x^2 } }. 
\end{equation}
\label{step1}
\end{subequations}
Next, the substitution of ${p_1}^2={p_2}^2={p_3}^2=0$ in Equation \eqref{step1} yields
\begin{subequations}
\begin{eqnarray}
\bar{\nu}_{\phi} =&&\frac{  a_1   + b_1  }{ 2 \pi \left(  a_2  +b_2  + c_2+ d_2 \right)}, \\ 
\bar{\nu}_{r}=&&\frac{1}{\pi \left(  a_2  +b_2  + c_2+ d_2 \right)}, \\
\bar{\nu}_{\theta}=&&\frac{ \bar{\nu}_{r} \mu^{1/2} \sqrt{\left( x^2+a^2+ 2a E x \right) } }{ \sqrt{ 1- 3\mu^2 x^2 } }. 
\end{eqnarray}
\label{step2}
\end{subequations}
By substituting $e=0$ in Equation (16) of \cite{RMarxiv2019}, we find that
\begin{subequations}
\begin{eqnarray}
a_1+b_1=&&\frac{2\mu^{1/2} \left( L_z -2 x \mu \right)}{\sqrt{1-3 \mu^2 x^2}\left(1- 2\mu + a^2 \mu^2 \right)}, \\
a_2 +b_2  + c_2+ d_2 =&&\frac{2 \left( E + E a^2 \mu^2 -2 a x \mu^3 \right)}{\mu^{3/2}\sqrt{1-3 \mu^2 x^2}\left(1- 2\mu + a^2 \mu^2 \right)}.
\end{eqnarray}
\label{sumconstants}
\end{subequations}
Now, by substituting Equation \eqref{sumconstants} in Equation \eqref{step2}, we get
\begin{subequations}
\begin{eqnarray}
\bar{\nu}_{\phi} =&&\frac{\mu^2 \left(L_z -2 x \mu \right)}{2 \pi \left( E + E a^2 \mu^2 -2 a x \mu^3 \right)}, \\
\bar{\nu}_{r}=&&\frac{\mu^{3/2}\sqrt{1-3 \mu^2 x^2}\left(1- 2\mu + a^2 \mu^2 \right)}{2 \pi \left( E + E a^2 \mu^2 -2 a x \mu^3 \right)}, \\
\bar{\nu}_{\theta}=&&\frac{\mu^{2}\left(1- 2\mu + a^2 \mu^2 \right) \sqrt{\left( x^2+a^2+ 2a E x \right) } }{  2 \pi \left( E + E a^2 \mu^2 -2 a x \mu^3 \right)}.
\end{eqnarray} 
\label{step3}
\end{subequations}
The expressions of $E$, $L_z$, and $x$ for $00$ orbits are given by \citep{Bardeen1972}
\begin{subequations}
\begin{eqnarray}
E=&& \dfrac{\left( r_{c}^{2} - 2r_{c} + a \sqrt{r_{c}}\right) }{r_{c}  \left(  r_{c}^{2} - 3 r_{c} + 2 a \sqrt{r_{c}} \right)^{1/2}}, \\
L_z=&& \dfrac{\sqrt{r_{c}} \left( r_{c}^{2} + a^{2} -  2 a \sqrt{r_{c}}\right) }{r_{c} \left(  r_{c}^{2} - 3 r_{c} + 2 a \sqrt{r_{c}} \right)^{1/2}}, \\
x=&& \frac{r_c\left( r_c^{1/2} -a\right) }{\left(  r_{c}^{2} - 3 r_{c} + 2 a \sqrt{r_{c}} \right)^{1/2}},
\end{eqnarray}
\label{ELx00}
\end{subequations}
where $r_c$ is the radius of the circular orbit. These expressions can be also be obtained by substituting \{$e=0$, $Q=0$, $\mu=1/r_c$\} in the more general expressions given by Equation (5) of \cite{RMCQG2019}. Finally, by substituting $E$, $L_z$, $x$, and $\mu=1/r_c$ from Equation \eqref{ELx00} into Equation \eqref{step3}, we recover the frequency formulae for $00$ orbits:
\begin{subequations}
\begin{eqnarray}
\bar{\nu}_{\phi}=&&\frac{1}{2 \pi \left( r_c^{3/2} + a\right)}, \\
\bar{\nu}_{r}=&& \bar{\nu}_{\phi}\left( 1- \dfrac{6}{r_c} - \dfrac{3 a^2}{r_c^2} + \dfrac{8a}{r_c^{3/2}}\right)^{1/2},\\
\bar{ \nu}_{\theta}=&& \bar{\nu}_{\phi} \left( 1+ \dfrac{3 a^2}{r_c^2} - \dfrac{4a}{r_c^{3/2}}\right)^{1/2},
\end{eqnarray}
\end{subequations}
as given by Equation \eqref{circfreq}.
\item \textit{Reduction from $Q0$ orbits}: We find that for circular orbits ($Q=0$), the expressions of $z_{\pm}$ [Equation (9d) of \cite{RMCQG2019}] reduce to
\begin{equation}
z_{-}=0, \ \ \ \ \ \ z_{+}=\frac{\sqrt{L_z^2 + a^2 \left(1- E^2 \right)}}{a \sqrt{1- E^2}}. \label{zpmcirc}
\end{equation}
The substitution of Equation \eqref{zpmcirc} in the frequency formulae of $Q0$ orbits, Equation (\ref{sphfreq}), yields
\begin{subequations}
\begin{eqnarray}
\bar{\nu}_{\phi}=&&\frac{ \left( -2L_z r_c + L_z r_c^2 + 2r_c aE \right)  }{2 \pi \left[ E \left( a^2 r_c^2 +r_c^4 +2a^2 r_c\right) -2 L_z a r_c \right] },  \\
\bar{\nu}_r =&&\frac{\sqrt{r_c^4 \left( 1- E^2\right) -2 x^2 r_c } \Delta}{2 \pi r_c \left[ E \left( a^2 r_c^2 +r_c^4 +2a^2 r_c\right) -2 L_z a r_c \right]  }, \\
\bar{\nu}_{\theta}=&&\frac{\sqrt{L_z^2 + a^2 \left(1- E^2 \right)  }\Delta}{2 \pi  \left[ E \left( a^2 r_c^2 +r_c^4 +2a^2 r_c\right) -2 L_z a r_c \right]}.
\end{eqnarray}
\label{step5}
\end{subequations}
Using the expressions of $E$, $L_z$, and $x$ from Equation \eqref{ELx00}, we find that
\begin{subequations}
\begin{eqnarray}
\left[ E \left( a^2 r_c^2 +r_c^4 +2a^2 r_c\right) -2 L_z a r_c \right]=&&\frac{r_c^{3/2} \Delta \left(r_c^{3/2} +a \right)}{\left( r_c^2 - 3 r_c +2 a r_c^{1/2}\right)^{1/2}}, \\
\left( -2L_z r_c + L_z r_c^2 + 2r_c aE \right)=&& \frac{r_c^{3/2} \Delta}{\left( r_c^2 - 3 r_c +2 a r_c^{1/2}\right)^{1/2}},  
\end{eqnarray}
\begin{eqnarray}
\sqrt{r_c^4 \left( 1- E^2\right) -2 x^2 r_c }=&& \frac{r_c^{3/2} \left( r_c^2 - 6 r_c - 3 a^2 + 8 a r_c^{1/2}\right)^{1/2}}{\left( r_c^2 - 3 r_c +2 a r_c^{1/2}\right)^{1/2}},  \\
\sqrt{L_z^2 + a^2 \left(1- E^2 \right)}=&&\frac{\sqrt{r_c^3+ 3 a^2 r_c -4 a r_c^{3/2}}}{ \left( r_c^2 - 3 r_c +2 a r_c^{1/2}\right)^{1/2}}.
\end{eqnarray} 
\label{sphconstants}
\end{subequations}
Finally, substituting these factors, given by Equation \eqref{sphconstants}, in Equation \eqref{step5}, we recover the expressions for $00$ orbits, which are given by
\begin{subequations}
\begin{eqnarray}
\bar{\nu}_{\phi}=&&\frac{1}{2 \pi \left( r_c^{3/2} + a\right)}, \\
\bar{\nu}_{r}=&& \bar{\nu}_{\phi}\left( 1- \dfrac{6}{r_c} - \dfrac{3 a^2}{r_c^2} + \dfrac{8a}{r_c^{3/2}}\right)^{1/2},\\
\bar{ \nu}_{\theta}=&& \bar{\nu}_{\phi} \left( 1+ \dfrac{3 a^2}{r_c^2} - \dfrac{4a}{r_c^{3/2}}\right)^{1/2},
\end{eqnarray}
\end{subequations}
as given in Equation \eqref{circfreq}.
\end{enumerate}

\section{Source History}
\label{sourcehistory}
We summarize the history of each BHXRB below:
\begin{enumerate}
\item M82 X-1: This is the brightest X-ray source in the M82 galaxy. This source is thought to harbor an intermediate-mass black hole because of its very high X-ray luminosity, average 2$-$10 keV luminosity $\sim 5 \times 10^{40}$ erg s$^{-1}$, and  variability characteristics \citep{Patruno2006MNRASL,Casella2008,Pasham2013ApJa}, although other models claim that it might contain a black hole of mass $\sim 20 M_{\odot}$ \citep{Okajima2006}. However, the discovery of twin-peak and stable QPOs at 3.32$\pm$0.06 Hz and 5.07$\pm$0.06 Hz in M82 X-1, which are nearly in 3:2 ratio, gave a shred of affirmative evidence that these QPOs are analogs of HFQPOs in stellar BHXRBs \citep{Pasham2014}. Following and extrapolating the inverse-mass scaling that holds for HFQPOs in stellar-mass BHXRBs \citep{2006csxs.book..157M}, it was found that the mass of the black hole in M82 X-1 could be 428$\pm$105$M_{\odot}$ \citep{Pasham2014}, making it an intermediate-mass black hole system.

\item GROJ 1655-40: This is one among the few BHXRBs in the Milky Way galaxy whose BH mass is known with good precision through the dynamical studies of the infrared and optical observations during the quiescent state \citep{Beer2002}. GROJ 1655-40 is also one of the BHXRBs known to produce relativistic radio jets having a double-lobed radio structure \citep{Mirabel1994}. The first detection of two simultaneous HFQPOs near $\sim 450$ and $300$ Hz in GROJ 1655-40 was reported by \cite{Strohmayer2001a}. The detection of 300Hz QPO was reported in BHXRB GROJ 1655-40 \citep{RemillardMorgan1999}, and later the detection of a simultaneous 450Hz QPO along with 300Hz in the same observations was confirmed \citep{Strohmayer2001a}. A systematic study of the LFQPOs and HFQPOs in 571 RXTE observations taken between the years 1996 and 2005 was carried out by \cite{Motta2014a}, who detected three simultaneous QPOs (two HFQPOs and one LFQPO) at 441$\pm$2 Hz, 298$\pm$4 Hz, and 17.3$\pm$0.1 Hz in one of these observations. Using these QPO frequencies, the mass, the spin of the black hole, and the radius of the equatorial circular orbit where these QPOs originated were estimated using Equations (\ref{nuphicirc})$-$(\ref{nuthetacirc}) assuming the RPM \citep{Motta2014a}.

\item XTEJ 1550-564: This BHXRB was first detected by ASM/RXTE on 1998 September 7. Since then, it has undergone four X-ray outbursts between the years 1998 and 2002 as observed by RXTE, among which the 1998 September to 1999 May outburst was the most luminous one. XTEJ 1550-564 is also among the few BHXRBs that have shown HFQPOs; for example, QPOs with frequencies in the range 185$-$237 Hz were detected during the 1998-1999 outburst \citep{Remillard1999XTE,Homan2001}. After a quiescent period of a few months, XTEJ 1550-564 again underwent a short X-ray outburst in the period 2000 April to May following a fast rise and an exponential decay of the X-ray luminosity. The simultaneous occurrence of two HFQPOs at 268$\pm$3 Hz and 188$\pm$3 Hz frequencies during the 2000 outburst was reported \citep{Miller2001}, indicating a resonance phenomenon. However, no LFQPOs were detected simultaneously with these two HFQPOs. A systematic study of all archival RXTE observations of XTEJ 1550-564 was carried out by \cite{Motta2014b}, who reported the detection of an HFQPO at $\sim$183 Hz along with a simultaneous type C LFQPO at $\sim$13 Hz and type B LFQPO at $\sim$5 Hz, but no second peak of HFQPO was detected during this observation.

\item 4U 1630-47: This soft X-ray transient was discovered by the Uhuru satellite \citep{Jones1976ApJL}, which is known to have an inclination of $\sim$ 60$^{\circ}-$75$^{\circ}$ \citep{Kuulkers1998ApJ}. This source is one among the few BHXRBs to show HFQPOs during its 1998 outburst in the frequency range $\sim$100-300 Hz, and also twin simultaneous HFQPOs with frequency ratio 1:4 \citep{KleinWolt2004}. It shows a regular X-ray outburst cycle after every $\sim$ 600-690 days \citep{Jones1976ApJL,Priedhorsky1986}. The QPO frequencies in this system during the 1998 X-ray outburst were observed to increase during the rising phase, followed by a phase where the frequencies were found to be stable near $\sim$180 Hz, and then a decrease in QPO frequencies was observed during the decay of the outburst.

\item GRS 1915+105: This BHXRB is known to be a very bright system during the whole RXTE period, showing its peculiar behavior and have also shown superluminal radio outflows \citep{Mirabel1994}. This is also the first BHXRB to show an HFQPO at a characteristic constant frequency of $\sim$67 Hz \citep{Morgan1997} in the RXTE observations taken during 1996 April to May. Later, simultaneous $\sim$67 Hz and $\sim$40 Hz QPOs were discovered in the RXTE observations taken during 1997 July and November \citep{Strohmayer2001b}. A systematic study of all RXTE observations of GRS 1915+105 discovered 51 observations that showed detection of HFQPOs, out of which 48 observations showed the centroid frequency of QPOs in the range 63$-$71 Hz \citep{BelloniAltamirano2013a}. Another pair of simultaneous HFQPOs was also discovered at $\sim$34 Hz and $\sim$68 Hz \citep{BelloniAltamirano2013b}.
\end{enumerate}

\section{Method for Errors Estimation of the Orbital Parameters}
\label{methodsection}
 Here, we describe a generic procedure that we have used to estimate errors in the orbital parameters. A flowchart of this method is provided in Figure \ref{methodflowchart}.
 
\begin{enumerate}
\item We assume that the QPO frequencies, $\nu_{1}$, $\nu_{2}$ and $\nu_{3}$, are Gaussian distributed  with mean values at $\nu_{10}$, $\nu_{20}$, and $\nu_{30}$ (with $\nu_{10}>\nu_{20}>\nu_{30}$), which are equal to the observed QPO centroid frequencies (see Table \ref{sourcelist}). For BHXRBs with two simultaneous QPOs, we only have $\nu_{1}$ and $\nu_{2}$. The joint probability density distribution of these frequencies will be given by
\begin{subequations}
\begin{equation}
P\left(\nu \right) = \prod_{i=1}^{l}  P_{i}\left( \nu_{i}\right), \label{jointP}
\end{equation}
where $l=3$ and $l=2$ for BHXRBs with three and two simultaneous QPOs, respectively. Here, $P_{i}\left( \nu_{i}\right)$ represents the Gaussian distribution of the $i$th QPO frequency, given by
\begin{equation}
P_{i}\left( \nu_{i}\right)=\dfrac{1}{\sqrt{2\pi \sigma_{i}^2}}\exp{\left[-\dfrac{\left( \nu_{i} -\nu_{i0}\right)^2}{2\sigma_{i}^2}\right]}.
\end{equation}
\end{subequations}
\item We find the Jacobian of the transformation from frequency to orbital parameter space using the formulae of fundamental frequencies, which are given by
\begin{subequations}
\begin{equation}
 \mathcal{J}_l=\frac{\partial \nu_{i} }{\partial x_j  }; \ \ \mathcal{J}_l=
\begin{cases}
   \mathcal{J}_2, & 2 \ \mathrm{simultaneous} \ \mathrm{QPOs}, \\
  \mathcal{J}_3, & 3 \ \mathrm{simultaneous} \ \mathrm{QPOs},
\end{cases}
\end{equation}
where \{$i, j$\}$=$1 to $l$ and $x_j$ represent the orbital parameters, and $\mathcal{J}$ is given by 
\begin{equation}
  \mathcal{J}_3 = \left[ {\begin{array}{ccc}
   \frac{\partial \nu_{1}}{\partial x_1} & \frac{\partial \nu_{1}}{\partial x_2} & \frac{\partial \nu_{1}}{\partial x_3} \\
   \frac{\partial \nu_{2}}{\partial x_1} & \frac{\partial \nu_{2}}{\partial x_2} & \frac{\partial \nu_{2}}{\partial x_3}\\
    \frac{\partial \nu_{3}}{\partial x_1} & \frac{\partial \nu_{3}}{\partial x_2} & \frac{\partial \nu_{3}}{\partial x_3} \\
  \end{array} } \right]
 \ \ \ \  \ \mathrm{and} \ \ \ \ \
 \mathcal{J}_2=  \left[ {\begin{array}{cc}
   \frac{\partial \nu_{1}}{\partial x_1} & \frac{\partial \nu_{1}}{\partial x_2}  \\
   \frac{\partial \nu_{2}}{\partial x_1} & \frac{\partial \nu_{2}}{\partial x_2} \\
  \end{array} } \right].
\end{equation}
\label{jacobian}
\end{subequations}
 For general eccentric trajectories ($Q\neq0$), which are implemented for BHXRBs with three QPOs, we have \{$x_1$, $x_2$, $x_3$\}$=$\{$e$, $r_p$, $a$\}, whereas for equatorial eccentric trajectories ($Q=0$), implemented for BHXRBs with two QPOs, we have \{$x_1$, $x_2$\}$=$\{$e$, $r_p$\}. Similarly, for the spherical orbit case, these parameters are \{$x_1$, $x_2$, $x_3$\}$=$\{$r_s$, $Q$, $a$\} or \{$x_1$, $x_2$\}$=$\{$r_s$, $Q$\}. The Jacobian is completely expressible in terms of the standard elliptic integrals and can be easily evaluated from Equation \eqref{jacobian} and using the frequency formulae, Equations \eqref{genfreq}, \eqref{eqfreq}, and \eqref{sphfreq}, where $\nu_{1}=\nu_{\phi}$, $\nu_{2}=\left( \nu_{\phi}-\nu_{r}\right) $, and $\nu_{3}=\left( \nu_{\phi}-\nu_{\theta}\right) $ according to the RPM and GRPM. The analytic expressions for the elements of the Jacobian are too long to reproduce here, but they are used to make our computations faster.

\item Next, we write the probability density distribution in the parameter space given by
\begin{equation}
P\left( [x]\right)= P\left(\nu \right)  \vert \mathcal{J}_l \vert, \label{Pera}
\end{equation}
where $[x]$ represent the set of parameters \{$x_1$, $x_2$, $x_3$\} for $l=3$ and \{$x_1$, $x_2$\} for $l=2$, and \{$\nu_{1}$, $\nu_{2}$, $\nu_{3}$\} or \{$\nu_{1}$, $\nu_{2}$\} are substituted in terms of parameters using our analytic formulae.

 We take $Q=\{0, 1, 2, 3, 4\}$ for the general \{$e$, $Q$\} trajectory solutions that are implemented for the sources M82 X-1 and GROJ 1655-40. For each fixed value of $Q$, we find the corresponding probability density distribution in the parameter space using Equation \eqref{Pera}.

\item We calculate the exact solutions for parameters by solving $\nu_{\phi}=\nu_{10}$, $\nu_{\rm pp}=\nu_{20}$, and $\nu_{\rm np}=\nu_{30}$ using Equations (\ref{nuphi}$-$\ref{nutheta}) for nonequatorial eccentric trajectories, Equations (\ref{eqnuphi}$-$\ref{eqnutheta}) for equatorial eccentric, and Equations (\ref{nuphisph2}$-$\ref{nuthetasph2}) for the spherical trajectories. We fix $\mathcal{M}$ for $l=3$, and both $\mathcal{M}$ and $a$ for $l=2$ to the previous values; see Table \ref{sourcelist}. We find 1$\sigma$ errors in the parameters by taking an appropriate parameter volume around the exact solution, and we generate sets of parameter combinations with resolution $\Delta x_j$ in this volume. The chosen parameter range, exact solutions, and corresponding resolutions are summarized in Tables \ref{3QPOresults}, \ref{2QPOresults}, and \ref{sphresults}. We then calculate the probability density using Equation \eqref{Pera}, for all of the generated parameter combinations and normalize the probability density by the normalization factor
\begin{subequations}
\begin{equation}
{\mathcal N}=\dfrac{ \sum_{k}  P\left( [x]_k\right) \Delta V_k}{V}, \ \ \Delta V_k=\prod_{j=1}^{l} \Delta x_{j , k}, \ \ V= \sum_k \Delta V_k, \label{normN}
\end{equation}
where $k$ varies from 1 to the number of total parameter combinations taken in the parameter volume, and $[x]_k$ is the $k$th combination of the parameters in the parameter volume. Hence, the normalized probability density is given by
\begin{equation}
\mathcal{P}\left([x]\right)=\dfrac{P\left( [x]\right)}{{\mathcal N}}. \label{normPera}
\end{equation}
\end{subequations}
\item The allowed parameter combinations for the bound orbits are governed by the condition Equation \eqref{boundcondition}. For a spherical orbit, we have $e=0$. Hence, we ensure that the parameters ($e$, $r_p$, $a$, $Q$) for eccentric and ($r_s$, $a$, $Q$) for spherical trajectories follow the bound orbit condition. If any parameter combination does not obey the bound orbit condition, then $\mathcal{P}\left([x]\right)$ is taken to be zero at that point in the parameter volume.

\item Next, we integrate the normalized probability density, $\mathcal{P}\left( [x ]\right)$, Equation \eqref{normPera}, in two dimensions to obtain the profile in the remaining third dimension of the parameters for BHXRBs with three simultaneous QPOs, and similarly by integrating in one dimension for the two QPO cases, we obtain the profile in the other dimension. So we finally obtain the one-dimensional distributions $\mathcal{P}_{1} \left( e\right)$, $\mathcal{P}_{1} \left( r_p\right)$, and $\mathcal{P}_1 \left( a\right)$.

\item Finally, we fit the normalized probability density profiles in each of the parameter dimensions to find the corresponding mean values, and quoted errors are obtained such that they contain a probability of 68.2\% about the peak value of the probability density. The results of these fits are given in Tables \ref{3QPOresults}, \ref{2QPOresults}, and \ref{sphresults}.

\item For BHXRBs M82 X-1 and GROJ 1655-40, we find various orbital solutions showing varying \{$a$, $Q$\} values. As the spin of the black hole should be fixed, we choose the most probable value of $a$, and then we estimate the remaining parameters \{$e$, $r_p$, $Q$\}, their profiles \{$\mathcal{P}_{1} \left( e\right)$, $\mathcal{P}_{1} \left( r_p\right)$, $\mathcal{P}_{1} \left( Q\right)$\}, and the corresponding errors using the same procedure given above in steps 1 to 6, where the orbital parameters are now given by \{$x_1$, $x_2$, $x_3$\}$=$\{$e$, $r_p$, $Q$\}. 

\item Although we have made accurate calculations described above, to obtain a rough and quick estimate of the errors, we may use the following procedure. Assuming that the probability density is Gaussian distributed independently in $e$, $r_p$ and $a$ parameters, the normalized joint probability density distribution is given by
\begin{subequations}
\begin{eqnarray}
\mathcal{P}\left(e, r_p, a \right) =\dfrac{1}{\left( 2 \pi \right) ^{3/2}\sigma_{e} \sigma_{r_p} \sigma_{a}} \exp{\left\lbrace - \dfrac{1}{2}\left[\left( \dfrac{e-e_0}{\sigma_{e}}\right) ^{2} + \left( \dfrac{r_p - r_{p0}}{\sigma_{r_p}}\right) ^{2} + \left( \dfrac{a - a_{0}}{\sigma_{a}}\right) ^{2} \right]\right\rbrace }, \nonumber \\
\end{eqnarray}
where the distribution is centered at the exact solution ($e_0, r_{p0}, a_0$), and $\sigma_{e}$, $\sigma_{r_p}$, and $\sigma_a$ are the corresponding 1$\sigma$ errors, derived using the method described above. The total probability contained in a volume $\mathcal V$ in ($e, r_p, a$) space is given by
\begin{eqnarray}
p =&& \dfrac{1}{\left( 2 \pi\right) ^{3/2} \sigma_{e} \sigma_{r_p} \sigma_{a}}\int \int \int_{\mathcal V} \exp{\left\lbrace - \dfrac{1}{2}\left[\left( \dfrac{e-e_0}{\sigma_{e}}\right) ^{2} + \left( \dfrac{r_p - r_{p0}}{\sigma_{r_p}}\right) ^{2} + \left( \dfrac{a - a_{0}}{\sigma_{a}}\right) ^{2} \right]\right\rbrace } {\rm d}e \cdot {\rm d}r_p \cdot {\rm d}a; \nonumber \\
\end{eqnarray}
so that  the total probability $p$ inside an ellipsoid  in ($e$, $r_p$, $a$) space specified by
\begin{equation}
\left[\left( \dfrac{e-e_0}{\sigma_{e}}\right) ^{2} + \left( \dfrac{r_p - r_{p0}}{\sigma_{r_p}}\right) ^{2} + \left( \dfrac{a - a_{0}}{\sigma_{a}}\right) ^{2} \right] =s_{3}^{2}, \label{errellip1}
\end{equation}
is given by
\begin{eqnarray}
p =&&\sqrt{\dfrac{2}{\pi}}\int^{s_3}_{0} \exp{\left( \dfrac{-s^2}{2}\right) } s^2 \ {\rm d}s= \dfrac{2}{\sqrt{\pi}} \gamma\left( \dfrac{3}{2}, \dfrac{s_{3}^2}{2}\right) , \label{ps3}
\end{eqnarray}
\end{subequations}
where $\gamma\left( \dfrac{3}{2}, \dfrac{s_{3}^2}{2}\right)$ is the incomplete gamma function. 

Similarly, for two QPO cases, the joint probability density distribution can be written as
\begin{subequations}
\begin{eqnarray}
\mathcal{P}\left(e, r_p \right)=&&\dfrac{1}{2 \pi \sigma_{e} \sigma_{r_p}} \exp{\left\lbrace - \dfrac{1}{2}\left[\left( \dfrac{e-e_0}{\sigma_{e}}\right) ^{2} + \left( \dfrac{r_p - r_{p0}}{\sigma_{r_p}}\right) ^{2} \right]\right\rbrace }.
\end{eqnarray}
The total probability contained in a surface $\mathcal S$ in ($e, r_p$) space is given by
\begin{eqnarray}
p =&& \dfrac{1}{ 2 \pi \sigma_{e} \sigma_{r_p} }\int \int_{\mathcal S} \exp{\left\lbrace - \dfrac{1}{2}\left[\left( \dfrac{e-e_0}{\sigma_{e}}\right) ^{2} + \left( \dfrac{r_p - r_{p0}}{\sigma_{r_p}}\right) ^{2} \right]\right\rbrace } {\rm d}e \cdot {\rm d}r_p.
\end{eqnarray}
The total probability inside an ellipse, specified by
\begin{equation}
\left[\left( \dfrac{e-e_0}{\sigma_{e}}\right) ^{2} + \left( \dfrac{r_p - r_{p0}}{\sigma_{r_p}}\right) ^{2} \right] =s_{2}^{2}, \label{errellip2}
\end{equation}
is given by
\begin{eqnarray}
p =&&\int^{s_2}_{0} \exp{\left( \dfrac{-s^2}{2}\right) } s \ {\rm d}s= 1 - \exp{\left( \dfrac{-s_{2}^2}{2}\right) }. \label{ps2}
\end{eqnarray}
\end{subequations}
\begin{figure}[h!]
\begin{center}
\includegraphics[width=8.0cm,height=5.0cm]{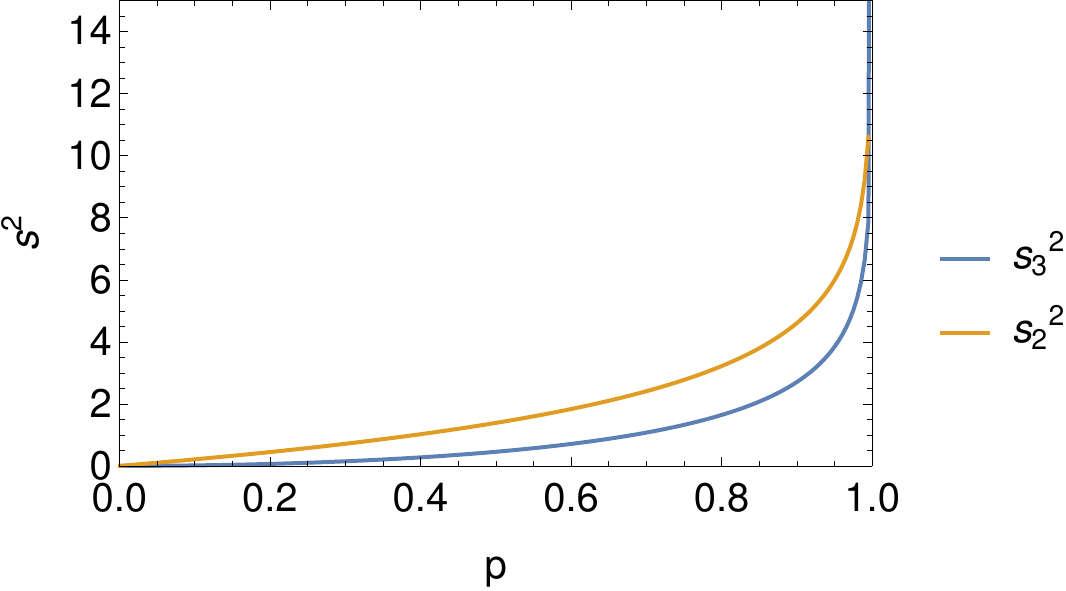} 
\end{center}
\caption{\label{s3s2p}Figure showing $s_{3}^2$ and $s_{2}^2$ as a function of probability $p$ given by Equations (\ref{ps3}) and (\ref{ps2}). }
\end{figure}
For a given $p$, we can calculate $s_{3}^2$ and $s_{2}^2$, and hence evaluate the error ellipsoid corresponding to $p$. ${s_2}^2$ and ${s_3}^2$ are shown as a functions of $p$ in Figure \ref{s3s2p}. This can be used to get rough estimates of the error distribution of the parameters. However, we calculate them exactly in \S \ref{method}.

\end{enumerate}

\bibliographystyle{aasjournal}
\bibliography{paper3_new}

\begin{thebibliography}{}
\expandafter\ifx\csname natexlab\endcsname\relax\def\natexlab#1{#1}\fi
\providecommand{\url}[1]{\href{#1}{#1}}
\providecommand{\dodoi}[1]{doi:~\href{http://doi.org/#1}{\nolinkurl{#1}}}
\providecommand{\doeprint}[1]{\href{http://ascl.net/#1}{\nolinkurl{http://ascl.net/#1}}}
\providecommand{\doarXiv}[1]{\href{https://arxiv.org/abs/#1}{\nolinkurl{https://arxiv.org/abs/#1}}}

\bibitem[{{Abramowicz} {et~al.}(2003){Abramowicz}, {Karas}, {Kluzniak}, {Lee},
  \& {Rebusco}}]{Abramowicz2003}
{Abramowicz}, M.~A., {Karas}, V., {Kluzniak}, W., {Lee}, W.~H., \& {Rebusco},
  P. 2003, Publications of the Astronomical Society of Japan, 55, 467,
  \dodoi{10.1093/pasj/55.2.467}

\bibitem[{{Bardeen} {et~al.}(1972){Bardeen}, {Press}, \&
  {Teukolsky}}]{Bardeen1972}
{Bardeen}, J.~M., {Press}, W.~H., \& {Teukolsky}, S.~A. 1972, \apj, 178, 347,
  \dodoi{10.1086/151796}

\bibitem[{{Beer} \& {Podsiadlowski}(2002)}]{Beer2002}
{Beer}, M.~E., \& {Podsiadlowski}, P. 2002, MNRAS, 331, 351,
  \dodoi{10.1046/j.1365-8711.2002.05189.x}

\bibitem[{{Belloni} {et~al.}(2002){Belloni}, {Psaltis}, \& {van der
  Klis}}]{Belloni2002ApJ}
{Belloni}, T., {Psaltis}, D., \& {van der Klis}, M. 2002, \apj, 572, 392,
  \dodoi{10.1086/340290}

\bibitem[{{Belloni} {et~al.}(2006){Belloni}, {Soleri}, {Casella}, {M{\'e}ndez},
  \& {Migliari}}]{Belloni2006}
{Belloni}, T., {Soleri}, P., {Casella}, P., {M{\'e}ndez}, M., \& {Migliari}, S.
  2006, MNRAS, 369, 305, \dodoi{10.1111/j.1365-2966.2006.10286.x}

\bibitem[{{Belloni} \&
  {Altamirano}(2013{\natexlab{a}})}]{BelloniAltamirano2013a}
{Belloni}, T.~M., \& {Altamirano}, D. 2013{\natexlab{a}}, MNRAS, 432, 10,
  \dodoi{10.1093/mnras/stt500}

\bibitem[{{Belloni} \&
  {Altamirano}(2013{\natexlab{b}})}]{BelloniAltamirano2013b}
---. 2013{\natexlab{b}}, MNRAS, 432, 19, \dodoi{10.1093/mnras/stt285}

\bibitem[{{Belloni} {et~al.}(2012){Belloni}, {Sanna}, \&
  {M{\'e}ndez}}]{Belloni2012}
{Belloni}, T.~M., {Sanna}, A., \& {M{\'e}ndez}, M. 2012, MNRAS, 426, 1701,
  \dodoi{10.1111/j.1365-2966.2012.21634.x}

\bibitem[{{Belloni} \& {Stella}(2014)}]{BelloniStella2014}
{Belloni}, T.~M., \& {Stella}, L. 2014, Space Science Reviews, 183, 43,
  \dodoi{10.1007/s11214-014-0076-0}

\bibitem[{{Carter}(1968)}]{Carter1968}
{Carter}, B. 1968, Physical Review, 174, 1559, \dodoi{10.1103/PhysRev.174.1559}

\bibitem[{{Casella} {et~al.}(2008){Casella}, {Ponti}, {Patruno}, {Belloni},
  {Miniutti}, \& {Zampieri}}]{Casella2008}
{Casella}, P., {Ponti}, G., {Patruno}, A., {et~al.} 2008, \mnras, 387, 1707,
  \dodoi{10.1111/j.1365-2966.2008.13372.x}

\bibitem[{{Dubus} {et~al.}(2001){Dubus}, {Hameury}, \& {Lasota}}]{Dubus2001}
{Dubus}, G., {Hameury}, J.~M., \& {Lasota}, J.~P. 2001, \aap, 373, 251,
  \dodoi{10.1051/0004-6361:20010632}

\bibitem[{{Fender} \& {Belloni}(2004)}]{FenderBelloni2004AA}
{Fender}, R., \& {Belloni}, T. 2004, \araa, 42, 317,
  \dodoi{10.1146/annurev.astro.42.053102.134031}

\bibitem[{{Fender} \& {Belloni}(2012)}]{Fender2012}
---. 2012, Science, 337, 540, \dodoi{10.1126/science.1221790}

\bibitem[{{Fender} {et~al.}(2004){Fender}, {Belloni}, \&
  {Gallo}}]{FenderBelloni2004}
{Fender}, R.~P., {Belloni}, T.~M., \& {Gallo}, E. 2004, \mnras, 355, 1105,
  \dodoi{10.1111/j.1365-2966.2004.08384.x}

\bibitem[{{Fujita} \& {Hikida}(2009)}]{Fujita2009}
{Fujita}, R., \& {Hikida}, W. 2009, Classical and Quantum Gravity, 26, 135002,
  \dodoi{10.1088/0264-9381/26/13/135002}

\bibitem[{{German{\`a}} {et~al.}(2009){German{\`a}}, {Kosti{\'c}},
  {{\v{C}}ade{\v{z}}}, \& {Calvani}}]{Germana2009}
{German{\`a}}, C., {Kosti{\'c}}, U., {{\v{C}}ade{\v{z}}}, A., \& {Calvani}, M.
  2009, in American Institute of Physics Conference Series, Vol. 1126, American
  Institute of Physics Conference Series, ed. J.~{Rodriguez} \& P.~{Ferrando},
  367--369

\bibitem[{{Glampedakis} \& {Kennefick}(2002)}]{Glampedakis2002}
{Glampedakis}, K., \& {Kennefick}, D. 2002, \prd, 66, 044002,
  \dodoi{10.1103/PhysRevD.66.044002}

\bibitem[{Gradshteyn \& Ryzhik(2007)}]{Grad}
Gradshteyn, I.~S., \& Ryzhik, I.~M. 2007, Table of integrals, series, and
  products, seventh edn. (Elsevier/Academic Press, Amsterdam), xlviii+1171

\bibitem[{{Homan} {et~al.}(2003){Homan}, {Klein-Wolt}, {Rossi}, {Miller},
  {Wijnands}, {Belloni}, {van der Klis}, \& {Lewin}}]{Homan2003}
{Homan}, J., {Klein-Wolt}, M., {Rossi}, S., {et~al.} 2003, \apj, 586, 1262,
  \dodoi{10.1086/367699}

\bibitem[{{Homan} {et~al.}(2005){Homan}, {Miller}, {Wijnands}, {van der Klis},
  {Belloni}, {Steeghs}, \& {Lewin}}]{Homan2005}
{Homan}, J., {Miller}, J.~M., {Wijnands}, R., {et~al.} 2005, \apj, 623, 383,
  \dodoi{10.1086/424994}

\bibitem[{{Homan} {et~al.}(2001){Homan}, {Wijnands}, {van der Klis}, {Belloni},
  {van Paradijs}, {Klein-Wolt}, {Fender}, \& {M{\'e}ndez}}]{Homan2001}
{Homan}, J., {Wijnands}, R., {van der Klis}, M., {et~al.} 2001, ApJS, 132, 377,
  \dodoi{10.1086/318954}

\bibitem[{{Ingram} \& {Done}(2011)}]{Ingram2011}
{Ingram}, A., \& {Done}, C. 2011, \mnras, 415, 2323,
  \dodoi{10.1111/j.1365-2966.2011.18860.x}

\bibitem[{{Ingram} \& {Done}(2012)}]{Ingram2012}
---. 2012, \mnras, 419, 2369, \dodoi{10.1111/j.1365-2966.2011.19885.x}

\bibitem[{{Ingram} {et~al.}(2009){Ingram}, {Done}, \& {Fragile}}]{Ingram2009}
{Ingram}, A., {Done}, C., \& {Fragile}, P.~C. 2009, \mnras, 397, L101,
  \dodoi{10.1111/j.1745-3933.2009.00693.x}

\bibitem[{{Jones} {et~al.}(1976){Jones}, {Forman}, {Tananbaum}, \&
  {Turner}}]{Jones1976ApJL}
{Jones}, C., {Forman}, W., {Tananbaum}, H., \& {Turner}, M.~J.~L. 1976, \apjl,
  210, L9, \dodoi{10.1086/182291}

\bibitem[{{Kato}(2004)}]{Kato2004b}
{Kato}, S. 2004, \pasj, 56, 905, \dodoi{10.1093/pasj/56.5.905}

\bibitem[{Kato(2008)}]{Kato2008}
Kato, S. 2008, Publications of the Astronomical Society of Japan, 60, 111,
  \dodoi{10.1093/pasj/60.1.111}

\bibitem[{{King} {et~al.}(2014){King}, {Walton}, {Miller}, {Barret}, {Boggs},
  {Christensen}, {Craig}, {Fabian}, {F{\"u}rst}, {Hailey}, {Harrison},
  {Krivonos}, {Mori}, {Natalucci}, {Stern}, {Tomsick}, \& {Zhang}}]{King2014}
{King}, A.~L., {Walton}, D.~J., {Miller}, J.~M., {et~al.} 2014, \apj, 784, L2,
  \dodoi{10.1088/2041-8205/784/1/L2}

\bibitem[{{Klein-Wolt} {et~al.}(2004){Klein-Wolt}, {Homan}, \& {van der
  Klis}}]{KleinWolt2004}
{Klein-Wolt}, M., {Homan}, J., \& {van der Klis}, M. 2004, Nuclear Physics B
  Proceedings Supplements, 132, 381, \dodoi{10.1016/j.nuclphysbps.2004.04.067}

\bibitem[{{Kuulkers} {et~al.}(1998){Kuulkers}, {Wijnands}, {Belloni},
  {M{\'e}ndez}, {van der Klis}, \& {van Paradijs}}]{Kuulkers1998ApJ}
{Kuulkers}, E., {Wijnands}, R., {Belloni}, T., {et~al.} 1998, \apj, 494, 753,
  \dodoi{10.1086/305248}

\bibitem[{{Levin} \& {Perez-Giz}(2009)}]{Levin2009}
{Levin}, J., \& {Perez-Giz}, G. 2009, \prd, 79, 124013,
  \dodoi{10.1103/PhysRevD.79.124013}

\bibitem[{{McClintock} \& {Remillard}(2006)}]{2006csxs.book..157M}
{McClintock}, J.~E., \& {Remillard}, R.~A. 2006, {Black hole binaries}, Vol.~39
  (Cambridge University Press), 157--213

\bibitem[{{Miller} {et~al.}(2001){Miller}, {Wijnands}, {Homan}, {Belloni},
  {Pooley}, {Corbel}, {Kouveliotou}, {van der Klis}, \& {Lewin}}]{Miller2001}
{Miller}, J.~M., {Wijnands}, R., {Homan}, J., {et~al.} 2001, \apj, 563, 928,
  \dodoi{10.1086/324027}

\bibitem[{{Miller} {et~al.}(2013){Miller}, {Parker}, {Fuerst}, {Bachetti},
  {Harrison}, {Barret}, {Boggs}, {Chakrabarty}, {Christensen}, {Craig},
  {Fabian}, {Grefenstette}, {Hailey}, {King}, {Stern}, {Tomsick}, {Walton}, \&
  {Zhang}}]{Miller2013}
{Miller}, J.~M., {Parker}, M.~L., {Fuerst}, F., {et~al.} 2013, \apj, 775, L45,
  \dodoi{10.1088/2041-8205/775/2/L45}

\bibitem[{{Miller} \& {Miller}(2015)}]{Miller2015}
{Miller}, M.~C., \& {Miller}, J.~M. 2015, PhysRep, 548, 1,
  \dodoi{10.1016/j.physrep.2014.09.003}

\bibitem[{{Mino}(2003)}]{Mino2003}
{Mino}, Y. 2003, \prd, 67, 084027, \dodoi{10.1103/PhysRevD.67.084027}

\bibitem[{{Mirabel} \& {Rodr{\'\i}guez}(1994)}]{Mirabel1994}
{Mirabel}, I.~F., \& {Rodr{\'\i}guez}, L.~F. 1994, \nat, 371, 46,
  \dodoi{10.1038/371046a0}

\bibitem[{{Mohan} \& {Mangalam}(2014)}]{2014ApJ...791...74M}
{Mohan}, P., \& {Mangalam}, A. 2014, \apj, 791, 74,
  \dodoi{10.1088/0004-637X/791/2/74}

\bibitem[{{Morgan} {et~al.}(1997){Morgan}, {Remillard}, \&
  {Greiner}}]{Morgan1997}
{Morgan}, E.~H., {Remillard}, R.~A., \& {Greiner}, J. 1997, \apj, 482, 993,
  \dodoi{10.1086/304191}

\bibitem[{{Motta}(2016)}]{Motta2016}
{Motta}, S.~E. 2016, Astronomische Nachrichten, 337, 398,
  \dodoi{10.1002/asna.201612320}

\bibitem[{{Motta} {et~al.}(2014{\natexlab{a}}){Motta}, {Belloni}, {Stella},
  {Mu{\~n}oz-Darias}, \& {Fender}}]{Motta2014a}
{Motta}, S.~E., {Belloni}, T.~M., {Stella}, L., {Mu{\~n}oz-Darias}, T., \&
  {Fender}, R. 2014{\natexlab{a}}, MNRAS, 437, 2554,
  \dodoi{10.1093/mnras/stt2068}

\bibitem[{{Motta} {et~al.}(2018){Motta}, {Franchini}, {Lodato}, \&
  {Mastroserio}}]{Motta2018}
{Motta}, S.~E., {Franchini}, A., {Lodato}, G., \& {Mastroserio}, G. 2018,
  \mnras, 473, 431, \dodoi{10.1093/mnras/stx2358}

\bibitem[{{Motta} {et~al.}(2014{\natexlab{b}}){Motta}, {Mu{\~n}oz-Darias},
  {Sanna}, {Fender}, {Belloni}, \& {Stella}}]{Motta2014b}
{Motta}, S.~E., {Mu{\~n}oz-Darias}, T., {Sanna}, A., {et~al.}
  2014{\natexlab{b}}, MNRAS, 439, L65, \dodoi{10.1093/mnrasl/slt181}

\bibitem[{{Okajima} {et~al.}(2006){Okajima}, {Ebisawa}, \&
  {Kawaguchi}}]{Okajima2006}
{Okajima}, T., {Ebisawa}, K., \& {Kawaguchi}, T. 2006, \apjl, 652, L105,
  \dodoi{10.1086/510153}

\bibitem[{{Orosz} {et~al.}(2011){Orosz}, {Steiner}, {McClintock}, {Torres},
  {Remillard}, {Bailyn}, \& {Miller}}]{Orosz2011}
{Orosz}, J.~A., {Steiner}, J.~F., {McClintock}, J.~E., {et~al.} 2011, \apj,
  730, 75, \dodoi{10.1088/0004-637X/730/2/75}

\bibitem[{{Pasham} \& {Strohmayer}(2013{\natexlab{a}})}]{Pasham2013ApJb}
{Pasham}, D.~R., \& {Strohmayer}, T.~E. 2013{\natexlab{a}}, \apj, 771, 101,
  \dodoi{10.1088/0004-637X/771/2/101}

\bibitem[{{Pasham} \& {Strohmayer}(2013{\natexlab{b}})}]{Pasham2013ApJa}
---. 2013{\natexlab{b}}, \apjl, 774, L16, \dodoi{10.1088/2041-8205/774/2/L16}

\bibitem[{{Pasham} {et~al.}(2014){Pasham}, {Strohmayer}, \&
  {Mushotzky}}]{Pasham2014}
{Pasham}, D.~R., {Strohmayer}, T.~E., \& {Mushotzky}, R.~F. 2014, Nature, 513,
  74, \dodoi{10.1038/nature13710}

\bibitem[{{Patruno} {et~al.}(2006){Patruno}, {Portegies Zwart}, {Dewi}, \&
  {Hopman}}]{Patruno2006MNRASL}
{Patruno}, A., {Portegies Zwart}, S., {Dewi}, J., \& {Hopman}, C. 2006, \mnras,
  370, L6, \dodoi{10.1111/j.1745-3933.2006.00176.x}

\bibitem[{{Penna} {et~al.}(2012){Penna}, {S{\k{a}}owski}, \&
  {McKinney}}]{2012MNRAS.420..684P}
{Penna}, R.~F., {S{\k{a}}owski}, A., \& {McKinney}, J.~C. 2012, \mnras, 420,
  684, \dodoi{10.1111/j.1365-2966.2011.20084.x}

\bibitem[{{Perez-Giz} \& {Levin}(2009)}]{Perez-Giz2009}
{Perez-Giz}, G., \& {Levin}, J. 2009, \prd, 79, 124014,
  \dodoi{10.1103/PhysRevD.79.124014}

\bibitem[{{Priedhorsky}(1986)}]{Priedhorsky1986}
{Priedhorsky}, W. 1986, \apss, 126, 89, \dodoi{10.1007/BF00644177}

\bibitem[{{Psaltis} {et~al.}(1999){Psaltis}, {Belloni}, \& {van der
  Klis}}]{PBK1999}
{Psaltis}, D., {Belloni}, T., \& {van der Klis}, M. 1999, \apj, 520, 262,
  \dodoi{10.1086/307436}

\bibitem[{{Rana} \& {Mangalam}(2019{\natexlab{a}})}]{RMCQG2019}
{Rana}, P., \& {Mangalam}, A. 2019{\natexlab{a}}, Classical and Quantum
  Gravity, 36, 045009, \dodoi{10.1088/1361-6382/ab004c}

\bibitem[{{Rana} \& {Mangalam}(2019{\natexlab{b}})}]{RMarxiv2019}
---. 2019{\natexlab{b}}, arXiv e-prints, arXiv:1901.02730

\bibitem[{{Rana} \& {Mangalam}(2020)}]{RMgalaxies2020}
---. 2020, Galaxies, 8, 67, \dodoi{10.3390/galaxies8030067}

\bibitem[{{Remillard} {et~al.}(2006){Remillard}, {McClintock}, {Orosz}, \&
  {Levine}}]{Remillard2006}
{Remillard}, R.~A., {McClintock}, J.~E., {Orosz}, J.~A., \& {Levine}, A.~M.
  2006, \apj, 637, 1002, \dodoi{10.1086/498556}

\bibitem[{{Remillard} {et~al.}(1999{\natexlab{a}}){Remillard}, {McClintock},
  {Sobczak}, {Bailyn}, {Orosz}, {Morgan}, \& {Levine}}]{Remillard1999XTE}
{Remillard}, R.~A., {McClintock}, J.~E., {Sobczak}, G.~J., {et~al.}
  1999{\natexlab{a}}, \apjl, 517, L127, \dodoi{10.1086/312038}

\bibitem[{{Remillard} {et~al.}(1999{\natexlab{b}}){Remillard}, {Morgan},
  {McClintock}, {Bailyn}, \& {Orosz}}]{RemillardMorgan1999}
{Remillard}, R.~A., {Morgan}, E.~H., {McClintock}, J.~E., {Bailyn}, C.~D., \&
  {Orosz}, J.~A. 1999{\natexlab{b}}, \apj, 522, 397, \dodoi{10.1086/307606}

\bibitem[{{Remillard} {et~al.}(2002){Remillard}, {Sobczak}, {Muno}, \&
  {McClintock}}]{Remillard2002}
{Remillard}, R.~A., {Sobczak}, G.~J., {Muno}, M.~P., \& {McClintock}, J.~E.
  2002, \apj, 564, 962, \dodoi{10.1086/324276}

\bibitem[{{Schmidt}(2002)}]{Schmidt2002}
{Schmidt}, W. 2002, Classical and Quantum Gravity, 19, 2743,
  \dodoi{10.1088/0264-9381/19/10/314}

\bibitem[{{Seifina} {et~al.}(2014){Seifina}, {Titarchuk}, \&
  {Shaposhnikov}}]{Seifina2014}
{Seifina}, E., {Titarchuk}, L., \& {Shaposhnikov}, N. 2014, \apj, 789, 57,
  \dodoi{10.1088/0004-637X/789/1/57}

\bibitem[{{Steeghs} {et~al.}(2013){Steeghs}, {McClintock}, {Parsons}, {Reid},
  {Littlefair}, \& {Dhillon}}]{Steeghs2013}
{Steeghs}, D., {McClintock}, J.~E., {Parsons}, S.~G., {et~al.} 2013, \apj, 768,
  185, \dodoi{10.1088/0004-637X/768/2/185}

\bibitem[{{Stella} \& {Vietri}(1999)}]{Stella1999a}
{Stella}, L., \& {Vietri}, M. 1999, Physical Review Letters, 82, 17,
  \dodoi{10.1103/PhysRevLett.82.17}

\bibitem[{{Stella} {et~al.}(1999){Stella}, {Vietri}, \&
  {Morsink}}]{Stella1999b}
{Stella}, L., {Vietri}, M., \& {Morsink}, S.~M. 1999, \apj, 524, L63,
  \dodoi{10.1086/312291}

\bibitem[{{Strohmayer}(2001{\natexlab{a}})}]{Strohmayer2001a}
{Strohmayer}, T.~E. 2001{\natexlab{a}}, \apj, 552, L49, \dodoi{10.1086/320258}

\bibitem[{{Strohmayer}(2001{\natexlab{b}})}]{Strohmayer2001b}
---. 2001{\natexlab{b}}, \apj, 554, L169, \dodoi{10.1086/321720}

\bibitem[{{Tagger} \& {Varni{\`e}re}(2006)}]{Tagger2006}
{Tagger}, M., \& {Varni{\`e}re}, P. 2006, \apj, 652, 1457,
  \dodoi{10.1086/508318}

\bibitem[{{T{\"o}r{\"o}k} {et~al.}(2005){T{\"o}r{\"o}k}, {Abramowicz},
  {Klu{\'z}niak}, \& {Stuchl{\'\i}k}}]{Torok2005}
{T{\"o}r{\"o}k}, G., {Abramowicz}, M.~A., {Klu{\'z}niak}, W., \&
  {Stuchl{\'\i}k}, Z. 2005, \aap, 436, 1, \dodoi{10.1051/0004-6361:20047115}

\bibitem[{{T{\"o}r{\"o}k} {et~al.}(2011){T{\"o}r{\"o}k}, {Kotrlov{\'a}}, {{\v
  S}r{\'a}mkov{\'a}}, \& {Stuchl{\'{\i}}k}}]{Torok2011}
{T{\"o}r{\"o}k}, G., {Kotrlov{\'a}}, A., {{\v S}r{\'a}mkov{\'a}}, E., \&
  {Stuchl{\'{\i}}k}, Z. 2011, A\&A, 531, A59,
  \dodoi{10.1051/0004-6361/201015549}

\bibitem[{{Varniere} {et~al.}(2019){Varniere}, {Casse}, \&
  {Vincent}}]{PeggyVarniere2019}
{Varniere}, P., {Casse}, F., \& {Vincent}, F.~H. 2019, \aap, 625, A116,
  \dodoi{10.1051/0004-6361/201935208}

\bibitem[{{Warburton} {et~al.}(2013){Warburton}, {Barack}, \&
  {Sago}}]{Warburton2013}
{Warburton}, N., {Barack}, L., \& {Sago}, N. 2013, \prd, 87, 084012,
  \dodoi{10.1103/PhysRevD.87.084012}

\bibitem[{{Wilkins}(1972)}]{Wilkins1972}
{Wilkins}, D.~C. 1972, \prd, 5, 814, \dodoi{10.1103/PhysRevD.5.814}

\end{thebibliography}

\end{document}